%% file: main.tex
\documentclass[
11pt, 
english, 
singlespacing, 
headsepline, 
]{MastersDoctoralThesis} 

\usepackage[utf8]{inputenc} 
\usepackage[T1]{fontenc} 
\usepackage{multirow}
\usepackage{mathpazo} 

\usepackage{mathtools,slashed}

\usepackage[autostyle=true]{csquotes}

\usepackage[
style=numeric-comp,
sorting=none,
backend=biber
]{biblatex}

\usepackage[colorlinks=true,pdfencoding=auto,psdextra]{hyperref}

\DeclareFieldFormat*{title}{#1}

\renewbibmacro{in:}{}

\DeclareSourcemap{
	\maps[datatype=bibtex]{
		\map{
			\step[fieldsource=doi, final]
			\step[fieldset=archiveprefix, null]
			\step[fieldset=arxivid, null]
			\step[fieldset=eprint, null]
		}
	}
}

\makeatletter
\def\blfootnote{\gdef\@thefnmark{}\@footnotetext}
\makeatother

\usepackage{mathtools}
\usepackage{lettrine}
\usepackage{amsmath,amssymb}
\usepackage{dsfont}
\usepackage{ stmaryrd }
\usepackage{pdfpages}
\usepackage{enumerate}
\usepackage{xurl} 
\usepackage{doi}  
\usepackage{braket}
\usepackage{cleveref}
\usepackage{subfigure}
\usepackage{microtype}

\usepackage[section]{placeins}
\usepackage{needspace}

\newcommand{\Trh}{T_\text{rh}}
\newcommand{\gs}{g_*}
\newcommand{\gss}{g_{* s}}
\newcommand{\arh}{a_\text{rh}}
\newcommand{\rp}{\rho_\phi}
\newcommand{\rR}{\rho_\text{rad}}
\newcommand{\lhs}{\lambda_{hs}}
\newcommand{\ls}{\lambda_s}

\usepackage{tikz}
\usepackage{tikz-feynman}
\tikzfeynmanset{compat=1.1.0}

\usepackage[most]{tcolorbox}

\newtcolorbox{chapterpublication}{
  colback=gray!5,
  colframe=black,
  boxrule=0.5pt,
  arc=2mm,
  left=6pt,
  right=6pt,
  top=6pt,
  bottom=6pt,
}

\thesistitle{Dynamics of Self-Interacting Dark Sectors} 
\supervisor{Dr. hab. Andrzej \textsc{Hryczuk}\\ Prof. NCBJ} 
\examiner{} 
\degree{Doctor of Philosophy} 
\author{Juan Esau \textsc{Cervantes-Hernandez}} 
\addresses{} 

\subject{Physics} 
\university{\href{http://www.ncbj.gov.pl}{National Centre for Nuclear Research}} 
\department{Department BP2} 
\faculty{{Factulty of Physics}} 

\AtBeginDocument{
\hypersetup{pdftitle=\ttitle} 
\hypersetup{pdfauthor=\authorname} 
\hypersetup{pdfkeywords=\keywordnames} 
}

\DeclareMathOperator{\Tr}{Tr}

\begin{document}

\frontmatter 

\pagestyle{plain} 


\begin{titlepage}
\begin{center}

{\scshape\LARGE \univname\par}\vspace{1.5cm} 
\textsc{\Large Doctoral Thesis}\\[0.5cm] 

\HRule \\[0.4cm] 
{\huge \bfseries \ttitle\par}\vspace{0.4cm} 
\HRule \\[1.5cm] 
 
\begin{minipage}[t]{0.4\textwidth}
\begin{flushleft} \large
\emph{Author:}\\
\authorname 
\end{flushleft}
\end{minipage}
\begin{minipage}[t]{0.4\textwidth}
\begin{flushright} \large
\emph{Supervisor:} \\
\supname\\ 
\end{flushright}
\end{minipage}\\[2cm]
 
\vspace{-1 cm}
\large \textit{A thesis submitted in fulfillment of the requirements\\ for the degree of \degreename}\\[0.3cm] 
\textit{in the}\\[0.4cm]
\deptname\\[-2cm] 
\includegraphics[width=0.5\textwidth]{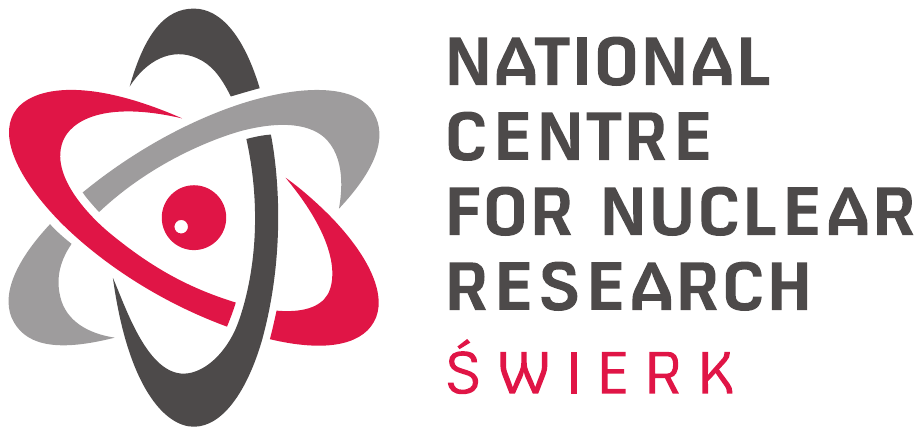}\\[-3cm]
\vfill
{\large April 28, 2026} 
 
\end{center}
\end{titlepage}


\begin{declaration}
\addchaptertocentry{\authorshipname} 
\noindent I, \authorname, declare that this thesis titled, \enquote{\ttitle} and the work presented in it are my own. I confirm that:

\begin{itemize} 
\item This work was done wholly or mainly while in candidature for a research degree at the National Centre for Nuclear Research.
\item Where any part of this thesis has previously been submitted for a degree or any other qualification at the National Centre for Nuclear Research or any other institution, this has been clearly stated.
\item Where I have consulted the published work of others, this is always clearly attributed.
\item Where I have quoted from the work of others, the source is always given. With the exception of such quotations, this thesis is entirely my own work.
\item I have acknowledged all main sources of help.
\item Where the thesis is based on work done by myself jointly with others, I have made clear exactly what was done by others and what I have contributed myself.\\
\end{itemize}
 
\noindent Signed:\\
\rule[0.5em]{25em}{0.5pt} 
 
\noindent Date:\\
\rule[0.5em]{25em}{0.5pt} 
\end{declaration}

\cleardoublepage





\begin{abstract}
\addchaptertocentry{\abstractname} 
This thesis investigates the dynamics of self-interacting dark sectors in the early Universe populated through the freeze-in mechanism. The main focus is on scenarios in which the dark sector contains particles undergoing self-number-changing reactions of the form \(2\leftrightarrow3\), and on how these interactions modify both the thermal history and the resulting phenomenology. Several realizations are studied. First, we consider a real scalar field in a spontaneously broken \(\mathbb{Z}_2\) theory, where symmetry breaking induces a cubic interaction and therefore allows for \(2\leftrightarrow3\) processes. Second, we study a complex scalar field stabilized by an exact \(\mathbb{Z}_3\) symmetry, for which $2\leftrightarrow3$ reactions are open in the symmetric phase. Third, we analyze scenarios in which the self-interacting dark matter interacts with an unstable mediator. In all these cases, the dark sector is assumed to be populated via freeze-in, mainly through Higgs decays. The thermal evolution is obtained by solving the coupled Boltzmann equations for the relevant number densities and temperatures, consistently accounting for both freeze-in production and cannibal interactions. The analysis reveals a rich phenomenology, which depends sensitively on the strength of the dark-sector self-interactions, the symmetry structure, and the particle content of the hidden sector. In particular, the resulting parameter space can be strongly constrained, effectively invisible, or potentially accessible to future searches, depending on the realization considered.
The thesis also extends this framework to non-standard cosmological histories by studying cannibal dark matter production during non-instantaneous reheating. We find that, if the temperature of reheating is low, there can be substantial differences between the portal and self-interaction coupling, opening regions of parameter space that may become accessible to future collider and long-lived particle searches.
Finally, we investigate the thermal evolution of a hidden U(1) gauge sector populated via freeze-in, where the fields under consideration do not constitute a dark matter candidate, but rather provide an example of a hidden sector with non-trivial phase structure. Dark sector particles are produced from the Standard Model bath and thermalize due to strong gauge interactions. The continuous injection of energy into the hidden sector can induce an inverse first-order phase transition and temporarily restore the symmetric phase. In this setup, standard transitions into the broken phase are accompanied by reheating of the hidden plasma, whereas the inverse transition does not exhibit an analogous reheating jump, since it is driven by energy injection rather than by latent-heat release. More broadly, the results suggest that the interactions required to generate a first-order barrier may also favor rapid chemical-equilibration within the hidden sector, non-trivially linking phase-transition dynamics and chemical equilibration.

\end{abstract}

\begin{extraAbstract}
	\addchaptertocentry{\extraAbstractname}

Niniejsza rozprawa bada dynamikę samo-oddziałujących ciemnych sektorów we wczesnym Wszechświecie, utworzonych poprzez mechanizm freeze-in. Główny nacisk położono na scenariusze, w których ciemny sektor zawiera cząstki podlegające reakcjom typu ($2\leftrightarrow3$), oraz na sposób, w jaki oddziaływania te modyfikują zarówno ewolucję we wczesnym Wszechświecie, jak i wynikającą z niej fenomenologię. Przeanalizowano kilka realizacji. Po pierwsze, zbadane zostało rzeczywiste pole skalarne w teorii ze spontanicznie łamaną symetrią ($\mathbb{Z}_2$), gdzie łamanie symetrii indukuje oddziaływanie umożliwiające procesy ($2\leftrightarrow3$). Po drugie, rozważono zespolone pole skalarne stabilizowane przez symetrię ($\mathbb{Z}_3$), dla którego reakcje ($2\leftrightarrow3$) są możliwe w fazie z zachowaną symetrią. Po trzecie, analizujemy scenariusze, w których samo-oddziałująca ciemna materia oddziałuje z niestabilnym polem pośredniczącym, tzw, mediatorem. We wszystkich tych przypadkach zakłada się, że ciemny sektor jest utworzony poprzez mechanizm freeze-in, głównie w wyniku rozpadów bozonu Higgsa. Ewolucję termiczną wyznaczono rozwiązując układ sprzężonych równań Boltzmanna dla odpowiednich gęstości liczbowych i temperatur, spójnie uwzględniając zarówno produkcję przez freeze-in, jak i oddziaływania kanibalistyczne. Analiza ujawnia bogatą fenomenologię, która silnie zależy od siły samo-oddziałowań w ciemnym sektorze, struktury symetrii oraz składu cząstkowego ukrytego sektora. W szczególności, wynikająca z tego przestrzeń parametrów może być silnie ograniczona, efektywnie niewidoczna lub potencjalnie dostępna dla przyszłych eksperymentów, w zależności od rozważanej realizacji. Rozprawa rozszerza również tę teorię o niestandardowe historie kosmologiczne, badając produkcję kanibalistycznej ciemnej materii podczas nie-natychmiastowego odgrzewania (reheating). Jeśli temperatura odgrzewania jest niska, mogą wystąpić znaczne różnice między sprzężeniem portalowym a samo-oddziałowaniem, co otwiera regiony przestrzeni parametrów, które mogą stać się dostępne dla przyszłych poszukiwań w akceleratorach oraz poszukiwań cząstek długożyciowych. Na koniec badana jest ewolucja termiczną ukrytego sektora cechowania U(1) utworzonego przez freeze-in, w którym rozważane pola nie stanowią opisu ciemnej materii, lecz służą jako przykład ukrytego sektora o nietrywialnej strukturze fazowej. Cząstki ciemnego sektora są produkowane z plazmy Modelu Standardowego i termalizują dzięki silnym oddziaływaniom cechowania. Ciągły przypływ energii do ukrytego sektora może indukować odwrotne przejście fazowe pierwszego rodzaju i tymczasowo przywrócić fazę symetryczną. W tym układzie standardowym przejściom do fazy z łamaną symetrią towarzyszy odgrzewanie ukrytej plazmy, podczas gdy przejście odwrotne nie wykazuje analogicznego skoku temperatury, ponieważ jest napędzane dopływem energii, a nie uwalnianiem ciepła utajonego. Szerzej mówiąc, wyniki sugerują, że oddziaływania wymagane do wygenerowania bariery pierwszego rodzaju mogą również sprzyjać szybkiemu ustalaniu równowagi chemicznej w ukrytym sektorze, co w nietrywialny sposób łączy dynamikę przejść fazowych z równowagą chemiczną.
	
\end{extraAbstract}

\begin{acknowledgements}
\setlength{\parindent}{0pt}
\addchaptertocentry{\acknowledgementname} 


First of all, I am grateful to my supervisor, Andrzej Hryczuk, for his excellent guidance, interesting research topics, and constructive feedback. Thank you for offering me the opportunity to pursue this research program in the first place, and for giving me enough intellectual freedom to follow my own curiosity and develop my ideas into concrete projects. I will certainly miss our discussions.

\vspace{0.15cm}

To all my collaborators and colleagues, thank you for the discussions, guidance, corrections, and insights. A special \textit{``Thank You''} goes to Nicolás Bernal and Kuldeep Deka, for the discussions, your (rapid!) feedback and for sharing with us your knowledge of reheating. Your expertise in this field has been, and continues to be, highly appreciated. We learned a great deal from you.

\vspace{0.15cm}

To my collaborators in Karlsruhe, a special mentioning goes to Felix Kahlhoefer, who kindly hosted me for an internship at KIT, where the \textit{inverse first order transition} project originated, and who encouraged me to pursue my curiosity further while providing feedback and insights into promising directions, this really made me grow as a physicist. To Santiago for coming up with interesting questions that made me rethink how to approach physics. To Jonas Matuszak, for your expertise and guidance on phase transition topics. I look forward to further discussions with you all.

\vspace{0.15cm}

I would also like to thank previous mentors who helped shape the path that brought me here, and from whom I learned a lot. To Saúl Ramos-Sánchez, for introducing me to dark matter phenomenology research during my bachelor’s studies and who also hosted me at UNAM during my visit in Mexico. To Kai Schmidt-Hoberg, Camilo García-Cely, and Marco Hufnagel, who guided me during my master’s studies and whose research direction I pursued further during this doctorate. Particularly to Marco Hufnagel, for explaining to me some of the BBN bound he derived and that appear in this thesis.

\vspace{0.15cm}

A dissertation reflects not only years of deep intellectual work, but also years of struggle in other spheres of life, especially for those far from home. I could not have managed this journey without the support of colleagues and friends, who made it much more joyful. A special \textit{¡Gracias!} goes to Andrés, for unearthing the latino inside me. To Joseph, Jonas, and Ludo, thanks for making the office hours more fun. I am also grateful with Abhishek for his comments on my CV and applications. And to Sven Mueller: for your insights on life, philosophy, and academia.

\vspace{0.15cm}

In a completely different category, I would like to express my deepest gratitude to my beloved girlfriend, Ania, who truly made Warsaw feel like home for me. Thank you for showing me the hidden gems of Poland, for your emotional support, for your patience, and for teaching me how to approach life and how to be more consistent. This work was greatly and positively influenced by you. I thank you for being present during such a significant moment in my life.

\vspace{0.15cm}

Un agradecimiento especial va para mi abuela, Blanca, que falleció mientras hacía el doctorado. Gracias por haber sido también mi madre, por haber sido el principal apoyo durante toda mi vida. Este trabajo va especialmente dedicado a ti. A mi familia, que siempre me mostró su apoyo incondicional; a mi padre Juan, que siempre creyó en mí y nunca dudó de mis capacidades. A mi hermano Isaí, por su trabajo duro. A mi hermana Alexa, por ser una fuente de inspiración para mí. Gracias a ustedes no me rendí y he logrado tanto. Los amo.

\vspace{0.15cm}

Al pueblo de México, por haberme provisto de educación gratuita y de calidad, a pesar de todos los problemas socioeconómicos que existen en mi país y en el mundo.

\vspace{0.15cm}

A special acknowledgment goes to people whom I have never met, but who nevertheless had a profound impact on my worldview and on the way I approach life. To Robert Sapolsky, for convincing me that there is no room for free will in a deterministic world. This helped me become more humble and forgiving. To Sam Harris and Dr.~Alok Kanojia (Dr. K), for teaching me that the ego is illusory, and that meditation is a way of seeing through that illusion.

\vspace{0.15cm}

And to any reader of this thesis, thank you for taking the time not only to review this document, which represents four years of work, but also to try to understand the world in a scientific and honest way.

\end{acknowledgements}


\tableofcontents 

\dedicatory{A la memoria de Blanca Clementina Cruz Manzanarez 
	\\ y Julio César Cervantes Cruz} 

\mainmatter 

\pagestyle{thesis} 

\clearpage
\thispagestyle{plain}
\begin{center}
	{\Large\bfseries Publications and Contributions}
\end{center}
\vspace{6em}
This thesis is based on the following published papers:

\vspace{1em}

\begin{itemize}
	\item
	N.~Bernal, E.~Cervantes, K.~Deka, and A.~Hryczuk,
	\emph{Freezing-in cannibals with low-reheating temperature},
	JHEP \textbf{09} (2025) 083,
	\href{https://arxiv.org/abs/2506.09155}{arXiv:2506.09155 [hep-ph]}.
	\vspace{1em}
	\item
	E.~Cervantes and A.~Hryczuk,
	\emph{Freezing-in cannibal dark sectors},
	JHEP \textbf{11} (2024) 050,
	\href{https://arxiv.org/abs/2407.12104}{arXiv:2407.12104 [hep-ph]},
\end{itemize}
\vspace{1em}
and publications in preparation
\vspace{1em}
\begin{itemize}
	\item
	E.~Cervantes, F.~Kahlhöfer, J. Matuszak and S. Roselon,
	\emph{First-order phase transitions induced via freeze-in}.
	
	\item E.~Cervantes, A.~Hryczuk, S.~Lederer,
	\emph{Light Dark Matter from Self-cooling Dark Sectors}.
\end{itemize}
%
\vspace{2em}
Other publications by the author not included in this thesis:
\vspace{1em}
\begin{itemize}
	\item
	E.~Cervantes, O.~Perez-Figueroa, R.~Perez-Martinez, and S.~Ramos-Sanchez,
	\emph{Higgs-portal dark matter from nonsupersymmetric strings},
	Phys.\ Rev.\ D \textbf{107} (2023) 115007,
	\href{https://arxiv.org/abs/2302.08520}{arXiv:2302.08520 [hep-ph]}.
\end{itemize}

\include{Chapters/Chapter1}

\include{Chapters/Chapter2}

\include{Chapters/Chapter3}

\include{Chapters/Chapter4}

\include{Chapters/Chapter5}
\include{Chapters/Chapter6}


\printbibliography[heading=bibintoc]



\appendix 


\include{Appendices/AppendixA}

\end{document}

%% file: Chapters/Chapter1.tex

\chapter{Introduction} 

\label{ch:1} 


\newcommand{\keyword}[1]{\textbf{#1}}
\newcommand{\tabhead}[1]{\textbf{#1}}
\newcommand{\code}[1]{\texttt{#1}}
\newcommand{\file}[1]{\texttt{\bfseries#1}}
\newcommand{\option}[1]{\texttt{\itshape#1}}



The Higgs boson, discovered at the Large Hadron Collider in 2012, is the only
fundamental scalar field currently known to exist in nature. While its discovery
completed the particle content of the Standard Model (SM), it also sharpened a
number of open conceptual and phenomenological questions. Despite its remarkable
success in describing collider and low-energy data, the Standard Model fails to
account for several well-established observations, including the existence of
dark matter, the nonzero masses of neutrinos, the observed baryon asymmetry of the
Universe, the absence of a solution to the strong CP problem, and the lack of a
consistent quantum description of gravity. These shortcomings strongly suggest
that the Standard Model should be regarded as an effective theory, valid up to
some cutoff scale, rather than as a complete description of fundamental
interactions. Within this broader context, the Higgs sector itself raises a particularly acute
theoretical puzzle. The Higgs mass,
$m_h \simeq 125\,\mathrm{GeV}$, and vacuum expectation value,
$v_h \simeq 246\,\mathrm{GeV}$, are many orders of magnitude smaller than the
Planck scale, $M_{\rm Pl} \sim 10^{19}\,\mathrm{GeV}$, at which gravitational
effects are expected to become strong. Understanding the stability of this large
hierarchy against quantum corrections constitutes the well-known hierarchy
problem. In the absence of additional structure, scalar masses are sensitive to
ultraviolet physics, rendering the electroweak scale unnaturally small compared
to the scale associated with gravity.

To address those shortcomings, a wide class of theoretical frameworks were proposed, extending
the Standard Model by enlarging its field content and symmetry structure. Such
extensions are commonly referred to as physics beyond the Standard Model (BSM).
Among the simplest and most widely studied BSM scenarios are those involving
additional scalar degrees of freedom. From a bottom-up perspective, scalar extensions are particularly minimal, as they can address experimental and observational shortcomings without enlarging the gauge sector or introducing new fermionic states. 
In
some cases, scalars may play an essential role, for example in the generation of
neutrino masses through seesaw mechanisms or radiative models, or in modifying
the nature of the electroweak phase transition in scenarios of baryogenesis. In
other contexts, scalar fields appear as part of a broader structure, such as in
supersymmetric or classically scale-invariant theories, where extended scalar
sectors accompany new symmetries or dynamics that can soften the hierarchy
problem by controlling radiative corrections to the Higgs mass. At the same time,
the introduction of additional scalar degrees of freedom generically gives rise
to new hierarchy problems of its own, unless their masses and couplings are
protected by symmetries or other organizing principles.

Scalar extensions are therefore typically stabilized or structured by additional 
discrete $\mathbb{Z}_N$ or continuous global or
gauge symmetries of the form $\mathrm{U}(N)$ or $\mathrm{SU}(N)$. These symmetries
may arise in top-down constructions, including supersymmetry or string-inspired
models, or be imposed in a purely bottom-up fashion to control couplings, protect
scalar masses, or forbid unwanted operators. Depending on the symmetry structure,
such scalar fields may be unstable and decay into Standard Model states, or they
may be stabilized and persist as long-lived or cosmologically stable relics.

In addition to their particle-physics implications, extended scalar sectors can
have profound consequences for early-Universe cosmology. Scalar fields may play a
central role in shaping the thermal history of the Universe: they may drive
inflation and the subsequent reheating of the Universe, they may govern the
structure and dynamics of cosmological phase transitions, while leaving observable imprints such as stochastic gravitational-wave backgrounds or modifications to the expansion history. As a
result, scalar extensions of the Standard Model provide a versatile framework in
which collider physics, cosmology, and astrophysical observations can be studied
at the same time. 

This thesis explores the cosmological implications of scalar extensions in two
complementary settings. In both cases, we focus on scenarios in which scalar fields
possess sizable self-interactions while coupling only feebly to the Standard Model
through suppressed portal interactions. As a result, these scalar sectors are
generically out of thermal equilibrium with the visible plasma, and their
cosmological evolution is controlled by an interplay between self-interactions,
cosmic expansion, and weak interactions with Standard Model degrees of freedom.

In the first setting, scalar fields appear as particle-like degrees of freedom whose relic abundance is determined by their initial production via Higgs decays and by the subsequent microscopic interactions among them, rather than by a departure from thermal equilibrium with the Standard Model. In this framework, the scalar particles may either constitute dark matter themselves or act as mediator particles coupling the dark sector to the Standard Model; in both cases, the scalar states exhibit self-interactions. 

In the second setting, scalar fields are treated as macroscopic condensates whose vacuum structure evolves dynamically and can undergo first-order phase transitions in the early Universe. While these realizations differ in their physical interpretation, both are governed by the same underlying principles: out-of-equilibrium dynamics in an expanding background, the role of self-interactions in shaping the evolution of the scalar
sector, and the use of Boltzmann equations to connect particle-physics parameters to cosmological observables.

We begin by considering the case in which a scalar extension gives rise to a stable particle species that contributes to the matter content of the Universe. This naturally leads to the dark matter problem, which provides one of the most direct and observationally driven motivations for physics beyond the Standard Model.

\section{The dark matter problem}

Throughout the 20th century, observations of galaxy rotation curves indicated that stars and galactic gas rotate faster than expected from the gravitational pull generated by visible matter alone. This picture has been reinforced by a wide range of independent astrophysical and cosmological probes pointing to the presence of a missing matter component~\cite{Bertone:2010zza}. The inferred gravitational dynamics therefore point either to a breakdown of the gravitational description on galactic and larger scales or to the existence of an additional non-luminous matter component.

The former possibility motivates theories of modified gravity. Among them, the most widely studied framework is \textit{Modified Newtonian Dynamics} (MOND)~\cite{Desmond:2025pmk}, which postulates a modification of Newtonian gravity at galactic scales and beyond. Although MOND can account for the observed rotation curves of galaxies, it faces a number of conceptual and phenomenological difficulties. One major challenge is the construction of a relativistic completion capable of simultaneously reproducing the observed cosmic microwave background (CMB) anisotropies and baryonic acoustic oscillations. In addition, MOND has difficulty explaining the gravitational lensing observed in the merging galaxy cluster 1E~0657--56~\cite{Randall:2008ppe}, where the lensing mass is spatially offset from the baryonic matter distribution.

The second possibility corresponds to the existence of a missing matter component. In this thesis, we adopt its particle interpretation and focus on the minimal scenario in which the observed dark matter abundance is accounted for by a single yet undiscovered elementary particle,\footnote{Dark matter could in principle be multicomponent. However, such scenarios generally involve a substantially enlarged and more model-dependent parameter space, due to the presence of additional masses, couplings, and conversion channels, and in many concrete realizations one component dominates the final relic abundance over broad regions of parameter space. For this reason, this thesis focuses on single-component dark matter as a minimal scenario.} under which the cosmological and cluster-scale observations mentioned above are well accommodated. 

On large scales, observations are well described by a cold dark matter component that behaves approximately collisionlessly. In particular, cluster mergers constrain the dark matter ($2\to2$) self-interaction cross section per unit mass to be at most of order
\(\sigma_{\rm self}/m_{\rm DM} \lesssim \mathcal{O}(1)\,\mathrm{cm^2\,g^{-1}}\),
with the Bullet Cluster yielding a representative bound
\(\sigma_{\rm self}/m_{\rm DM} \lesssim 1.25\,\mathrm{cm^2\,g^{-1}}\) and, under additional assumptions, stronger bounds such as \(\sigma_{\rm self}/m_{\rm DM} \lesssim 0.7\,\mathrm{cm^2\,g^{-1}}\)~\cite{Randall:2008ppe}. This still corresponds to a microscopically sizeable interaction strength, of order \(1\,\mathrm{barn/GeV}\), i.e. roughly a nuclear scale, so there is no contradiction between allowing appreciable dark-sector self-interactions and satisfying current astrophysical bounds. Thus, any self-interactions relevant for dark-matter phenomenology must remain weak enough to preserve the approximately collisionless behavior inferred on cluster scales.

In the particle DM picture, the dark matter states are excitations of a quantum field associated with an additional 
(gauge or global) symmetry group extending the $\text{SU}(3)_c\times \text{SU}(2)_L\times U(1)_Y$ gauge structure of the Standard Model. 
Such a hidden symmetry (or set of symmetries) would assist in the stabilization of DM, preventing it from decaying and ensuring that it remains the dominant matter component in the Universe. 
If allowed by the underlying charge assignments, the matter content of hidden
sectors may interact with the Standard Model through gauge-invariant operators,
commonly referred to as \emph{portals}. At the renormalizable level, the most
frequently considered possibilities include the Higgs portal~\cite{Arcadi:2019lka}, the vector
portal~\cite{Baek:2012se}, and the neutrino portal~\cite{Blennow:2019fhy}, which couple hidden-sector fields to the Higgs
doublet, to the hypercharge gauge field, or to Standard Model leptons,
respectively. In addition, interactions may arise through higher-dimensional
operators, such as purely gravitational couplings~\cite{Cata:2016epa} or effective baryonic
operators often referred to as neutron portals~\cite{Lonsdale:2018xwd}. 

Among the most studied particle DM realizations containing a portal or gauge interaction with SM, 
one finds Weakly Interacting Massive Particles (WIMPs)~\cite{Roszkowski:2017nbc}, whose
predicted annihilation cross section into Standard Model states is naturally set
by weak-scale masses and couplings (hence the term \textit{weakly} interacting).
The WIMP prediction contained three main ingredients:
\begin{itemize}
    \item interactions with the Standard Model enforced thermal equilibrium between the two sectors in the early Universe, until the expansion of the Universe became too rapid for dark matter and Standard Model particles to maintain chemical equilibrium. This epoch is known as decoupling, or freeze-out.
    \item after freeze-out, the annihilation rate of dark matter particles falls below the
Hubble expansion rate, and the Boltzmann equation no longer supports efficient
number-changing processes. The dark matter number density then simply redshifts
with the expansion, implying a constant comoving abundance;
    \item if the dark matter mass lies in the electroweak range
($10\,\mathrm{GeV}$--$1\,\mathrm{TeV}$) and its interactions with Standard Model
particles are characterized by couplings of order $\mathcal{O}(0.01)$--$\mathcal{O}(1)$, 
yielding a comoving dark matter abundance roughly consistent with
observations.
\end{itemize}
This observation became known as the WIMP \textit{miracle}. Its appeal lay in the fact
that the observed relic abundance emerges naturally from electroweak-scale masses
and couplings within a clear dynamical picture. The chemical
decoupling occurs when the dark matter particles become non-relativistic, at
temperatures comparable to their mass, ensuring consistency with the formation
of large-scale structure. 

These features motivated an extensive experimental
program, including numerous direct and indirect detection (DD/ID)
experiments~\cite{XENON:2017lvq, LZ:2019sgr, PandaX-II:2017hlx, SuperCDMS:2017mbc,DAMA:2008bis, 2009ApJ...697.1071A}, 
aimed at probing WIMP interactions with Standard
Model particles. At the time of writing this thesis (2026), the sensitivity of leading direct
detection experiments has reached unprecedented levels and, in some regions of parameter space, is beginning to approach the regime in which neutrino-induced recoils become a relevant background for conventional recoil-based searches~\cite{Blanco-Mas:2024ale}. 
This introduces an important experimental challenge, since neutrino-induced nuclear recoils can mimic a dark matter signal when only rate and spectral information are used. As a
result, simple and highly predictive WIMP realizations, such as those involving tree-level 
$Z$-boson exchange, are now strongly constrained. 
Viable regions of parameter space persist only in non-minimal constructions, 
for instance involving inelastic dark matter with vector mediators~\cite{Foguel:2024lca}.
 
At the same time,
models in which dark matter couples predominantly through the Higgs portal remain
actively probed, including scenarios with dark matter masses extending into the
TeV scale~\cite{EscuderoAbenza:2025cfj}. Taken together, these developments indicate that while the original, minimal
WIMP paradigm is under increasing experimental pressure, it is not excluded.
Rather, present constraints motivate a reassessment of the simplest
implementations, favoring either weaker couplings to visible matter or more
secluded dark sectors, potentially interacting only gravitationally, over a wide
range of dark matter masses. 

It is worth emphasizing, however, that retaining electroweak-scale interactions with the Standard Model does not preclude viable WIMP dark matter. Rather, such realizations may persist in less minimal BSM frameworks, where the enlarged particle content and parameter space permit compatibility with existing constraints, albeit with reduced predictivity. This occurs, for example, in models with extended dark sectors, including scenarios with multiple dark matter candidates~\cite{Bhattacharya:2013hva}, as well as top-down approaches like in supersymmetric theories, whose rich spectrum and coupling structure continue to admit viable WIMP parameter regions.


This thesis explores alternatives to the minimal WIMP picture by considering dark matter production mechanisms that lead to distinct cosmological histories and phenomenological signatures. Two broad classes of mechanisms are particularly relevant for the scenarios studied here. The first is the \textit{freeze-in} mechanism, in which the dark matter abundance is gradually built up through extremely feeble interactions with the Standard Model plasma, so that the dark sector never attains thermal equilibrium with the visible sector. The second is \textit{cannibal} dark matter, in which sufficiently strong self-interactions give rise to number-changing reactions within the dark sector, thereby determining its final relic abundance. Although these mechanisms are logically independent and can be realized separately, in this thesis we focus on scenarios in which freeze-in production populates a self-interacting dark sector that subsequently undergoes cannibal dynamics. A detailed introduction to freeze-in and cannibal dark matter, as well as to their combined realization, is presented in~\Cref{ch:3}.



\section{Cosmological phase transitions}

In 2016, the Laser Interferometer Gravitational-Wave Observatory (LIGO) collaboration
reported the first direct detection of gravitational waves (GWs) originating from
the merger of a binary black-hole system~\cite{LIGOScientific:2016aoc}, almost exactly
one hundred years after Einstein had realized that his field equations admit wave
solutions in the weak-field, linearized regime. This discovery provided yet another
spectacular confirmation of General Relativity and marked the beginning of
gravitational-wave astronomy, motivating intense experimental and theoretical
efforts to identify and characterize new GW sources. 
The sources detected so far are astrophysical in nature, involving compact objects
such as black holes and neutron stars. Future ground- and space-based interferometers
are expected to significantly expand this catalog. However, gravitational waves may also have been produced in the early Universe,
long before the formation of any astrophysical structures. In particular,
violent processes associated with cosmological phase transitions could have
generated a stochastic gravitational-wave background that persists to the
present day. Other sources of a stochastic gravitational-wave background in
cosmology include inflation, during which quantum fluctuations of the metric
are stretched to macroscopic scales, and the thermal emission of gravitons from
a hot primordial plasma.

Cosmological phase transitions arise when the vacuum structure of one or more scalar quantum
fields changes as the Universe cools. If the transition is sufficiently abrupt, most
notably in the case of a first-order phase transition (FOPTs), it proceeds through the
nucleation of bubbles of a new vacuum phase within a metastable background. In such
scenarios, a scalar field typically forms a condensate characterized by a vacuum
expectation value (VEV). At high temperatures, interactions of the scalar field with
itself and with other degrees of freedom modify its free-energy density through
thermal corrections, often restoring a symmetric phase in which the VEV vanishes.
As the Universe expands and cools, thermal effects become subdominant, allowing a
new vacuum configuration to emerge. If the transition between these vacua is of first order, the system can become
temporarily trapped in a metastable state before transitioning to the true vacuum,
during which, some regions of space-time start transitioning into the true vacuum, 
forming bubbles with the new phase inside.
The subsequent nucleation, expansion, and collision of bubbles release latent heat
into the surrounding plasma and drive bulk motions and turbulence. A fraction of the
energy stored in the scalar sector and the plasma is converted into gravitational
radiation, giving rise to a stochastic gravitational-wave background whose spectral
properties encode information about the underlying particle physics.

Cosmological phase transitions are macroscopic events whose
observable consequences are determined by microscopic properties of the theory,
such as the field content, interaction strengths, and thermal corrections to the
effective potential. Their characteristic features, including the order of the
transition, its duration, the amount of supercooling, and the released latent
heat, are therefore sensitive probes of physics beyond the Standard Model. 
Phase transitions may occur over a wide range of energy scales. Well-established
examples include the QCD transition (100~MeV~scale), associated with the confinement of quarks into hadrons, and the electroweak phase transition (100~GeV~scale), during which electroweak symmetry is
spontaneously broken and Standard Model particles acquire mass. While the Standard
Model electroweak transition is a crossover, many extensions predict a strongly
first-order transition, which could simultaneously source gravitational waves and
provide the out-of-equilibrium conditions required for electroweak baryogenesis.
Importantly, upcoming space-based gravitational-wave detectors are expected to be
sensitive to cosmological sources. In particular, the Laser Interferometer Space
Antenna (LISA)~\cite{2018AdOT....7..395E,LISACosmologyWorkingGroup:2022jok}
will probe the millihertz frequency band, which is well suited to detecting
stochastic gravitational-wave backgrounds originating from first-order phase
transitions in a wide class of extensions of the Standard Model. 

Complementarily, pulsar timing array experiments, such as those conducted by the
\textit{NANOGrav} Collaboration~\cite{NANOGrav:2025blx}, are sensitive to
gravitational waves in the nanohertz frequency range and have recently reported
evidence for a stochastic gravitational-wave background. In this context, the 
\textit{stochastic} refers to a signal that is not resolved into a small number of
individual sources, but instead arises as the superposition of many independent
and typically unresolved emitters, giving rise to an effectively random background
characterized statistically through its spectral shape and angular
correlations. In the astrophysical interpretation, the leading candidate is the
population of inspiralling supermassive black-hole binaries distributed throughout
the Universe, whose combined emission naturally generates a nanohertz stochastic
background. However, pulsar timing arrays are also sensitive to gravitational waves
of cosmological origin, produced in the early Universe by mechanisms such as
first-order phase transitions, cosmic strings, or enhanced primordial
perturbations~\cite{Gangui:2001wc}.

The cosmological interest of such a signal lies in the fact that gravitational
waves propagate essentially freely once produced and therefore retain information
about the physical conditions of the epoch in which they were generated. In this
sense, a stochastic gravitational-wave background can act as a fossil relic of the
early Universe, probing energy scales and out-of-equilibrium dynamics that are
otherwise inaccessible to direct observation. In particular, the characteristic
frequency and amplitude of the signal are linked to the Hubble scale and the
microphysics of the source at the time of production, so that a detection in the
nanohertz band could provide information about processes such as symmetry-breaking
phase transitions or topological defects in the primordial plasma. For this
reason, the detection of a stochastic gravitational-wave background of cosmological
origin would open a unique observational window onto high-energy particle physics
and the thermal history of the early Universe, complementing collider experiments
and other cosmological probes~\cite{Figueroa:2023zhu}.

The thesis is organized as follows. In Chapter~\ref{ch:2}, we introduce the theoretical tools required to describe dark matter production in the early Universe, including a review of the $\Lambda$CDM model in cosmology, Friedmann–Lemaître–Robertson–Walker (FLRW) metric, the Boltzmann equation, and an overview of thermal field theory and the parametrization of cosmological first-order phase transitions. Chapter~\ref{ch:3}, based on~\cite{Cervantes:2024ipg}, is devoted to the phenomenology of cannibal dark matter produced via the freeze-in mechanism, and we will address the details of the freeze-in production mechanisms and cannibal dynamics. Chapter~\ref{ch:4}, building on a completed and published collaboration~\cite{Bernal:2025osg}, generalizes the framework of the previous chapter to scenarios with continuous reheating histories and provides a careful assessment of detection prospects.~\Cref{ch:5} presents work carried out as part of an ongoing collaboration in the Karlsruhe Institute of Technology, initiated during a research internship in the group of Prof.~Felix Kahlhoefer, and investigates whether an inverse first-order phase transition can be induced through freeze-in dynamics. Finally, Chapter~\ref{ch:conclusions} summarizes the main conclusions of this research program and discusses its broader implications and outlook.

%% file: Chapters/Chapter2.tex
\chapter{Background}\label{ch:2}

Throughout this thesis, we work within the cosmological framework provided by
the Friedmann--Lemaître--Robertson--Walker (FLRW) metric, which describes a
homogeneous and isotropic expanding Universe on large scales. Together with the
$\Lambda$CDM model, it provides the standard geometrical and dynamical setting
for modern cosmology, in which the present-day energy budget is dominated by a
cosmological constant $\Lambda$ and cold dark matter (CDM).

This chapter collects the background material needed in the remainder of the
thesis. We first review the main observational evidence for dark matter. We
then discuss the basic cosmological framework, relevant thermodynamic
quantities, and the Boltzmann equation. Finally, we introduce the elements of
thermal field theory needed to describe thermal corrections to the free energy
density, as well as the basic concepts of cosmological first-order phase
transitions and their associated gravitational-wave signatures.

\section{Observational evidence for dark matter}
\label{sec:dm_evidence}

Before introducing the theoretical tools to compute
the dark matter relic abundance, we begin this chapter by discussing the
empirical evidence supporting the existence of dark matter. The purpose of this section is to
summarize the key astrophysical and cosmological observations that motivate the
introduction of a non-luminous, non-baryonic matter component.

The evidence for DM is cumulative and arises from a wide range of
independent probes, spanning galactic~\cite{2017PhyU...60....3Z}, cluster~\cite{Einasto:1999dy}, 
and cosmological scales~\cite{Giesen:2012rp}. Despite
their very different physical origins and systematic uncertainties, these
observations consistently point toward the presence of a cold, gravitationally
interacting matter component that dominates the matter budget of the Universe.
Importantly, the inferred properties of this component are largely insensitive
to the microscopic nature of dark matter and rely primarily on its gravitational
effects.

Historically, some of the earliest indications for dark matter emerged from
observations of galaxy rotation curves. In spiral galaxies, the circular
velocities of stars and interstellar gas are inferred from Doppler shifts of
spectral lines as a function of galactocentric radius. Under the assumption that
the gravitational potential is generated predominantly by the observed luminous
matter, Newtonian dynamics predicts a Keplerian fall-off of the rotational
velocity, $v(r) \propto r^{-1/2}$, at distances beyond the bulk of the stellar
disk. Instead, observations reveal that rotation curves remain approximately
flat out to the largest measured radii, implying an enclosed mass that continues
to grow roughly linearly with radius~\cite{Corbelli:1999af,Mistele:2024hfh,Lelli:2016zqa,Wechsler:2018pic}. 
This behavior is naturally explained by
the presence of an extended, approximately spherical dark matter halo that
dominates the gravitational potential at large distances from the galactic
center.

\begin{figure}[t!]
	\centering
	\includegraphics[width=0.85\textwidth]{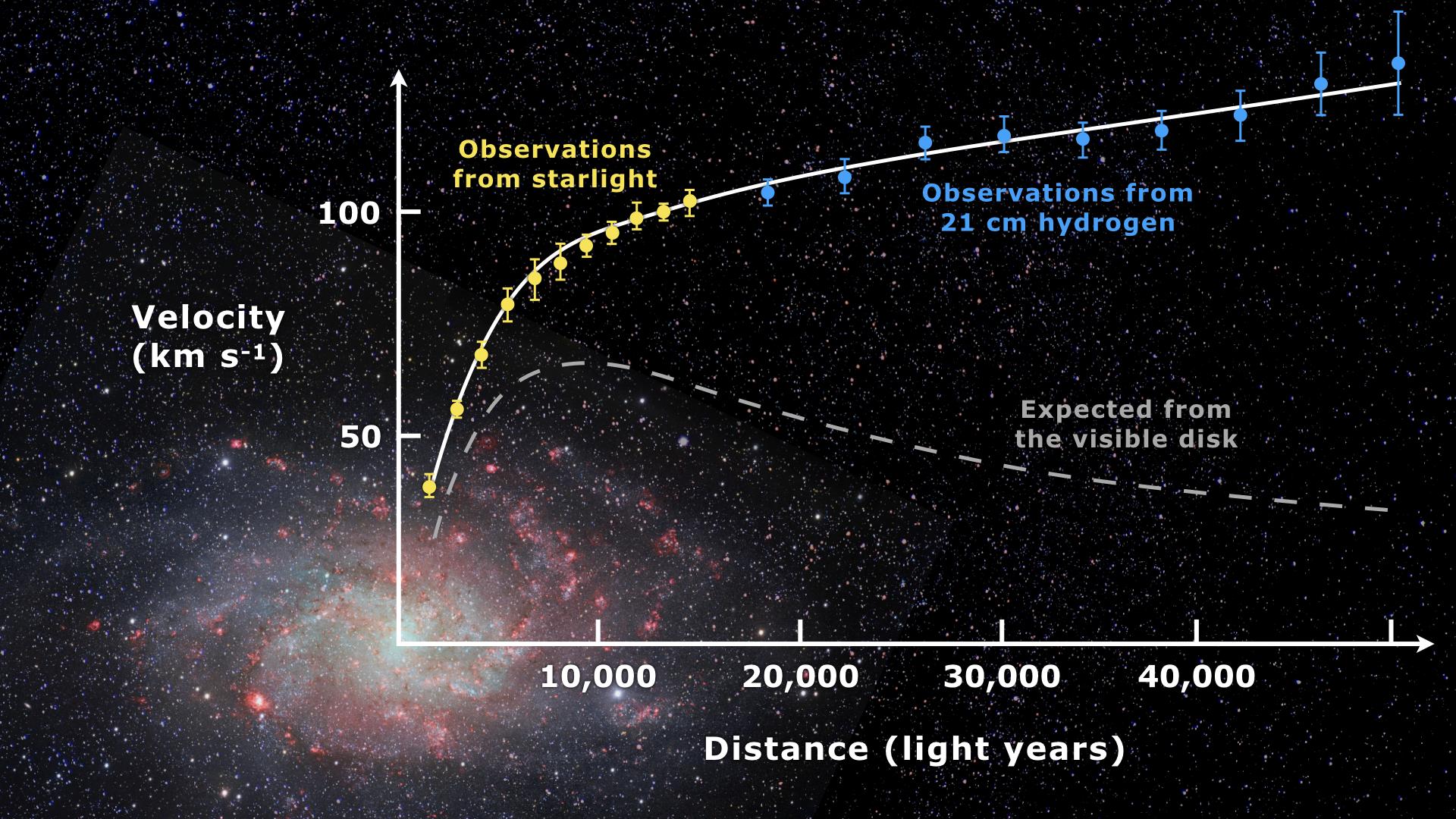}
	\caption{Observed rotation curve of the Messier~33 galaxy. The gray dashed line shows the
		expected rotational velocity inferred from the luminous matter distribution,
		while the solid line corresponds to a model including an extended dark matter
		halo. The yellow and blue points denote the measured stellar and gas velocities,
		respectively. Figure from~\cite{Corbelli:1999af}.}
	\label{fig:rot_gal}
\end{figure}

On larger scales, galaxy clusters provide independent and complementary evidence
for dark matter. The total mass of clusters can be inferred through several
observational channels, including the velocity dispersion of member galaxies,
X-ray measurements of the hot intracluster gas assuming hydrostatic equilibrium,
and gravitational lensing of background sources. These methods yield
consistent mass estimates that exceed the observed baryonic mass by factors of
several. In particular, gravitational lensing directly probes the projected mass
distribution irrespective of its composition, and has revealed substantial
amounts of matter distributed throughout clusters and in their outskirts, well
beyond the visible galaxy population.

A particularly instructive case is provided by the merging galaxy cluster
1E~0657--56 (also referred to as 1E~0657--558), commonly known as the
\emph{Bullet Cluster}~\cite{Clowe:2006eq,Randall:2008ppe, Cha2025Bullet}. This system consists of
two galaxy clusters that have undergone a high-velocity collision, in
which different cluster components respond differently depending on their
interaction properties. The baryonic mass is dominated by hot, ionized
intracluster gas, which behaves as a collisional fluid and is decelerated by ram
pressure and shocks during the merger. 
\begin{figure}[t!]
	\centering
	\includegraphics[width=0.85\textwidth]{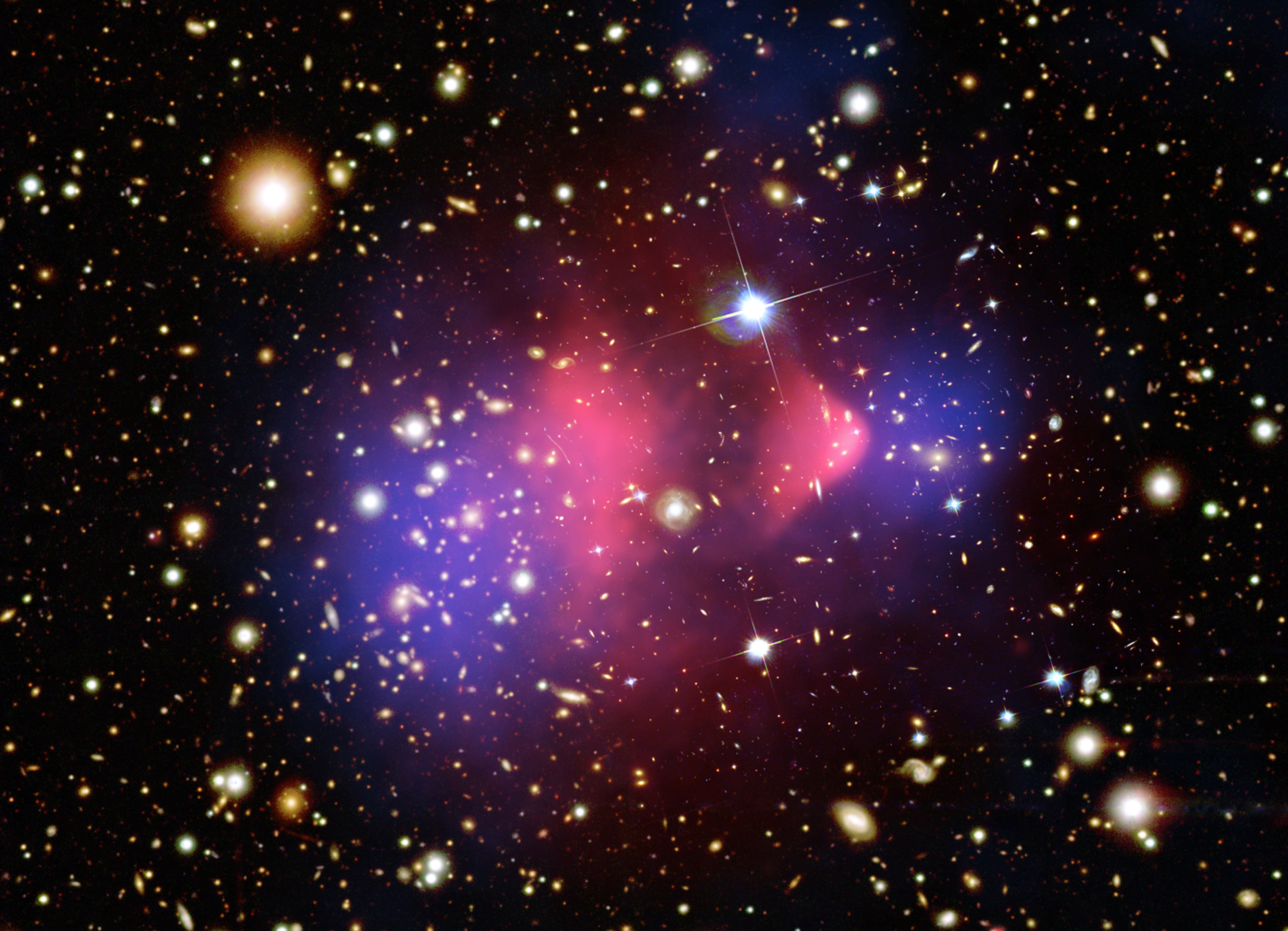}
	\caption{Composite optical, X-ray (pink), and weak gravitational lensing (blue) image of
		the Bullet Cluster (1E~0657--558). The X-ray emission traces the hot intracluster
		gas, while the lensing map reconstructs the projected total mass distribution.
		Image from~\cite{AstronomyBulletCluster2024}.}
	\label{fig:bullet}
\end{figure}
In contrast, galaxies are effectively
collisionless on cluster scales and pass through the collision with little
direct interaction. Gravitational lensing maps of the Bullet Cluster reveal that the dominant mass
peaks are spatially aligned with the galaxy distributions rather than with the
X-ray emitting gas, which contains most of the baryons. This offset indicates
that the gravitational potential is dominated by a collisionless component that
remains bound to the galaxies during the merger, naturally interpreted as a cold gas
of dark matter particles. Such a configuration is difficult to reconcile with modified-gravity
scenarios in which the gravitational field is expected to closely follow the
baryonic mass distribution. 

Beyond supporting the particle dark matter interpretation, the Bullet Cluster
also constrains possible dark matter self-interactions. If dark matter exhibited
frequent self-scattering, its distribution would experience drag during the
collision, leading to observable offsets or broadening relative to the galaxy
population. The absence of such effects implies that dark matter must be
sufficiently weakly self-interacting on cluster scales, placing upper limits on
the self-interaction cross section per unit mass,
$\sigma_{2\to2}/m_\text{DM} \lesssim \mathcal{O}(1)\,\mathrm{cm^2\,g^{-1}}$~\cite{Randall:2008ppe}, consistent with
dark matter behaving approximately collisionlessly in high-velocity cluster
mergers. Beyond establishing the approximately collisionless behavior of dark matter,
merging clusters also provide sensitive probes of dark matter self-interactions.
Even finite-range, coherent forces would induce observable distortions or offsets
in cluster dynamics, leading to stringent constraints across a wide range of
interaction lengths~\cite{Bogorad:2023wzn}. Notably, 
the elastic cross section for DM, $\sigma_{2\to2}/m_\text{DM}$, is of the order of roughly the (baryonic) nuclear scale, 
so there is no contradiction between allowing the appreciable dark-sector self-interactions in the upcoming chapters 
and satisfying current astrophysical bounds. For a review on self interacting dark matter see~\cite{Tulin:2017ara}.
 
\begin{figure}[t!]
	\centering
	\includegraphics[width=0.85\textwidth]{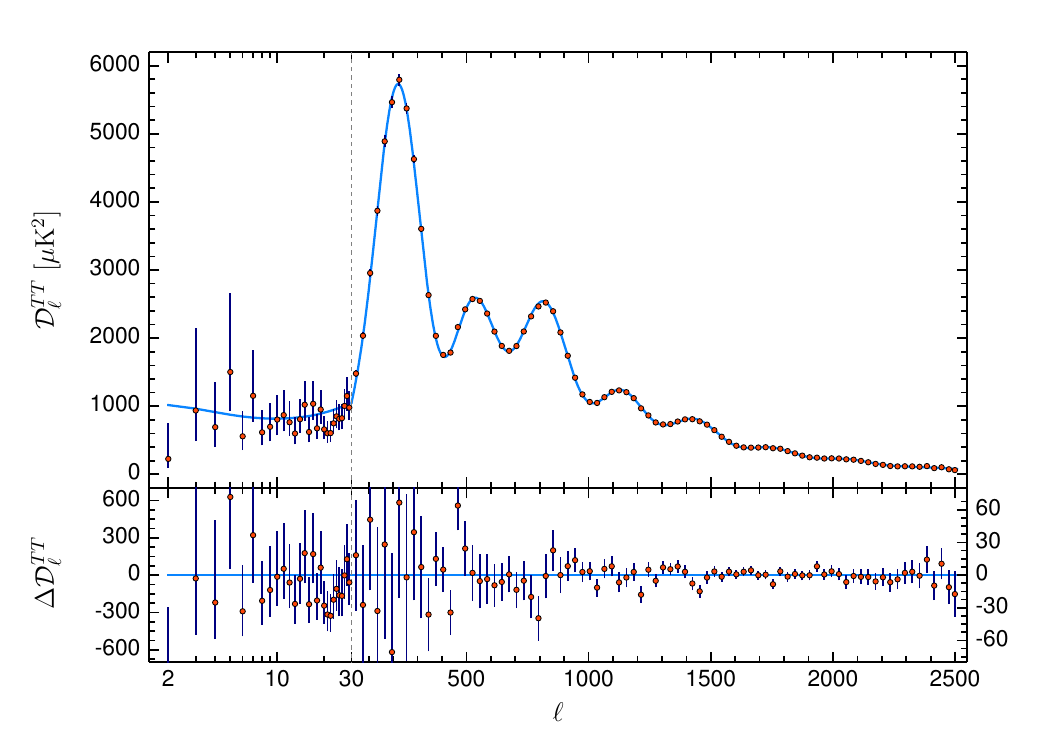}
	\caption{Temperature anisotropy power spectrum of the cosmic microwave background
		measured by the Planck satellite. The data points show the angular power spectrum
		of temperature fluctuations over a wide range of multipoles $\ell$, corresponding
		to angular scales on the sky from large to small. The solid curve represents the
		best-fit theoretical prediction within the base-$\Lambda$CDM cosmological model.
		The lower panel displays the residuals between the measured spectrum and the
		best-fit model. Figure from the Planck 2018 results~\cite{Planck:2018vyg}.
	}
	\label{fig:planck}
\end{figure}

At the largest observable scales, the existence of dark matter is firmly
established by precision cosmological measurements, most notably observations of
the cosmic microwave background~\cite{Planck:2018vyg,2013ApJS..208...19H,DESI:2024mwx}. The CMB originates from the epoch of
recombination, when the Universe was approximately $380{,}000$ years old and had
cooled sufficiently for electrons and protons to combine into neutral hydrogen,
allowing photons to decouple from the primordial plasma and propagate freely.
The resulting temperature anisotropies encode detailed information about the
composition and dynamics of the Universe at that time. In particular, the
relative heights and positions of the acoustic peaks in the CMB angular power
spectrum are sensitive to the total matter density, the baryon fraction, and the
presence of a non-baryonic dark matter component. Dark matter contributes to the
gravitational potential wells in which the baryon--photon fluid oscillates prior
to recombination, shaping the observed anisotropy pattern.
These features are illustrated in Fig.~\ref{fig:planck}, which shows the
temperature anisotropy power spectrum measured by the Planck Collaboration.
 
Despite the challenges to the $\Lambda$CDM model, 
its best fit predictions remain to large extent in agreement with the observed spectrum, which
demonstrates the consistency of a cosmological model containing cold dark matter
over a wide range of angular scales~\cite{Perivolaropoulos:2021jda}. 
The same early gravitational potential perturbations subsequently seed the
growth of large-scale structure, as dark matter density fluctuations begin to
collapse well before baryons decouple from radiation. The observed distribution
of galaxies and clusters at late times is therefore closely connected to the dark
matter imprint in the CMB, providing a consistent and self-contained picture
across vastly different cosmic epochs. At the same time, these observations posit constraints on the
allowed characteristics of dark matter~\cite{Cang:2020exa,Xu:2024vdn,Kawasaki:2021etm,Nygaard:2020sow,Slatyer:2016qyl}.

\section{Cosmological framework}\label{sec:cosmo_framework}

The best description of the large-scale evolution of the Universe is provided
by the $\Lambda$CDM model, which assumes General Relativity as the theory of
gravity and a matter content composed of radiation, baryonic matter, cold dark
matter (CDM), and a cosmological constant $\Lambda$. The
Universe is observed to be homogeneous and isotropic on sufficiently large
scales, an assumption that is well supported by measurements of the cosmic
microwave background and large-scale structure. These symmetries allow the
spacetime geometry to be described by the Friedmann--Lemaître--Robertson--Walker
(FLRW) metric. In spherical coordinates and natural units, it takes the form
\begin{equation}
	\label{FLRW}
	(g_{\mu\nu}) =
	\begin{pmatrix}
		1 & 0 & 0 & 0 \\
		0 & -\dfrac{a^2(t)}{1-k r^2} & 0 & 0 \\
		0 & 0 & -a^2(t) r^2 & 0 \\
		0 & 0 & 0 & -a^2(t) r^2 \sin^2\theta
	\end{pmatrix},
\end{equation}
where $a(t)$ is the dimensionless \textit{scale factor}. The curvature parameter
$k$ characterizes the spatial geometry and may take the values $k=+1$ (closed),
$k=0$ (flat), or $k=-1$ (open).\footnote{Observations of the cosmic microwave
	background favor a spatially flat Universe,
	$k \simeq 0$~\cite{2013ApJS..208...19H}.}

The dynamics of the scale factor are determined by Einstein's field equations,
\begin{equation}
	R_{\mu\nu} - \frac{1}{2} g_{\mu\nu} R = 8\pi G T_{\mu\nu},
\end{equation}
where $R_{\mu\nu}$ is the Ricci tensor, $R \equiv g^{\mu\nu}R_{\mu\nu}$ is the
Ricci scalar, $T_{\mu\nu}$ is the energy-momentum tensor of the cosmological
fluid and $G$ is Newton’s gravitational constant, which can be expressed in terms of the reduced Planck mass as
\begin{equation}
G \equiv \frac{1}{8\pi M_\text{Pl}^2}\,,
\end{equation}
with $M_\text{Pl} \simeq 2.4 \times 10^{18}\,\text{GeV}$.

In the following, we focus on post-inflationary epochs and therefore neglect the
contribution of the cosmological constant and the associated late-time
accelerated expansion, which are irrelevant for the early-Universe processes
considered in this thesis. The matter content of the Universe is modeled as a perfect fluid with energy
density $\rho$ and pressure $P$, whose thermodynamic properties are encoded in an
equation-of-state parameter $\omega$,
\begin{equation}
P = \omega \rho,
\end{equation}
where $\omega$ takes characteristic values for different components. In
particular,
\begin{equation}
\omega =
\begin{cases}
\dfrac{1}{3}, & \text{radiation}, \\
0, & \text{matter}, \\
-1, & \text{vacuum energy}.
\end{cases}
\end{equation}
The corresponding energy--momentum tensor is
\[
(T^\mu{}_\nu) = \mathrm{diag}(\rho,-P,-P,-P)
= \rho\,\mathrm{diag}(1,-\omega,-\omega,-\omega).
\]
The $(0,0)$ component of Einstein’s equations yields
\begin{equation}
\label{friedman1}
\frac{\dot a^2}{a^2} + \frac{k}{a^2} = \frac{8\pi G}{3}\rho,
\end{equation}
which motivates the definition of the \textit{Hubble parameter}
\begin{equation}
	\label{Hubble_definition}
	H(t) \equiv \frac{\dot a(t)}{a(t)} = \sqrt{\frac{8\pi G}{3}\rho}\,,
\end{equation}
where \(\rho\) denotes the total energy density of the Universe. It includes the SM radiation energy density, and may also receive contributions from other components, such as inflaton field(s) and dark sectors. Eq.~\eqref{friedman1} can then be written as
\begin{equation}
\label{friedman2}
H^2 + \frac{k}{a^2} = \frac{8\pi G}{3}\rho,
\end{equation}
commonly referred to as the \textbf{Friedmann equation}. Energy--momentum conservation, $\nabla^\mu T_{\mu\nu}=0$, leads to the
\textbf{continuity equation}
\begin{equation}
\label{continuity}
\dot\rho + 3H(\rho + P)
= \dot\rho + 3H(1+\omega)\rho = 0.
\end{equation}
Integrating~\eqref{continuity} gives
\begin{equation}
\label{sol_rho}
\rho(t) = \rho_0\left(\frac{a}{a_0}\right)^{-3(1+\omega)},
\end{equation}
which implies
\[
\rho \propto
\begin{cases}
a^{-4}, & \text{radiation},\\
a^{-3}, & \text{matter},\\
\text{constant}, & \text{dark energy}.
\end{cases}
\]

A dark sector follows the same scaling behavior as any other component
of the cosmic fluid, with its contribution to the total energy density determined by
its equation of state. Consider $m_\text{ds}$ to be the heaviest particles within the dark sector. When the dark sector temperature satisfies
$T_{\rm ds} \gg m_{\rm ds}$,\footnote{In general, the dark sector temperature $T_{\rm ds}$ need not coincide with the visible-sector temperature $T$, and, in fact, the dark sector may be completely out of thermal equilibrium if self interactions are negligible or absent.
} the dark sector particles are relativistic and their energy density
scales as $\rho_{\rm ds}\propto a^{-4}$, corresponding to an effective equation-of-state
parameter $\omega_{\rm ds}\simeq 1/3$. In this regime, the dark sector behaves as an additional radiation component and contributes to the expansion rate in the same way as the Standard Model plasma.

As the Universe expands and the dark sector temperature drops below the particle mass, $T_{\rm ds}\ll m_{\rm ds}$, the particles become non-relativistic and their energy density transitions to $\rho_{\rm ds}\propto a^{-3}$, characteristic of pressureless matter ($\omega_{\rm ds}\simeq 0$). The dark sector then behaves as dust. Since non-relativistic energy density redshifts more slowly than radiation, even a subdominant dark sector component at early times may become cosmologically important at later times.

Let us now stress again that the Hubble expansion rate is determined by the \emph{total} energy
density of the Universe, independently of the microscopic nature of its
constituents. In a spatially flat Universe, the Friedmann equation is
\begin{equation}\label{eq:Hubble}
	H^2 = \frac{8\pi G}{3}\,\rho
	= \frac{8\pi G}{3}\left(\rho_{\rm SM} + \rho_{\rm ds} + \cdots \right),
\end{equation}
where the dots denote any additional contribution beyond the Standard Model (SM)
and the dark sector. Then, any component---relativistic or
non-relativistic---affects the expansion history whenever it contributes
non-negligibly to $\rho$.

At key epochs in the thermal history of the Universe, such as Big Bang
Nucleosynthesis (BBN) and recombination, the expansion rate is tightly
constrained by observations. Additional energy density present at these times
modifies $H(t)$ and can therefore be bounded. If the extra component behaves as
radiation, its effect is commonly parametrized in terms of an effective number
of relativistic degrees of freedom, $\Delta N_{\rm eff}$. More general expansion
histories, for instance those involving matter-like components arising from
coherent scalar-field condensates or periods of early matter domination, are
constrained through their impact on the expansion rate and, in some cases,
through entropy injection into the visible sector. As a result, any dark sector
must either be sufficiently dilute or sufficiently cold relative to the SM plasma,
to remain consistent with observational bounds at the
onset of these epochs.

\subsection{Thermodynamic quantities}\label{subsec:thermo_quant}

The kinetic evolution of particle species in the early Universe can be described
within the framework of relativistic phase-space dynamics. Rather than following
individual particle trajectories, the state of a system of particles is encoded in a
phase-space distribution function $f(x^\mu,p^\nu)$, which specifies the density
of particles at spacetime position $x^\mu$ with four-momentum $p^\nu$. Throughout
this discussion we assume homogeneity and isotropy of the cosmological background,
so that the distribution function depends only on cosmic time and on the magnitude
of the physical momentum, $f(x^\mu,p^\nu) \to f(p,t)$, and here often denoted simply as $f$.

All thermodynamic properties of a particle species can be derived from its
phase-space distribution function. In particular,
the number density, energy density, and pressure of a species $i$ are given by~\cite{Kolb:1990vq}
\begin{equation}
	\label{eq:thermoquantities}
	n_i = g \int_p f_i \,, \qquad
	\rho_i = g \int_p E\, f_i \,, \qquad
	P_i = g \int_p \frac{p^2}{3E}\, f_i\,,
\end{equation}
where $\int_p \equiv \int d^3p/(2\pi)^3 = (2\pi^2)^{-1}\int dp\,p^2$, $g$ denotes the
number of internal degrees of freedom, and
$E = \sqrt{p^2 + m^2}$ is the single-particle energy.

Let us assume that the species $i$ consists of a single particle type with mass
$m_i$ and that its distribution is well approximated by a Maxwell-Boltzmann form,
$f_i = z_i\, e^{-E/T_i}$. Here $T_i$ denotes the effective temperature of the species
and $z_i = e^{\mu_i/T_i}$ is the fugacity. Under these assumptions, the integrals in
Eq.~\eqref{eq:thermoquantities} can be evaluated analytically, yielding
\begin{equation}
	\label{eq:MBthermo}
	\begin{split}
		n_i &=
		\frac{g}{2\pi^2}\, z_i\, m_i^2 T_i\,
		K_2\!\left(\frac{m_i}{T_i}\right) , \\[4pt]
		\rho_i &=
		\frac{g}{2\pi^2}\, z_i\, m_i^3 T_i
		\left[
		K_1\!\left(\frac{m_i}{T_i}\right)
		+ 3\,\frac{T_i}{m_i}\,
		K_2\!\left(\frac{m_i}{T_i}\right)
		\right] , \\[4pt]
		P_i &=
		\frac{g}{2\pi^2}\, z_i\, m_i^2 T_i^2\,
		K_2\!\left(\frac{m_i}{T_i}\right) ,
	\end{split}
\end{equation}
where $K_n$ denote modified Bessel functions of the second kind.

In practice, it is often convenient to trade the fugacity for the
number density $n_i$ or for the comoving abundance
\begin{equation}
	Y_i \equiv \frac{n_i}{s} ,
\end{equation}
where \(s\) is the entropy density of the Standard Model plasma. The comoving abundance is particularly useful because it remains constant after chemical decoupling. In the case where the species \(i\) constitutes dark matter, it is directly related to the present-day relic density,
\begin{equation}
	\label{eq:OmegaY}
	\Omega_c h^2
	=
	\frac{m_{\rm DM}\, s_0\, Y_i^\infty}{\rho_c/h^2} ,
\end{equation}
where \(Y_i^\infty \equiv Y_i(t\to \infty)\), \(m_{\rm DM}\) is the dark matter mass, \(s_0 \simeq 2891\,{\rm cm^{-3}}\) is the current entropy density, and \(\rho_c/h^2 \simeq 1.05\times10^{-5}\,{\rm GeV\,cm^{-3}}\) is the present critical density. Equivalently, this relation can be expressed numerically as
\begin{equation}
	\Omega_c h^2 \simeq
	2.74 \times 10^8
	\left( \frac{m_{\rm DM}}{\rm GeV} \right)
	Y_\infty ,
\end{equation}
which provides the direct link between the distribution function of the DM (and therefore its energy and number density) and
the observed dark matter abundance. Importantly, the relic abundance constitutes
the only precisely measured quantitative property of the dark sector to date.
As such, it provides a decisive constraint on BSM physics, allowing the parameter space of dark matter models to be
systematically tested and restricted through their predicted cosmological
history.

We now briefly review the thermodynamics of the early Universe and its connection
to the Hubble expansion rate. During the
radiation-dominated era, and at temperatures well above the masses of the
relevant SM particles, the energy density of the Universe is
dominated by relativistic degrees of freedom. Particle species with masses
$m_i \ll T$ behave as radiation, while heavier species with $m_i \gtrsim T$ are
Boltzmann suppressed and contribute negligibly to the total energy density.\footnote{\textit{Boltzmann suppression} denotes the non-relativistic limit of the distribution in equilibrium, which is suppressed by the factor $e^{-m_i/T}$.}
Around the epoch of Big Bang Nucleosynthesis (BBN), corresponding to temperatures
$T \sim \mathcal{O}(\text{MeV})$~\cite{Fields:2014uja, Fields:2019pfx}, the relativistic plasma consists primarily of
photons, neutrinos, and, prior to their annihilation, electron--positron pairs.
Assuming thermal and chemical equilibrium of the Standard Model plasma, the
energy density of relativistic bosons and fermions follows from
Eq.~\eqref{eq:thermoquantities} using Bose--Einstein and Fermi--Dirac statistics.

\begin{equation}
	\rho_{\rm rad} =
	\begin{cases}
		\dfrac{\pi^2}{30}\, g\, T^4 \,, & \text{bosons}, \\[6pt]
		\dfrac{7}{8}\,\dfrac{\pi^2}{30}\, g\, T^4 \,, & \text{fermions}.
	\end{cases}
\end{equation}
Absorbing the fermionic factor into $g_*$, the total radiation energy density can be written as
\begin{equation}\label{eq:rho_rad}
	\rho_{\rm rad} = \frac{\pi^2}{30}\, g_*\, T^4 ,
\end{equation}
where $g_*$ denotes the effective number of relativistic degrees of freedom. Inserting this expression into the Friedmann equation yields the characteristic
scaling relations of a radiation-dominated Universe,
\begin{equation}
	H \propto T^2 \,, \qquad
	t \propto T^{-2} \,, \qquad
	a \propto T^{-1} \,,
\end{equation}
which can be written explicitly as
\begin{equation}
	\label{eq:HubbleT}
	H = 1.66\, g_*^{1/2}\, \frac{T^2}{M_{\rm Pl}} \,,
	\qquad
	t = 0.301\, g_*^{-1/2}\, \frac{M_{\rm Pl}}{T^2} \,.
\end{equation}
%

\subsection{Inflation and the $\Lambda$CDM model}
\label{sec:reheating}

The radiation-dominated Universe derived above implicitly assume that
the Universe has already reached a state of thermal equilibrium dominated by
relativistic Standard Model degrees of freedom. However, our direct observational
access to the thermal history of the Universe begins only at the onset of BBN. 
At earlier times, the expansion history is less
constrained, allowing for the possibility that cosmic expansion was driven by a matter content other than radiation.

An example is provided by cosmic inflation, which was originally
introduced to address conceptual problems of the standard Big Bang cosmology,
such as the observational absence of magnetic monopoles~\cite{Lazarides:1984bh}, the flatness and horizon problems~\cite{Guth:1980zm}, and to provide a mechanism for the generation of primordial density
perturbations~\cite{Starobinsky:1980te}. In typical inflationary scenarios, the
accelerated expansion is driven by the nearly constant potential energy of a
scalar field, the inflaton~\cite{Linde:1981mu}, which ``rolls'' slowly along the flat potential. Slow-roll inflation 
can be approached from a bottom-up perspective (adopted in this work), and also
top-down perspectives like in string theory~\cite{Cicoli:2023opf}. 
During slow-roll inflation, the dynamics are dominated by the potential energy
of a scalar field $\phi$ (the inflaton), whose potential is sufficiently flat that
the field evolves slowly compared to the Hubble expansion rate. As a result, the
inflaton energy density is approximately constant and given by
$\rho_\phi \simeq V(\phi)$, while its kinetic contribution is subdominant.
The Friedmann equation then implies an approximately constant Hubble rate,
\begin{equation}
	H^2 \simeq \frac{8\pi G}{3}\, V(\phi) \simeq \text{const.},
\end{equation}
leading to an exponential expansion of the scale factor,
\begin{equation}
	a(t) \propto e^{H t}.
\end{equation}
Note that the potential of the inflaton field need not be flat. Among different potential shapes, ones leading to a first order phase transition with bubble nucleation and collision has been explored in~\cite{Watkins:1991zt}.

\begin{figure}[t!]
	\centering
	\includegraphics[width=0.8\textwidth]{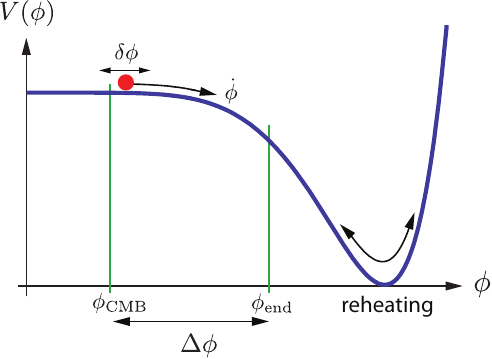}
	\caption{Schematic illustration of an inflaton field $\phi$ rolling along a flat
		potential, corresponding to a period of exponential inflation. The figure shows the end of inflation
		($\phi_\text{end}$) and the period of time when the cosmological fluctuations are created ($\phi_\text{CMB}$). 
		Once inflation
		ends, the inflaton undergoes coherent oscillations around the minimum of the potential, 
		in the process decaying into SM particles and reheating the Universe. Figure from~\cite{Baumann:2009ds}.
	}
	\label{fig:rot_gal}
\end{figure}

Such accelerated expansion dynamically drives the Universe towards spatial
flatness setting the initial conditions required for the subsequent
cosmological evolution. The late time dynamics of this nearly flat Universe are
well described by the standard cosmological model, commonly referred to as
$\Lambda$CDM, which is based on a homogeneous and isotropic FLRW spacetime whose
energy budget is dominated at late times by a cosmological constant $\Lambda$
and cold dark matter (CDM), in addition to baryonic matter and radiation.

In its minimal form, $\Lambda$CDM is specified by a small set of parameters,
including the present-day Hubble rate, the physical densities of baryons and dark
matter, the properties of the primordial scalar perturbations generated
during inflation (acoustic scale and scalar spectral index), and the optical depth 
to reionization~\cite{Weinberg:1972kfs,Kolb:1990vq,Dodelson:2003ft,Weinberg:2008zzc}.
The $\Lambda$CDM model arises as the simplest cosmological framework obtained by combining General Relativity with the assumptions of large-scale homogeneity and isotropy, together with a matter content consisting of radiation, baryons, cold dark matter, and a cosmological constant. Its success lies in the fact that this minimal setup provides a consistent description of a wide range of observations, including cosmic microwave background anisotropies, baryon acoustic oscillations, large-scale structure, and primordial light-element abundances, using only a small number of parameters.
In fact, various alternative cosmological scenarios have been proposed, involving, for
example, among others 
dynamical dark energy components~\cite{Rezaei:2023xkj} or departures from large-scale homogeneity or isotropy~\cite{Saadeh:2016sak}.
However, none of these frameworks currently matches the level of simultaneous
agreement with cosmological observations achieved by $\Lambda$CDM despite it having its own tension
with observations~\cite{Perivolaropoulos:2021jda}.


After inflation ends, the inflaton does not immediately decay into radiation. Instead, once the slow-roll conditions break down, the inflaton typically undergoes coherent oscillations about the minimum of its potential~\cite{Albrecht:1982mp}, behaving approximately as a damped classical condensate while gradually transferring its energy to Standard Model states through decays or annihilations. This process is denoted as \textit{reheating} and in this period the thermal radiation bath is gradually
populated and only then it become meaningful to
characterize the cosmic plasma by a well-defined temperature. 
For common choices of the potential, such as a
quadratic minimum, the oscillating inflaton condensate behaves on average as
pressureless matter, with an energy density scaling as
$\rho_\phi \propto a^{-3}$. During this reheating phase, the expansion history
therefore differs from that of a radiation-dominated Universe. The transition to
radiation domination is commonly characterized by the reheating temperature
$T_{\rm rh}$. Requiring successful BBN imposes a robust lower bound,
$T_{\rm rh} \gtrsim \mathcal{O}(\text{MeV})$, ensuring that the Universe is
radiation dominated by the onset of nucleosynthesis~\cite{Hannestad:2004px,
Hasegawa:2019jsa}. For temperatures above $T_{\rm rh}$, however, the expansion
rate need not follow the standard radiation-dominated scaling. 


Such nonstandard
expansion histories can significantly affect the production and evolution of
dark matter, and will be discussed in~\Cref{ch:4} in the context
of freeze-in dark matter production during reheating.

\section{The Boltzmann equation}
\label{sec:Boltzmanneq}

The evolution of particle species in an expanding background is commonly
described by the \textbf{Boltzmann equation}, which governs the time evolution of
phase-space distribution functions in the presence of interactions and cosmic
expansion. From a more fundamental perspective, this kinetic description can be
derived as an effective limit of nonequilibrium quantum field theory formulated
on the Schwinger--Keldysh closed time path, although the Boltzmann approach
provides a sufficiently accurate and practical framework for the purposes of
this thesis~\cite{Kolb:1990vq}.

In curved spacetime, the evolution of the distribution function, $f$, along particle trajectories is governed
by Liouville's theorem, which states that in the absence of interactions the
distribution function is conserved along geodesics. Including interactions,
this statement generalizes to the relativistic Boltzmann equation,
\begin{equation}
\frac{d f}{d \lambda} = C[f] \, ,
\end{equation}
where $\lambda$ denotes an affine parameter along the particle worldline and
$C[f]$ is the collision operator accounting for microscopic interactions, i.e., it incorporates the matrix element provided by QFT of the reaction of interest.

To make this equation explicit, we express the total derivative in terms of
spacetime and momentum derivatives,
\begin{equation}
\frac{d f}{d \lambda}
=
\frac{d x^\mu}{d \lambda}\,\frac{\partial f}{\partial x^\mu}
+
\frac{d p^\mu}{d \lambda}\,\frac{\partial f}{\partial p^\mu} \, .
\end{equation}
The first term describes transport in spacetime, while the second accounts for
the evolution of the four-momentum along the trajectory.

In an FLRW spacetime, the particle motion between collisions follows geodesics,
and the four-momentum evolves according to
\begin{equation}
\frac{d p^\mu}{d \lambda}
+
\Gamma^\mu_{\alpha\beta}\,
p^\alpha p^\beta
= 0 \, ,
\end{equation}
where $\Gamma^\mu_{\alpha\beta}$ are the Christoffel symbols associated with the
FLRW metric. For a homogeneous and isotropic background, the only non-vanishing
Christoffel symbols relevant for the momentum evolution are those involving the
scale factor $a(t)$. Evaluating the geodesic equation then yields the familiar
result that the physical three-momentum redshifts with the expansion,
\begin{equation}
\frac{d p}{d t} = - H\, p \, ,
\end{equation}
where $H$ is the Hubble rate defined in Eq.~\eqref{Hubble_definition}.

Using this relation, and identifying the affine parameter with cosmic time for
on-shell particles, the total derivative of the distribution function can be
written as
\begin{equation}
\frac{d f}{d t}
=
\frac{\partial f}{\partial t}
-
H\, p\,\frac{\partial f}{\partial p} \, .
\end{equation}
Including interactions through the collision operator, one arrives at the
cosmological Boltzmann equation in its unintegrated form~\cite{Kremer:2012dd}
,
\begin{equation}
\label{eq:fBE}
\left(
\partial_t
-
H\, p\,\partial_p
\right) f(p,t)
=
C[f] \, ,
\end{equation}
which we denote as \textit{full} Boltzmann equation (fBE). In this form, the fBE equation can be used to describe various aspects of the early Universe cosmology, 
like the freeze-out and freeze-in of dark matter, the formation of light elements and the decoupling
of neutrinos. Such a description is possible by implementing the corresponding collision operator $C[f]$
for the respective scenario of interest. 

The collision term $C[f]$ encodes all microphysical processes that change the
occupation of phase-space cells, including elastic scatterings, inelastic
reactions, and number-changing processes. For a given reaction channel
$r : (i + j + \ldots \leftrightarrow k + \ldots)$, the contribution to the collision
operator acting on the distribution of species $i$ can be written as
\begin{equation}
\label{eq:intCO_compact}
C_r[f_i] =
- \frac{1}{2E_ig_i}
\int
\left(
\prod_{\alpha = j,k,\ldots} d\Pi_\alpha
\right)
\,
\bigl|
\tilde{\mathcal M}_r
\bigr|^2
\,
\Bigl[
f_i f_j \ldots
\bigl(1 \pm f_k\bigr)\ldots
-
f_k \ldots
\bigl(1 \pm f_i\bigr)
\bigl(1 \pm f_j\bigr)\ldots
\Bigr],
\end{equation}
where $g_\alpha$ corresponds to the degrees of freedom of the species $\alpha$ and we have introduced the shorthand
\begin{equation}\label{eq:Msq}
\bigl|
\tilde{\mathcal M}_r
\bigr|^2
\equiv
(2\pi)^4
\delta^{(4)}
\!\left(
\sum_{f} p_f - \sum_{i} p_i
\right)
\bigl|
\overline{\mathcal M}_r
\bigr|^2 .
\end{equation}
Here $\lvert \overline{\mathcal M}_r \rvert^2$ denotes the squared matrix element,
summed (averaged) over final (initial) internal degrees of freedom, and
$d\Pi_\alpha = S_\alpha d^3p_\alpha / \bigl[(2\pi)^3 2E_\alpha\bigr]$ 
is the
Lorentz-invariant phase-space measure including all appropriate symmetry factors, $S_\alpha$. The $(\pm)$ signs account for Bose
enhancement and Pauli blocking, respectively.

In the remainder of this thesis we focus on scenarios in which dark matter is
produced out of equilibrium and remains a highly dilute component of the cosmic
plasma. We therefore approximate all distribution functions by the MB
form and neglect quantum statistical factors,
$(1 \pm f) \simeq 1$. The impact of quantum statistics on freeze-in production has
been studied in the literature and leads to modest numerical
corrections to the dark matter abundance~\cite{Heeba:2018wtf}. However, such effects do
not qualitatively modify the parametric dependence or the overall dynamics of
out-of-equilibrium dark matter production. Including quantum statistical
corrections would considerably increase the technical complexity of the collision
integrals without qualitatively affecting the results relevant for the present
analysis.

\subsection{Coupled Boltzmann equations}\label{subsec:cBE}

Under the assumption of a fixed functional form for the distribution function, the
integro--partial differential Boltzmann Eq.~\eqref{eq:fBE} can be
significantly simplified. Rather than solving for the full phase-space
distribution $f_i(p,t)$, one may instead track the time evolution of a reduced
set of parameters, such as the effective temperature $T_i$ and the fugacity
$z_i$. In other words, the fBE is re-casted as a set ordinary differential equations for both $z_i$ and $T_i$. In this work we denote this set as \textit{coupled} Boltzmann equations (cBE).

To obtain an equation for the number density of a species $i$, we integrate Eq.~\eqref{eq:fBE} over phase-space momenta $g\int_p$:
\begin{equation}\label{eq:nBE}
	g\int_p(\partial_t f_i - H\,p\partial_p f_i) = g\int_p C[f_i]\equiv C_0 \implies \dot n_i + 3Hn_i =C_0\,,
\end{equation}
which we will denote as \textit{number density} Boltzmann equation (short nBE). Note that this equation can be expressed in terms of the comoving number density, $N_i = a^3 n_i$,
\begin{equation}
	\dot n_i = \frac{d}{dt}(a^{-3} N_i) = a^{-3}\dot N_i - 3a^{-4}\,\dot a\,N_i=a^{-3}\dot N_i - 3a^{-3} H N_i\,,
\end{equation}
therefore 
\begin{equation}
	a^{-3}\dot N_i=\dot n_i + 3 H n_i \implies \dot N_i = a^3 C_0\,.
\end{equation}
It is also convenient to cast the nBE in terms of the scale factor $a$ using $\frac{dN_i}{dt}= aH\frac{dN_i}{da}$, which leads to
\begin{equation}\label{eq:N}
	\frac{dN_i}{da} = \frac{a^2}{H}C_0\,.
\end{equation}
A different useful parametrization that often appears in the literature is $Y_i =n_i/s$ already introduced in the last subsection. This parametrization of the comoving number density is useful when the SM plasma is completely isolated, such that the entropy is conserved along the evolution, $a^3s=S_0=\text{constant}$,\footnote{Note that this assumption may breakdown for temperatures above BBN if there is injection of entropy into the SM as, e.g., during reheating.} then
\begin{equation}
	\dot n_i = s\dot Y_i + Y_i \dot s = s\dot Y_i + Y_i (-3 H s)\implies \dot n_i + 3Hn_i = s\dot Y_i\,.
\end{equation}
Additionally, since $a\propto 1/T$ when the universe is dominated by radiation, the quantity $x=m_i/T$ serves a good proxy of time. To trade cosmic time for the dimensionless inverse-temperature variable $x$ we use the chain rule
\begin{equation}
	\frac{dY_i}{dx}=\frac{1}{\dot x}\,\dot Y_i\,.
\end{equation}
Since $x=m_i/T$, we have
\begin{equation}
	\dot x = -\frac{m_i}{T^2}\dot T = -x\,\frac{\dot T}{T}\,.
\end{equation}
Assuming that the visible sector evolves adiabatically, the entropy density
$s(T)=(2\pi^2/45)\,g_{*s}(T)\,T^3$ satisfies $\dot s+3Hs=0$,\footnote{Here \(g_{*s}(T)\) accounts for the effective number of relativistic degrees of freedom contributing to the entropy density of the Standard Model.} which implies
\begin{equation}
	\frac{\dot T}{T}
	= -H\left(1+\frac{1}{3}\frac{d\ln g_{*s}}{d\ln T}\right)^{-1}
	\equiv -\frac{H}{\tilde g(T)}\,,
	\qquad
	\tilde g(T) \equiv 1+\frac{1}{3}\frac{d\ln g_{*s}}{d\ln T}\,.
\end{equation}
Therefore
\begin{equation}
	\dot x = \frac{xH}{\tilde g(T)}\,,
\end{equation}
and the yield equation becomes
\begin{equation}
	\label{eq:Yeq_x}
	\frac{dY_i}{dx} = \frac{\tilde g(T)}{x\,H\,s}\,C_0\,.
\end{equation}
A precise determination of the abundances and energy densities of particle
species in the early Universe is essential for confronting cosmological models
with observations. In particular, the public code \texttt{AlterBBN}~\cite{Arbey:2011nf}
provides numerical routines to compute key Big Bang Nucleosynthesis observables,
such as the baryon-to-photon ratio and the effective number of relativistic
degrees of freedom.

In the context of dark matter relic abundance calculations, robust and numerically
stable routines for solving Eq.~\eqref{eq:Yeq_x} are available in several
well-established public codes. These include \texttt{MicrOMEGAs}~\cite{Alguero:2023zol},
\texttt{MadDM}~\cite{Ambrogi:2018jqj}, \texttt{SuperISO}~\cite{Arbey:2018msw},
\texttt{DarkSUSY}~\cite{Bringmann:2018lay}, and \texttt{DRAKE}~\cite{Binder:2021bmg}
which are widely used for computing thermal and non-thermal dark matter
abundances in a variety of particle physics models. In addition, the code
\texttt{DarkKROME}~\cite{Ryan:2021tgw} provides dedicated tools for computing the
abundance of compact objects.

For freeze-out calculations, some of the codes above reduce the complexity of the full Boltzmann equation not only by assuming a thermal momentum distribution, but also by enforcing kinetic equilibrium with the Standard Model plasma throughout
the entire evolution, including the period following chemical decoupling.
However, the number-density equation alone is insufficient whenever the momentum distribution departs from a single-temperature form tied to the SM bath like in the absence of strong portal interactions. In that case one must evolve (at least) an additional moment of the distribution, e.g. the energy density, which can be mapped to an effective temperature within the MB ansatz~\cite{Binder:2017rgn}. 

To derive the temperature equation, we first note that the dark-sector temperature is defined implicitly by the equilibrium distribution. Its time evolution may then be inferred from the equation of motion for the energy density. The corresponding continuity equation follows from integrating the fBE with the measure \(g\int_p E\),
\begin{equation}
	g\int_p\,E(\partial_t f_i-H\,p\,\partial_p\,f_i) = g\int_p E C[f_i]\equiv C_E[f_i]\,,
\end{equation}
and if $\dot m_i=0$, then 
\begin{equation}
	\dot \rho_i + 3H(\rho_i+P_i) = C_E[f_i]\,,
\end{equation}
or equivalently, 
\begin{equation}
	\frac{d\rho_i}{da} = \frac{C_E}{a\,H}-\frac{3}{a}(\rho_i + P_i)
\end{equation}
and using
\begin{equation}
	\frac{d\rho_i}{da} = \frac{\partial \rho_i}{\partial T_i}\frac{dT_i}{da} + \frac{\partial \rho_i}{\partial N_i}\frac{dN_i}{da} + \frac{\partial \rho_i}{\partial a}\,, 
\end{equation}
the equation for the temperature is
\begin{equation}\label{eq:dT_da}
	\frac{dT_i}{da} = \frac{ C_E/(aH) - \frac{3}{a} \left(\rho_i + P_i \right) - \frac{\partial\rho_i}{\partial N_i} \frac{dN_i}{da} - \frac{\partial\rho_i}{\partial a} }{ \frac{\partial \rho_i}{\partial T_i} }\,.
\end{equation}

We can close the last equation using $\rho_i = \rho_i(T_i, N_i)$ in Eq.~\eqref{eq:MBthermo}, identifying the fugacity as $z = n/n^\text{eq} = N/N^\text{eq} = Y/Y^\text{eq}$, where $n^\text{eq} = n(z=1)$ (c.f. Eq.~\eqref{eq:MBthermo}). Hence, Eqs.~\eqref{eq:N} and~\eqref{eq:dT_da} form a set of cBE. It is useful to emphasize that the equation for the temperature can be also cast in terms of $x = m_i/T$,
\begin{equation}\label{eq:dTi_dx}
	\frac{dT_i}{dx}
	=
	\frac{
		\frac{\tilde g}{xH}C_E
		-\frac{3\tilde g}{x}(\rho_i+P_i)
		-\frac{\partial\rho_i}{\partial Y_i}\frac{dY_i}{dx}
	}{
		\frac{\partial\rho_i}{\partial T_i}
	}\,,
\end{equation}
with
\begin{equation}
	dY_i/dx = \tilde g\,C_0/(xHs)
\end{equation}
form a closed set of coupled Boltzmann
equations within the Maxwell-Boltzmann ansatz.

\subsection{Temperature from the pressure moment}
\label{subsec:cBE_pressure}

We have derived an evolution equation for the dark sector's
temperature by taking the energy-density moment. An equivalent and often convenient route is to work with the pressure
moment. The starting point is the definition of the (isotropic) pressure of a
species $i$,
\begin{equation}
	P_i \;\equiv\; g \int_p \frac{p^2}{3E}\, f_i\,,
	\qquad
	E \equiv \sqrt{p^2+m_i^2}\,,
\end{equation}
together with the number density $n_i = g \int_p f_i$. If the distribution is
well described by a Maxwell-Boltzmann form with some temperature $T_i$ and
arbitrary chemical potential (or fugacity), then the ideal-gas relation
\begin{equation}
	P_i = n_i\,T_i
\end{equation}
holds, and one may identify the temperature directly as $T_i = P_i/n_i$.
This relation is not true in general for Bose--Einstein or Fermi--Dirac statistics:
for quantum distributions the equation of state depends nontrivially on the
occupation numbers (e.g. through polylogarithms in the nonrelativistic limit),
so that the ratio $P_i/n_i$ is no longer equal to the physical temperature $T_i$.
In equilibrium, the temperature is instead defined by the functional form of the
distribution,
$f_i = [\exp((E-\mu_i)/T_i)\mp 1]^{-1}$, and one can only infer ($T_i$, $\mu_i$) from moments
once a distribution shape is specified.

Nevertheless, even away from perfect local thermal equilibrium (LTE) it is useful to define
an effective \emph{kinetic temperature} as the second moment of the distribution,
\begin{equation}
	\label{eq:Tkin_def}
	T_i \;\equiv\; \frac{1}{n_i}\, g \int_p \frac{p^2}{3E}\, f_i
	\;=\; \frac{P_i}{n_i}\,,
\end{equation}
which can be interpreted as a velocity-dispersion parameter encoding the typical
kinetic energy per particle. When the system is in local thermal equilibrium with a Maxwellian shape this
definition coincides with the physical temperature, while for more general
non-thermal distributions it remains a useful characterization of the momentum
scale.

To obtain an evolution equation for $T_i$, we take the pressure moment of the
unintegrated fBE, multiply by $g\,p^2/(3E)$, and integrate over $\int_p$. Defining the
corresponding collision moment
\begin{equation}
	\label{eq:C2_def}
	C_2 \;\equiv\; g \int_p \frac{p^2}{E}\, C[f_i]\,,
\end{equation}
the left-hand side becomes
\begin{equation}
	\label{eq:pressure_moment_step}
	\dot P_i \;-\; H\, g \int_p \frac{p^2}{3E}\,p\,\partial_p f_i
	\;=\; \frac{1}{3}C_2\,.
\end{equation}
Integrating by parts one obtains the identity
\begin{equation}
	\label{eq:identity_pp}
	g \int_p \frac{p^2}{3E}\,p\,\partial_p f_i
	=
	g \int_p
	\left(
	\frac{p^4}{3E^{3}} - 5\,\frac{p^2}{3E}
	\right) f_i
	=
	n_i \left\langle \frac{p^4}{3E^{3}} \right\rangle - 5 P_i\,,
\end{equation}
where we introduced the average
$\langle X\rangle \equiv (1/n_i)\,g\int_p X f_i$. Inserting this into
Eq.~\eqref{eq:pressure_moment_step} yields
\begin{equation}
	\label{eq:P_evolution}
	\dot P_i + 5 H P_i - H\, n_i\left\langle \frac{p^4}{3E^{3}} \right\rangle
	= \frac{1}{3}C_2\,.
\end{equation}
At this point we combine Eq.~\eqref{eq:P_evolution} with the number-density Eq.~\eqref{eq:nBE} and use $P_i = n_i T_i$ with $T_i$ defined by Eq.~\eqref{eq:Tkin_def}. Differentiating
$P_i=n_iT_i$ gives $\dot P_i = n_i \dot T_i + T_i \dot n_i$, and one arrives at
\begin{equation}
	\label{eq:T_evolution_time}
	n_i \dot T_i
	=
	-2H\, n_i T_i
	+ \frac{1}{3}C_2
	- T_i C_0
	+ H\, n_i \left\langle \frac{p^4}{3E^{3}} \right\rangle .
\end{equation}
Finally, rewriting the time derivative in terms of the scale factor and using $N_i \equiv a^3 n_i$,
we obtain an evolution equation in the form
\begin{equation}
	\label{eq:T_evolution_a}
	\frac{dT_i}{da}
	=
	-\frac{2T_i}{a} + \frac{1}{3a}\left\langle \frac{p^4}{E^{3}} \right\rangle
	+\frac{a^2}{3H N_i}C_2
	-\frac{a^2 T_i}{H N_i}\, C_0 .
\end{equation}
It is instructive to briefly comment on the physical interpretation of the first
two terms on the right-hand side of
Eq.~\eqref{eq:T_evolution_a} in the absence of
collisions, i.e.~setting $C_0=C_2=0$.
In this limit, the temperature evolution is entirely driven by the Liouville
term of the Boltzmann equation and therefore reflects pure redshifting due to
cosmic expansion. The first term, $-2T_i/a$, corresponds to the adiabatic cooling of a
non-relativistic gas. Indeed, for freely streaming non-relativistic particles
one has $p\propto a^{-1}$ and hence $p^2\propto a^{-2}$, implying
$T_i\propto \langle p^2\rangle \propto a^{-2}$. This term therefore encodes the
expected scaling of the kinetic temperature once the species has become
non-relativistic. The second term,
$\frac{1}{a}\langle p^4/(3E^3)\rangle$, captures relativistic corrections to this
scaling. When the particles are relativistic, $E\simeq p$ and the moment reduces
to $\langle p^4/E^3\rangle\simeq \langle p\rangle \sim T_i$, so that the two terms
partially cancel and one recovers the relativistic scaling
$T_i\propto a^{-1}$.
As the population becomes increasingly non-relativistic, $E\simeq m_i$ and this
term becomes parametrically suppressed, leaving the $T_i\propto a^{-2}$
behavior dictated by the first term. Taken together, these two contributions interpolate smoothly between the
relativistic regime, where the temperature redshifts as $T_i\sim a^{-1}$, and the
non-relativistic regime, where the kinetic temperature scales as $T_i\sim a^{-2}$.
The remaining terms proportional to $C_0$ and $C_2$ quantify deviations from
pure redshifting induced by particle-number–changing and momentum-exchanging
interactions, respectively.

In this formulation, the system is closed once $C_0$ and $C_2$ are specified for
the interactions under consideration. If the Maxwell-Boltzmann ansatz holds, $T_i$ defined
through Eq.~\eqref{eq:Tkin_def} coincides with the physical temperature and the
average $\langle p^4/E^3\rangle$ can be evaluated using the corresponding
equilibrium distribution. If local thermal equilibrium is not guaranteed, one may
either keep $\langle p^4/E^3\rangle$ as an additional moment (leading to a
hierarchy). 

As in the previous subsection, let us emphasize that it 
is often convenient to rewrite the temperature evolution equation in terms of
the dimensionless inverse SM temperature variable
$x \equiv m_i/T$, rather than the scale factor. Starting from Eq.~\eqref{eq:T_evolution_time}, 
we rewrite the time derivative using the chain rule with
$x \equiv m_i/T$ (with $m_i$ constant),
\begin{equation}
	\dot T_i = \frac{dT_i}{dx}\,\dot x,
	\qquad
	\dot x = \frac{d}{dt}\left(\frac{m_i}{T}\right)
	= -\frac{m_i}{T^2}\,\dot T
	= -\frac{x}{T}\,\dot T.
\end{equation}
Substituting $\dot T_i = (dT_i/dx)\dot x$ into
Eq.~\eqref{eq:T_evolution_time} and dividing by $n_i\,\dot x$ yields
\begin{equation}
	\label{eq:T_evolution_x_from_time}
	\frac{dT_i}{dx}
	=
	\frac{1}{H\,x\,\tilde g(T)}
	\left[
	-2H\,T_i
	+\frac{C_2}{3n_i}
	-\frac{T_i\,C_0}{n_i}
	+H\left\langle \frac{p^4}{3E^{3}} \right\rangle
	\right].
\end{equation}
Equivalently,
\begin{equation}
	\label{eq:T_evolution_x_compact_Y}
	\frac{dT_i}{dx}
	=
	-\frac{2T_i}{x\,\tilde g(T)}
	+\frac{1}{H\,x\,\tilde g(T)}
	\left[
	\frac{C_2}{3s\,Y_i}
	-\frac{T_i\,C_0}{s\,Y_i}
	\right]
	+\frac{1}{x\,\tilde g(T)}
	\left\langle \frac{p^4}{3E^{3}} \right\rangle ,
	\qquad
	T=\frac{m_i}{x}.
\end{equation}

Numerical tools such as
\texttt{DarkSUSY}~\cite{Bringmann:2018lay} and \texttt{DRAKE}~\cite{Binder:2021bmg} 
provide solutions to the coupled Boltzmann equations,
Eqs.~\eqref{eq:Yeq_x} and~\eqref{eq:T_evolution_x_compact_Y},
within the framework of a standard cosmological history assuming radiation domination.
In particular, \texttt{DRAKE} and more recently \texttt{BEST}~\cite{Yoon:2026rce} also include routines for solving the full,
unintegrated Boltzmann equation~\eqref{eq:fBE}, going beyond the
moment-based approach discussed here.
Such a fully phase-space-resolved treatment is numerically more demanding, but becomes
necessary in scenarios where elastic self-interactions are not efficient enough
to maintain local thermal equilibrium.




In a different context, cosmological Boltzmann codes such as
\texttt{CLASS}~\cite{2011arXiv1104.2932L} and \texttt{CAMB}~\cite{Storchi:2025who}
solve the Boltzmann equation for linear perturbations around homogeneous
background distributions in order to compute CMB anisotropies and large-scale
structure observables. These codes are not designed to treat out-of-equilibrium
dark-sector and are therefore not applicable to the scenarios studied
here.

We emphasize that the explicit form of the collision operators
$C_0$, $C_2$, and $C_E$ has not been specified at this stage.
In practice, $C_0$ is commonly expressed in terms of thermally averaged reaction
rates, such as $\langle\sigma_{\rm ann}\,v\rangle$, while analogous
definitions apply to $C_2$ and $C_E$, weighted by appropriate powers of the
momentum. Since the precise structure of these integrated collision terms depends on the underlying particle-physics realization, we will present their explicit forms on a case-by-case basis in the subsequent chapters, where we introduce concrete examples involving freeze-in production and self-number-changing (cannibal) interactions.

Finally, the results presented in this thesis are obtained using numerical routines developed for the specific dynamical regimes and interaction structures studied in the following chapters, rather than through a direct application of public codes.

\section{Thermal corrections to the free energy density}
\label{sec:Thermal_corr}

While the observational evidence reviewed above establishes the presence of dark
matter on cosmological and astrophysical scales, many proposed particle
realizations involve new degrees of freedom whose dynamics are governed by the
thermal conditions of the early Universe. In particular, scalar fields, which
play a central role in symmetry breaking and phase
transitions, are generically affected by finite-temperature effects.

The dynamics of scalar fields in the early Universe cannot be understood solely
from their zero-temperature potentials. At finite temperature, interactions with
the surrounding plasma modify the effective free energy density of the system,
altering both the location and the stability of its extrema. These thermal
corrections play a central role in determining the pattern of symmetry breaking,
the order of cosmological phase transitions, and the presence or absence of
metastable states.

In thermal equilibrium, the relevant quantity governing the macroscopic evolution
of a scalar field is the finite-temperature effective potential, or equivalently
the free energy density, obtained by integrating out quantum and thermal
fluctuations around a background field configuration. As the temperature evolves
due to cosmic expansion, the competition between vacuum contributions and thermal
effects can qualitatively change the structure of the potential, inducing
phase transitions in the scalar sector. Before defining the effective potential
explicitly, we begin by recalling how expectation values are defined in thermal
field theory. This section is largely based on the standard textbooks~\cite{Kapusta:2006pm,Laine:2016hma}.
For a detailed study of thermal field theory we refer the interested reader to these materials.

In the canonical ensemble, the partition function is
\begin{equation}
	\label{eq:Zcanonical}
	Z(\beta_i) \equiv \Tr\!\left[e^{-\beta_i \hat H}\right],\qquad
	\beta_i \equiv \frac{1}{T_i},
\end{equation}
from which the free energy follows as $F = -T_i \ln Z$, where $\hat H$ denotes the
Hamiltonian operator. Thermal expectation values of an operator
$\hat{\mathcal O}$ are then defined as
\begin{equation}
	\label{eq:thermal_expval_trace}
	\langle \hat{\mathcal O} \rangle_{T_i}
	\equiv
	\frac{1}{Z}\,
	\Tr\!\left(e^{-\beta_i \hat H}\,\hat{\mathcal O}\right).
\end{equation}
In particular, for a scalar field operator $\hat\phi(x)$, the equilibrium background
(or condensate) is defined as
\begin{equation}
	\label{eq:phi_background_def}
	\phi_b(x) \equiv \langle \hat\phi(x)\rangle_{T_i} .
\end{equation}
In a homogeneous and isotropic setting, the background field $\phi_b$ can depend
at most on time, and in thermal equilibrium it reduces to a constant determined
by the minimum of the free-energy density. The temperature $T_i$ denotes the
temperature of the plasma associated with the thermal fluctuations being
integrated out, and coincides with the temperature introduced in the previous
section. It is often convenient to rewrite the canonical partition function
Eq.~\eqref{eq:Zcanonical} as a Euclidean path integral. This is achieved by
introducing imaginary time $\tau = it$ and compactifying it on the interval
$\tau \in [0,\beta_i]$, with $\beta_i = 1/T_i$. In this formulation,
finite-temperature field theory is defined on a compact Euclidean time
direction. The trace over physical states in the thermal partition function then
enforces specific boundary conditions along the imaginary-time direction:
bosonic fields are periodic, while fermionic fields are anti-periodic. For a
bosonic scalar field, this implies
\begin{equation}
\phi(\tau+\beta_i,\mathbf{x}) = \phi(\tau,\mathbf{x}) \,,
\end{equation}
whereas fermionic fields satisfy
\begin{equation}
\psi(\tau+\beta_i,\mathbf{x}) = -\,\psi(\tau,\mathbf{x}) \,.
\end{equation}
These boundary conditions encode the quantum statistics of the fields and ensure
the correct implementation of the thermal trace in the path-integral
representation.

With these boundary conditions, the partition function can be written as
\begin{equation}
	\label{eq:Zpath}
	Z(\beta_i)
	=
	\int_{\phi(\beta_i,\mathbf{x})=\phi(0,\mathbf{x})}
	\!\!\mathcal D\phi\;
	e^{-S_E[\phi]} ,
\end{equation}
where $S_E$ denotes the Euclidean action. In the presence of an external source
$J(x)$, the generating functional generalizes to
\begin{equation}
	\label{eq:ZJ}
	Z[J] =
	\int \mathcal D\phi\;
	\exp\!\left[
	- S_E[\phi]
	+ \int_0^{\beta_i} d\tau \int d^3x\; J(x)\,\phi(x)
	\right].
\end{equation}
The corresponding background field is obtained by functional differentiation,
\begin{equation}
	\label{eq:phi_from_W}
	\phi_b(x)
	\equiv
	\langle \phi(x)\rangle_J
	=
	\frac{\delta W[J]}{\delta J(x)},
	\qquad
	W[J]\equiv \ln Z[J].
\end{equation}
and the finite-temperature effective action is defined as the Legendre transform
\begin{equation}
	\label{eq:Gamma_def}
	\Gamma[\phi_b]
	\equiv
	- W[J]
	+ \int_0^{\beta_i} d\tau \int d^3x\; J(x)\,\phi_b(x),
\end{equation}
with $J$ understood as a functional of $\phi_b$. For a spacetime-independent
background, the effective action reduces to
\begin{equation}
	\Gamma[\phi_b] = \beta_i V\, V_{\rm eff}(\phi_b,T_i),
\end{equation}
where $V=a^3$ denotes the spatial volume. The effective potential
$V_{\rm eff}(\phi_b,T_i)$ can therefore be interpreted as the free energy density,
and the equilibrium value of the condensate is obtained by minimizing it,
\begin{equation}
	\left.
	\frac{\partial V_{\rm eff}(\phi_b,T_i)}{\partial \phi_b}
	\right|_{\phi_b=\phi_{\rm min}(T_i)} = 0 .
\end{equation}

To make this construction explicit, it is instructive to consider first a free
scalar theory. In this case the Euclidean action is quadratic, and the path
integral in Eq.~\eqref{eq:Zpath} can be evaluated exactly. Since the imaginary
time direction is compact, $\tau \in [0,\beta_i]$, and the scalar field satisfies
periodic boundary conditions, the field can be expanded in a discrete Fourier
series along the imaginary-time direction. Writing
\begin{equation}
	\phi(\tau,\mathbf{x})
	=
	T_i \sum_{n=-\infty}^{\infty}
	\int_k\,
	e^{i(\omega_n \tau + \mathbf{k}\cdot\mathbf{x})}
	\,\tilde\phi(\omega_n,\mathbf{k}),
	\qquad
	\omega_n = 2\pi n T_i,
\end{equation}
the frequencies $\omega_n$ label the discrete Fourier modes associated with the
compact imaginary-time direction. These modes are referred to as
\emph{bosonic Matsubara modes}. With this expansion, the finite-temperature
partition function factorizes into a product over Matsubara modes. As a result,
the effective potential naturally separates into a zero-temperature contribution
and a purely thermal correction, the latter encoding the effects of thermally
excited bosonic fluctuations.

In practice the quantity of interest is not the partition function itself,
but the corresponding free energy density evaluated in the presence of a background
field $\phi_b$. At one-loop order, this amounts to integrating out Gaussian
fluctuations around $\phi_b$, which results in a functional determinant. For a
particle species $i$ (bosonic or fermionic) whose field-dependent mass in the
background is $m_i(\phi_b)$, the resulting contribution to the effective potential
can be written as
\begin{equation}
	\label{eq:V1loop_general}
	V_{\rm eff}(\phi_b,T_i)
	=
	V_0(\phi_b)
	+
	\sum_i g_i
	\left[
	V_i^{(0)}(\phi_b)
	+
	V_i^{(T)}(\phi_b,T_i)
	\right],
\end{equation}
where $V_0$ denotes the tree-level potential and $g_i$ counts the number of internal
degrees of freedom of species $i$, including spin, color, and particle--antiparticle
multiplicities. The one-loop contribution naturally separates into a
zero-temperature vacuum part $V_i^{(0)}(\phi_b)$, corresponding to the
Coleman--Weinberg potential, and a purely thermal part $V_i^{(T)}(\phi_b,T_i)$
arising from thermally excited modes. For fermionic fields, the contribution
carries an additional overall minus sign, reflecting Fermi--Dirac statistics and
the Grassmann nature of the functional integral.

The thermal contribution arises from the Matsubara sum over discrete frequencies
$\omega_n = 2\pi n T_i$ for bosons and $\omega_n = (2n+1)\pi T_i$ for fermions. After
performing the sum, one obtains
\begin{equation}
	\label{eq:Vthermal_general}
	V_i^{(T)}(\phi_b,T_i)
	=
	\pm\, T_i
	\int_p
	\ln\!\left(1 \mp e^{-\beta_i E}\right),
	\qquad
	E \equiv \sqrt{p^2+m_i^2(\phi_b)} .
\end{equation}
where the upper (lower) sign corresponds to bosons (fermions). This expression has a transparent thermodynamic interpretation: it is precisely the free energy density of a relativistic gas of excitations with dispersion relation $E$,
evaluated in the background field $\phi_b$.

The dependence of $m_i$ on $\phi_b$ originates from interaction terms in the
Lagrangian. In the presence of a nonvanishing background, such interactions
generate effective masses for fluctuations around $\phi_b$. For example, scalar
interactions at leading order of the form $\lambda\,\phi^2\chi^2$ lead to
$m_\chi^2(\phi_b)=m_{\chi,0}^2+\lambda\,\phi_b^2$, while Yukawa couplings
$y\,\phi\,\bar\psi\psi$ induce $m_\psi(\phi_b)=y\,\phi_b$. As a result, thermal
corrections to the free energy density directly probe the field-dependent spectrum
of the theory.

As expected from thermodynamics, the thermal contribution to the effective
potential is related to the pressure of the corresponding particle species,
$V_i^{(T)}(\phi_b,T_i) = -P_i(\phi_b,T_i)$, providing a useful consistency check and
clarifying the interpretation of $V_{\rm eff}$ as a free energy density.

\subsection{Infrared sensitivity and ring resummation}
\label{subsec:ring}


Introducing a finite temperature fundamentally modifies the infrared structure
of quantum field theory and, in certain regimes, leads to a breakdown of the
naive perturbative expansion. For detailed reviews of this issue in the context
of cosmological phase transitions, see e.g.~\cite{Curtin:2016urg,Senaha:2020mop}.
The origin of this breakdown can be traced to the behavior of bosonic fields at
finite temperature. In the imaginary-time formalism, bosonic fields obey periodic boundary
conditions and therefore possess a Matsubara zero mode. This mode plays a
distinguished role in the infrared and is absent at zero temperature. As a
result, long-wavelength bosonic fluctuations can become strongly coupled even
when the underlying zero-temperature theory is weakly interacting.

The one-loop finite-temperature contribution to the effective potential for a
bosonic degree of freedom with field-dependent mass $m_i^2(\phi_b)$ is given
by~\cite{Laine:2016hma}
\begin{equation}
	V^{(1)}_{T_i}(\phi_b)
	=
	\frac{T_i}{2}
	\sum_{n=-\infty}^{\infty}
	\int_p
	\ln\!\left(\omega_n^2 + p^2 + m_i^2(\phi_b)\right),
	\qquad
	\omega_n = 2\pi n T_i .
\end{equation}
All nonzero Matsubara modes are separated from the infrared by a thermal gap
$\omega_n^2 \sim (2\pi T_i)^2$ and therefore do not generate long-distance
singularities. The potentially problematic contribution originates entirely from
the zero mode $n=0$,
\begin{equation}
	V^{(1)}_{n=0}
	=
	\frac{T_i}{2}
	\int_p
	\ln\!\left(p^2 + m_i^2(\phi_b)\right),
\end{equation}
which describes a three-dimensional scalar field.\footnote{This is due to the zero Matsubara mode ($n=0$) being independent of Euclidean time and therefore has no time-like kinetic term; as a result, its contribution involves only a three-dimensional spatial momentum integral, so that it is equivalent to a scalar field propagating purely in a three dimensional space, which is why the infrared dynamics at finite temperature is effectively described by a three-dimensional Euclidean field theory.}

In an interacting theory, bosonic fields acquire thermal self-energy corrections
$\Pi_i$ from loop effects involving particles present in the plasma.
To make the connection to distribution functions explicit, consider a scalar
degree of freedom $\phi$ with a quartic interaction.\footnote{At this stage we keep the
	discussion schematic and encode model-dependent combinatorics and internal degrees
	of freedom in an overall coefficient $c_i$.}
The static (zero external Matsubara frequency) self-energy at vanishing external
3-momentum is given by
\begin{equation}
	\Pi_i
	=
	\;
	\raisebox{0.15em}{\begin{tikzpicture}[baseline={(b.base)}]
		\begin{feynman}
			\diagram[scale=0.8] {
				a -- [scalar] b -- [scalar] c,
				b -- [scalar, loop, min distance=1.6cm] b,
			};
		\end{feynman}
	\end{tikzpicture}}
	=
	c_i\lambda\;
	T_i\sum_{n=-\infty}^{\infty}\int_{ q}
	\frac{1}{\omega_n^2 + q^2+m_i^2(\phi_b)} .
	\label{eq:Pi_with_diagram}
\end{equation}
We can see that the self-energy is related to the free-energy density by 
differentiating the latter with respect to the squared mass, making explicit the appearance of the
thermal propagator,
\begin{equation}
	\frac{\partial V^{(1)}_{T_i}}{\partial m_i^2}
	=
	\frac{T_i}{2}\sum_{n=-\infty}^{\infty}\int_{ p}
	\frac{1}{\omega_n^2+ p^2+m_i^2(\phi_b)}.
	\label{eq:dVdm2_prop}
\end{equation}
Isolating the $n=0$ term in \eqref{eq:dVdm2_prop} shows that the same Matsubara zero mode
responsible for the infrared sensitivity of $V^{(1)}_{n=0}$ also controls the infrared
behavior of the tadpole.
The Matsubara sum can be evaluated in closed form (standard contour methods), yielding
\begin{equation}
	T_i \sum_{n=-\infty}^{\infty}\frac{1}{\omega_n^2 + E_q^2}
	=
	\frac{1}{2E_q} + \frac {f_{\rm BE}(E_q)}{E_q},
	\qquad
	f_{\rm BE}(E)\equiv \frac{1}{e^{E/T_i}-1}.
	\label{eq:matsubara_sum_BE}
\end{equation}
The $1/(2E_q)$ term is temperature-independent and
ultraviolet divergent; it is renormalised by the usual zero-temperature counterterms.
The \emph{thermal} contribution is finite and is therefore isolated as
\begin{equation}
	\Pi_i
	=
	c_i\,\lambda\;
	\int_{q}
	\frac{f_{\rm BE}(E_{q})}{E_{q}}
	\;\equiv\;
	c_i\,\lambda\;\mathcal{I}_i\,,
	\label{eq:Pi_equals_I_def}
\end{equation}
which makes explicit why $\Pi_i$ can be written in terms of a Bose--Einstein
distribution: it originates from the \emph{thermal part} of the Matsubara sum. In contrast, 
in the real-time formalism, the finite-temperature scalar propagator contains a thermal contribution proportional to $2\pi\,\delta(q^2-m^2)\,f(E_q)$, i.e., the distribution function appears explicitly. The thermal self-energy then arises directly from this term, so that the appearance of $f_{\rm BE}(E_q)$ in Eq.~\eqref{eq:Pi_equals_I_def} reflects the fact that the loop correction is sourced by the population of on-shell excitations in the plasma. In fact, the real-time formalism is the appropriate framework when the system is out of equilibrium, and we will use it in~\Cref{ch:5}. 

If the distribution corresponds to a Bose-Einstein one, we can expand as 
$f_\text{BE} = (e^{E/T_i}-1)^{-1} = \sum_{k} \,e^{-k\,E/T_i}$. Then the integral $\mathcal{I}_i$ admits a useful representation in terms of modified Bessel
functions,
\begin{align}
	\mathcal{I}_i
	&=
	\int_{q}
	\frac{f_{\rm BE} }{E}
	=
	\frac{1}{2\pi^2}
	\int_0^\infty dq\,\frac{q^2}{\sqrt{q^2+m_i^2}}
	\frac{1}{e^{\sqrt{q^2+m_i^2}/T_i}-1}
	\nonumber\\
	&=
	\frac{1}{2\pi^2}\,m_i T_i
	\sum_{k=1}^{\infty}\frac{1}{k}\,
	K_1\!\left(k\,\frac{m_i}{T_i}\right),
	\label{eq:I_bessel}
\end{align}
\begin{figure}[t]
	\centering
	\begin{tikzpicture}
		\def\R{1.25}   
		\def\rp{0.28}  
		
		\coordinate (O) at (0,0);
		
		\draw[dashed] (O) circle (\R);
		
		\foreach \ang in {30,90,150,210}{
			\coordinate (Pc) at ($(O)+(\ang:{\R+\rp})$);
			\draw[dashed] (Pc) circle (\rp);
		}
		
		\def\Rdot{1.45} 
		\foreach \ang in {330,345,0}{
			\fill ($(O)+(\ang:\Rdot)$) circle (0.035);
		}

		\coordinate (t) at ($(O)+(0,-\R)$);
		\coordinate (a) at (-2.2,-\R);
		\coordinate (c) at ( 2.2,-\R);
		
		\draw[dashed](a) -- (t) -- (c);
		
	\end{tikzpicture}
	\caption{Schematic daisy (ring) diagram with multiple self-energy insertions. The dots denoted an arbitrary number of insertions.}
	\label{fig:daisy_petals}
\end{figure}
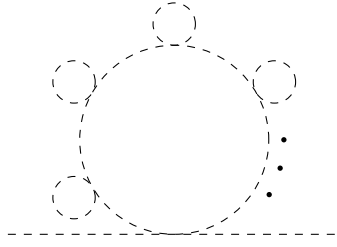
where $m_i\equiv m_i(\phi_b)$.
In the relativistic limit $m_i \ll T_i$, one obtains
$\mathcal{I}_i \to T_i^2/12$, reproducing the familiar scaling
$\Pi_i \sim c_i\,\lambda\,T_i^2$ (and similarly $\sim g^2 T_i^2$ for gauge interactions).
Near a phase transition, the curvature $m_i^2(\phi_b)$ around the relevant field region
can become small, so that $\Pi_i(T_i)$ is no longer a small perturbation. In this regime,
the infrared dynamics is governed by an effective mass
\begin{equation}
	m_{i,\rm eff}^2(\phi_b;T_i)
	=
	m_i^2(\phi_b) + \Pi_i(T_i)\,,
\end{equation}
and insertions of $\Pi_i(T_i)$ on the zero-mode propagator must be treated
non-perturbatively. We can 
resum the propagator by adding an arbitrary number of insertions into the loop contribution
as shown in Fig.~\ref{fig:daisy_petals}, where the resummed propagator is given by
\begin{equation}\label{eq:resummed_propagator}
	\frac{1}{p^2-m_i^2}+\frac{\Pi_i}{(p^2-m_i^2)^2} + \frac{\Pi_i^2}{ (p^2-m_i^2)^3 } + \dots = \frac{1}{p^2-m_i^2 - \Pi_i}\,.
\end{equation}

Diagrams with an arbitrary number of such insertions, the so-called ring or daisy
diagrams, contribute at the same parametric order. The contribution with $k$ self-energy
insertions reads
\begin{equation}\label{eq:Vk}
	V_k
	=
	\frac{T_i}{2}\,
	\frac{(-1)^k}{k}
	\int_p
	\left(
	\frac{\Pi_i(T_i)}{p^2 + m_i^2(\phi_b)}
	\right)^k ,
\end{equation}
where the factor $1/k$ accounts for the cyclic symmetry of the ring. Summing over
all such diagrams yields
\begin{align}
	V_{\text{ring}}
	=
	\frac{T_i}{2}
	\int_p
	\sum_{k=1}^{\infty}
	\frac{(-1)^k}{k}
	\left(
	\frac{\Pi_i}{p^2 + m_i^2}
	\right)^k
	\nonumber
	=
	-\frac{T_i}{2}
	\int_p
	\ln\!\left(
	1 + \frac{\Pi_i}{p^2 + m_i^2}
	\right),
\end{align}
and therefore
\begin{equation}
	V_{\text{ring}}
	=
	-\frac{T_i}{2}
	\int_p
	\left[
	\ln\!\left(p^2 + m_i^2 + \Pi_i \right)
	-
	\ln\!\left(p^2 + m_i^2 \right)
	\right].
\end{equation}
The three-dimensional momentum integral is elementary,
\begin{equation}
	\int_p
	\ln(p^2 + M^2)
	=
	\frac{M^3}{6\pi}
	\,+\,(\text{UV terms}),
\end{equation}
where ultraviolet contributions cancel in the difference. One therefore obtains
the standard ring-improved contribution
\begin{equation}
	\label{eq:Vring_final}
	V_{\text{ring}}
	=
	-\frac{T_i}{12\pi}
	\left[
	\bigl(m_i^2 + \Pi_i\bigr)^{3/2}
	-
	\bigl(m_i^2\bigr)^{3/2}
	\right],
\end{equation}
summed over all bosonic degrees of freedom.

Physically, this result corresponds to thermal screening of long-wavelength bosonic
modes, implemented through the replacement $m_i^2 \to m_i^2 + \Pi_i$ in the infrared.
Explicit expressions for $\Pi_i$ in the case of a U(1) dark gauge symmetry will be
derived in Chapter~\ref{ch:5}.
%

\section{Cosmological first-order phase transitions and gravitational waves}
\label{sec:FOPT_GW}

In the previous section, we discussed how thermal corrections modify the
effective potential of scalar fields in the early Universe, with particular
emphasis on infrared effects and the necessity of daisy resummation. These
thermal effects can qualitatively change the structure of the potential,
including the appearance of barriers separating local minima. Such barriers are
a necessary condition for a cosmological first-order phase transition, which is 
intrinsically a non-equilibrium process. As
the Universe cools, the system may become trapped in a metastable (false)
vacuum, even after a lower-energy phase has appeared. The transition to the
true vacuum then proceeds through the nucleation and expansion of bubbles,
rather than through a smooth crossover. The dynamics of this process are
controlled by the thermal effective potential and its temperature dependence~\cite{Linde:1981zj}.
For a complete review see~\cite{Mazumdar:2018dfl,Hindmarsh:2020hop}.

A cosmological first-order phase transition is characterized by several
distinct temperatures, reflecting different stages of its dynamics.
In the following, we will make use of three temperatures in particular:
\begin{itemize}
	\item the \emph{critical temperature} $T_i^c$, at which the symmetric and
	broken minima of the thermal effective potential become degenerate;
	\item the \emph{nucleation temperature} $T_i^n$, at which the rate of bubble
	nucleation becomes comparable to the Hubble expansion rate and the phase
	transition effectively begins;
	\item the \emph{percolation temperature} $T_i^p$, at which a substantial
	fraction of the Universe has transitioned to the true vacuum and expanding
	bubbles start to overlap and collide.
\end{itemize}
When the field associated with $\phi_b$ is in thermal contact with the SM,
one typically finds $T_i^p \lesssim T_i^n < T_i^c$. This relation need not hold
in an inverse FOPT, as we will see in Chapter~\ref{ch:5}.

\begin{figure}[t!]
	\def\sepf{0.49}
	\centering
	\includegraphics[width=0.3\textwidth]{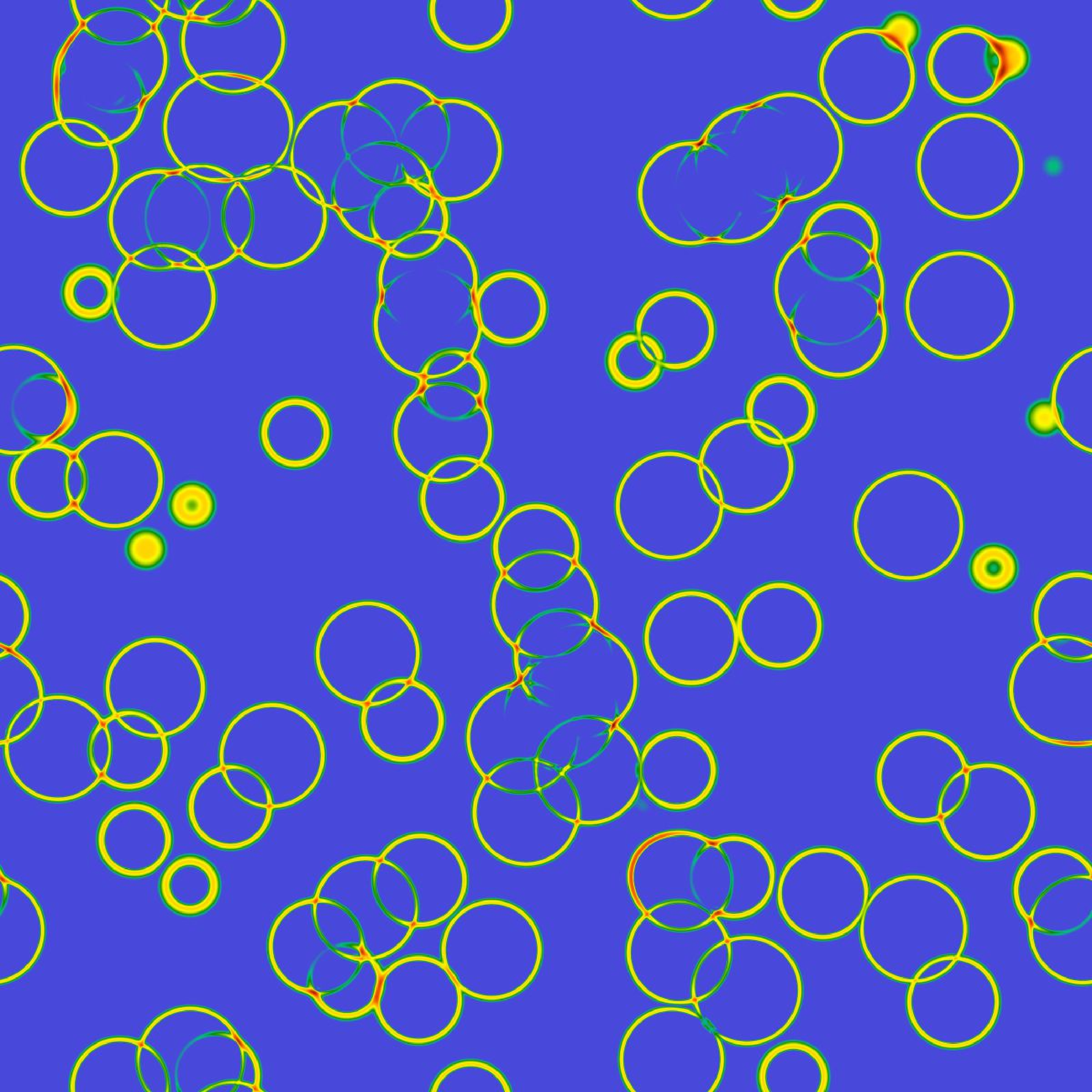}
	\includegraphics[width=0.3\textwidth]{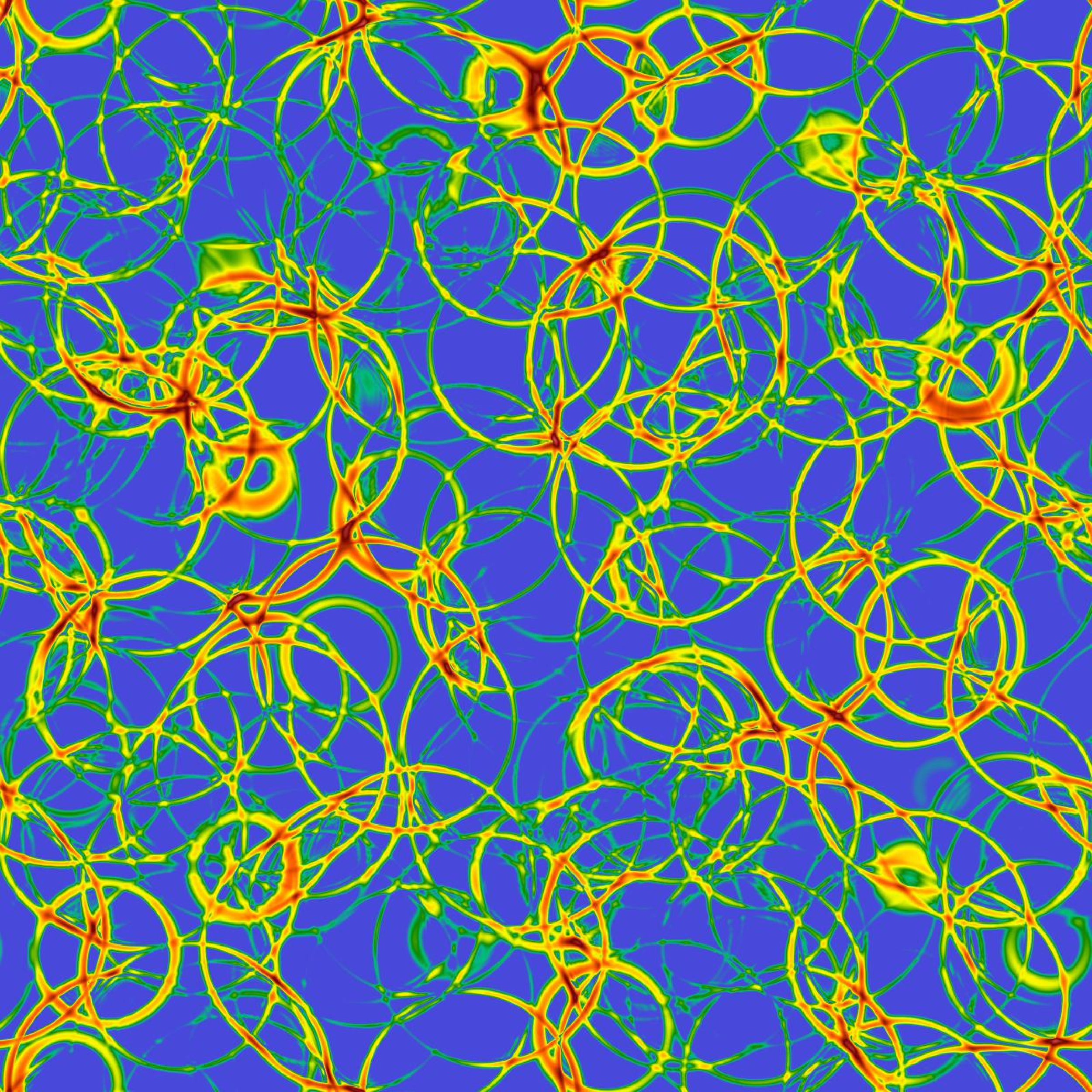}
	\includegraphics[width=0.3\textwidth]{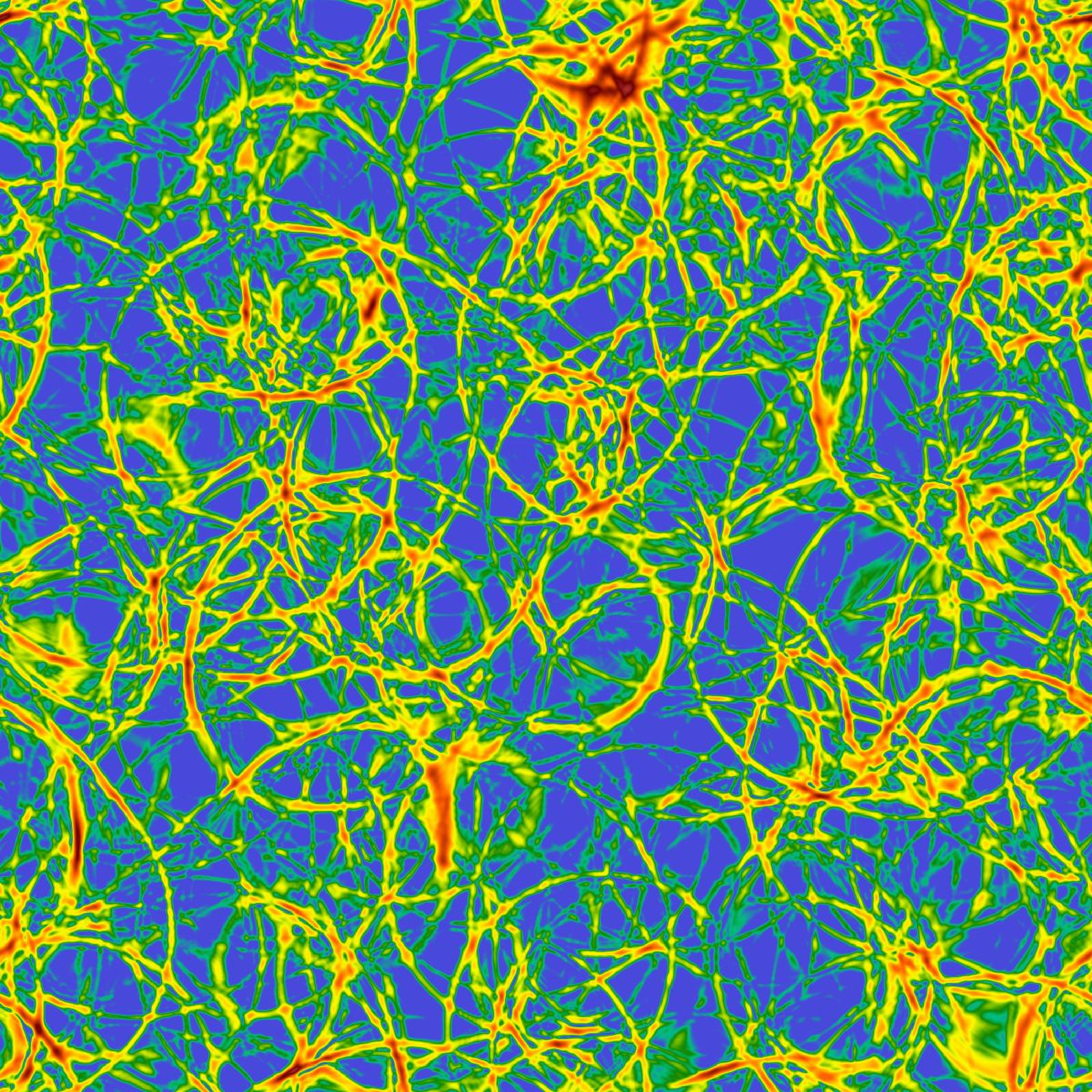}
	\caption{Numerical simulation of nucleation, expansion and percolation of bubbles. Images from~\cite{Hindmarsh:2015qta}.}
	\label{fig:bubbles}
\end{figure}

The nucleation, growth, and subsequent collisions of expanding bubbles, as shown in~\Cref{fig:bubbles}, involve
large energy densities and relativistic bulk motion of the surrounding plasma~\cite{Steinhardt:1981ct}.
As a result, first-order phase transitions in the early Universe can act as
powerful sources of stochastic gravitational-wave backgrounds. This was first noted by Witten and 
studied in the context of QCD phase transition~\cite{Witten:1984rs} and
subsequently studied in a more general context in~\cite{Hogan:1986qda}. The properties of
the resulting signal encode detailed information about the underlying scalar
potential and the thermal history of the transition.

In general the transition does not complete at $T_i^c$: the Universe can supercool
until bubble nucleation becomes sufficiently efficient. In this context, \textit{supercooling} 
refers to the period during which the Universe cools below the critical temperature 
$T_i^c$ while remaining trapped in the metastable (false vacuum) phase, until the bubble nucleation rate becomes sufficiently large to complete the transition.
For a thermal transition, the bubble nucleation rate per unite volume was originally derived by Coleman in~\cite{Coleman:1977py}
and improved by Weinberg in~\cite{Weinberg:1994tk},
\begin{equation}
	\label{eq:Gamma_nucl}
	\Gamma =  A\,e^{-S_3/T_i},
	\qquad
	A\approx T_i^4\left( \frac{S_3}{2\pi T_i} \right)^{3/2},
\end{equation}
where $S_3$ is the three dimensional Euclidean bounce action.
We emphasise that $\Gamma$ denotes a nucleation rate \emph{per unit volume}, i.e.\ a probability per unit time and per unit spatial volume. In natural units it therefore carries mass dimension four. This reflects the fact that bubble nucleation is a local process in spacetime and is naturally expressed as a probability per unit four-volume.

For a single scalar
field,
\begin{equation}
	\label{eq:S3_def}
	S_3
	=
	4\pi\int_0^\infty dr\,r^2
	\left[
	\frac{1}{2}\left(\frac{d\phi_b}{dr}\right)^2
	+ V_{\rm eff}(\phi_b) - V_{\rm eff}(\phi_{\rm false})
	\right],
\end{equation}
with boundary conditions $\phi_b'(0)=0$ and $\phi_b(r\to\infty)=\phi_{\rm false}$.
The bounce action represents the free-energy cost of forming a critical bubble of true vacuum inside the metastable phase. The $\frac{d\phi_b}{dr}$ term in Eq.~\eqref{eq:S3_def} corresponds to the gradient energy associated with the bubble wall (surface tension), while the potential difference $V_{\rm eff}(\phi_b)-V_{\rm eff}(\phi_{\rm false})$ accounts for the free-energy gain from converting the interior of the bubble to the true vacuum. Small bubbles are energetically disfavoured because the surface term dominates, whereas sufficiently large bubbles gain free energy from the volume term. The bounce solution $\phi_b(r)$ extremises $S_3$ and determines the critical bubble configuration for which these two contributions balance. The exponential factor $e^{-S_3/T_i}$ therefore reflects the Boltzmann suppression for thermally activating such a critical fluctuation in the plasma.

A practical definition of the \emph{nucleation temperature} \(T_n\) is obtained by
requiring that roughly one bubble nucleates per Hubble volume~\cite{Turner:1992tz,Athron:2023xlk},
\begin{equation}
	\label{eq:Tn_condition}
	\frac{4\pi}{3}\int_{t(T_n)}^{t(T_c)} dt\,\frac{\Gamma}{H^3} = 1.
\end{equation}
\noindent
Here $\Gamma/H^3$ is a nucleation rate per Hubble volume and the condition is dimensionless.
In Chapter~\ref{ch:5} we will be interested in FOPT of a dark sector that is not necessarily in kinetic equilibrium with the SM, we therefore reformulate the previous formula
in terms of the scale factor $a$,
\begin{equation}
	\label{eq:Tn_condition_a}
	\frac{4\pi}{3}\int_{a(T_n)}^{a(T_c)} da\,
	\frac{\Gamma}{a\,H^4} = 1.
\end{equation}
%
%
%

While the nucleation temperature $T_n$ characterizes the onset of the phase
transition, it does not by itself guarantee that the transition completes.
Whether the Universe successfully converts from the false vacuum to the true
vacuum depends on the subsequent growth and overlap of bubbles, i.e. on
\emph{percolation}. This distinction is particularly relevant in strongly
supercooled transitions, where bubble nucleation may occur but fail to fill a
sufficient fraction of space.

A convenient way to quantify the progress of the transition is through the
fraction of the Universe that remains in the false vacuum. Assuming homogeneous
nucleation, this fraction can be written as
\begin{equation}
	\label{eq:false_vacuum_fraction}
	P(t)
	=
	\exp\!\left[
	- \frac{4\pi}{3}
	\int_{t_c}^{t} dt'\,
	\Gamma(t')\,
	a^3(t')\,
	R^3(t,t')
	\right],
\end{equation}
where $\Gamma(t')$ is the bubble nucleation rate per unit physical volume and
$R(t,t')$ denotes the physical radius at time $t$ of a bubble nucleated at time
$t'$. For relativistically expanding walls, one may approximate
\begin{equation}
	R(t,t') \simeq \int_{t'}^{t} dt''\, \frac{v_w}{a(t'')},
\end{equation}
so that bubble growth is controlled by both the wall velocity and the expansion of
the Universe. The \emph{percolation time} $t_p$ is then defined as the time at which the false-vacuum fraction drops below a
critical value,
\begin{equation}
	\label{eq:percolation_condition}
	P(t_p) \equiv P_* \simeq 0.7,
\end{equation}
corresponding to the point at which bubbles begin to form a connected network and
the transition can complete efficiently. The precise numerical value of $P_*$
is not fundamental; values in the range $P_*\sim 0.7$ are commonly adopted in the
literature as a practical criterion for percolation and lead to only mild
quantitative differences in the resulting GW predictions~\cite{Enqvist:1991xw,Turner:1992tz,Ellis:2018mja,1971AdPhy..20..325S}.

A convenient definition of the false-vacuum volume fraction is~\cite{Athron:2023xlk}
\begin{equation}
	\label{eq:Pf_def}
	P_f(t)=e^{-I(t)},
	\qquad
	I(t)\equiv \frac{4\pi}{3}\int_{t_c}^{t} dt'\,\Gamma(t')\,a^3(t')\,R^3(t,t')\,,
\end{equation}
where $\Gamma(t')$ is the nucleation rate per unit \emph{physical} volume and
$R(t,t')$ is the physical radius at time $t$ of a bubble nucleated at $t'$.
Assuming a wall speed $v_w$ (taken constant for simplicity in this section), the bubble radius
follows from integrating the physical growth rate $dR/dt=v_w$ in comoving
coordinates,
\begin{equation}
	\label{eq:R_ttprime}
	R(t,t') = a(t)\int_{t'}^{t} dt''\,\frac{v_w}{a(t'')}\,.
\end{equation}
Rewriting the time integrals in terms of the scale factor,
one obtains
\begin{equation}
	\label{eq:R_aa}
	R(a,a')
	=
	a\int_{a'}^{a} da''\,\frac{v_w}{a''^{2}H(a'')}\,,
\end{equation}
and therefore
\begin{equation}
	\label{eq:I_of_a}
	I(a)
	=
	\frac{4\pi}{3}
	\int_{a_c}^{a} da'\,
	\frac{\Gamma(a')}{a'H(a')}\,
	a'^3
	\left[
	a\int_{a'}^{a} da''\,\frac{v_w}{a''^{2}H(a'')}
	\right]^3.
\end{equation}
This makes the dependence on the Hubble rate explicit: percolation is controlled
not only by the microscopic nucleation rate $\Gamma$, but also by the expansion
history through $H(a)$, which governs both the available four-volume for nucleation
and the bubble growth in an expanding background. 

The general expression for the false-vacuum fraction,
Eq.~\eqref{eq:false_vacuum_fraction}, simplifies considerably in the case of a
standard radiation-dominated cosmology with slowly varying relativistic degrees
of freedom. In this regime one has $T\propto a^{-1}$ and
$H(a)\propto a^{-2}$, such that the combination $a^2H(a)$ is constant. As a
result, the physical radius of a bubble nucleated at scale factor $a'$ and
observed at $a$ can be evaluated analytically, yielding
\begin{equation}
	\label{eq:R_RD}
	R(a,a') =
	\frac{v_w}{aH(a)}\left(a-a'\right)
	=
	\frac{v_w}{H(a)}\left(1-\frac{a'}{a}\right).
\end{equation}
Substituting this expression into Eq.~\eqref{eq:false_vacuum_fraction}, the false-vacuum fraction can be written in the compact form
\begin{equation}
	\label{eq:Pf_RD}
	P_f(a)
	=
	\exp\!\left[
	-\,\frac{4\pi}{3}\,
	\frac{v_w^{\,3}}{a^{2}H^{4}(a)}
	\int_{a_i}^{a} da'\,
	\Gamma(a')\,a'^{4}
	\left(1-\frac{a'}{a}\right)^{3}
	\right].
\end{equation}
In this approximation the percolation problem reduces
to a single integral over the nucleation history, while the effects of cosmic
expansion enter explicitly through the Hubble rate evaluated at the observation
time $a$.


\subsection{Thermal parameters}\label{subsec:therm_param_gws}

Gravitational wave production is dominated by the late stages of the transition,
around the percolation temperature $T_i^p$, when expanding bubbles begin to
overlap and collide and a connected network of true-vacuum regions forms.
The GW signal from a FOPT is commonly expressed in terms of a small set of macroscopic
parameters evaluated at a characteristic temperature $T^*_i$, typically taken as $T_i^n$
(or sometimes a percolation temperature). In what follows we keep $T^*_i$ explicit.
A standard measure of the released vacuum energy relative to the background radiation is~\cite{Athron:2023xlk}
\begin{equation}
	\label{eq:alpha_def}
	\alpha
	\equiv
	\frac{\Delta\rho}{\rho_{\rm rad}}
	\bigg|_{T_i^*},
	\qquad
	\rho_{\rm rad}(T)=\frac{\pi^2}{30}g_*(T)\,T^4.
\end{equation}
We recall that the energy density of the dark sector 
can be directly obtained from the corresponding distribution functions via 
$\rho = \sum_j g_j\int_pE\,f_j$ (c.f. Eq.~\eqref{eq:thermoquantities}). We can also rely on the useful identity
\begin{equation}
	\rho = -P + T_i\frac{\partial P}{\partial T_i} = V_\text{eff} - T_i \frac{\partial V_\text{eff}}{\partial T_i}
\end{equation}
to compute the energy density. Hence,
\begin{equation}
	\label{eq:Drho_def}
	\Delta\rho
	\equiv
	\Delta V_{\rm eff}
	- T_i\,\frac{\partial}{\partial T_i}\Delta V_{\rm eff},
	\qquad
	\Delta V_{\rm eff}\equiv
	V_{\rm eff}(\phi_{\rm true})-V_{\rm eff}(\phi_{\rm false}).
\end{equation}
This captures latent heat effects in a way that is well suited to an effective-potential computation.
%
%
%
%
Another useful quantity that characterizes the time scale of the transition is commonly encoded in the parameter \(\beta\), defined as the time derivative of the logarithm of the nucleation rate,
\begin{equation}
	\label{eq:beta_def_general}
	\beta \;\equiv\; \left.\frac{d}{dt}\ln \Gamma(t)\right|_{t_*} =\left.a\,H\,\frac{d}{da}\ln \Gamma(a)\right|_{a_*}\,,
\end{equation}
where \(t_*\) (equivalently \(a_*\)) denotes the time at the characteristic transition epoch.
%
%
%
%
%
The parameters $\alpha$ and $\beta/H$ characterize, respectively, the strength and
the time scale of the phase transition, but they do not fully determine its
dynamical evolution. In a first-order phase transition, the conversion of the
false vacuum into the true vacuum proceeds through the nucleation and subsequent
expansion of bubbles of the broken phase. Once nucleated, a bubble grows because
the pressure difference between the two phases accelerates the bubble wall
outwards, converting vacuum energy into kinetic energy of the wall and of the
surrounding plasma.

The dynamics of this expansion is governed by a balance between the driving force,
set by the released vacuum energy $\Delta\rho$, and friction forces arising from
interactions between the bubble wall and particles in the plasma. These
interactions transfer energy and momentum from the scalar field configuration
forming the wall to the surrounding medium, limiting the acceleration of
the wall. As a result, the bubble wall typically reaches a terminal velocity
$v_w$, which can range from subsonic to ultra-relativistic values depending on the
particle content of the theory and the strength of the couplings to the plasma. In fact,
the bubble wall velocity plays a central role in the generation of gravitational
waves~\cite{Dorsch:2018pat}: a rapidly expanding wall deposits energy efficiently into bulk fluid
motions of the plasma, leading to long-lasting acoustic waves that constitute one
of the dominant sources of GWs~\cite{Esposito:2018sdc}, which typically is the main source 
of a GW background when the dark sector's FOPT occurs in thermal equilibrium with the SM, 
as the friction with SM particles becomes substantial. In contrast, if friction is weak and the wall
continues to accelerate (the runaway regime), a significant fraction of
the vacuum energy remains stored in the scalar field gradients themselves, and GW
production from bubble collisions can become important~\cite{Hawking:1982ga,Jinno:2016vai}. The wall velocity
therefore controls not only the overall amplitude of the GW signal, but also the
relative importance of the different production mechanisms.

To quantify how the released vacuum energy is distributed among these channels, it
is customary to introduce efficiency factors $\kappa$, which encode the fraction
of $\Delta\rho$ transferred into different components:
\begin{equation}
	\kappa_\phi:\ \Delta\rho \;\to\; \rho_\phi,
	\qquad
	\kappa_{\rm sw}:\ \Delta\rho \;\to\; \rho_{\rm bulk\ fluid},
	\qquad
	\kappa_{\rm turb}:\ \Delta\rho \;\to\; \rho_{\rm turbulence}.
\end{equation}
Here $\rho_\phi$ denotes energy stored in the scalar field gradients of the bubble
walls, $\rho_{\rm bulk\ fluid}$ the kinetic energy of coherent plasma motions
(sound waves), and $\rho_{\rm turbulence}$ the energy in magnetohydrodynamic
turbulence generated at later times. In phenomenological treatments one often
assumes $\kappa_{\rm turb}\sim \epsilon\,\kappa_{\rm sw}$ with
$\epsilon\sim\mathcal{O}(10^{-2}\!-\!10^{-1})$, reflecting the expectation that only
a small fraction of the bulk fluid energy cascades into fully developed
turbulence. Together, the parameters $(\alpha,\beta/H,v_w,\kappa)$ provide a macroscopic
description of the phase transition that directly maps onto predictions for the
resulting stochastic gravitational-wave background. While their precise values
depend on the microscopic particle-physics model and on the thermal history of the
plasma, their physical interpretation is universal and largely independent of the
details of the underlying theory.

The present-day spectrum of a GW signal from a FOPT is conventionally written as a sum of contributions
associated with the different sources of anisotropic stress active during and
after the transition,
\begin{equation}
	\label{eq:OmegaGW_sum}
	\Omega_{\rm GW}(f)
	=
	\Omega_{\rm col}(f)
	+
	\Omega_{\rm sw}(f)
	+
	\Omega_{\rm turb}(f),
\end{equation}
corresponding, respectively, to scalar-field gradients during bubble collisions,
acoustic (sound) waves in the plasma, and magnetohydrodynamic turbulence generated
at later times. Note that the collision and turbulence contributions
were already discussed as early as~\cite{Kamionkowski:1993fg}, 
while characterization of the sound-waves spectrum
has been studied in~\cite{Hindmarsh:2013xza,Ellis:2020awk}. 
Which of these contributions dominates depends sensitively on the
bubble wall velocity and on how efficiently the released vacuum energy is
transferred to the surrounding plasma. In most non-runaway transitions, the sound
wave contribution provides the leading signal, while scalar-field collisions
become relevant mainly if friction is sufficiently weak for the wall to continue
accelerating.

The GW spectrum is usually expressed in terms of the present-day energy density per
logarithmic frequency interval,
\begin{equation}
	\Omega_{\rm GW}(f)
	\equiv
	\frac{1}{\rho_{c}}\,
	\frac{d\rho_{\rm GW}}{d\ln f},
\end{equation}
The
characteristic frequency scale of the signal is set by the typical bubble size at
the time of collision, $R_*\sim v_w/\beta$, redshifted from the transition epoch to
the present. Assuming a standard radiation-dominated expansion after the phase
transition and a dark sector in kinetic equilibrium with SM model ($T_i =T$), the peak frequency scales parametrically as
\begin{equation}
	\label{eq:f_scaling_generic}
	f_{\rm peak}
	\propto
	\left(\frac{\beta}{H_*}\right)
	\left(\frac{T_*}{100~{\rm GeV}}\right)
	\left(\frac{g_*}{100}\right)^{1/6},
\end{equation}
so that transitions occurring near the electroweak scale naturally peak in the milli-Hertz range, making them prime targets for space-based GW observatories. Here \(H_* \equiv H(t_*)\) is the Hubble rate evaluated at the transition epoch.

In phenomenological studies, the dominant contribution for non-runaway walls is
typically associated with long-lived acoustic waves in the plasma. A commonly
used fit for this contribution is~\cite{Hindmarsh:2013xza}
\begin{equation}
	\label{eq:Omega_sw_fit_thesis}
	\Omega_{\rm sw}(f)\,h^2
	\simeq
	2.65\times 10^{-6}
	\left(\frac{H_*}{\beta}\right)
	\left(\frac{\kappa_{\rm sw}\,\alpha}{1+\alpha}\right)^2
	\left(\frac{100}{g_*}\right)^{1/3}
	v_w\,
	S_{\rm sw}(f),
\end{equation}
with spectral shape
\begin{equation}
	\label{eq:S_sw}
	S_{\rm sw}(f)
	=
	\left(\frac{f}{f_{\rm sw}}\right)^3
	\left[
	\frac{7}{4+3\left(f/f_{\rm sw}\right)^2}
	\right]^{7/2},
\end{equation}
and peak frequency
\begin{equation}
	\label{eq:f_sw_thesis}
	f_{\rm sw}
	\simeq
	1.9\times 10^{-5}\,{\rm Hz}\,
	\frac{1}{v_w}
	\left(\frac{\beta}{H_*}\right)
	\left(\frac{T_*}{100~{\rm GeV}}\right)
	\left(\frac{g_*}{100}\right)^{1/6}.
\end{equation}
These expressions capture the parametric dependence of the GW amplitude and peak
frequency on the strength and time scale of the transition, and provide a reliable
baseline for order-of-magnitude forecasts.

At later times, part of the bulk fluid motion may cascade into magnetohydrodynamic
turbulence, sourcing an additional GW component. A widely used phenomenological
fit for this contribution is~\cite{Caprini:2009yp}
\begin{equation}
	\label{eq:Omega_turb_fit_thesis}
	\Omega_{\rm turb}(f)\,h^2
	\simeq
	3.35\times 10^{-4}
	\left(\frac{H_*}{\beta}\right)
	\left(\frac{\kappa_{\rm turb}\,\alpha}{1+\alpha}\right)^{3/2}
	\left(\frac{100}{g_*}\right)^{1/3}
	v_w\,
	S_{\rm turb}(f),
\end{equation}
with
\begin{equation}
	\label{eq:S_turb}
	S_{\rm turb}(f)
	=
	\frac{(f/f_{\rm turb})^3}
	{\left[1+(f/f_{\rm turb})\right]^{11/3}\,
		\left(1+8\pi f/h_*\right)},
\end{equation}
and a peak frequency parametrically similar to that of the sound-wave signal. In
practice, this contribution is typically subdominant and is often estimated using
the relation $\kappa_{\rm turb}\sim\epsilon\,\kappa_{\rm sw}$ with
$\epsilon\sim\mathcal{O}(10^{-2}\!-\!10^{-1})$.

If the bubble walls enter a runaway regime, a non-negligible fraction of the
released vacuum energy can remain stored in scalar-field gradients, giving rise to
an additional contribution from bubble collisions. In the envelope approximation
this contribution may be written as~\cite{Huber:2008hg}
\begin{equation}
	\label{eq:Omega_col_fit_thesis}
	\Omega_{\rm col}(f)\,h^2
	\simeq
	1.67\times 10^{-5}
	\left(\frac{H_*}{\beta}\right)^2
	\left(\frac{\kappa_{\phi}\,\alpha}{1+\alpha}\right)^2
	\left(\frac{100}{g_*}\right)^{1/3}
	\left(\frac{0.11\,v_w^3}{0.42+v_w^2}\right)
	S_{\rm col}(f),
\end{equation}
with
\begin{equation}
	\label{eq:S_col}
	S_{\rm col}(f)
	=
	\frac{3.8\,(f/f_{\rm col})^{2.8}}
	{1+2.8\,(f/f_{\rm col})^{3.8}}.
\end{equation}
Although this term is not expected to dominate in the scenarios explored in this
thesis, it is included here for completeness and to illustrate how different
microscopic realizations of a phase transition can lead to qualitatively distinct
GW signatures.


It is worth emphasizing that, in practice, the computation of these parameters
for a given model is often performed using dedicated numerical tools developed
for first-order phase transitions. 
A prominent example is \texttt{CosmoTransitions}~\cite{Wainwright:2011kj},
which provides a flexible framework for multi-scalar effective potentials,
including automated minimization at finite temperature and numerical solutions
of the bounce equations governing vacuum decay. Such packages have been
instrumental in enabling systematic studies of GW signals from phase
transitions and is used as a benchmark code in the literature~\cite{Camargo-Molina:2013qva,Caprini:2019egz}. 
An important limitation of this package is its formulation
under the assumption of full thermal equilibrium with the SM plasma.
In this setting, the bounce action is computed using the standard finite-temperature
formalism, with boundary conditions corresponding to a homogeneous thermal bath
and a background field configuration approaching a fixed vacuum expectation
value at spatial infinity. While this treatment is well justified for transitions
occurring in equilibrium, it becomes inadequate when the dark sector is not in
thermal contact with the SM and its thermodynamic properties evolve
independently. We will study such a case carefully in Chapter~\ref{ch:5} for the case
of an inverse FOPT induced via the freeze-in mechanism.



%% file: Chapters/Chapter3.tex

\chapter{Origin of cannibal dark sectors}\label{ch:3} 


\begin{chapterpublication}
This chapter is largely based on the publication titled
\textit{Freezing-in cannibal dark sectors}~\cite{Cervantes:2024ipg} by E.~Cervantes and A.~Hryczuk,
DOI~[\href{https://link.springer.com/article/10.1007/JHEP11(2024)050}{10.1007/JHEP11(2024)050}].
The chapter also incorporates elements of the publication in preparation with A.~Hryczuk and S.~Lederer on
\textit{Light Dark Matter from Self-cooling Dark Sectors}.

\vspace{0.5em}

The author of this thesis contributed to this body of work in three main directions, under the guidance and feedback of A.~Hryczuk throughout the development of the project.
First, the author participated in the conceptual development of the beyond-the-Standard-Model scenarios studied in this chapter, in particular in the introduction of a decaying mediator into the framework.
Second, the author carried out the numerical implementation and phenomenological analysis presented here, including the development of the numerical solvers, the parameter scans, the implementation of experimental constraints, and the production of the plots and figures.
Third, the author contributed to the drafting of the manuscript and to the presentation and interpretation of the main results.

For the \textit{Light Dark Matter from Self-cooling Dark Sectors} project, the author contributed through active discussions on the direction of the work, the further development of the numerical routines describing the cannibal-sector dynamics, and the interpretation of the relevant thermodynamic aspects. 
\end{chapterpublication}

\section{Production mechanisms}

\subsection{Freeze-in}

An alternative to thermal freeze-out for generating the observed dark matter abundance is the freeze-in (FI) mechanism~\cite{Hall:2009bx}, which we analyse in detail below. This scenario rests on two essential assumptions. First, the dark sector is taken to be either empty or negligibly populated at the end of reheating. Second, interactions linking the dark and visible sectors are extremely weak, such that dark matter is slowly produced from the Standard Model plasma while never attaining thermal equilibrium with it. 
In the realisations considered here, the interaction responsible for populating the dark sector is assumed to be renormalizable. In this case, production is typically dominated at temperatures comparable to the masses of the particles that decay or annihilate populating the dark sector, and the resulting relic abundance is largely insensitive to the reheating temperature. We note, however, that freeze-in can also arise from higher-dimensional operators (as in gravitino or axino production), in which case the relic density may instead be sensitive to the highest temperatures attained in the early Universe~\cite{Bernal:2025fdr}. Once freeze-in production becomes inefficient, the visible and dark sectors evolve independently, and the dynamics of the dark sector are thereafter entirely determined by the energy density accumulated during the production stage.

A characteristic feature of freeze-in scenarios is the feeble portal coupling, which calls for a justification. Given a UV completion,\footnote{Specifying the full UV completion is not required for the FI mechanism itself, which can consistently be formulated within an effective field theory. Rather, the UV framework provides a possible explanation for the origin of the feeble coupling.} such a small coupling may naturally arise from the symmetry breaking of a heavy state whose vacuum expectation value is associated with the UV theory. In this case, the portal coupling may scale parametrically as $\lambda_p \sim 1/\Lambda_{\text{UV}} \ll 1$, where $\Lambda_\text{UV}$ denotes the VEV or mass of a heavy state associated with the UV theory. However, detection prospects remain highly challenging. Current DD experiments probe only a limited subset of scenarios involving feeble portals, typically requiring the dark sector particle to lie at or below the MeV mass scale~\cite{Hambye:2018dpi,Bhattiprolu:2023akk}. 

\subsection{Internal freeze-out}

\begin{figure}[t!]
	\centering
	\includegraphics[width=0.49\textwidth]{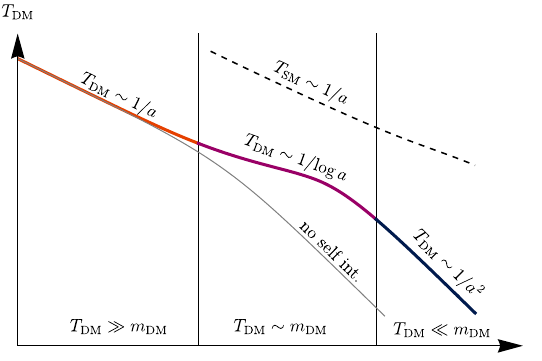}
	\hfill
	\includegraphics[width=0.48\textwidth]{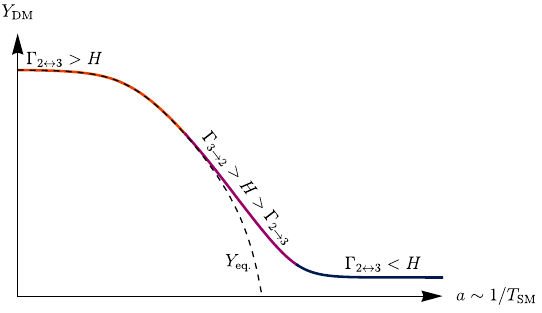}
	\caption{Schematic temperature (left) and yield (right) evolution of a cannibal DM candidate as a function of the scale factor. The sketch highlights the three relevant phases of the evolution. When the system is relativistic, the DM behaves as radiation and its temperature scales with the scale factor as $T_\text{DM}\propto 1/a$, while $\Gamma_{3\leftrightarrow 2}>H$ enforce $Y=Y_\text{eq}$ (red). As the system cools, the DM enters the cannibal phase where $T_\text{DM}\propto 1/\log(a)$ and $\Gamma_{3\to 2}>H>\Gamma_{2\to 3}$ mark the departure from chemical equilibrium (purple). Finally, in the last stage all the rates decouple, the yield is constant and the system turns into a non-relativistic gas, $T_\text{DM}\propto1/a^2$ (blue). The temperature of the SM is shown (dashed line) to highlight that $T_\text{SM}>T_\text{DM}$ in all relevant stages to ensure successful structure formation (see main text).}
	\label{fig:evol_DM_sketched}
\end{figure}

Yet another production mechanism at the core of this work is provided by models of self-interacting dark matter (also known as \textit{cannibal} dark matter), in which the relic abundance is determined by a freeze-out process occurring entirely within the dark sector itself. Here both sectors are (up to gravity) typically secluded and evolve independently,\footnote{As we will see below, this is not a strict requirement. However, the production mechanism was originally proposed under this premise.} and it is the number-changing self-interactions among dark sector particles that ensure chemical equilibrium ($\mu_\text{ds} = 0$) and set the final dark matter relic abundance after decoupling, that is, once self-number-changing reactions occur at a rate exceeding the Hubble expansion rate, the system may eventually reach chemical equilibrium, followed by a phase denoted \textit{cannibalism} (internal freeze-out). The three relevant stages of this evolution are sketched in~\Cref{fig:evol_DM_sketched}. In this sense, cannibalization arises once a sufficiently populated and self-interacting dark sector is present, and does not rely on special assumptions about the ultraviolet origin of the dark sector or its coupling to the Standard Model. This observation is particularly relevant for freeze-in, where the feeble portal interaction ensures that the dark sector evolves autonomously once populated, making self-thermalization and cannibal dynamics a generic possibility rather than a model-dependent choice.

This mechanism was first proposed in~\cite{1992ApJ...398...43C} and has since been explored in a wide range of cosmological contexts. In cannibal dark matter scenarios, the leading number-changing processes are most often \(4\to2\) and \(3\to2\). Such interactions arise when dark matter stability is protected by a \(\mathbb{Z}_2\) or \(\mathbb{Z}_3\) symmetry. In particular, the latter permits cubic self-couplings~\cite{Choi:2015bya,Bernal:2015bla,Bernal:2015lbl,Ko:2014nha,Choi:2017mkk,Chu:2017msm}. By contrast, if dark matter is stabilized by a \(\mathbb{Z}_2\) symmetry, all \(3\to2\) processes are forbidden, and the relic abundance is instead controlled by higher-order \(4\to2\) reactions~\cite{Bernal:2015xba,Heikinheimo:2016yds,Bernal:2017mqb,Heikinheimo:2017ofk,Bernal:2018hjm}. Note that if the $\mathbb{Z}_2$ symmetry is spontaneously broken, $3\to 2$ annihilation channels may also open, as we will see in the following section. Cannibalism is not exclusive to scalars charged under discrete or gauge symmetries, but can also occur, for example, in dark-pion-like theories~\cite{Beauchesne:2019ato}, as well as in theories involving fermions or vector bosons, where the external particles are not necessarily all identical.

A variety of mechanisms have been proposed to account for the origin of cannibal DM in the early Universe, including purely gravitational particle production~\cite{Chung:2001cb}, decay processes involving the inflaton~\cite{Takahashi:2007tz,Moroi:2020has}, asymmetric reheating~\cite{Feng:2008mu}, and vacuum decay dynamics~\cite{Asadi:2021pwo}. In fact, during the production of cannibal dark matter, its evolution can exhibit a nontrivial feedback between the energy stored in the dark sector and its redistribution through number-changing self-interactions. These processes allow the dark sector to convert kinetic energy into particle number while simultaneously cooling, leading to a dynamical enhancement of dark matter production relative to naive freeze-in expectations~\cite{Bernal:2015xba,Heeba:2018wtf,March-Russell:2020nun,Bernal:2020gzm}. Capturing this effect reliably requires a self-consistent treatment of both the energy (or temperature) evolution and the number density of the dark sector.

The absence of any definitive experimental identification has motivated the exploration of a wide range of particle candidates with suppressed interactions with the visible sector. Examples include sterile neutrinos~\cite{Boyarsky:2018tvu}, ultralight scalar fields such as fuzzy dark matter~\cite{Anchordoqui:2023tln}, and axions~\cite{OHare:2024nmr}. These frameworks differ primarily in their field content, symmetry structure, and mass scales, and may be realized within either top-down or bottom-up approaches. Cannibal dark matter, by contrast, does not correspond to a distinct particle candidate, but rather to a dynamical regime that may arise in self-interacting dark sectors once they are sufficiently populated. In particular, scalar dark matter models, including axion-like or fuzzy dark matter extensions, can in principle exhibit cannibal dynamics if efficient number-changing self-interactions are present. From this perspective, cannibalization should be viewed not as a property of a specific candidate, but as a generic dynamical possibility in theories containing self-number-changing reactions.

\subsection{Bounds on freeze-in and cannibalism}

\subsubsection{Small-scale-structure and Lyman-\texorpdfstring{$\alpha$}{alpha} constraints}

During the cannibal phase, number-changing reactions convert rest mass into kinetic energy, heating the dark sector as its particle number decreases. As a consequence, dark matter can remain warm for an extended period and acquire an enhanced effective sound speed. This delays the onset of efficient gravitational clustering and suppresses the growth of density perturbations below a characteristic scale. In Fourier space, this appears as a damping of the matter power spectrum at large wavenumbers. The impact of such self-heating dark matter on structure formation was already noted in~\cite{1992ApJ...398...43C} and has been extensively revisited in later studies~\cite{deLaix:1995vi,Irsic:2017ixq,Chatterjee:2019olf,Buen-Abad:2018mas,Erickcek:2020wzd}.

This suppression may lead to significant constraints from small-scale-structure observables, in particular Lyman-$\alpha$ forest measurements and galaxy clustering data. Since CMB lensing and related secondary anisotropies are sensitive to the integrated matter distribution, they can also provide complementary bounds~\cite{Heimersheim:2020aoc}. Several mechanisms have been proposed to alleviate these constraints. One possibility is to assume that the dark sector is initially colder than the visible sector, such that the heating induced during cannibalization remains compatible with structure formation~\cite{Hufnagel:2022aiz,Farina:2016llk,Ghosh:2022asg,Arcadi:2019oxh}. Another possibility is to introduce a feeble but non-negligible coupling to the Standard Model, allowing the excess kinetic energy generated during the cannibal phase to be partially transferred to the visible sector. This realization, commonly referred to as the SIMP mechanism, was introduced in~\cite{Hochberg:2014dra,Hochberg:2014kqa} and further explored in~\cite{Bernal:2015bla,Kuflik:2015isi,Pappadopulo:2016pkp,Kuflik:2017iqs}. Related variations consider unstable mediators whose decay can reduce the kinetic energy stored in the dark sector during cannibalization~\cite{Yang:2023xgk}.



In the scenarios considered in this work, however, the connection between cannibal dynamics and Lyman-$\alpha$ constraints is more subtle. The suppression of the matter power spectrum depends on the dark matter momentum distribution and, in particular, on the typical particle velocities at the onset of structure formation. While dark matter never thermalizes with the SM bath, elastic self-scatterings within the dark sector may remain efficient well after number-changing reactions have frozen out. 
As long as these elastic processes satisfy $\Gamma_{\rm el}\gg H$, the dark sector behaves as a collisional fluid rather than as a collisionless gas. In that regime, the mean free path remains short and the onset of ordinary collisionless free streaming is delayed, even when the particles still carry sizeable velocities. The impact on structure formation is then controlled not by standard free streaming alone.


By the time elastic self-scatterings decouple and dark matter becomes effectively collisionless, it is already sufficiently non-relativistic, yielding a parametrically small free-streaming length~\cite{Decant:2021mhj}. The resulting suppression of the matter power spectrum on Lyman-$\alpha$ scales is therefore expected to be much weaker than in the standard thermal-relic warm-dark-matter benchmark, making such bounds subdominant over most of the parameter space considered here. 


A different situation arises when the mediator undergoes cannibalization, but the dark matter produced in its decays is effectively collisionless. In that case, the dark matter no longer behaves as a self-interacting fluid, yet its phase-space distribution still carries the imprint of the mediator thermal history. Because number-changing reactions modify the mediator temperature evolution, they also affect the momentum spectrum inherited by the dark matter. The final distribution therefore generally differs from the thermal or quasi-thermal forms usually assumed in standard warm-dark-matter analyses. The corresponding Lyman-\(\alpha\) constraints must then be obtained from the actual phase-space distribution, rather than imported from standard thermal-WDM bounds. We carry out this analysis in~\Cref{sec:wDM}.



For reference, present Lyman-$\alpha$ analyses typically constrain the mass of a thermal-relic warm-dark-matter particle to be at least of order a few~keV~\cite{Kamada:2019kpe,Villasenor:2022aiy,Irsic:2023equ}. However, these limits cannot be straightforwardly translated to the present scenario, since dark matter produced from mediator decays in a cannibal background generally exhibits a non-thermal and model-dependent phase-space distribution. Establishing the corresponding bound would require a dedicated treatment of the resulting matter-power-spectrum suppression, or an appropriate mapping to an equivalent thermal-relic description~\cite{Ballesteros:2020adh}.

\subsubsection{Primordial isocurvature constraints}

Distinct from constraints associated with the late-time evolution of density perturbations, cosmological observations are also sensitive to the primordial initial conditions of the dark matter distribution. In particular, isocurvature modes correspond to fluctuations in the relative abundances of different species at fixed total energy density on superhorizon scales, and are therefore determined by primordial initial conditions rather than by the subsequent microphysical evolution of the dark sector. In the absence of a primordial isocurvature component, freeze-in production does not generate isocurvature perturbations on cosmological scales.\footnote{This statement assumes that the primordial perturbations of the radiation bath are purely adiabatic and that dark matter is produced locally from Standard Model particles through freeze-in interactions whose rates depend only on local thermodynamic quantities. Under these conditions, freeze-in production does not generate isocurvature perturbations, but merely transfers the pre-existing adiabatic fluctuations of the visible sector to the dark matter. If, instead, a fraction of the dark matter abundance, or its parent field, is already present during inflation, or if the production rate depends on an additional fluctuating field, a primordial isocurvature component may arise and persist due to the feeble nature of freeze-in interactions.}

If a fraction of the dark matter abundance is already present at the end of inflation and carries an initial isocurvature perturbation, the feeble and out-of-equilibrium nature of freeze-in interactions prevents its efficient erasure, in contrast to the case of thermal dark matter~\cite{Bellomo:2022qbx}. Cannibal interactions, which redistribute energy and particle number internally within the dark sector while conserving its total energy density, do not modify this conclusion and neither generate nor erase primordial isocurvature modes. Any residual isocurvature component is therefore constrained by CMB temperature, polarization, and lensing anisotropies, independently of the cannibal dynamics, which affect only the subsequent evolution of perturbations rather than their primordial origin.

\section{An unstable real scalar candidate}\label{sec:real_scalar}

We begin with the simplest realization of particle dark matter: a real scalar singlet charged under a discrete $\mathbb{Z}_2$ symmetry. We denote this candidate by $\varphi$ and emphasize that it is a singlet under all gauge groups of the SM. The interaction between $\varphi$ and the SM proceeds through the Higgs portal, described by the renormalizable operator
\begin{equation}\label{eq:HP_par}
	V_{\rm HP}(H,\varphi)=\frac{1}{2}\,\lambda_{h\varphi}\,\varphi^2\,H^\dagger H \, ,
\end{equation}
where $H$ is the Higgs doublet of the SM and it is given by $H=(0,h+v_h)^\intercal/\sqrt{2}$ in unitary gauge after the electroweak phase transition (EWPT), i.e., here
$h$ corresponds to the real mode, and $v_h$ to its VEV.

The phenomenological impact of the Higgs portal is largely determined by how efficiently it mediates the exchange of energy and chemical potential between the visible and dark sectors in the early universe. For sufficiently large values of $\lambda_{h\varphi}$, Higgs-mediated scattering and decay processes occur rapidly compared to the Hubble expansion rate, enforcing both kinetic and chemical equilibrium between $\varphi$ and the SM plasma. In this regime, the cosmological evolution of the singlet closely resembles that of a conventional WIMP relic, with its abundance set by thermal freeze-out once interaction rates fall below the expansion rate.

As the portal coupling is reduced, this picture gradually changes. Below a critical strength, Higgs-mediated interactions become too inefficient to equilibrate the two sectors, and $\varphi$ instead evolves as an out-of-equilibrium species. In this limit, the dark matter population is generated predominantly through rare production events, such as Higgs decays or annihilations. When these processes dominate, reproducing the observed dark matter density typically selects portal couplings of order $\lambda_{h\varphi}\sim 10^{-9}$ for singlet masses in the GeV scale~\cite{Lebedev:2019ton}. Even smaller couplings render the Higgs portal cosmologically irrelevant, leaving gravitational particle production as the dominant source of $\varphi$.

On the other hand, the dynamics of the singlet field are governed by the potential
\begin{equation}\label{eq:Vself_par}
	V_{\rm self}(\varphi)=\frac{1}{2}\,\mu^2\varphi^2+\frac{\lambda}{4!}\,\varphi^4 \, .
\end{equation}
In the following, we also consider the possibility that the $\mathbb{Z}_2$ symmetry is explicitly or spontaneously broken on the cosmological time scales relevant for our analysis. This assumption serves two purposes. First, symmetry breaking induces a cubic self-interaction term in the Lagrangian, opening up $3\to2$ processes, which occur with higher frequency than $4\to2$ reactions due both to the lower power of the self-coupling and to the reduced phase-space suppression. Second, although often undesirable, the resulting instability of the DM candidate can have phenomenologically relevant implications for experiments and observations. Since the Higgs portal constitutes the only non-gravitational interaction with the SM, the singlet can remain cosmologically long-lived provided the portal coupling is sufficiently small. In earlier studies this issue was circumvented by setting the portal coupling exactly to zero~\cite{Hufnagel:2022aiz}. Here, we relax this assumption and allow for a nonvanishing Higgs portal, focusing on the case in which symmetry breaking arises from a nonzero vacuum expectation value of $\varphi$,
\begin{equation}\label{eq:vevphi_par}
	\langle \varphi \rangle \equiv w
	= \pm \sqrt{\frac{3}{\lambda}}\,
	\sqrt{v_h^2\lambda_{h\varphi}-2\mu^2} \, .
\end{equation}

Once $\varphi$ acquires a VEV, mixing between the singlet and the Higgs boson is induced. As a result, a Higgs-like scalar state appears whose decay width into SM particles scales as $\sin^2\theta$, implying a lifetime $\tau_\varphi\propto \sin^{-2}\theta$. The mixing angle is
\begin{equation}\label{eq:sinth}
	\sin 2\theta  =\frac{4\,v_h\,w\,\lambda_{h\varphi}}{m_\varphi^2-m_h^2}\,,
\end{equation} 
and observations demand this mixing angle to be suppressed, which can be naturally accommodated in the limit $\lambda_{h\varphi}\ll 1$. 
This condition can be translated into a constraint on the portal coupling, first we expanding $w$ as
\begin{equation}
	w=\sqrt{\frac{3}{\lambda}}
	\left(
	m_\varphi
	+\frac{3m_\varphi v_h^2}{\lambda(2m_h^2-2m_\varphi^2)}\,\lambda_{h\varphi}^2
	\right)
	+\mathcal{O}(\lambda_{h\varphi}^3)\, .
\end{equation}
In the limit $m_\varphi\ll m_h$, the leading-order mixing angle is then given by
\begin{equation}\label{eq:DM_Higgs_mixing_par}
	\theta\simeq
	\frac{2\sqrt{3}\,\lambda_{h\varphi}\,m_\varphi\,v_h}
	{(m_\varphi^2-m_h^2)\sqrt{\lambda}}
	\qquad \Rightarrow \qquad
	\Gamma_{\varphi\to {\rm SM\,SM}}
	\propto
	\frac{\lambda_{h\varphi}^2\,m_\varphi^2\,v_h^2}
	{m_h^4\,\lambda}\, .
\end{equation}
In other words, the DM is long-lived in the FI regime. 

A well-known consequence of spontaneously breaking a discrete symmetry in the early universe is the emergence of topological defects. In the present context, the breaking of the $\mathbb{Z}_2$ symmetry generically induces a dark-sector phase transition, during which domain walls may form. If sufficiently long-lived, such defects can quickly come to dominate the total energy density, leading to severe cosmological inconsistencies, as discussed in~\cite{Saikawa:2017hiv}. Avoiding this outcome requires the domain walls to be sufficiently light, which translates into an upper bound on their surface tension $\sigma$. Cosmological considerations impose $\sigma \lesssim \mathcal{O}(\mathrm{MeV})$, and using the parametric estimate $\sigma \sim \sqrt{\lambda}\,w^3$~\cite{Zeldovich:1974uw}, this condition implies $\sqrt{\lambda}\,w^3 \lesssim \mathrm{MeV}$. Taken at face value, such a bound appears to severely restrict (if not exclude) the phenomenologically accessible parameter space of the model.

It is important to stress, however, that these constraints rely on assumptions about the thermal history and stability of the dark sector that are not universal. There exist well-motivated scenarios in which the standard domain-wall bounds are substantially weakened or entirely avoided, for instance if the dark sector reheats to a temperature below the symmetry-breaking scale, or if the domain walls are unstable and decay sufficiently early~\cite{Hufnagel:2022aiz}. An explicit, albeit small, breaking of the $\mathbb{Z}_2$ symmetry provides another viable way to circumvent these arguments. Since the present analysis does not hinge on the specific resolution of the domain-wall problem, we do not pursue this issue further. 

\begin{figure}[t!]
	\centering
	\begin{tabular}{c@{\hspace{1.2cm}}c@{\hspace{1.2cm}}c}
		\begin{tikzpicture}[baseline=(b)]
			\begin{feynman}
				\vertex (a1);
				\vertex [below right=1.1cm of a1] (b);
				\vertex [below=0.7cm of b](a2);
				\vertex [right=1.5cm of b](c);
				\vertex [above right=1.2cm of c](d1);
				\vertex [below right=1.2cm of c](d2);
				\vertex [above left=1.1cm of a2](e1);
				\vertex [below left=1.1cm of a2](e2);
				\diagram* {
					(a1) -- [scalar] (b),
					(a2) -- [scalar] (b) --[scalar](c),
					(c) -- [scalar](d1),
					(c) -- [scalar](d2),
					(a2) -- [scalar](e1),
					(a2) -- [scalar](e2),
				};
			\end{feynman}
		\end{tikzpicture} &
		
		\begin{tikzpicture}[baseline=(in)]
			\begin{feynman}
				\vertex [dot] (a);
				\vertex [below =0.7cm of a] (in);
				\vertex [below =1cm of in] (b);
				\vertex [left =0.7cm of a] (i1);
				\vertex [left =1.5cm of b] (i2) ;
				\vertex [right =1.5cm of b] (f1) ;
				\vertex [right =1.5cm of a] (f2) ;
				\vertex [below left=1cm of i1] (ii2) ;
				\vertex [left=0.8cm of i1] (ii1) ;
				\diagram* {
					(i1) -- [scalar](a)
					-- [scalar] (f2),
					(a) -- [scalar] (in),
					(in) -- [scalar] (b),
					(i2) -- [scalar] (b),
					(b) -- [scalar] (f1),
					(ii1) -- [scalar] (i1),
					(ii2) -- [scalar] (i1),
				};
			\end{feynman}
		\end{tikzpicture} &
		
		\begin{tikzpicture}[baseline=(in)]
			\begin{feynman}
				\vertex [dot] (a);
				\vertex [below =0.8cm of a] (in);
				\vertex [below =0.8cm of in] (b);
				\vertex [left =1.5cm of in] (i3);
				\vertex [left =1.5cm of a] (i1);
				\vertex [left =1.5cm of b] (i2);
				\vertex [right =3.2cm of i2] (f1);
				\vertex [right =3.2cm of i1] (f2);
				\diagram* {
					(i1) -- [scalar](a)
					-- [scalar] (f2),
					(a) -- [scalar] (b),
					(i2) -- [scalar] (b),
					(b) -- [scalar] (f1),
					(i3) -- [scalar] (in),
				};
			\end{feynman}
		\end{tikzpicture} \\ \\ 
		
		\begin{tikzpicture}[baseline=(g)]
			\begin{feynman}
				\vertex (a1);
				\vertex [below right=1.2cm of a1] (b);
				\vertex [left =1.2cm of b] (g);
				\vertex [below left=1.2cm of b](a2);
				\vertex [right=of b](c);
				\vertex [above right=1.1cm of c](d1);
				\vertex [below right=1.1cm of c](d2);
				\diagram* {
					(a1) -- [scalar] (b),
					(a2) -- [scalar] (b) -- [scalar] (c),
					(c) -- [scalar](d1),
					(c) -- [scalar](d2),
					(g) -- [scalar](b),
				};
			\end{feynman}
		\end{tikzpicture} &
		
		\begin{tikzpicture}[baseline=(g)]
			\begin{feynman}
				\vertex (a1);
				\vertex [below right=1.1cm of a1] (b);
				\vertex [left =1.1cm of b] (g);
				\vertex [below left=1.1cm of b](a2);
				\vertex [right=of b](c);
				\vertex [above right=1.1cm of c](d1);
				\vertex [below right=1.1cm of c](d2);
				\diagram* {
					(a1) -- [scalar] (c),
					(a2) -- [scalar] (b) -- [scalar] (c),
					(c) -- [scalar](d1),
					(c) -- [scalar](d2),
					(g) -- [scalar](b),
				};
			\end{feynman}
		\end{tikzpicture} &
		
		\begin{tikzpicture}[baseline=(in)]
			\begin{feynman}
				\vertex [dot] (a);
				\vertex [below=0.6cm of a] (in);
				\vertex [below=1cm of in] (b);
				\vertex [left =1.5cm of a] (i1);
				\vertex [below=1cm of i1] (ii2);
				\vertex [left =1.5cm of b] (i2);
				\vertex [right=1.5cm of b] (f1);
				\vertex [right=1.5cm of a] (f2);
				\diagram* {
					(i1) -- [scalar](a)
					-- [scalar] (f2),
					(a) -- [scalar] (b),
					(i2) -- [scalar] (b),
					(b) -- [scalar] (f1),
					(ii2) -- [scalar] (a),
				};
			\end{feynman}
		\end{tikzpicture}
	\end{tabular}
	\caption{Diagrams for the self-number changing reactions.  All initial, virtual and final states are $\varphi$.}
	\label{fig:feynman3-2}
\end{figure}
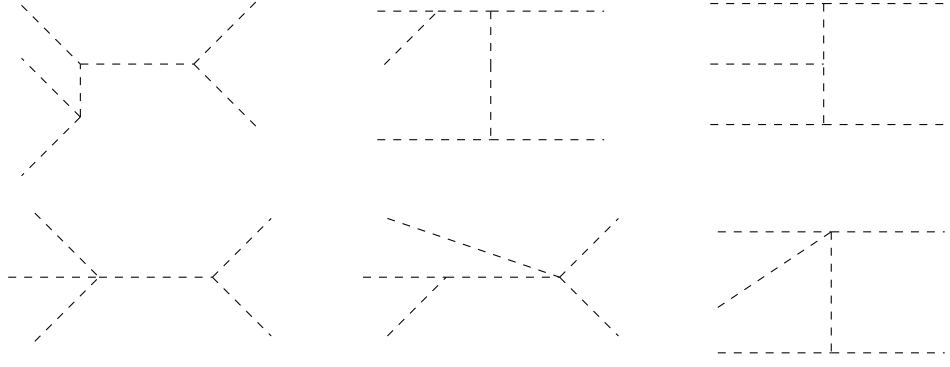

Once the discrete symmetry is broken, the scalar potential acquires additional interaction terms and can be rewritten as
\begin{equation}\label{eq:Z2_broken_phase_potential}
	V_{\rm self}
	= \frac{1}{2}m_\varphi^2 \varphi^2
	+ \frac{g}{3!}\varphi^3
	+ \frac{\lambda}{4!}\varphi^4 \, .
\end{equation}
The parameters appearing in this expression are not independent. In particular, the cubic coupling is fixed by the quartic self-interaction and the physical mass according to $g=\sqrt{3\lambda}\,m_\varphi$ if the symmetry is spontaneously broken, while the mass itself is given by $m_\varphi^2=2|\mu|^2=\lambda v_h^2/3$~\cite{Hufnagel:2022aiz}. As mentioned above, the presence of the cubic term plays a central role in the dynamics of the dark sector, as it enables efficient number-changing reactions already at leading order in the self-coupling. Indeed, the dominant process controlling the chemical evolution of the singlet population is the $3 \leftrightarrow 2$ interaction. This reaction arises at the lowest nontrivial order in $\lambda$ and provides the main mechanism through which the dark sector can self-thermalize or reduce its number density. The corresponding matrix element is given in Eq.~\eqref{eq:mat_el_real}, while the associated tree-level Feynman diagrams are shown in Fig.~\ref{fig:feynman3-2}. 

The evolution of the DM is described by the cBE provided in the subsection~\ref{subsec:cBE_pressure}. Note that in this case we consider the second moment stemming from the pressure instead of the energy density, equivalent to~\cite{Cervantes:2024ipg}. The collision term associated with the $3\varphi\leftrightarrow2\varphi$ process takes the form
\begin{equation}\label{eq:C_3to2_realscalar}
	\begin{split}
		C_{3\varphi\leftrightarrow 2\varphi}
		=\frac{1}{2E_\varphi g_\varphi}
		\int \Bigg(
		&- f_\varphi(p)\,
		\vert\tilde{\mathcal{M}}_{\underline{\varphi}2\to345}\vert^2\,
		f_2\, d\Pi_2
		\left(\frac{1}{3!} d\tilde\Pi_3 d\tilde\Pi_4 d\tilde\Pi_5\right)
		\\
		&+(1+f_\varphi(p))\,
		\vert\tilde{\mathcal{M}}_{12\to\underline{\varphi}45}\vert^2
		\left(\frac{1}{2!} d\Pi_1 d\Pi_2\, f_1 f_2\right)
		\left(\frac{1}{2!} d\tilde\Pi_4 d\tilde\Pi_5\right)
		\\
		&- f_\varphi(p)\,
		\vert\tilde{\mathcal{M}}_{12\leftarrow\underline{\varphi}45}\vert^2
		\left(\frac{1}{2!} d\Pi_4 d\Pi_5\, f_4 f_5\right)
		\left(\frac{1}{2!} d\tilde\Pi_1 d\tilde\Pi_2\right)
		\\
		&+(1+f_\varphi(p))\,
		\vert\tilde{\mathcal{M}}_{\underline{\varphi}2\leftarrow345}\vert^2
		\left(\frac{1}{3!} d\Pi_3 d\Pi_4 d\Pi_5\, f_3 f_4 f_5\right)
		d\tilde\Pi_2
		\Bigg)\, ,
	\end{split}
\end{equation}
where momenta are labeled as $12\leftrightarrow345$, and we have defined
$\vert\tilde{\mathcal{M}}\vert^2$ in Eq.~\eqref{eq:Msq} as well as $d\tilde\Pi_i=d\Pi_i(1+f_i)$, with $d\Pi_i\equiv\frac{d^3p_i}{(2\pi)^3 2E_i}$. The underlined momentum indicates the external $\varphi$ state whose production or annihilation is being tracked, and symmetry factors are shown explicitly. Notice that we incorporated the Bose-enhancement effects, and as we pointed out in the previous chapter, we will in practice approximate $(1+f)\simeq1$, since the DM is dilute at early times and approximately non-relativistic during cannibalization. The zeroth and second moments of the collision operator are given in Equations~\eqref{eq:C0_self} and~\eqref{eq:C2_self}, respectively.

As initial conditions we set the dark-sector number density to its equilibrium value at the initial DM temperature,
\begin{equation}
	n_\varphi^i = n_\varphi^{\rm eq}(T_\varphi^i)\,,
\end{equation}
and parametrize the relative coldness of the dark sector by the free parameter
\begin{equation}
	\xi_\infty \equiv \frac{T_\varphi^i}{T^i}<1\,.
\end{equation}
This choice of initial condition is not merely a technical detail, but a crucial ingredient for the phenomenological viability of the model. In particular, $\xi=T_\varphi/T<1$ throughout the evolution is a requirement to the successful formation of large structures.
Additionally, we take the initial SM temperature to coincide with the electroweak phase transition scale,
\begin{equation}
	T^i \equiv T_{\rm EWPT}=150~{\rm GeV}\,.
\end{equation}

Two remarks are useful at this point. First, the cBE employed here track production \emph{after} the EWPT and neglect contributions from \emph{before} and \emph{during} the transition. For sub-GeV dark matter, and hence in the limit $m_\varphi\ll m_h$, the dominant freeze-in channel is Higgs decay, $h\to\varphi\varphi$, which is most efficient around $T\sim 40~{\rm GeV}$~\cite{Heeba:2018wtf}. During the EWPT the Higgs mass briefly vanishes and then increases as the temperature drops. Then, there can be a moment when $m_h(T)\simeq m_\varphi$, for which the mixing angle in Eq.~\eqref{eq:sinth} becomes resonantly enhanced and $\varphi$ can also be produced via Higgs oscillations. The resulting contribution to the yield can be estimated as~\cite{Heeba:2018wtf}
\begin{equation}
	Y_{\varphi}^{\rm EWPT}
	= \left(1.93\times 10^5~{\rm GeV}^{-4}\right)\lambda_{h\varphi}^2\, m_\varphi^2\,w^2
	= \left(1.93\times 10^5~{\rm GeV}^{-4}\right)\lambda_{h\varphi}^2\,\frac{3m_\varphi^4}{\lambda}\,.
\end{equation}
For $m_\varphi=100~{\rm MeV}$ and $\lambda=10^{-2}$ this gives
$Y_\varphi^{\rm EWPT}\simeq 5.79\times 10^{-15}$, which is subdominant compared to the contribution after EWPT (cf.\ Fig.~\ref{fig:BM1_Z2}).
Second, production from $h\to\varphi\varphi$ is largely insensitive to the detailed EWPT dynamics (and, in practice, also to the precise value of $T_{\rm EWPT}$). The contribution of Higgs-decay into DM to the collision operator is
\begin{equation}\label{eq:C_higgs_decay}
	 C^{h\to\varphi\varphi}
	=
	\frac{(\lambda_{h\varphi} v_h)^2}{8\pi}\,\frac{T}{E_\varphi p_\varphi g_\varphi}\,\left(e^{-E_h^-(p_\varphi)/T} -e^{-E_h^+(p_\varphi)/T} \right)\,,
\end{equation}
where $E_h^\pm(p_\varphi) = \left(m_h^2E_\varphi \pm m_hp_\varphi\sqrt{m_h^2 - 4m_\varphi^2}\right)/(2m_\varphi^2)$. The corresponding zeroth and second moment collision terms are presented in~\Cref{subsec:Higgs_decay} .

The combined effect of the FI mechanism and number-changing self-interactions can lead to qualitatively different thermal histories than pure FI without self-interactions. A convenient classification is in terms of the efficiency of the cannibal reactions, where we identify
three cases: 
\begin{itemize}
\item 1) Number-changing reactions remain always inefficient, i.e., they never compete with the expansion rate, $\Gamma_{2\leftrightarrow 3}\ll H$ at all times. This is the standard freeze-in regime.

\item  2) Number-changing reactions are initially negligible, but become efficient at later times. The DM then attempts to establish chemical equilibrium internally; during this stage, $2\to 3$ reactions rapidly increase the particle number while simultaneously cooling the dark sector~\cite{Bernal:2020gzm,March-Russell:2020nun}.

\item 3) Number-changing reactions remain fast throughout the evolution and the DM stays in chemical equilibrium with itself, and freeze-in effectively proceeds through the internal process $2\to 3$ without a distinct ``re-equilibration'' phase.
\end{itemize}
%
\subsection{Results}\label{sec:Z2_results}

\begin{figure}[t!]
	\centering
	\includegraphics[width=0.55\textwidth]{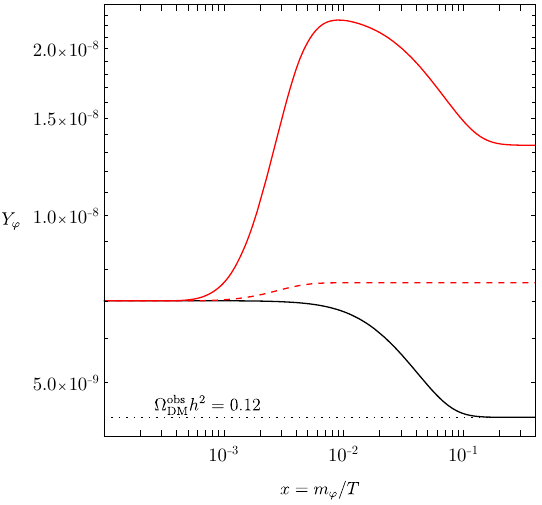}
	\hfill
	\includegraphics[width=0.55\textwidth]{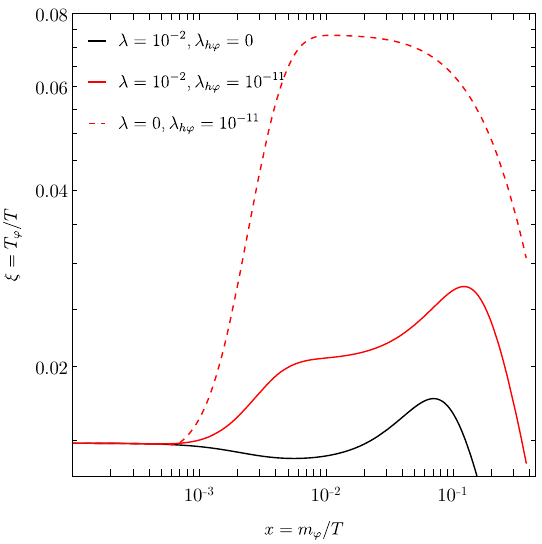}
	\caption{Evolution of the dark matter yield (top) and temperature (down) for a benchmark point with
		$m_\varphi = 100\,\mathrm{MeV}$ and $\xi_\infty = 0.014$. The secluded case $\lambda_{h\varphi}=0$
		(black curves) reproduces the observed relic abundance through dark-sector freeze-out.
		Activating a small Higgs portal coupling, $\lambda_{h\varphi}=10^{-11}$ (red curves), injects
		energetic $\varphi$ particles from Higgs decays. If number-changing self-interactions are
		inefficient (red dashed curves), this injection leads only to a mild enhancement of the
		abundance. By contrast, when $2\leftrightarrow3$ reactions remain efficient (red solid curves),
		the injected energy is converted into additional particle number, resulting in a significantly
		larger yield and a cooler dark sector.}
	\label{fig:BM1_Z2}
\end{figure}

\begin{figure}[t!]
	\centering
	\includegraphics[width=0.7\textwidth]{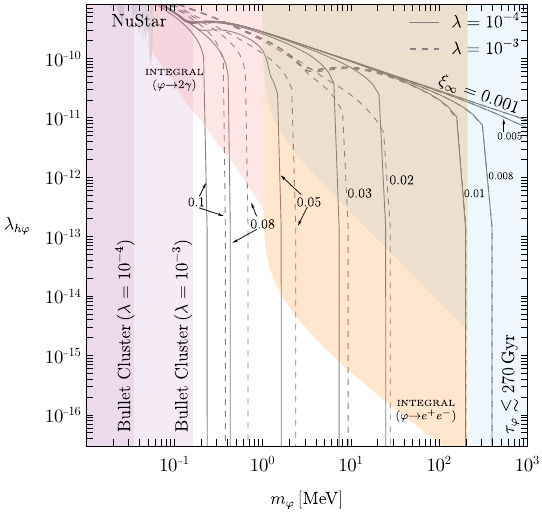}
	\caption{Summary of viable and excluded regions in the $(\lambda_{h\varphi},\,m_\varphi)$ parameter
		space of the scalar singlet dark matter model with a spontaneously broken $\mathbb{Z}_2$
		symmetry. Results are shown for two values of the quartic self-coupling $\lambda$. Gray solid
		and dashed curves denote combinations yielding the observed relic abundance for different
		initial temperature ratios $\xi_\infty$. Shaded regions indicate constraints from dark matter
		self-interactions (Bullet Cluster) and from indirect searches for decay products with INTEGRAL,
		\textit{NuSTAR}, and CMB observations.}
	\label{fig:resultsZ2}
\end{figure}

Despite the restricted viability of this minimal realization, the model offers a transparent setting in which to visualize the non-trivial interplay between freeze-in production and number-changing self-interactions. By solving Eqs.~\eqref{eq:Yeq_x}~and~\eqref{eq:T_evolution_x_compact_Y}, we illustrate this in Fig.~\ref{fig:BM1_Z2}, where we show the time evolution of both the dark matter yield and temperature for a representative benchmark point. 

A brief comment is in order regarding the interpretation of the red curves in
Fig.~\ref{fig:BM1_Z2}. Strictly speaking, these trajectories do not correspond to
a ``pure'' freeze-in production, since the initial conditions assume a finite dark
matter abundance, $n_\varphi^i = n_\varphi^{\rm eq}(T_\varphi^i)$, with
$T_\varphi^i < T^i$. Nevertheless, the subsequent evolution remains within the
freeze-in regime in the sense that the dark sector is never in thermal equilibrium
with the SM plasma. In other words, this parametrization of the initial condition
is equivalent to $T_\varphi^i =0$ and $n_\varphi = 0$ as long as $\xi_\infty\ll 1$, and we quantify deviations from this limit in the plot shown in~\Cref{fig:resultsZ2}.

Although the initial condition is assumed to satisfy $z_\varphi^i = 1$,\footnote{Recall that the fugacity is defined as $z_\varphi \equiv e^{\mu/T_\varphi}$ and in the Maxwell-Boltzmann limit can be identified with the number density as $z_\varphi = n_\varphi/n_\varphi^\text{eq}$.} the FI mechanism pushes the DM into a dilute regime ($z_\varphi \ll 1$), unless number-changing
self-interactions are sufficiently efficient to restore chemical equilibrium (red-thick line),
where initially the $2\to 3$ reaction produces more DM (rising yield) in exchange of kinetic energy
(c.f. red thick and dashed lines).

A broader view of the phenomenological viability of the model is provided in Fig.~\ref{fig:resultsZ2}, which summarizes the allowed and excluded regions in the $(\lambda_{h\varphi},\,m_\varphi)$ plane. The figure combines relic-density requirements with constraints from dark matter stability and self-interactions, and is shown for two representative choices of the quartic coupling $\lambda$. The gray contours indicate parameter combinations that reproduce the observed relic abundance for different assumptions about the initial temperature ratio $\xi_\infty$ (i.e., quantifying deviations from the FI initial condition assumption). Once the Higgs portal interaction is present, the finite lifetime of the dark matter particle becomes a dominant constraint. For masses above the $\mu^+\mu^-$ threshold, decays into charged leptons severely restrict the allowed values of $\lambda_{h\varphi}$, as cosmological observations require lifetimes exceeding $\tau_\varphi \sim 270\,\mathrm{Gyr}$~\cite[Table~2]{Nygaard:2020sow}. In the intermediate mass range, $2m_e < m_\varphi < 2m_\mu$, the strongest limits arise from indirect searches for electromagnetic decay products, with INTEGRAL and \textit{NuSTAR} excluding lifetimes shorter than $\tau_\varphi \sim 10^{27}\,\mathrm{s}$~\cite{Laha:2020ivk,Slatyer:2016qyl}. Below the $e^+e^-$ threshold the bounds weaken, as decays proceed dominantly into photons. 

A comment is in order regarding the origin of the lower bound on the dark matter
lifetime, which in this work is taken to be $\tau_\varphi \gtrsim 270\,\mathrm{Gyr}$,
significantly longer than the age of the Universe, $t_0 \simeq 13.8\,\mathrm{Gyr}$.
At first sight, one might expect that requiring dark matter to survive until the
present epoch would be sufficient. However, much stronger constraints arise from
cosmological observations, in particular from measurements of the Cosmic Microwave
Background. The bound adopted here corresponds to the limits reported in~\cite{Nygaard:2020sow}, assuming that the decay products carry the full dark
matter energy density (\(f=1\)) and behave as dark radiation. The physical origin
of these constraints is discussed in detail in Ref.~\cite{Slatyer:2016qyl}. Even if
dark matter decays occur well after recombination (which takes place at
$t \sim 4\times10^5\,\mathrm{yr}$), they can still leave observable imprints on the
CMB through their impact on the evolution of large-scale gravitational potentials.
In particular, dark matter decays modify the growth of cosmic structures and the
time dependence of gravitational potential wells, which in turn affects CMB photons
via late-time effects such as the Integrated Sachs--Wolfe (ISW) effect and CMB
lensing. 

As a consequence, the CMB does not merely constrain energy injection at early times,
but is also sensitive to the cumulative impact of dark matter decays occurring over
cosmological timescales. Even a lifetime as long as $\tau_\varphi \sim 270\,\mathrm{Gyr}$
implies that a non-negligible fraction of dark matter (of order a few percent) has
already decayed by today. Such a depletion is sufficient to alter the gravitational
potential landscape in a way that is incompatible with the observed CMB anisotropies
and large-scale structure. This explains why CMB-based limits on decaying dark matter
are substantially stronger than the naive requirement that the lifetime exceed the
age of the Universe.

Additional constraints from Big Bang Nucleosynthesis could in principle arise from late-time energy injection at temperatures $T\sim0.1$--$1\,\mathrm{MeV}$. However, such effects become relevant only for lifetimes $\tau_\varphi \lesssim 10^{12}\,\mathrm{s}$~\cite{Depta:2020zbh}, which are already far below those excluded by indirect detection. For this reason, BBN bounds do not further restrict the parameter space shown in Fig.~\ref{fig:resultsZ2}.

Independent of decay constraints, elastic self-scattering of dark matter is bounded by observations of galaxy clusters. In particular, Bullet Cluster data impose an upper limit on the transfer cross section per unit mass, $\sigma_T/m_\varphi < 1\,\mathrm{cm}^2/\mathrm{g}$ at velocities $v\sim10^{-4}$~\cite{Randall:2008ppe},\footnote{We quote the Bullet Cluster bound as it provides a particularly clean and robust upper limit on DM self-interactions at cluster velocities, inferred from the observed offset between the collisional baryonic gas and the gravitational lensing mass peaks in this high-velocity merger. Other cluster systems yield comparable constraints.} where
\begin{equation}\label{eq:sigmaT}
	\sigma_T = \int d\Omega\,(1-\cos\alpha)\,
	\frac{d\sigma_{2\to2}}{d\Omega}\,.
\end{equation}
In the present model this translates into the approximate condition
$m_\varphi/16.32\,\mathrm{MeV} \gtrsim \lambda^{2/3}$, explicitly linking the exclusion region to the
quartic self-coupling. As illustrated in Fig.~\ref{fig:resultsZ2}, increasing $\lambda$ shifts the
self-interaction bound to higher masses and simultaneously enhances cannibal depletion, which
must be compensated by stronger freeze-in production.

Overall, these findings suggest that in this model a dark matter particle with a cosmologically
viable lifetime generally requires a sizable initial abundance generated by mechanisms beyond
freeze-in alone. In the regions of parameter space that remain phenomenologically allowed, the
Higgs portal coupling must be sufficiently small that its influence on both particle production and
thermal energy exchange becomes effectively negligible. While these constraints are admittedly
severe, they are a natural consequence of the substantial progress achieved by indirect detection
experiments and astrophysical observations. Finally, although self-interactions can give rise to
non-trivial dynamical behavior, their quantitative impact on the phenomenology of this particular
scenario remains marginal.

\section{A $\mathbb{Z}_3$ model}\label{sec:Z3}

The strong constraints encountered when the stabilizing $\mathbb{Z}_2$ symmetry is broken motivate the consideration of alternative discrete symmetries that preserve DM stability while retaining the essential features of cannibal dynamics. A particularly economical option is provided by a $\mathbb{Z}_3$ symmetry, which ensures absolute stability while still allowing renormalizable number-changing interactions. Implementing such a symmetry requires promoting the dark matter candidate from a real to a complex scalar field. Such a scalar DM stabilized by a $\mathbb{Z}_3$ symmetry was first introduced in a different context in~\cite{Ma:2007gq}. Its WIMP phenomenology was subsequently studied in~\cite{Belanger:2012zr} and further discussed in~\cite{Hektor:2019ote}. Here we are instead interested in the regime where sizeable self-interactions~\cite{Ghosh:2022asg} coexist with feeble portal couplings, i.e.\ a FIMP-like production history~\cite{Choi:2016tkj}. A key structural advantage of the $\mathbb{Z}_3$ symmetry is that it permits a cubic invariant, and therefore $3\leftrightarrow2$ cannibal reactions already at the renormalizable level. In contrast to the broken-$\mathbb{Z}_2$ case, where cubic terms typically jeopardize stability unless additional structure is imposed, a $\mathbb{Z}_3$ symmetry allows cubic self-interactions while keeping the lightest $\mathbb{Z}_3$-charged state exactly stable.
We denote the complex scalar as $S$, and assume it transforms under the discrete symmetry as $S \rightarrow e^{2\pi i/3} S$. 

The most general renormalizable scalar potential consistent with this symmetry can be written as
\begin{equation}\label{HP_Z3}
	V(H,S) = V_s(S) + \lambda_{hs}\,|H|^2 |S|^2\,,
\end{equation}
note that the portal interaction gives rise to three types of production processes: direct Higgs decay ($h\to S^* S$), direct $2\leftrightarrow 2$ ($hh\to S^*S$) and Higgs-mediated production from other SM states populating the plasma. Out of these, whenever kinematically possible and as long as Higgs bosons are abundant in the plasma, the decay mode dominates. 
Also
\begin{equation}\label{eq:VS}
	V_s(S) =
	\mu_s^2 |S|^2
	+ \frac{g_s}{3!}\left(S^3 + (S^\ast)^3\right)
	+ \frac{\lambda_s}{4} |S|^4
\end{equation}
encodes the self-interactions of the dark scalar.
We require $\mu_s^2>0$ and $\lambda_s>0$ to guarantee that the potential is bounded from below. 
Note that any complex phase in the vacuum expectation value can be absorbed by a redefinition of the field, allowing the extrema of the potential to be obtained by solving
$\partial_S V_s = 0$.
This yields one trivial solution,
$v_s^0 = 0$,
and two additional extrema. Requiring that the symmetry remains unbroken implies that the discriminant is negative, or equivalently
$g_s^2 < 8\lambda_s\mu_s^2$.
It is convenient to express this condition in terms of the dimensionless ratio
$k \equiv g_s^2/(3\lambda_s\mu_s^2)$,
which must satisfy $k < 8/3$.
In this regime the physical mass of the scalar is simply
$m_s^2 = \mu_s^2$.

Charge conservation severely restricts the number-changing processes that can
occur within the dark sector. In the present realization, the only allowed
reactions are
$SSS \leftrightarrow S^{*}S$ and $S^{*}S^{*}S \leftrightarrow SS$, together with
their complex conjugates. The corresponding Feynman diagrams are shown in
\Cref{fig:feynman3-2_Z3,fig:feynman3-2_Z3_}.
We first consider the process $S^{*}S \leftrightarrow SSS$ and treat $S$ and
$S^{*}$ as distinct particle species. The collision operator for $S^{*}$, written
in its most general form, reads
\begin{equation}
	\begin{split}
		C_{S^{*}S\leftrightarrow SSS}[S^{*}]
		=
		\frac{1}{2E_{S^{*}}\,g_{S^{*}}}
		\int \Bigg[
		&(1+f_{S^{*}})
		\left|\tilde{\mathcal{M}}_{345\to\underline{S^{*}}2}\right|^{2}
		\left(\frac{1}{3!}\,d\Pi_{3} d\Pi_{4} d\Pi_{5}\, f_{3} f_{4} f_{5}\right)
		d\tilde{\Pi}_{2}
		\\
		&-
		f_{S^{*}}
		\left|\tilde{\mathcal{M}}_{\underline{S^{*}}2\to345}\right|^{2}
		d\Pi_{2}\, f_{2}
		\left(\frac{1}{3!}\,d\tilde{\Pi}_{3} d\tilde{\Pi}_{4} d\tilde{\Pi}_{5}\right)
		\Bigg] .
	\end{split}
\end{equation}
Similarly, the collision operator for the species $S$ associated with the same
process is given by
\begin{equation}
	\begin{split}
		C_{S^{*}S\leftrightarrow SSS}[S]
		=
		\frac{1}{2E_{S}\,g_{S}}
		\int \Bigg[
		&-
		f_{S}
		\left|\tilde{\mathcal{M}}_{1\underline{S}\to345}\right|^{2}
		d\Pi_{1}\, f_{1}
		\left(\frac{1}{3!}\,d\tilde{\Pi}_{3} d\tilde{\Pi}_{4} d\tilde{\Pi}_{5}\right)
		\\
		&+
		(1+f_{S})
		\left|\tilde{\mathcal{M}}_{12\to\underline{S}45}\right|^{2}
		\left(d\Pi_{1} d\Pi_{2}\, f_{1} f_{2}\right)
		\left(\frac{1}{2!}\,d\tilde{\Pi}_{4} d\tilde{\Pi}_{5}\right)
		\\
		&-
		f_{S}
		\left|\tilde{\mathcal{M}}_{\underline{S}45\to12}\right|^{2}
		\left(\frac{1}{2!}\,d\Pi_{4} d\Pi_{5}\, f_{4} f_{5}\right)
		\left(d\tilde{\Pi}_{1} d\tilde{\Pi}_{2}\right)
		\\
		&+
		(1+f_{S})
		\left|\tilde{\mathcal{M}}_{345\to1\underline{S}}\right|^{2}
		\left(\frac{1}{3!}\,d\Pi_{3} d\Pi_{4} d\Pi_{5}\, f_{3} f_{4} f_{5}\right)
		d\tilde{\Pi}_{1}
		\Bigg] .
	\end{split}
\end{equation}

\begin{figure}[t!]
	\centering
	\begin{tabular}{c@{\hspace{1.2cm}}c@{\hspace{1.2cm}}c@{\hspace{1.2cm}}}
		\begin{tikzpicture}[baseline=(in)]
			\begin{feynman}
				\vertex [dot] (a);  
				\vertex [below=0.6cm of a] (in);
				\vertex [below=1cm of in] (b);
				\vertex [left =1.5cm of a] (i1) {\(S\)};  
				\vertex [below=0.82cm of i1] (ii2){\(S\)};
				\vertex [left =1.5cm of b] (i2){\(S\)};
				\vertex [right=1.5cm of b] (f1){\(S^*\)};
				\vertex [right=1.5cm of a] (f2){\(S\)};
				\diagram* {
					(i1) -- [scalar](a)
					-- [scalar] (f2),
					(a) -- [scalar, edge label=\(S^*\)] (b),
					(i2) -- [scalar] (b),
					(b) -- [scalar] (f1),
					(ii2) -- [scalar] (a),
				};
			\end{feynman}
		\end{tikzpicture} &
		
		\begin{tikzpicture}[baseline=(in)]
			\begin{feynman}
				\vertex [dot] (a) ; 
				\vertex [below =1.6cm of a] (b);
				\vertex [left =0.85cm of a] (i1);  
				\vertex [left =1.5cm of b] (i2){\(S\)};
				\vertex [right =1.3cm of b] (f1){\(S^*\)};
				\vertex [right =1.3cm of a] (f2){\(S\)};  
				\vertex [below left=0.9cm of i1] (ii2) {\(S\)};
				\vertex [left=0.75cm of i1] (ii1) {\(S\)};
				\diagram* {
					(i1) --[scalar, edge label=\(S\)](a)
					-- [scalar] (f2),
					(a) -- [scalar,edge label=\(S\)] (b),
					(i2) -- [scalar] (b),
					(b) -- [scalar] (f1),
					(ii1) -- [scalar] (i1),
					(ii2) -- [scalar] (i1),
				};
			\end{feynman}
		\end{tikzpicture} &
		\begin{tikzpicture}[baseline=(g)]
			\begin{feynman}
				\vertex (a1) {\(S\)};  
				\vertex [below right=1.1cm of a1] (b);
				\vertex [left =1.1cm of b] (g){\(S\)};
				\vertex [below left=1.1cm of b](a2){\(S\)};
				\vertex [right=of b](c);
				\vertex [above right=1cm of c](d1){\(S\)};
				\vertex [below right=1cm of c](d2){\(S^*\)};
				\diagram* {
					(a1) -- [scalar] (c),
					(a2) -- [scalar] (b) -- [scalar,edge label'=\(S^*\)] (c),
					(c) -- [scalar](d1),
					(c) -- [scalar](d2),
					(g) -- [scalar](b),
				};
			\end{feynman}
		\end{tikzpicture}
	\end{tabular}
	\caption{Diagrams for $SSS\leftrightarrow S^*S$.}
	\label{fig:feynman3-2_Z3}
\end{figure}
\begin{figure}[t!]
	\centering
	\begin{tabular}{c@{\hspace{1.2cm}}c@{\hspace{1.2cm}}c@{\hspace{1.2cm}}}
		\begin{tikzpicture}[baseline=(in)]
			\begin{feynman}
				\vertex [dot] (a);  
				\vertex [below=1cm of in] (b);
				\vertex [left =1.35cm of a] (i1) {\(S\)};  
				\vertex [below=0.82cm of i1] (ii2){\(S^*\)};
				\vertex [left =1.2cm of b] (i2){\(S^*\)};
				\vertex [right=1.2cm of b] (f1){\(S\)};
				\vertex [right=1.2cm of a] (f2){\(S\)};
				\diagram* {
					(i1) -- [scalar](a)
					-- [scalar] (f2),
					(a) -- [scalar, edge label=\(S\)] (b),
					(i2) -- [scalar] (b),
					(b) -- [scalar] (f1),
					(ii2) -- [scalar] (a),
				};
			\end{feynman}
		\end{tikzpicture} &
		
		\begin{tikzpicture}[baseline=(b)]
			\begin{feynman}
				\vertex (a1){\(S\)};
				\vertex [below right=1.1cm of a1] (b);
				\vertex [below=0.8cm of b](a2);
				\vertex [right=1.2cm of b](c);
				\vertex [above right=1cm of c](d1){\(S\)};
				\vertex [below right=1cm of c](d2){\(S\)};
				\vertex [above left=0.8cm of a2](e1){\(S^*\)};
				\vertex [below left=0.8cm of a2](e2){\(S^*\)};
				\diagram* {
					(a1) -- [scalar] (b),
					(a2) -- [scalar, edge label'=\(S^*\)] (b) --[scalar, edge label=\(S^*\)](c),
					(c) -- [scalar](d1),
					(c) -- [scalar](d2),
					(a2) -- [scalar](e1),
					(a2) -- [scalar](e2),
				};
			\end{feynman}
		\end{tikzpicture} 
		&
		\begin{tikzpicture}[baseline=(g)]
			\begin{feynman}
				\vertex (a1) {\(S\)};  
				\vertex [below right=1.1cm of a1] (b);
				\vertex [left =1.1cm of b] (g){\(S^*\)};
				\vertex [below left=1.1cm of b](a2){\(S^*\)};
				\vertex [right=of b](c);
				\vertex [above right=1.1cm of c](d1){\(S\)};
				\vertex [below right=1.1cm of c](d2){\(S\)};
				\diagram* {
					(a1) -- [scalar] (c),
					(a2) -- [scalar] (b) -- [scalar,edge label'=\(S^*\)] (c),
					(c) -- [scalar](d1),
					(c) -- [scalar](d2),
					(g) -- [scalar](b),
				};
			\end{feynman}
		\end{tikzpicture}
	\end{tabular}
	\\[0.3cm]
	\centering
	\begin{tabular}{c@{\hspace{1.2cm}}c@{\hspace{1.2cm}}}
		\begin{tikzpicture}[baseline=(in)]
			\begin{feynman}
				\vertex  (a);
				\vertex [below =0.8cm of a] (in);
				\vertex [below =0.8cm of in] (b);
				\vertex [left =1.2cm of in] (i3){\(S\)};
				\vertex [left =1.2cm of a] (i1){\(S^*\)};
				\vertex [left =1.2cm of b] (i2){\(S^*\)};
				\vertex [right =3.2cm of i2] (f1){\(S\)};
				\vertex [right =3.2cm of i1] (f2){\(S\)};
				\diagram* {
					(i1) -- [scalar](a)
					-- [scalar] (f2),
					(a) -- [scalar,edge label=\(S^*\)] (in),
					(in) --[scalar,edge label=\(S^*\)] (b),
					(i2) -- [scalar] (b),
					(b) -- [scalar] (f1),
					(i3) -- [scalar] (in),
				};
			\end{feynman}
		\end{tikzpicture} &
		\begin{tikzpicture}[baseline=(g)]
			\begin{feynman}
				\vertex (a1){\(S\)};
				\vertex [below right=1.4cm of a1] (b);
				\vertex [left =1.1cm of b] (g){\(S^*\)};
				\vertex [below left=1.1cm of b](a2){\(S^*\)};
				\vertex [right=of b](c);
				\vertex [above right=1.1cm of c](d1){\(S\)};
				\vertex [below right=1.1cm of c](d2){\(S\)};
				\diagram* {
					(a1) -- [scalar] (b),
					(a2) -- [scalar] (b) -- [scalar,edge label=\(S^*\)] (c),
					(c) -- [scalar](d1),
					(c) -- [scalar](d2),
					(g) -- [scalar](b),
				};
			\end{feynman}
		\end{tikzpicture}
	\end{tabular}
	\caption{Diagrams for $S^*S^*S\leftrightarrow SS$.}
	\label{fig:feynman3-2_Z3_}
\end{figure}
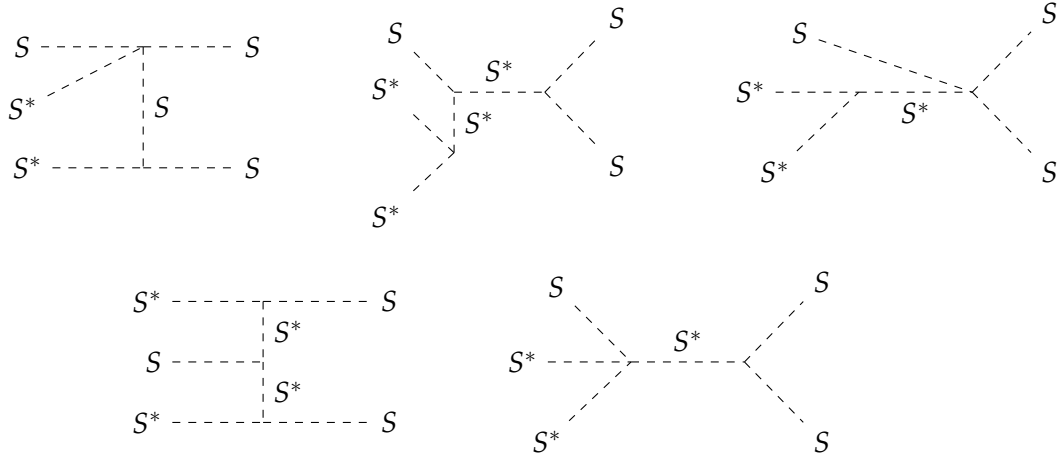

The freeze-in production contribution can be treated analogously to the
discussion in the previous section. In this model DM is assumed to be composed of the
two degrees of freedom of the complex scalar field, such that the total number
density is given by $n = n_{S} + n_{S^{*}}$, with $n_{S} = n_{S^{*}}$.\footnote{In principle CP may be violated within the dark sector. In that case there can
be phenomenological implications for asymmetric cannibal DM annihilating in neutron stars~\cite{Dey:2025atz}. Albeit interesting, we do not investigate this case here, as we focus on the simplest realization of complex cannibal DM.} This
follows from the absence of an initial particle--antiparticle asymmetry in the
dark sector and from the assumption that the model does not violate CP. The
initial conditions are taken to be identical to those adopted in the previous
section.

%
%
\subsection{Results}\label{sec:Z3_results}

\begin{figure}[t!]
	\centering
	\includegraphics[width=0.55\textwidth]{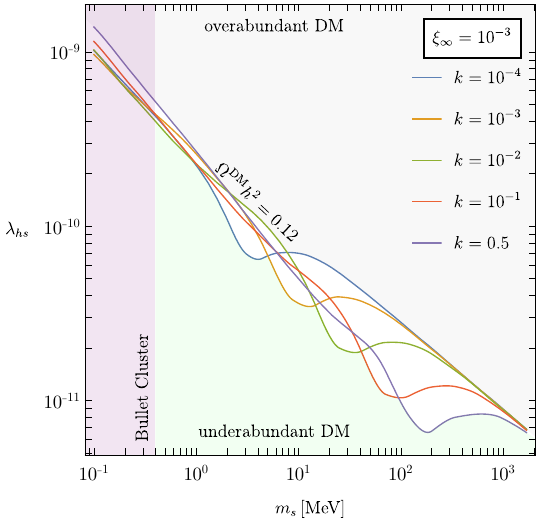}
	\hfill
	\includegraphics[width=0.55\textwidth]{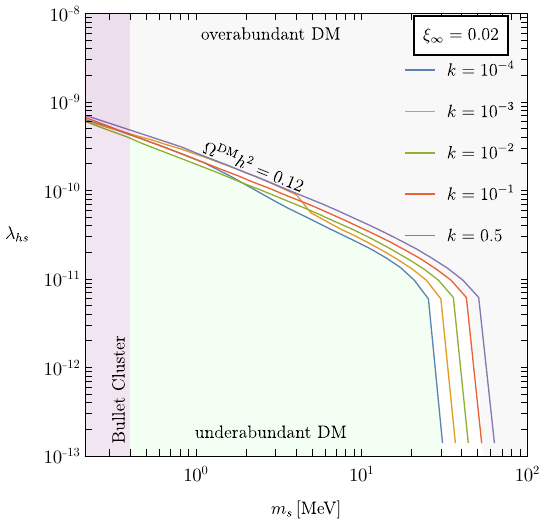}
	\caption{Colored curves show parameter combinations yielding the observed dark matter relic abundance~\cite{Planck:2018vyg} for different values of $k$ in the $\lambda_{hs}$--$m_s$ plane, fixing $\lambda_s = 10^{-2}$. The top (down) panel corresponds to $\xi_\infty = 10^{-3}$ ($\xi_\infty = 0.02$). The green (gray) regions lead to an under- (over-) abundance of dark matter and depend on the assumed initial condition $\xi_\infty$. The purple shaded area is excluded by constraints on dark matter self-interactions. The solution curves exhibit non-trivial structure and do not follow a simple power-law scaling; in particular, regions with enhanced $2\to3$ self-interactions require smaller values of the portal coupling $\lambda_{hs}$ to avoid dark matter overproduction.}
	\label{fig:resultsZ3}
\end{figure}

To investigate the interplay between FI production and cannibal reactions, and to determine the values of the Higgs portal coupling $\lambda_{hs}$ compatible with the observed relic abundance, 
we perform a parameter scan in the $\lambda_{hs}$--$m_s$ plane for several choices of $k \equiv \frac{g_s^2}{3\lambda_s m_s^2}$,
as shown in Fig.~\ref{fig:resultsZ3}. The colored curves correspond to parameter points reproducing the observed DM abundance, with $\lambda_s$ fixed to $10^{-2}$. 
The purple shaded region is excluded by excessive dark matter self-scattering, using Eq.~\eqref{eq:sigmaT}, where $\sigma_T$ includes both $SS\to SS$ and $SS^*\to SS^*$ processes. The resulting constraint depends on $m_s$, $\lambda_s$, and $k$. For instance, for $m_s = 400\,\mathrm{keV}$ and $\lambda_s = 10^{-2}$, one finds $\sigma_T/m_s = 0.54\,\mathrm{cm}^2/\mathrm{g}$ for $k=10^{-4}$ and $\sigma_T/m_s = 0.64\,\mathrm{cm}^2/\mathrm{g}$ for $k=0.5$. The exclusion shown corresponds to the most stringent bound over the scanned parameter space.

We first examine the role of the cannibal reactions by comparing solution curves with different values of $k$, focusing on the top panel of Fig.~\ref{fig:resultsZ3}. For $m_s \lesssim 1\,\mathrm{MeV}$, the dark sector reaches chemical equilibrium at early times and undergoes a standard cannibalization phase. Increasing the strength of self-interactions (larger $k$) enhances dark matter depletion during cannibalization. 
To maintain the relic abundance fixed, a larger initial freeze-in yield is required, which translates into a larger value of $\lambda_{hs}$.
The non-linear behaviour of the solution curves in this region arises because self-interactions decouple \emph{before} chemical equilibrium is achieved. Therefore, the cannibal phase does not occur, and no significant depletion takes place.

We now turn to the dependence on the initial dark sector abundance by comparing the top and bottom panels of Fig.~\ref{fig:resultsZ3}. For $\xi_\infty = 10^{-3}$, corresponding to a very suppressed initial dark matter population, slightly larger values of $\lambda_{hs}$ are required compared to the case $\xi_\infty = 0.02$. The interplay between FI production and cannibal interactions, manifested as deviations from simple power-law behaviour, is more pronounced for smaller $\xi_\infty$. This is because, for larger initial abundances, the energy injected into the dark sector during FI is comparable to its pre-existing energy density, reducing the relative impact of subsequent cannibal dynamics.
In summary, these results illustrate the rich dynamics of the FI production and self interactions. Allowing for $\xi_\infty < 1$ enables a realization of self-interacting DM in the $\mathbb{Z}_3$ model that is compatible with both the observed relic abundance and structure formation constraints. This possibility was shown to be excluded in the case $\xi_\infty = 1$~\cite{Choi:2016tkj}. However, this viable region remains beyond the reach of current experimental probes.

If the dark sector consists of a single particle species that is a singlet under the Standard Model gauge group, detection prospects are inherently challenging, since all non-gravitational interactions proceed exclusively through the Higgs portal. Nevertheless, a variety of ongoing and proposed experiments aim to probe sub-GeV dark matter. In particular, dark matter--electron scattering experiments are sensitive to masses in the MeV range~\cite{Essig:2011nj,Derenzo:2016fse,Essig:2017kqs}, while astrophysical observations also target similar mass scales~\cite{Kumar:2018heq}. DM with masses in the GeV scale are already accessible to nuclear recoil direct detection experiments~\cite{Hambye:2018dpi}. In this work, however, our primary focus is on cannibal sectors, which favor dark matter with $m_{\rm DM}\sim\mathcal{\text{MeV}}$. It is worth noting that current electron recoil experiments are not yet sensitive to FI portal couplings relevant for these masses.\footnote{For the Higgs-portal model considered here, the non-relativistic dark matter--electron scattering cross section can be estimated as $\sigma_{Se\to Se} \sim \lambda_{hs}^2 m_e^2/m_h^4 \sim 4\lambda_{hs}^2 \times 10^{-43}\,\mathrm{cm}^2$. The sensitivity of silicon-based electron recoil experiments is approximately $\sigma_{\mathrm{DM}\,e\to\mathrm{DM}\,e} \sim 10^{-42}\,\mathrm{cm}^2$ for $m_{\mathrm{DM}} \sim 10\,\mathrm{MeV}$~\cite{Ramanathan:2020fwm}.}

\section{SM $+$ $\mathbb{Z}_3$ complex scalar $+$ scalar mediator}
\label{sec:Z3_w_mediator}

In this section we introduce a model that combines the key dynamical features of the simpler scenarios discussed previously while allowing for a broader phenomenology. 
The model is obtained by extending the scalar potential in Eq.~\eqref{HP_Z3} with an additional real singlet field $\phi$. 
This mediator couples both to the SM Higgs doublet and to the complex $\mathbb{Z}_3$-charged dark matter field $S$.

\subsection{The model}

The most general renormalizable scalar potential consistent with the assumed symmetries is
\begin{equation}\label{eq:full_potential}
	V(H,S,\phi)=V_s + V_\phi + V_{\phi s} + V_\text{HP}\,,
\end{equation}
where
\begin{equation}
	\begin{split}
		V_{\phi} &= \lambda_1\phi + \frac{\mu_\phi^2}{2}\phi^2
		+ \frac{g_\phi}{3!}\phi^3 + \frac{\lambda_\phi}{4!}\phi^4\,,
		\\[0.5ex]
		V_{\phi s} &= \frac{\lambda_{\phi s}}{2}\,\phi^2 |S|^2
		+ A_{\phi s}\,\phi |S|^2
		+ \kappa_{\phi s}\,\phi S^3 + \text{c.c.}\,,
		\\[0.5ex]
		V_\text{HP} &= \left(
		\lambda_{hs}\,|S|^2
		+ \lambda_{h\phi}\,\phi^2
		+ B_{h\phi}\,\phi
		\right)|H|^2\,,
	\end{split}
\end{equation}
and $V_s$ is given in Eq.~\eqref{eq:VS}. The interactions contained in $V_{\phi s}$ allow the dark matter and mediator sectors to exchange energy and entropy (see e.g.~\cite{Ghosh:2020lma}). In particular, the cubic term proportional to $\kappa_{\phi s}$ can induce freeze-in semi-production of dark matter through $\phi S \to SS$~\cite{Bringmann:2021tjr,Hryczuk:2021qtz}. In the present analysis we set $\kappa_{\phi s}=0$ and focus on the thermal coupling between $S$ and $\phi$ controlled by $\lambda_{\phi s}$.

To further simplify the setup, we set $\lambda_{hs}=0=\lambda_{h\phi}$ and generate the dark sector population through the linear Higgs portal term proportional to $B_{h\phi}$. Throughout the following discussion we assume that the scalar $S$ is stable, which requires $k<8/3$, where $k=g_s^2/(3\lambda_s \mu_s^2)$. After electroweak symmetry breaking, the mediator mixes with the Higgs boson via $B_{h\phi}$. This mixing is parameterized by the angle $\theta$, defined in Eq.~\eqref{eq:mixing_angle}. A detailed discussion of the vacuum stability conditions for the full scalar potential is provided in Appendix~\ref{ap:vac_stab}.

After the electroweak phase transition, mixing between the scalar mediator and the Higgs boson implies the field redefinition
\[
\phi \to \cos\theta\,\phi + \sin\theta\,h \simeq \phi + \theta\,h\,,
\]
which induces effective Higgs--DM interactions through the mediator couplings,
\begin{equation}\label{eq:terms_in_Vphis}
	\begin{split}
		V_{\phi s} \supset \lambda_{\phi s}\theta\,\phi h |S|^2
		+ A_{\phi s}\theta\,h|S|^2
		+ \mathcal{O}(\theta^2)\,.
	\end{split}
\end{equation}
The first term gives rise to the three-body decay $h\to\phi SS^*$ when kinematically allowed. Since we work in the regime $m_\phi,m_s\ll m_h$, this decay is evaluated in the massless limit for both the dark matter and the mediator within the collision term. Mixing also generates interactions between $\phi$ and the Higgs boson from the Higgs potential,\footnote{For simplicity, we neglect the additional interactions induced by \(V_\phi\) through \(g_\phi\) and \(\lambda_\phi\). Our goal here is to isolate the minimal scenario driven by Higgs-induced production and mixing, rather than to study the extra dark-sector dynamics that these self-interactions may generate.}
\begin{equation}\label{eq:scalar_higgsinteract}
	\lambda_h (H^\dagger H)^2
	= \frac{\lambda_h}{4}(h-\theta\phi+v_h)^4
	= -\lambda_h\,\theta\,h^{3}\phi
	+ \mathcal{O}(\theta^2)\,.
\end{equation}
We restrict to the parameter region $m_\phi<2m_s$ so that mediator decays into dark matter are kinematically forbidden. Under this assumption, the coupled Boltzmann equations governing the number densities and temperatures of $S$ and $\phi$ are
\begin{align}\label{eq:system_med_DM}
	\frac{Y'_S}{Y_S}&=\frac{1}{x\tilde H}\left(\braket{C_{h\to \phi SS^*}} + \braket{C_{h\to SS^*}}+\braket{C_{\phi\phi\leftrightarrow SS^*}} +  \braket{C_{3\leftrightarrow 2}} \right)\,,\nonumber
	\\-\frac{x_S'}{x_S}&=
	\begin{aligned}[t]&\frac{1}{x\tilde H}\big(\braket{C_{h\to \phi SS^*}}_2+\braket{C_{h\to SS^*}}_2+\braket{C_{\phi S\leftrightarrow \phi S}}_2 +  \braket{C_{3\leftrightarrow 2}}_2 \big)
		\\&-\frac{Y'_S}{Y_S} + \frac{H}{x\tilde H}\frac{\braket {p^4/E^3}}{3T_S} + \frac{2s'}{3s}\,,\end{aligned}
	\\\frac{Y'_\phi}{Y_\phi}&=\frac{1}{x\tilde H}\left(\braket{C_{h\to \phi SS^*}}+\braket{C_{\text{SM SM}\to \text{SM}\,\phi}}
	+\braket{C_{\phi\phi\leftrightarrow SS^*}}\right)\,,\nonumber
	\\-\frac{x_\phi'}{x_\phi}&=\frac{1}{x\tilde H}\left(\braket{C_{h\to \phi SS^*}}_2+\braket{C_{\text{SM SM}\to \text{SM}\,\phi}}_2+\braket{C_{\phi S\leftrightarrow \phi S}}_2  \right) -\frac{Y'_\phi}{Y_\phi} + \frac{H}{x\tilde H}\frac{\braket {p^4/E^3}}{3T_\phi} + \frac{2s'}{3s}\,.\nonumber
\end{align}
Here we define $x_i=m_i/T_i$ for $i=S,\phi$, $x = m_s/T$ and $\tilde H=H/(1+3T(dg_{*s}/dT)/g_{*s})$. Also $\braket{C}= C_0/n_i$ and $\braket{C}_2= C_2/(3n_iT_i) $. The collision operators involving $S$--$\phi$ interactions are given in Appendix~\ref{ap:C_DM_med}, while the collision term for the three-body Higgs decay is presented in Appendix~\ref{ap:triple_higgs_decay}.

\subsection{Results}\label{sec:Z3_w_mediator_results}

\begin{figure}[t!]
	\centering
	\begin{subfigure}
		\centering
		\includegraphics[width=0.46\textwidth]{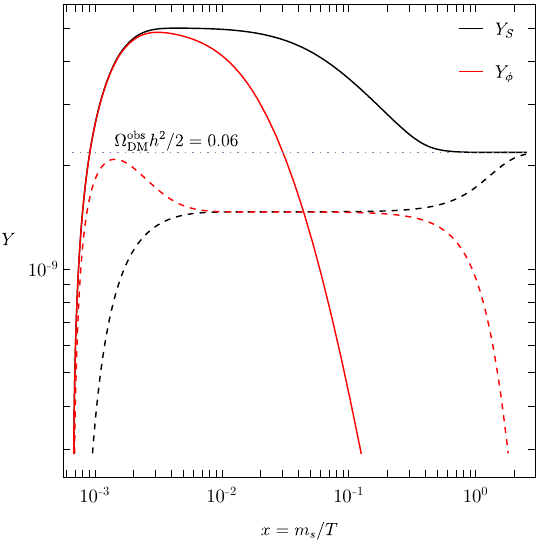}
	\end{subfigure}
	\hfill
	\begin{subfigure}
		\centering
		\includegraphics[width=0.46\textwidth]{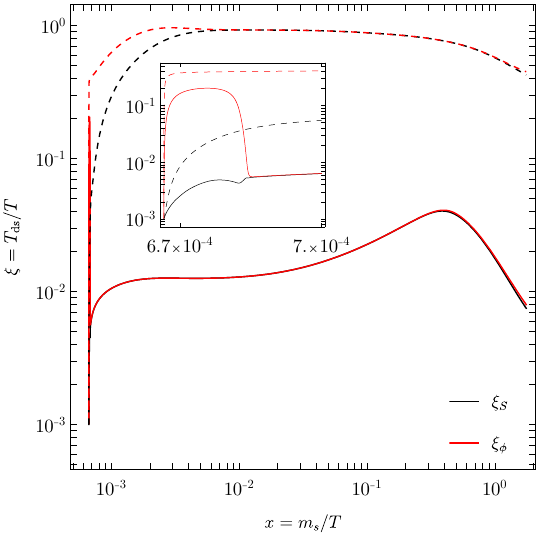}
	\end{subfigure}
	\\
	\begin{subfigure}
		\centering
		\includegraphics[width=0.46\textwidth]{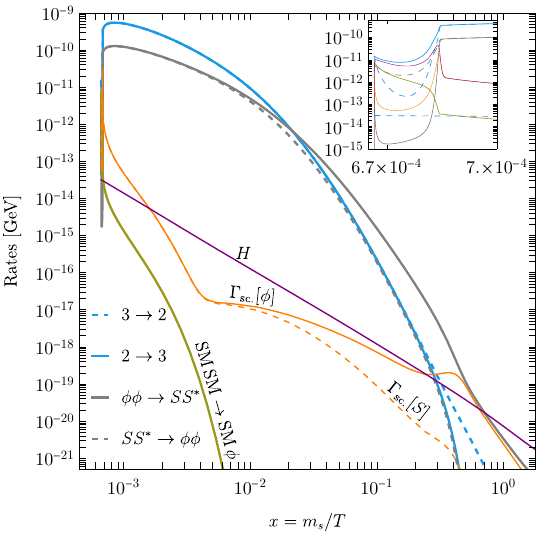}
	\end{subfigure}
	\hfill
	\begin{subfigure}
		\centering
		\includegraphics[width=0.46\textwidth]{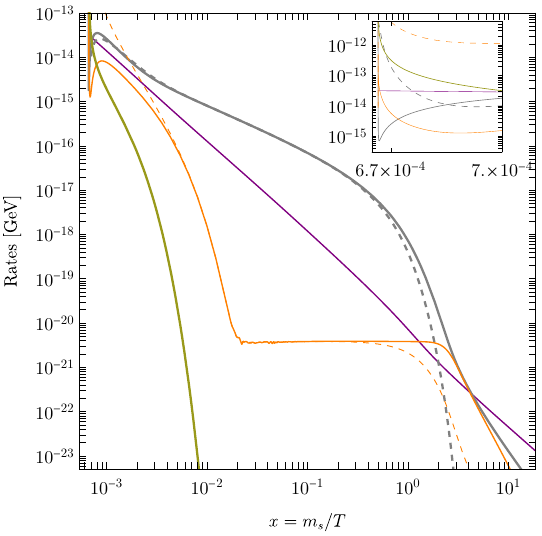}
	\end{subfigure}
	\caption{Evolution of the dark sector for $m_s=100\,\mathrm{MeV}$ and $m_\phi=160\,\mathrm{MeV}$, with $\lambda_{\phi s}=10^{-3}$, $A_{\phi s}=0$, $k=0.5$, and $\lambda_s=0.05$. The mixing angle $\theta$ is fixed to reproduce the observed relic abundance (dotted line). 
		Top panels show the dark matter yield (left) and the temperature ratio $\xi=T_{\rm ds}/T$ (right). 
		Bottom panels display the relevant interaction rates, including self-interactions (left) and with self-interactions switched off (right). 
		Black (red) curves correspond to evolution without (with) self-interactions, with $\theta\simeq2.2\times10^{-10}$ ($\theta\simeq5.2\times10^{-11}$). 
		Inset panels highlight the early-time evolution around the electroweak phase transition, where a rapid injection of particles occurs.
	}
	\label{fig:evol1}
\end{figure}

\begin{figure}[t!]
	\centering
	\begin{subfigure}
		\centering
		\includegraphics[width=0.46\textwidth]{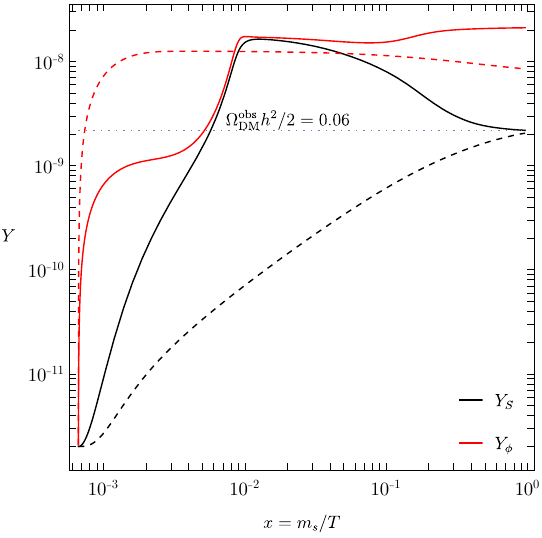}
	\end{subfigure}
	\hfill
	\begin{subfigure}
		\centering
		\includegraphics[width=0.46\textwidth]{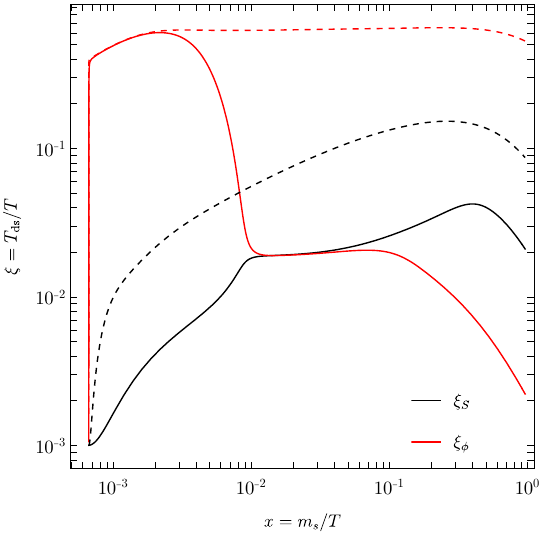}
	\end{subfigure}
	\\
	\begin{subfigure}
		\centering
		\includegraphics[width=0.46\textwidth]{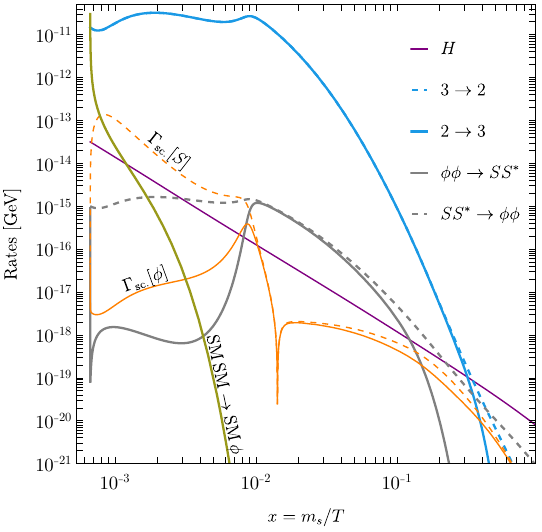}
	\end{subfigure}
	\hfill
	\begin{subfigure}
		\centering
		\includegraphics[width=0.46\textwidth]{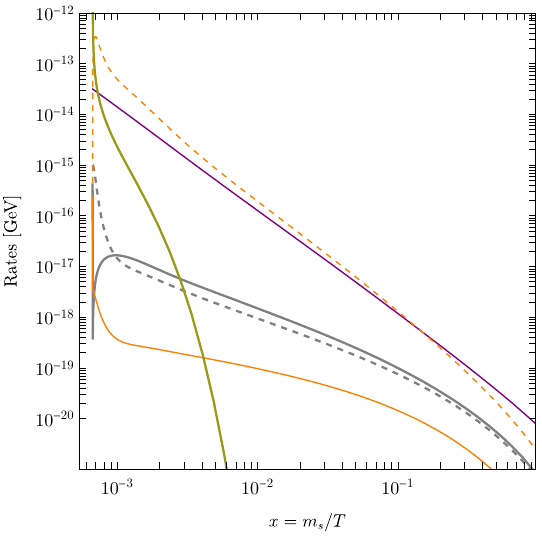}
	\end{subfigure}  
	\caption{Same as Figure~\ref{fig:evol1}, but for an inverted mass hierarchy with $m_\phi=80\,\mathrm{MeV}<m_s=100\,\mathrm{MeV}$ and a weaker DM--mediator coupling $\lambda_{\phi s}=10^{-5}$. 
		The required mixing angle is $\theta\simeq3.73\times10^{-10}$ without self-interactions and $\theta\simeq1.12\times10^{-10}$ when self-interactions are included.
	}
	\label{fig:evol2}
\end{figure}

\begin{figure}[t!]
	\centering
	\begin{subfigure}
		\centering
		\includegraphics[width=0.46\textwidth]{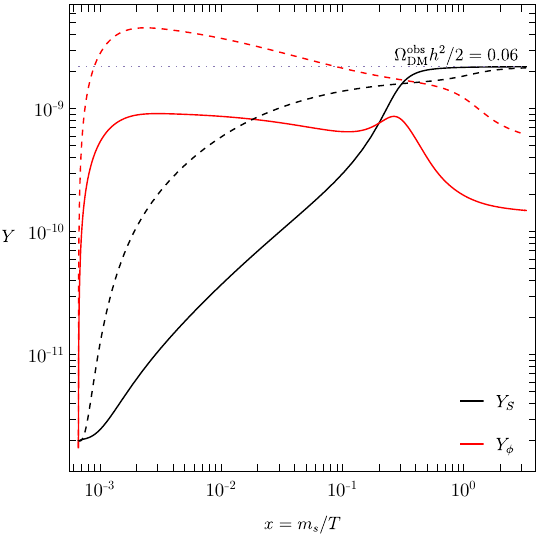}
	\end{subfigure}
	\hfill
	\begin{subfigure}
		\centering
		\includegraphics[width=0.46\textwidth]{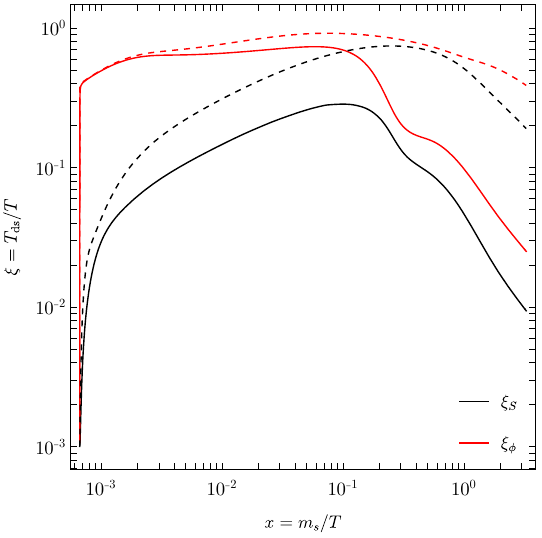}
	\end{subfigure}
	\\
	\begin{subfigure}
		\centering
		\includegraphics[width=0.46\textwidth]{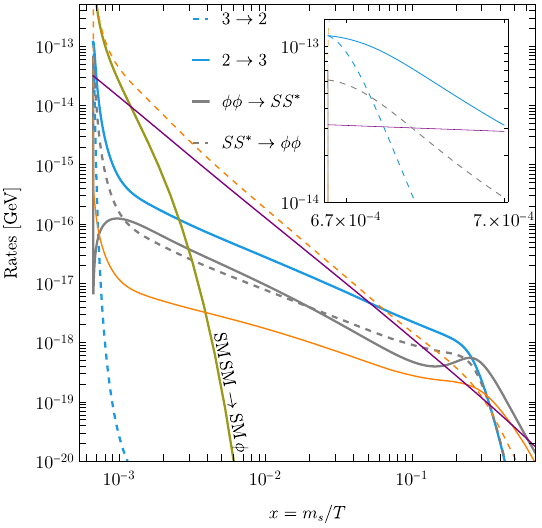}
	\end{subfigure}
	\hfill
	\begin{subfigure}
		\centering
		\includegraphics[width=0.46\textwidth]{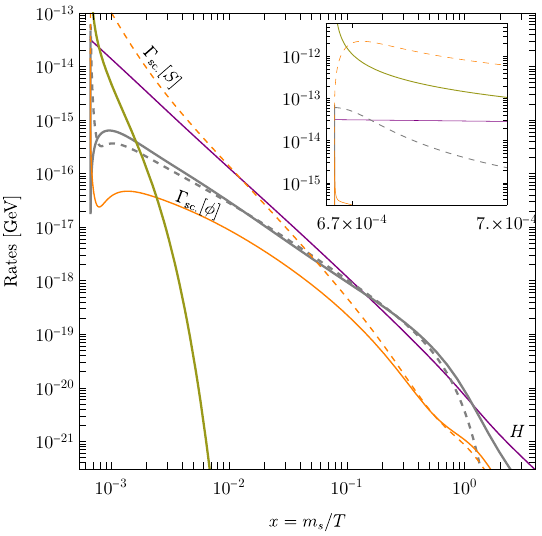}
	\end{subfigure}
	\caption{Same as Figure~\ref{fig:evol1}, but with intermediate DM--mediator coupling $\lambda_{\phi s}=10^{-4}$ and weaker DM self-interactions, $\lambda_s=10^{-2}$. 
		The corresponding mixing angle is $\theta\simeq2.3\times10^{-10}$ without self-interactions and $\theta\simeq1.0\times10^{-10}$ when self-interactions are included.
	}
	\label{fig:evol3}
\end{figure}

\begin{figure}[t!]
	\centering
	\begin{subfigure}
		\centering
		\includegraphics[width=0.46\textwidth]{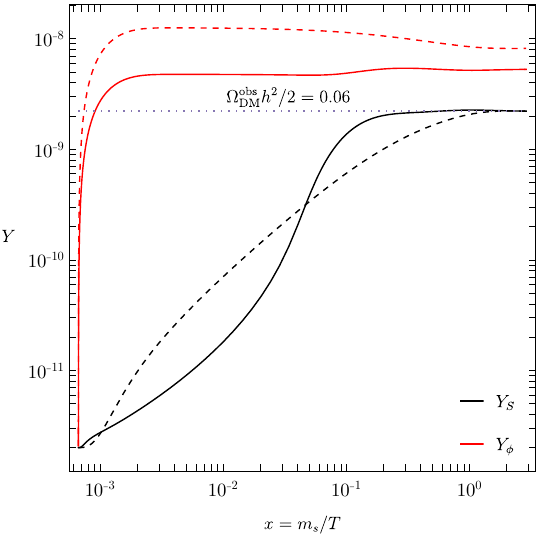}
	\end{subfigure}
	\hfill
	\begin{subfigure}
		\centering
		\includegraphics[width=0.46\textwidth]{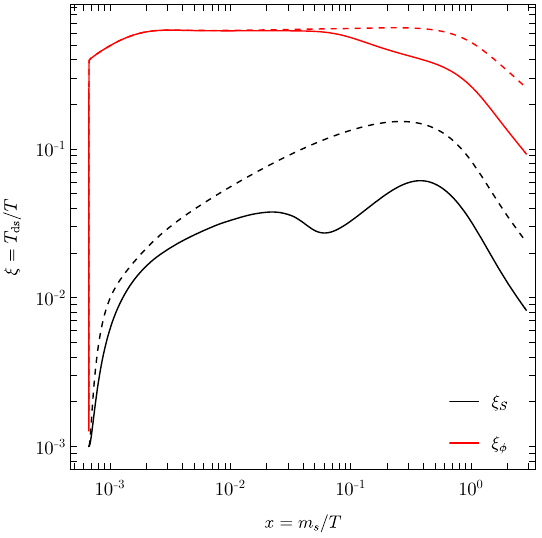}
	\end{subfigure}
	\\
	\begin{subfigure}
		\centering
		\includegraphics[width=0.46\textwidth]{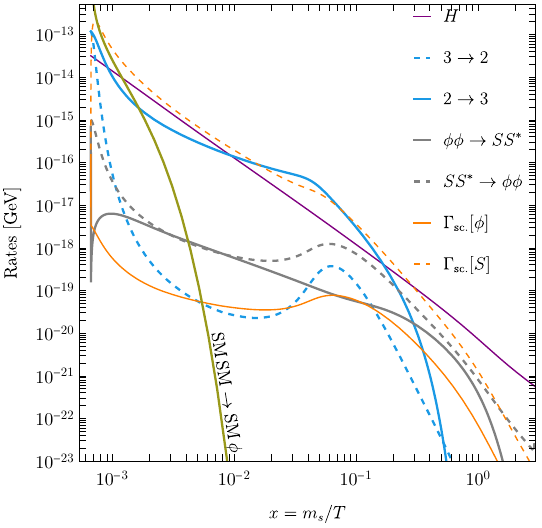}
	\end{subfigure}
	\caption{Same as Figure~\ref{fig:evol2}, but with stronger DM self-interactions, $\lambda_s=10^{-2}$. 
		The mixing angle required to reproduce the relic abundance is $\theta\simeq3.73\times10^{-10}$ without self-interactions and $\theta\simeq2.29\times10^{-10}$ when self-interactions are included. 
		For reference, the interaction rates in the absence of self-interactions coincide with those shown in the bottom-right panel of Figure~\ref{fig:evol2}.
	}
	\label{fig:evol4}
\end{figure}

Compared to the models discussed previously, the presence of a scalar mediator leads to a wider range of dynamical behaviors. 
We begin by illustrating these effects with four benchmark examples, which show how dark matter cannibalization can be either enhanced or suppressed by the mediator, depending on the mass hierarchy between $S$ and $\phi$ and on the strength of number-changing self-interactions.
We then examine the phenomenology associated with the mediator itself, focusing on constraints related to its lifetime. Unlike the secluded $\mathbb{Z}_3$ scenario of Section~\ref{sec:Z3}, this model gives rise to observable signals in astrophysical searches and in experiments targeting long-lived particles. For mediator masses in the MeV range, decays can occur during cosmologically relevant epochs and modify the thermal history of the Universe. In particular, decays around $T\sim 0.1$--$1\,\mathrm{MeV}$ may affect the primordial abundances of light elements produced during BBN. The resulting bounds depend sensitively on the mediator abundance at the onset of, and during BBN.

Mediator decays at later times inject energy into the SM plasma and can leave imprints on CMB. Decays producing energetic electrons and positrons may also generate X-ray photons through inverse Compton scattering. Below we identify regions of parameter space that remain consistent with these cosmological and astrophysical constraints. We also comment on the prospects for detecting the mediator in future long-lived particle searches with emphasis on the MATHUSLA experiment~\cite{Curtin:2024xxo}.

\subsubsection{Benchmarks solutions}
\label{sec:Z3_w_mediator_bms}

We consider four benchmark solutions that reproduce the observed dark matter relic abundance and display qualitatively different dynamical regimes, shown in Figures~\ref{fig:evol1}--\ref{fig:evol4}. The first two benchmarks correspond to relatively strong self-interactions, for which the dark sector approaches equilibrium efficiently, while the latter two illustrate cases with weaker self-interactions. Each figure shows the evolution of the yield (top left), the temperature ratio $\xi=T_\text{ds}/T$ (top right), the interaction rates including self-interactions (bottom left), and the rates obtained by setting $\lambda_s=0$ (bottom right).

The interaction rates are encoded within the zeroth moment collision integral, 
$\Gamma_\text{rate}\equiv \langle C_\text{int.}\rangle$, with the exception of the thick and dashed orange curves, which denote the DM--mediator scattering rates,
\begin{equation}
	\begin{split}
		\Gamma_\text{sc.}[\phi] &\equiv \big|\langle C_\text{scatter}[\phi]\rangle_2\big|\,,\\
		\Gamma_\text{sc.}[S] &\equiv \big|\langle C_\text{scatter}[S]\rangle_2\big|\,,
	\end{split}
\end{equation}
where $\langle C_\text{scatter}\rangle_2$ is defined in Eq.~\eqref{eq:S_med_scatter}.

For $m_s<m_\phi$ with strong interactions, shown in Figure~\ref{fig:evol1}, the evolution begins with entropy injection at $T\simeq150\,\mathrm{GeV}$ ($x\simeq6.6\times10^{-4}$, dark green curves in the rate panels). Via elastic scattering, mediators transfer energy to the dark matter, driving the system toward kinetic equilibrium. For $\lambda_{\phi s}=10^{-3}$, kinetic and chemical equilibrium between $S$ and $\phi$ are established at early times. DM self-interactions ($\lambda_s=0.05$) maintain chemical equilibrium within the $S$ population. As temperatures drop, mediators annihilates into DM (thick gray curve), delaying the onset of a standard cannibal phase.
The second benchmark, shown in Figure~\ref{fig:evol2}, corresponds to $m_s>m_\phi$ with weaker dark-sector interactions. In this case dark matter depletion proceeds through both $3\to2$ and $SS^*\to\phi\phi$ annihilations. Self-interactions set $\mu_s = 0$, while $\Gamma_\text{sc.}[S]$ (orange dashed) quantifies heat transfer into the DM, while $\Gamma_\text{sc.}[\phi]$ (orange solid) describes the corresponding heat loss from the mediators.
Figure~\ref{fig:evol3} illustrates a case with $m_s<m_\phi$ where interactions in the dark sector are insufficient to establish equilibrium. As the Universe expands, the $2\to3$ reaction becomes efficient around $x\simeq3\times10^{-2}$, increasing the DM yield. The resulting high-energy tail allows mediator production, but since the mediator is heavier this channel quickly shuts off, and the inverse process dominates. The system decouples before reaching chemical equilibrium.
The final benchmark, shown in Figure~\ref{fig:evol4}, also corresponds to weak self-interactions with $m_s>m_\phi$. Here dark matter production is entirely given by $2\to3$, whose efficiency is sustained by entropy injection from mediators into DM. During periods where $\Gamma_{2\to3}>\Gamma_\text{sc}[S]$, self-interactions cool the DM while its abundance grows. At later times, increased scattering reheats the dark matter, and once sufficient kinetic energy is available, $\Gamma_{2S\to2\phi}$ grows, leading to a final substantial abundance of $\phi$.

\subsubsection{Numerical scan setup}\label{sec:Z3_w_mediator_scan}
We perform a random scan over the parameter ranges listed in Table~\ref{tab:params_scan}, sampling each parameter independently with uniform distributions in logarithmic space. For each point we solve the coupled system and retain only those that reproduce the observed relic abundance within $1\%$, i.e.\ $\Omega_\text{DM}^\text{obs}/\Omega_\text{DM}=1\pm 0.01$, consistent with the 68\% intervals from the TT,TE,EE+lowE+lensing Planck data~\cite{Planck:2018vyg}. 

To probe different strengths of dark matter self-interactions, we fix $k=0.5$\footnote{Both $k$ and $\lambda_s$ control the strength of self-interactions and can partially compensate each other. The choice $k=0.5$ also satisfies the stability bound $k<8/3$.} and scan $\log_{10}\lambda_s$ on a grid from $-4$ to $-1$ in steps of $0.5$.
\begin{table}[t!]
	\begin{center}
		{
			\begin{tabular}{|c|c|c|c|}
				\cline{1-4}
				Parameter  & min. & max. & Impact on \\
				\hline\hline
				\raisebox{-0.5ex}{$m_s$} & \raisebox{-0.5ex}{$500\,\text{keV}$} & \raisebox{-0.5ex}{$10\,\text{GeV}$} & \raisebox{-0.5ex}{ID and DD}  \\[5pt]
				\cline{1-4}
				\raisebox{-0.5ex}{$r=\frac{m_\phi}{2m_s}$} & \raisebox{-0.5ex}{0} & \raisebox{-0.5ex}{1}  & \raisebox{-0.5ex}{ID and DD} \\ [5pt]
				\cline{1-4}
				\raisebox{-0.5ex}{ $\tilde A=A_{\phi s}/m_\phi$}  & \raisebox{-0.5ex}{$10^{-7}$} & \raisebox{-0.5ex}{$1$}  &  \raisebox{-0.5ex}{ID and FI }    \\ [5pt]
				\cline{1-4}
				\raisebox{-0.5ex}{$\lambda_{\phi s}$} & \raisebox{-0.5ex}{$10^{-10}$} &  \raisebox{-0.5ex}{$10^{-1}$}  &  \parbox[t]{8cm}{ID, DD, FI and equilibrium between $\phi$ and $S$}   \\ 
				\cline{1-4}
				\raisebox{-0.5ex}{$\log_{10}\lambda_{s}$} & \raisebox{-0.5ex}{$-4$} &  \raisebox{-0.5ex}{$-1$}  &  \raisebox{-0.5ex}{DM self-interactions}   \\ 
				\cline{1-4}
			\end{tabular}
			\caption{Relevant parameters, their ranges in the numerical scan and phenomenological impact. The choice of the DM mass scale is motivated by $3\leftrightarrow 2$ processes and that for $m_s\lesssim \mathcal{O}(100\,\text{keV})$ the bounds on self-scatterings (Eq.~\eqref{eq:sigmaT}) are very stringent. The choice of values for $\tilde A$ and $\lambda_s$ is to preserve unitarity and perturbativity~\cite{Schuessler:2007av}. The listed parameters are scanned with $\log_{10}$ scaling, with the exception of $r$, which is scanned with a linear scaling. DD and ID denote direct and indirect detection, respectively.}
			\label{tab:params_scan}
		}
	\end{center}
\end{table}

\begin{figure}[t!]
	\centering
	\begin{subfigure}
		\centering
		\includegraphics[width=0.65\textwidth]{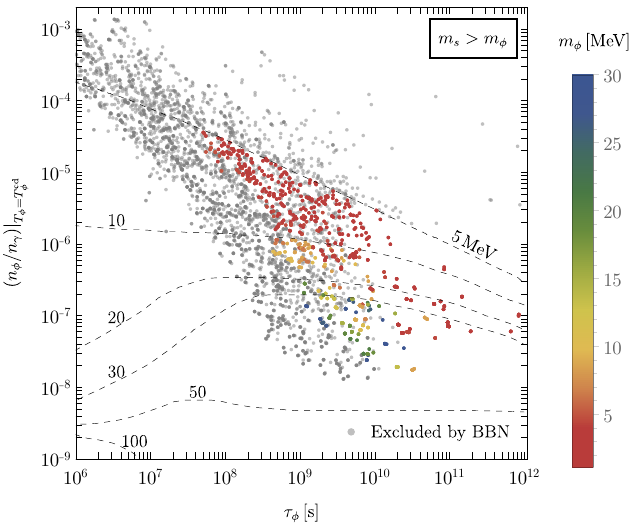}
	\end{subfigure}
	\hfill
	\begin{subfigure}
		\centering
		\includegraphics[width=0.65\textwidth]{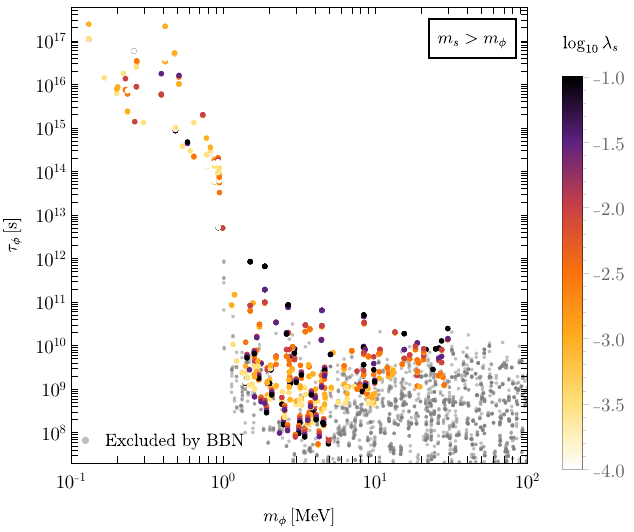}
	\end{subfigure}
	\caption{Points reproducing the observed relic projected onto the planes $\left.n_\phi/n_\gamma\right\vert_{T_\phi=T_\phi^\text{cd}}\,\text{vs}\,\tau_\phi$ (top) and $\tau_\phi\,\text{vs}\,m_\phi$ (down) for masses up to $2m_\mu$, fixing $\xi_\infty = 10^{-3}$ and $k=0.5$. The dashed lines in the top plot indicate the BBN limits for each scalar mass and are adapted from~\cite[Figure 6]{Depta:2020zbh}, while the color bar indicates the mediator mass. There is an overabundance of mediators arising from $S^*S\to 2\phi$, thus mediator masses in the interval $30\,\text{MeV}\lesssim m_\phi<2m_\mu$ are excluded. The allowed points can lead to signals in indirect detection via scalar decay (cf. Figure~\ref{fig:cross_ID}). The bottom plot includes points with $m_\phi < 1\,\text{MeV}$, which are not in conflict with BBN because their lifetimes exceed $\tau_\phi > 10^{12}\,\text{s}$ with the color bar displaying the values of DM self-interacting coupling $\log_{10}\lambda_s$.
	}
	\label{fig:tauphi_vs_abundance_ms>mphi}
\end{figure}
\begin{figure}[t!]
	\centering
	\begin{subfigure}
		\centering
		\includegraphics[width=0.65\textwidth]{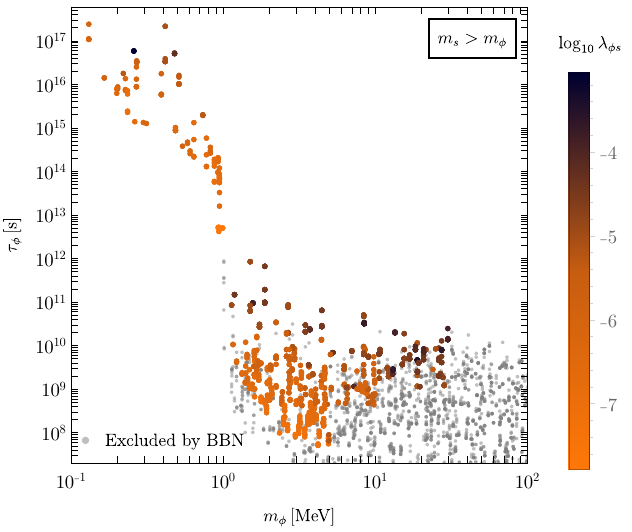}
	\end{subfigure}
	\hfill
	\begin{subfigure}
		\centering
		\includegraphics[width=0.65\textwidth]{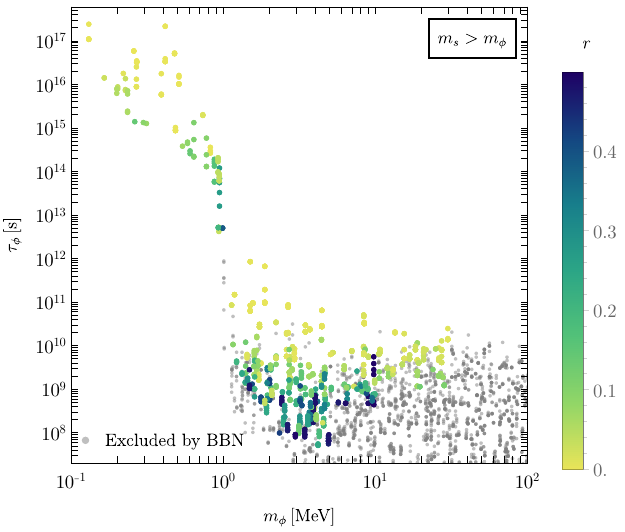}
	\end{subfigure}
	\caption{The same points as in Figure~\ref{fig:tauphi_vs_abundance_ms>mphi} projected on the $\tau_\phi\,\text{vs}\,\,m_\phi$ plane with the color bar displaying the values for $\log_{10}\lambda_{\phi s}$ (top) and $r=m_\phi/(2m_s)$ (down).}
	\label{fig:tauphi_vs_abundance_ms>mphi2}
\end{figure}

\begin{figure}[t!]
	\centering
	\begin{subfigure}
		\centering
		\includegraphics[width=0.65\textwidth]{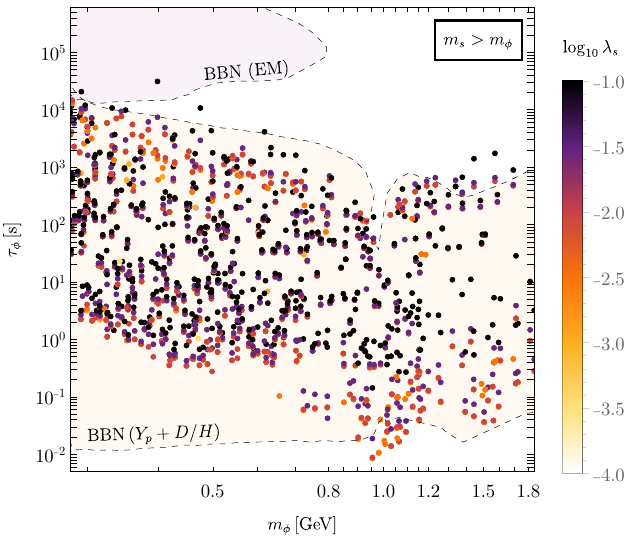}
	\end{subfigure}
	\hfill
	\begin{subfigure}
		\centering
		\includegraphics[width=0.65\textwidth]{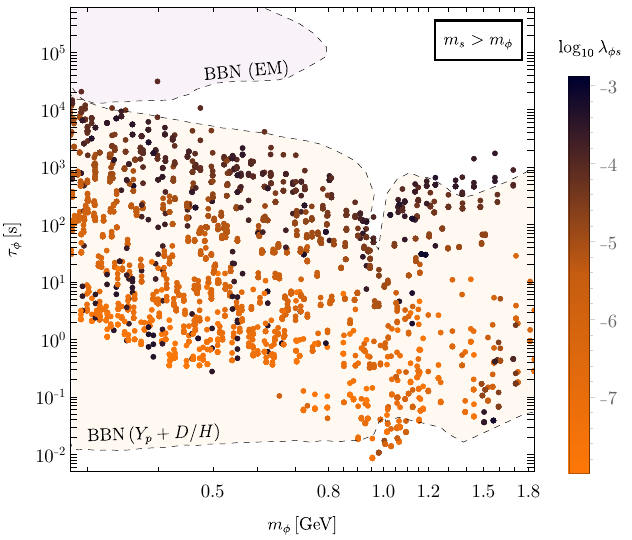}
	\end{subfigure}
	\caption{Same as Figure~\ref{fig:tauphi_vs_abundance_ms>mphi} for mediator masses in the GeV scale with the superimposed BBN limits adapted from~\cite[Figure 13]{Fradette:2018hhl}. Note that most of the points are excluded by BBN observations of the primordial mass fraction $Y_p$ and the $D/H$ ratio.}
	\label{fig:tauphi_vs_abundance_ms>mphiGeV}
\end{figure}

\begin{figure}[t!]
	\centering
	\includegraphics[width=0.65\textwidth]{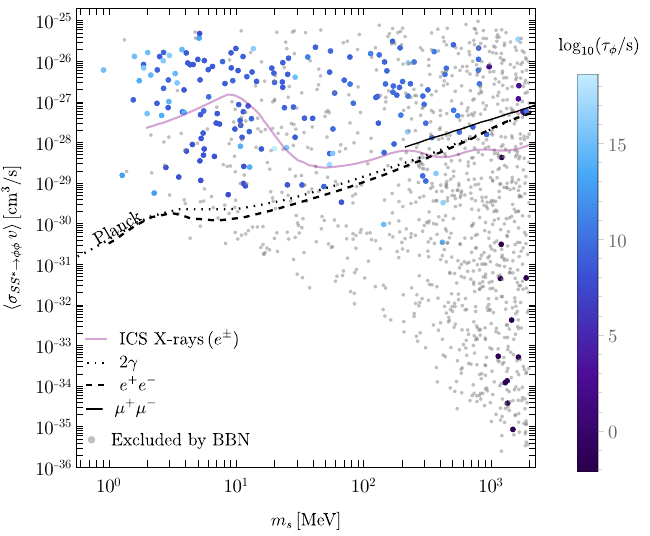}
	\caption{Present day cross section for $SS^*\to 2\phi$ with the superimposed limits of PLANCK~\cite{Planck:2018vyg} (black lines) and from Inverse Compton Scattering (ICS) X-rays produced by energetic electron/positron pairs~\cite{Cirelli:2024kph} (purple line).}
	\label{fig:cross_ID}
\end{figure}

\begin{figure}[t!]
	\centering
	\begin{subfigure}
		\centering
		\includegraphics[width=0.65\textwidth]{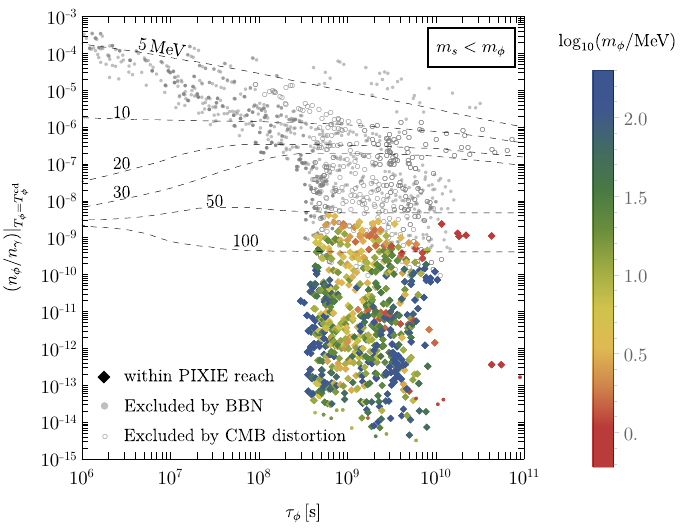}
	\end{subfigure}
	\hfill
	\begin{subfigure}
		\centering
		\includegraphics[width=0.65\textwidth]{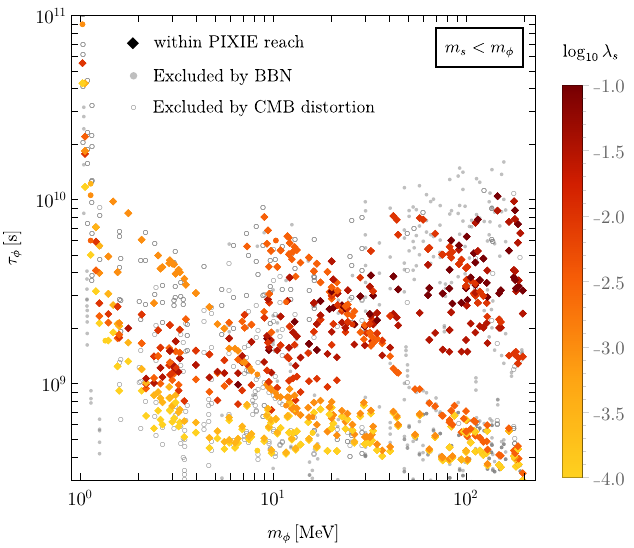}
	\end{subfigure}
	\caption{Same as Figure~\ref{fig:tauphi_vs_abundance_ms>mphi} for the case $m_s<m_\phi$. Empty circles denote points excluded by CMB spectral distortion constraints, while rhomboids indicate points within the projected sensitivity of a PIXIE-like experiment~\cite{Kogut:2024vbi}.}
	\label{fig:tauphi_vs_abundance_ms<mphi}
\end{figure}

\begin{figure}[t!]
	\centering
	\begin{subfigure}
		\centering
		\includegraphics[width=0.65\textwidth]{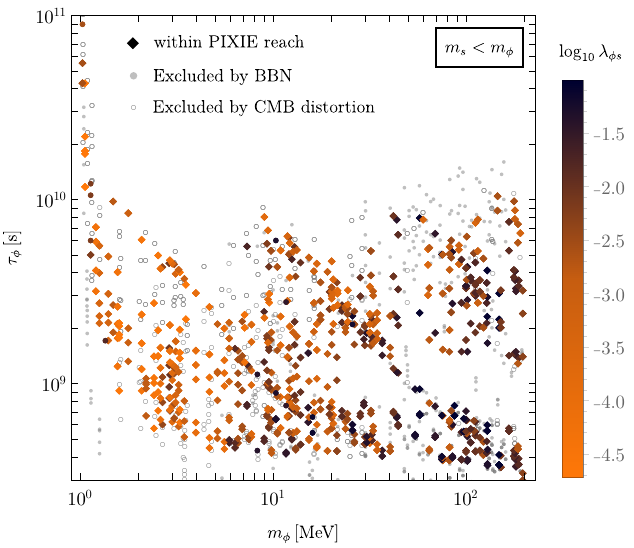}
	\end{subfigure}
	\begin{subfigure}
		\centering
		\includegraphics[width=0.65\textwidth]{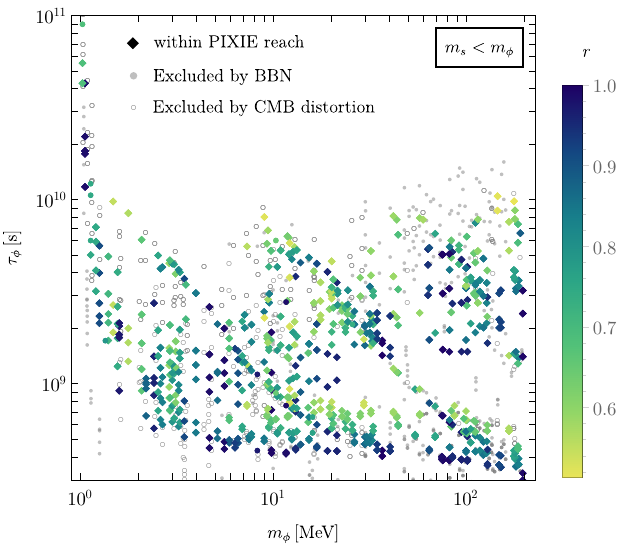}
	\end{subfigure}
	\caption{Same as Figure~\ref{fig:tauphi_vs_abundance_ms>mphi2} for the hierarchy $m_s<m_\phi$.}
	\label{fig:tauphi_vs_abundance_ms<mphi2}
\end{figure}

The phenomenology depends strongly on the mass ordering of $S$ and $\phi$. Observable annihilation signals arise mainly for $m_s>m_\phi$. In this regime the Higgs–mediator mixing is too small for direct annihilation $SS\to\mathrm{SM}$ to be relevant, while $SS\to\phi\phi$ followed by mediator decay can yield an observable signal. The mass hierarchy also affects the cosmological evolution of the dark sector, as it determines both the final dark matter abundance and the abundance of $\phi$. The latter controls the total energy injected into the Standard Model plasma by mediator decays, which is a key ingredient in setting BBN constraints.

Although Big Bang Nucleosynthesis occurs at temperatures $T\sim0.1$--$1\,\mathrm{MeV}$, the resulting light nuclei can be modified at much later times by the decay of long-lived particles. In particular, mediator decays with lifetimes in the range $10^{-2}\,\mathrm{s}\lesssim\tau_\phi\lesssim10^{12}\,\mathrm{s}$ can alter the primordial abundances through both hadronic and electromagnetic channels. For $\tau_\phi\lesssim10^{5}\,\mathrm{s}$, hadronic decay products such as mesons interact efficiently with nucleons and can modify light-element yields, leading to constraints from the observed abundances of $^4\mathrm{He}$, D, $^6\mathrm{Li}$, and $^7\mathrm{Li}$.

At later times, $\tau_\phi\gtrsim10^{5}\,\mathrm{s}$, electromagnetic decays dominate. The injected photons and $e^\pm$ initiate electromagnetic cascades, generating a non-thermal photon spectrum with a time-dependent cutoff energy. As the Universe expands and cools, this cutoff increases, reaching $E\simeq2.2\,\mathrm{MeV}$ at $\tau_\phi\sim10^{5}\,\mathrm{s}$ and $E\simeq19.8\,\mathrm{MeV}$ at $\tau_\phi\sim10^{7}\,\mathrm{s}$. Once these energies exceed the photodisintegration thresholds of deuterium and helium, the cascade photons can destroy nuclei that were produced during BBN, modifying the final light-element abundances~\cite{Jedamzik:2006xz}. A quantitative treatment requires solving the Boltzmann network for nuclear reactions~\cite{Hufnagel:2020nxa}; here we estimate the resulting bounds using~\cite[Fig.~6]{Depta:2020zbh}.

\subsubsection{Scan results for $m_s>m_\phi$}\label{subsubsec:mphi>ms}

For the hierarchy $m_s>m_\phi$, BBN constraints become more restrictive because the channel $SS^*\to2\phi$ can temporarily boost the mediator abundance.~\Cref{fig:tauphi_vs_abundance_ms>mphi,fig:tauphi_vs_abundance_ms>mphi2} show scan points that reproduce the observed relic abundance. Points shaded in gray are excluded by BBN, following the exclusion curves in the top panel of Figure~\ref{fig:tauphi_vs_abundance_ms>mphi}. The remaining points are color-coded to display the dependence on selected parameters. The bottom panel of Figure~\ref{fig:tauphi_vs_abundance_ms>mphi} and the projections in Figure~\ref{fig:tauphi_vs_abundance_ms>mphi2} present the viable solutions in the $(m_\phi,\tau_\phi)$ plane. These plots highlight the role of $\lambda_s$, $\lambda_{\phi s}$, and the mass ratio $r=m_\phi/(2m_s)$. The mediator lifetime increases with $\lambda_{\phi s}$, while varying $\lambda_s$ does not generate a simple trend in these projections. The correlation between $r$ and $\lambda_{\phi s}$ can be understood from the efficiency of $SS^*\to2\phi$: for a strongly hierarchical spectrum ($r\ll1/2$), mediator production from dark matter annihilation is efficient. Avoiding excessive depletion of $S$ in that regime favors smaller $\lambda_{\phi s}$.
Figure~\ref{fig:tauphi_vs_abundance_ms>mphiGeV} shows the same scan projected onto the GeV mediator-mass range, together with BBN exclusion limits adapted from~\cite[Figure 13]{Fradette:2018hhl}. Even when $m_\phi>2m_\mu$ and the mediator decays rapidly, the injected energy can still modify the light-element abundances, and most of the relic-density solutions are excluded by constraints on $Y_p$ and $D/H$. The remaining allowed region lies at $m_\phi\gtrsim1\,\mathrm{GeV}$, with a preference for $\log_{10}\lambda_s\lesssim-2.5$ (top left), and with $\lambda_{\phi s}$ clustered around $\log_{10}\lambda_{\phi s}\lesssim-5$ and $\log_{10}\lambda_{\phi s}\sim-3$ (top right).

This mass ordering is also constrained by late-time energy injection from present-day annihilation. In Figure~\ref{fig:cross_ID} we compare the predicted $SS^*\to2\phi$ cross sections to CMB limits from PLANCK~\cite{Planck:2018vyg}, recast using~\cite{Slatyer:2015jla}, and to X-ray limits from inverse Compton scattering~\cite{Cirelli:2024kph}. The scan points shown here satisfy the relic density constraint; the color bar indicates the mediator lifetime. The limits are displayed assuming a 100\% branching fraction of $\phi$ into photons, electrons, or muons; the differences among these idealized channels do not affect the qualitative conclusions.
To compare with the CMB and X-ray bounds we assume that $SS^*\to2\phi$ occurs at rest, so that each mediator carries energy $E_\phi=m_s$, and we account for the two mediators produced per annihilation as well as the conventional rescaling of the effective dark matter mass by a factor of $1/2$. In the MeV range the predicted cross sections can lie within the sensitivity of telescope observations, while at larger $m_s$ the cross section decreases and the signal typically falls below current reach, especially for GeV-scale dark matter.

In summary, for $m_s>m_\phi$ the combination of relic-density requirements with BBN, CMB, and X-ray constraints removes most of the parameter space, leaving only narrow windows at mediator masses in the GeV range as viable targets for future searches.

\subsubsection{Scan results for $m_s<m_\phi$}\label{subsubsec:mphi>ms}

\begin{figure}[t!]
	\centering
	\begin{subfigure}
		\centering
		\includegraphics[width=0.65\textwidth]{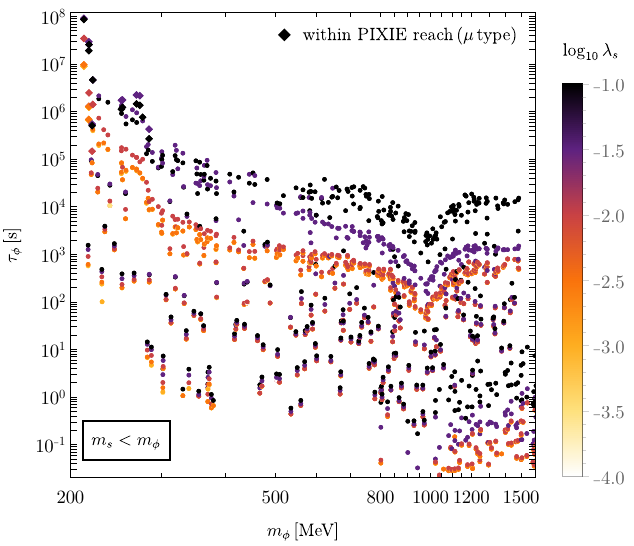}
	\end{subfigure}
	\hfill
	\begin{subfigure}
		\centering
		\includegraphics[width=0.65\textwidth]{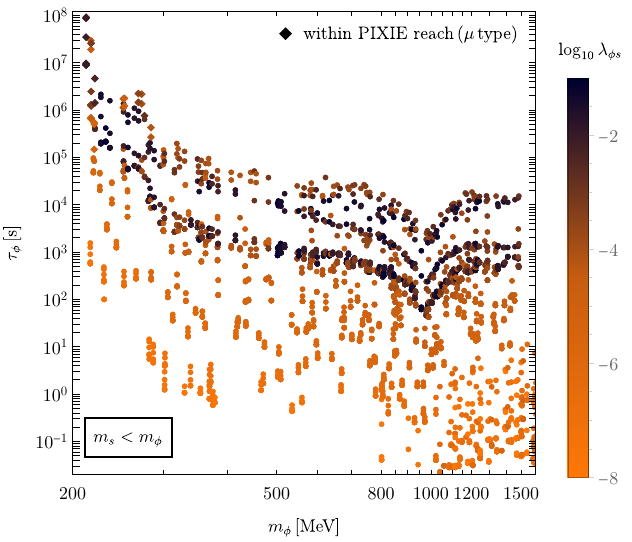}
	\end{subfigure}
	\caption{Same as Figure~\ref{fig:tauphi_vs_abundance_ms<mphi} for $m_s<m_\phi$, restricted to solutions above the muon threshold. Viable points occur for mediator masses in the range $212$--$284.35\,\mathrm{MeV}$ with $\log_{10}\lambda_{\phi s}\in[-5.27,-2]$.}
	\label{fig:tauphi_vs_abundance_ms<mphi_GeV}
\end{figure}

In the inverted mass hierarchy, $m_s<m_\phi$, mediators are efficiently depleted through annihilation into dark matter, which significantly relaxes cosmological constraints. This behavior is illustrated in the top panel of Figure~\ref{fig:tauphi_vs_abundance_ms<mphi}, where a large fraction of the viable points lies below the BBN exclusion curves.\footnote{For mediator masses above the muon threshold (cf. Figure~\ref{fig:tauphi_vs_abundance_ms<mphi_GeV}), BBN limits from the primordial helium mass fraction and the $D/H$ ratio require dedicated computations for this specific scenario, which are beyond the scope of this work.}

Unlike the $m_s>m_\phi$ case, the impact of mediator decays on the CMB spectrum in this scenario requires an explicit treatment of spectral distortions, due to the lower abundance of $\phi$. The photon spectrum prior to recombination is tightly constrained by COBE/FIRAS measurements~\cite{Fixsen:1996nj}. Energy injection from $\phi$ decays around or after recombination can distort the blackbody spectrum and modify the ionization history near last scattering ($z\simeq1000$), affecting the CMB temperature and anisotropy spectra. These distortions are commonly parameterized by the $\mu$- and $y$-type parameters~\cite{Chluba:2016bvg},
\begin{equation}
	\begin{split}
		y&=\frac{1}{4}\int_{z_\text{rec}}^{z_{\mu y}}\frac{d(Q/\rho_\gamma)}{dz'}\,dz'\,,\\
		\mu&=1.401\int_{z_{\mu y}}^{\infty}e^{-\left( \frac{z'}{z_\mu}\right)^{5/2}}\frac{d(Q/\rho_\gamma)}{dz'}\,dz'\,,
	\end{split}
\end{equation}
where $z_\text{rec}=1000$, $z_{\mu y}\simeq5\times10^{4}$, and $z_\mu=2\times10^{6}$. The COBE/FIRAS bounds~\cite{Fixsen:1996nj},
\begin{equation}
	\begin{split}
		|\mu|_\text{CF}&<9\times10^{-5}\,,\\
		|y|_\text{CF}&<1.5\times10^{-5}\,,
	\end{split}
\end{equation}
exclude the points marked with empty circles in~\Cref{fig:tauphi_vs_abundance_ms<mphi,fig:tauphi_vs_abundance_ms<mphi2}. The proposed PIXIE mission~\cite{Kogut:2024vbi}, with a sensitivity roughly three orders of magnitude stronger, would probe distortions down to
\begin{equation}
	\begin{split}
		|\mu|_\text{Pixie}&<10^{-9}\,,\\
		|y|_\text{Pixie}&<2\times10^{-9}\,.
	\end{split}
\end{equation}
Points predicting distortions at this level are indicated by rhomboids in~\Cref{fig:tauphi_vs_abundance_ms<mphi,fig:tauphi_vs_abundance_ms<mphi2,fig:tauphi_vs_abundance_ms<mphi_GeV}. PIXIE is sensitive to most solutions in the MeV range below the muon threshold, except for those with $\left.n_\phi/n_\gamma\right|_{\rm cd}\lesssim1.63\times10^{-13}$ (filled colored circles in Figure~\ref{fig:tauphi_vs_abundance_ms<mphi}). When the mediator can decay into muons, its lifetime is shortened and the resulting distortions fall below PIXIE sensitivity; nevertheless, mediator masses in the interval $212$--$284\,\mathrm{MeV}$ remain accessible through $\mu$-type distortions (cf. Figure~\ref{fig:tauphi_vs_abundance_ms<mphi_GeV}).

In this mass hierarchy, an additional correlation between $m_\phi$ and $\tau_\phi$ emerges due to dark matter self-interactions, as seen in the bottom panel of Figure~\ref{fig:tauphi_vs_abundance_ms<mphi}. Larger values of $\lambda_s$ lead to longer mediator lifetimes, since efficient cannibalization converts kinetic energy into number density and reduces the required freeze-in contribution. Moreover, points consistent with both BBN and CMB constraints typically require stronger DM--mediator interactions than in the $m_s>m_\phi$ case. This is evident from Figure~\ref{fig:tauphi_vs_abundance_ms<mphi2} (top), where viable solutions satisfy $-4.5\lesssim\log_{10}\lambda_{\phi s}\le -1$, compared to $-8\lesssim\log_{10}\lambda_{\phi s}\lesssim-3$ in the opposite hierarchy. Smaller $\lambda_{\phi s}$ suppress mediator annihilation into dark matter and reduce heat transfer to the DM sector, necessitating enhanced freeze-in production, which in turn strengthens cosmological constraints. Then, larger values of $\lambda_{\phi s}$ are favored.

\begin{figure}[t!]
	\centering
	\begin{subfigure}
		\centering
		\includegraphics[width=0.65\textwidth]{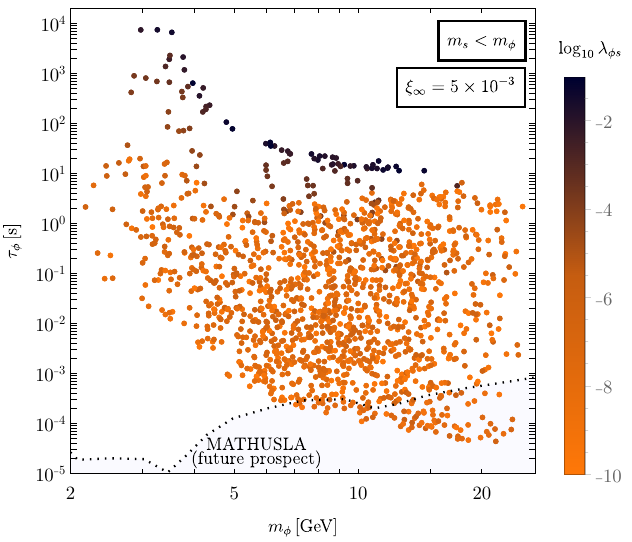}
	\end{subfigure}
	\hfill
	\begin{subfigure}
		\centering
		\includegraphics[width=0.65\textwidth]{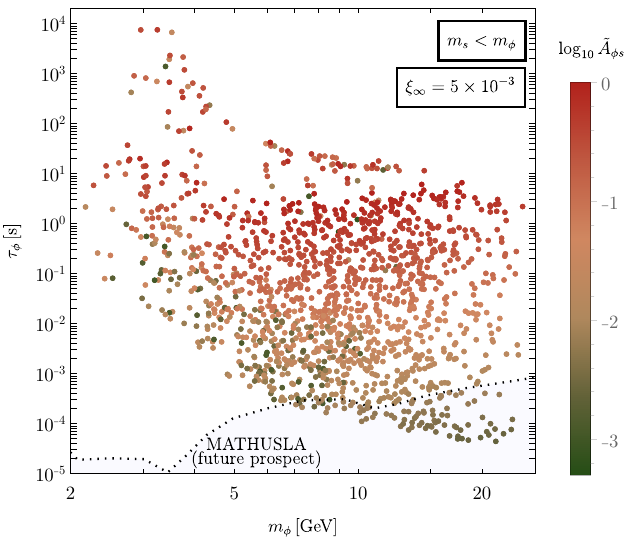}
	\end{subfigure}
	\caption{Results in the GeV mass range with projected sensitivity of the MATHUSLA experiment~\cite{Curtin:2024xxo}. Potentially detectable points satisfy $\log_{10}\lambda_{\phi s}\lesssim-6.15$, corresponding to a regime in which the mediator and dark matter sectors are effectively decoupled and populated independently.}
	\label{fig:Mathusla}
\end{figure}

The influence of $\lambda_s$ is visible through discrete bands in the scan results, most clearly in Figure~\ref{fig:tauphi_vs_abundance_ms<mphi_GeV}. These features are a numerical artifact of the discrete grid used to sample $\log_{10}\lambda_s$ and carry no intrinsic physical significance. The visible bands correspond to $\log_{10}\lambda_s=-2,-1.5$, and $-1$. Physically, smaller $\lambda_s$ leads to shorter mediator lifetimes, as weaker self-interactions reduce the efficiency of cannibalization and require stronger freeze-in production, which in turn increases the mediator coupling (cf.~\Cref{fig:evol1,fig:evol3}). The coupling $\lambda_{\phi s}$ further separates the solutions into two regimes, $\log_{10}\lambda_{\phi s}\lesssim-2.5$ and $\log_{10}\lambda_{\phi s}\gtrsim-2.5$ (Figure~\ref{fig:tauphi_vs_abundance_ms<mphi_GeV}, right). The resonance-like feature around $m_\phi\simeq1\,\mathrm{GeV}$ arises from a sharp decrease in the mediator lifetime due to a resonance in the $\Gamma_{\phi\to\pi\pi}$ decay width~\cite{Winkler:2018qyg}.
Finally, the $m_s<m_\phi$ scenario can also be probed by searches for long-lived particles at MATHUSLA. The experiment is designed to detect neutral particles produced at the LHC with lifetimes up to $\mathcal{O}(10^{-4}\,\mathrm{s})$~\cite{Curtin:2024xxo}, placing GeV-scale mediators in this model within reach. The projected sensitivity is shown in Figure~\ref{fig:Mathusla}, highlighting the role of $\lambda_{\phi s}$ and $\tilde A_{\phi s}=A_{\phi s}/m_\phi$. Detectable points satisfy $\log_{10}\lambda_{\phi s}\lesssim-6.15$ and $-2.9\lesssim\log_{10}\tilde A_{\phi s}\lesssim-2$, corresponding to a regime where the two sectors are effectively decoupled: $\phi$ is produced directly through Higgs mixing, while $S$ originates predominantly from Higgs decays. Smaller $\lambda_{\phi s}$ suppress mediator-induced DM production and require larger mixing angles, shortening the mediator lifetime, whereas larger $\tilde A_{\phi s}$ enhances DM production from Higgs decay and must be compensated by smaller $\theta$, pushing $\tau_\phi$ beyond the detector sensitivity.

In summary, the hierarchy $m_s<m_\phi$ significantly enlarges the viable parameter space in the MeV range and provides strong motivation for future searches targeting CMB spectral distortions and long-lived particles, such as PIXIE and MATHUSLA.

\clearpage

\section{Fermionic DM with cannibalising mediator}
\label{sec:wDM}

In this section we consider a simple scenario in order to illustrate how cannibal dynamics can quantitatively modify the phase-space distribution of frozen-in dark matter, with potential implications for Lyman-$\alpha$ constraints on dark matter. The goal is to use the model introduced below as a minimal benchmark in which an unstable scalar mediator undergoes self-number-changing reactions before decaying into collisionless dark matter. This provides a controlled framework to isolate the impact of cannibalization on the dark-sector temperature evolution and on the resulting dark-matter momentum distribution. Similar effects may also arise more generally in hidden-sector models with unstable mediators and out-of-equilibrium dark-matter production. 
It should be noted that the possibility of weakening Lyman-\(\alpha\) constraints through self-number-changing reactions was already pointed out in~\cite{Garny:2018byk}, where DM itself undergoes cannibalization so that its own thermal history directly determines its free-streaming properties. 

Cannibalism has been discussed in the previous sections in the context of the dark-matter itself undergoing number-changing reactions. Here we focus instead on a different possibility, in which cannibal dynamics takes place in an intermediate component of the dark sector~\cite{Pappadopulo:2016pkp}. Concretely, the model introduced in~\Cref{sec:real_scalar} is extended such that the real scalar field in the broken phase acts as a mediator, while the dark matter is identified with a Dirac fermion \(\chi\). In this case, the scalar mediator undergoes cannibal evolution, whereas \(\chi\) is produced from mediator decays and does not participate directly in the number-changing reactions.

The scalar potential corresponds to the one introduced in Eq.~\eqref{eq:Z2_broken_phase_potential}, supplemented by the mediator--dark-matter Yukawa interaction
\begin{equation}\label{eq:med_wdm_interactions}
	V
	= \frac{1}{2}m_\varphi^2 \varphi^2
	+ \frac{g}{3!}\varphi^3
	+ \frac{\lambda}{4!}\varphi^4
	+ \frac{\lambda_{h\varphi}}{2}\varphi^2\vert H\vert^2
	+ y \varphi \bar{\chi}\chi \,,
\end{equation}
where \(y\ll 1\) is assumed, so that \(\chi\) is populated through the decay process \(\varphi \to \chi\bar{\chi}\). Since \(\varphi\) is not required to be stable, no stabilizing symmetry is imposed on the scalar potential. As a result, \(2\leftrightarrow 3\) self-number-changing reactions are allowed in the mediator sector and are responsible for the cannibal phase.

This scenario differs qualitatively from scenarios in which the dark-matter particle itself participates in the cannibal dynamics. In such cases, the final dark-matter momentum distribution is shaped directly by the thermodynamics of the cannibal phase. Here, by contrast, \(\chi\) is produced via freeze-in and does not scatter with itself. Its phase-space distribution is inherited indirectly from the decay of a mediator whose abundance and temperature evolution are modified by cannibalization. The resulting warmness of the dark-matter spectrum is therefore controlled by the interplay between mediator thermodynamics and decay kinematics, rather than by dark-matter number-changing reactions acting directly on \(\chi\).


The phase-space distribution of \(\chi\) is governed by the unintegrated Boltzmann equation~\eqref{eq:fBE},
\begin{equation}\label{eq:fbe_for_chi}
	\left( \partial_t - H p\,\partial_p \right) f_\chi(t,p)
	= C[f_\chi] \,,
\end{equation}
which must be solved at the phase-space level because the dark-matter self-interactions are too weak to maintain local thermal equilibrium. Indeed, the only self-scattering channel is \(\chi\chi \to \varphi \to \chi\chi\), whose amplitude is suppressed by the feeble Yukawa coupling \(y\). As a consequence, the momentum distribution of \(\chi\) is not reshuffled after production, but instead retains memory of the microscopic production process. While the total dark-matter abundance generated via freeze-in can be obtained without \emph{a priori} solving for the full distribution function \(f_\chi\), quantities relevant for structure formation, such as the velocity dispersion and free-streaming length, depend on the detailed shape of \(f_\chi\). For this reason, a moment-based treatment is not sufficient if one wishes to assess the warmness of the produced dark matter.

Let us now address the processes in the collision operator. In the present freeze-in regime, the DM production occurs via mediator decays, so that the collision term in Eq.~\eqref{eq:fbe_for_chi} is determined primarily by this process, while inverse decays and quantum-statistical corrections can be safely neglected. The corresponding phase-space evolution has been studied, for example, in Ref.~\cite{DEramo:2025fvy} in the context of non-standard reheating histories.

For the decay process \(\varphi \to \chi\bar{\chi}\), the collision term reads
\begin{equation}\label{eq:C_for_chi}
	C[f_\chi]
	= 
	\frac{1}{2E_\chi g_\chi}\int d\Pi_\varphi\, d\Pi_{\bar\chi}\, \vert \tilde{\mathcal M}\vert^2 f_\varphi(p_\varphi)\,,
\end{equation}
where
\begin{equation}
	|{\mathcal M}|^2 \equiv 2 y^2 \left(m_\varphi^2 - 4 m_\chi^2\right)\,,
\end{equation}
already includes the sum over spins. We track the distribution of a single particle species, \(\chi\). Note that Eq.~\eqref{eq:C_for_chi} results in Eq.~\eqref{eq:C_higgs_decay} with the corresponding trilinear coupling and distribution being $f_\varphi = z_\varphi\,e^{-E/T_\varphi}$. We assume that both \(\chi\) and \(\bar\chi\) contribute equally to the final dark-matter relic abundance, and that there is neither a mass splitting nor an asymmetry within the dark sector. The final abundance is therefore given by \(n = n_{\chi} + n_{\bar\chi} = 2n_\chi\).

With this collision operator, it is possible to integrate Eq.~\eqref{eq:fbe_for_chi} in terms of the comoving momentum \(q = a\,p\) as
\begin{equation}\label{eq:f_chi_sol}
	f_\chi(q,t) = f^0_\chi+\int_{a_i}^a \frac{da'}{a'H}\,C[\chi]\,,
\end{equation}
where we used \((\partial_t - H\,p\,\partial_p)f_\chi(t,p) = \partial_t f_\chi(t,q) = aH\,\partial_a f_\chi(a,q)\), and in the freeze-in regime we take \(f_\chi^0=0\). In practice, the integral in Eq.~\eqref{eq:f_chi_sol} is evaluated numerically, similarly to~\cite{Bae:2017dpt}.

From the perspective of \(\varphi\), its collision operator includes the cannibal reactions in Eq.~\eqref{eq:C_3to2_realscalar}, the Higgs-decay source term in Eq.~\eqref{eq:C_higgs_decay}, and in the present case also the term in Eq.~\eqref{eq:C_for_chi} with a minus sign, which depletes the mediator abundance while sourcing the dark matter.

%

We now need the moments of the collision operator \(C^{\varphi\to\bar\chi\chi}\) that enter the cBE for the mediator. Since the mediator remains in local thermal equilibrium within the dark sector and is described by a Maxwell-Boltzmann distribution, the zeroth moment results in
\begin{equation}
	C_0^{\varphi \to \chi\bar{\chi}}
	=
	\Gamma_{\varphi \to \chi\bar{\chi}}
	\frac{m_\varphi^2 T_\varphi}{2\pi^2}
	K_1\!\left(\frac{m_\varphi}{T_\varphi}\right),
\end{equation}
for a single real scalar degree of freedom.

For a Dirac fermion \(\chi\) with interaction \(y \varphi \bar{\chi}\chi\), the mediator decay width is
\begin{equation}
	\Gamma_{\varphi \to \chi\bar{\chi}}
	=
	\frac{y^2 m_\varphi}{8\pi}
	\left(1-\frac{4m_\chi^2}{m_\varphi^2}\right)^{3/2}\,.
	\label{eq:phi_to_chichi_width}
\end{equation}

Upon weighting the Boltzmann equation by the particle energy and integrating over phase space, one obtains the corresponding source term for the dark-matter energy density,
\begin{equation}
	C_E^{\varphi\to\chi\bar\chi}
	=
	\Gamma_{\varphi\to\chi\bar\chi}
	\frac{m_\varphi^3 T_\varphi}{2\pi^2}
	K_2\!\left(\frac{m_\varphi}{T_\varphi}\right)\,.
\end{equation}

We can then utilise the cBEs in Eqs.~\eqref{eq:Yeq_x} and~\eqref{eq:dTi_dx},
\begin{equation}
	\begin{split}
		\frac{dY_\varphi}{dx} &= \frac{\tilde g(T)}{x\,H\,s}\,C_0[\varphi]\,,
		\\
		\frac{dT_\varphi}{dx}
		&=
		\frac{
			\frac{\tilde g}{xH}C_E[\varphi]
			-\frac{3\tilde g}{x}(\rho_\varphi+P_\varphi)
			-\frac{\partial\rho_\varphi}{\partial Y_\varphi}\frac{dY_\varphi}{dx}
		}{
			\frac{\partial\rho_\varphi}{\partial T_\varphi}
		}\,,
	\end{split}
\end{equation}
to obtain the thermal evolution of \(\varphi\). Note that here the cannibal reactions satisfy \(C_E^{3\varphi\leftrightarrow2\varphi} = 0\) due to conservation of energy.

The momentum distribution of \(\chi\) is not determined solely by the decay kinematics, but also by the thermal history of the mediator sector. During the cannibal phase, number-changing reactions maintain chemical equilibrium in the mediator bath and modify the scaling of its temperature with the scale factor. As a result, the decay injects dark matter with a non-thermal spectrum that differs from the standard thermal or quasi-thermal warm-dark-matter templates. In this sense, the impact of cannibalization is encoded directly in the final phase-space distribution \(f_\chi\), which is the quantity relevant for any subsequent assessment of structure-formation constraints.

In a more complete analysis, one would use the resulting distribution to extract quantities such as the average velocity or the free-streaming horizon~\cite{Garzilli:2019qki,Coy:2021ann},
\begin{equation}\label{eq:lambda_FS}
	\lambda_{\rm FS}(a)
	=
	\int_{a_{\rm prod}}^{a}
	\frac{\langle v(a') \rangle}{a'^2 H(a')}\, da' \,,
\end{equation} 
and ultimately to infer the corresponding suppression of the matter power spectrum. In the present work, we focus on how cannibal dynamics modifies \(f_\chi\) itself, and use the resulting spectral distortion as the main diagnostic of the warmness of the produced dark matter.

\subsection{Results}
\label{subsec:results_wDM}

\begin{figure}[t!]
	\centering
	\includegraphics[width=0.55\textwidth]{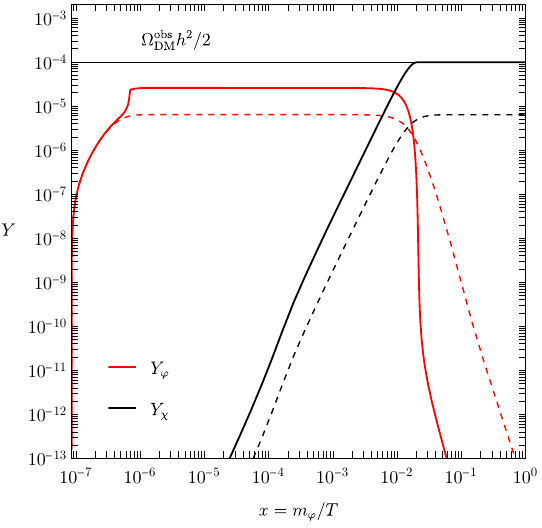}
	\hfill
	\includegraphics[width=0.55\textwidth]{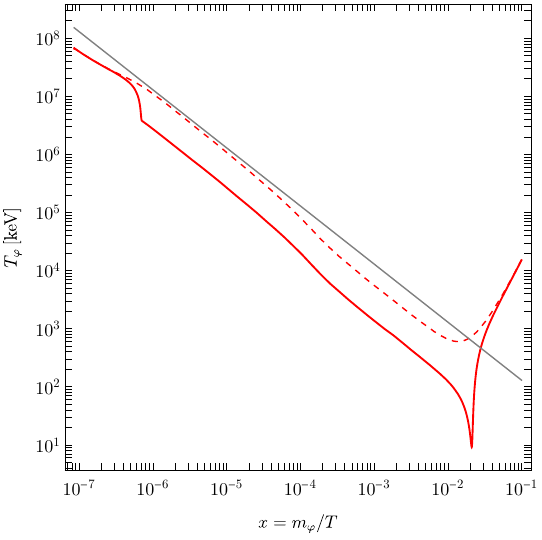}
	\caption{Yield (top) and temperature evolution of the mediator (bottom) for the parameter point in Eq.~\eqref{eq:BMch3}. The red lines correspond to the cannibal mediator, while the black lines correspond to the fermionic dark matter. For reference, the gray line corresponds to the temperature of the Standard Model bath in the bottom plot.}
	\label{fig:BM_cann_med}
\end{figure}

As a benchmark point to illustrate the qualitative impact of mediator self-interactions, we fix
\begin{equation}\label{eq:BMch3}
	\begin{split}
		m_\varphi &= 12.93\,\text{keV}\,,\qquad m_\chi = 2.22\,\text{keV}\,,
		\\
		\lambda &= 8.9\times 10^{-4}\qquad\text{and}\qquad y = 8\times 10^{-9}\,,
	\end{split}
\end{equation}
while \(\lambda_{h\varphi} = 10^{-9}\) is chosen such that the observed DM relic abundance is reproduced. We focus on a single benchmark, rather than a broader parameter scan, because our aim is to make transparent the new qualitative effect induced by cannibal dynamics in the mediator sector: namely, the modification of the dark-sector thermal history and the resulting distortion of the freeze-in dark-matter momentum distribution. The chosen point lies in a regime where mediator self-number-changing reactions are sufficiently efficient to produce a visible effect, allowing a direct comparison between the solutions with and without cannibalization. In~\Cref{fig:BM_cann_med} we show the corresponding thermal evolution.

Note that at later times, once \(\varphi\to\chi\bar\chi\) becomes the dominant process, the mediator yield drops rapidly while its temperature increases. This behaviour is a consequence of slower mediators decaying more efficiently than highly boosted ones. As a result, the surviving \(\varphi\) population becomes increasingly dominated by the energetic tail of the distribution. The average energy per remaining mediator therefore increases, which is reflected as an increase in the effective mediator temperature even though the total number density is being depleted.

An apparently odd feature of the yield evolution is that, when self-number-changing reactions are efficient, the peak value of the single-species dark-matter yield \(Y_\chi\) can exceed the initial mediator yield \(Y_\varphi\) by a factor of a few (compare black and red thick lines on the yield evolution of~\Cref{fig:BM_cann_med}). This does not signal a violation of particle conservation in the decay \(\varphi\to\chi\bar\chi\). Rather, it reflects the fact that particle number in the mediator sector is not conserved prior to the decay stage when \(2\leftrightarrow3\) reactions are active. After the initial freeze-in injection, the approximately conserved quantity is the comoving energy density of the dark sector, not the mediator number density. As the mediator bath approaches chemical equilibrium, part of this energy is converted into additional mediator quanta, increasing the total number of \(\varphi\) particles while lowering their average energy. These extra mediators subsequently decay into \(\chi\bar\chi\), so that the final dark-matter yield can become larger than the initial mediator yield, even when tracking only a single species \(\chi\).

\begin{figure}[t!]
	\centering
	\includegraphics[width=0.65\textwidth]{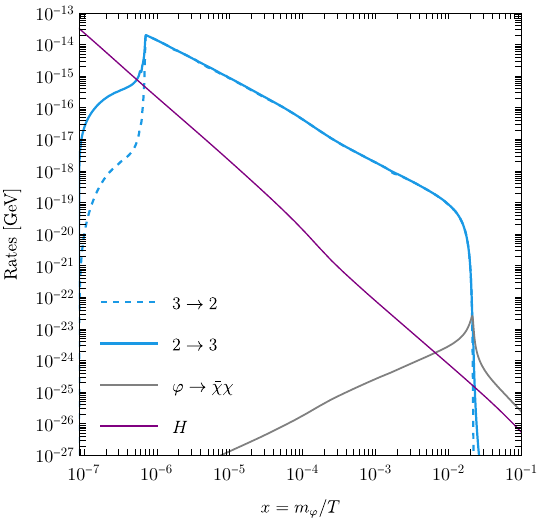}
	\caption{Relevant rates for the benchmark point.}
	\label{fig:cannmed_rates}
\end{figure}

\begin{figure}[t!]
	\centering
	\includegraphics[width=0.65\textwidth]{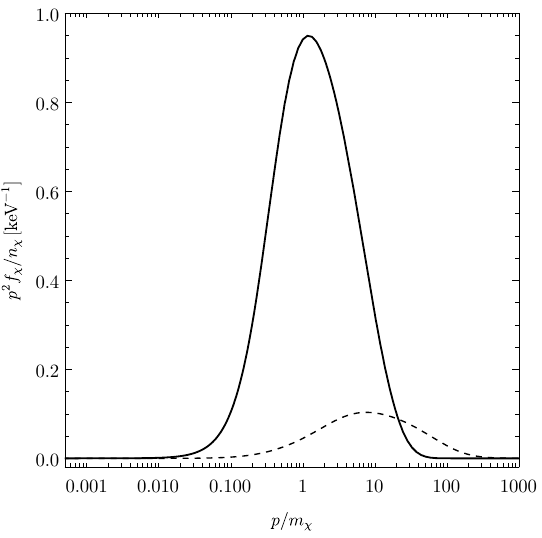}
	\caption{Solutions of Eq.~\eqref{eq:f_chi_sol} normalized with the number density for the cases with (thick) and without (dashed) self-number-changing reactions corresponding to $x = 0.05$, i.e., $T = 258.686\,\text{keV}$.}
	\label{fig:dist_func_chi}
\end{figure}

\begin{figure}[t!]
	\centering
	\includegraphics[width=0.55\textwidth]{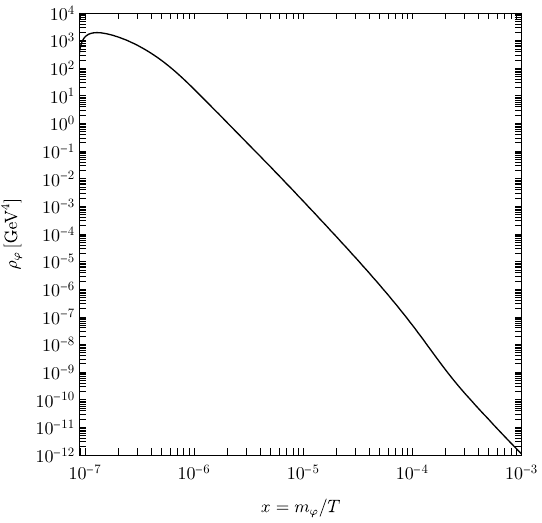}
	\hfill
	\includegraphics[width=0.55\textwidth]{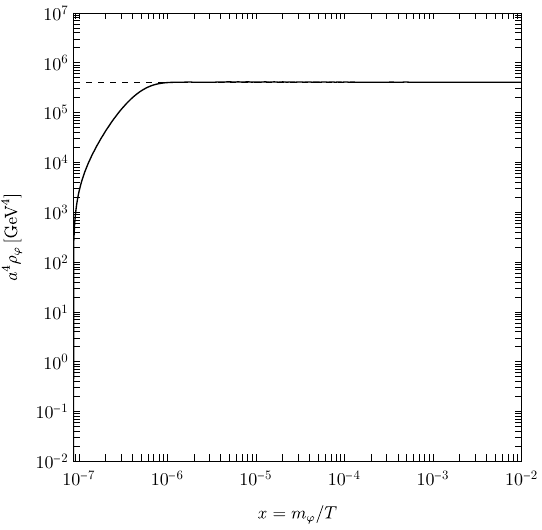}
	\caption{Energy density (top) and comoving energy density (down) of \(\varphi\). After the energy injection from Higgs decay ends, the comoving energy density remains approximately constant, \(a^4\rho_\varphi \simeq 4\times10^5\,\text{GeV}^4\), fixing the maximum possible cooling of the mediator sector.}
	\label{fig:rho_evol}
\end{figure}

Another interesting feature appears close to decoupling around $x(\equiv m_\varphi/T) = 0.02$, where the red line in the temperature evolution drops momentarily. This occurs due to the production process \(2\to3\) becoming more efficient than \(3\to2\) just before decoupling, contrary to the standard cannibal freeze-out. In this scenario, the decay \(\varphi\to\chi\bar\chi\) depletes the mediator abundance, suppressing the \(3\to2\) rate, which is more sensitive to the population of $\varphi$'s than the \(2\to3\) process, temporarily cooling the mediators. The rates are shown in~\Cref{fig:cannmed_rates}.

After evaluating Eq.~\eqref{eq:f_chi_sol}, we obtain the phase-space distributions shown in~\Cref{fig:dist_func_chi}. The two solutions differ both in normalization and in shape. In the absence of self-number-changing reactions, the mediator sector remains chemically underpopulated, \(z_\varphi \ll 1\), so that fewer mediator particles are available to decay into dark matter. This suppresses the overall normalization of the \(\chi\) distribution. At the same time, however, the mediator sector is hotter, such that the average energy per mediator is larger. The daughter particles produced in \(\varphi\to\chi\bar\chi\) therefore inherit a broader and harder momentum distribution, which explains why the resulting \(f_\chi\) leans toward higher momenta and leads to a larger effective dark-matter temperature. 

Quantitatively, around the time of decoupling, \(x=0.05\), the case without self-interactions gives \(T_\chi^0 = 84.07\,\text{keV}\), while the case with self-interactions gives \(T_\chi = 6.57\,\text{keV}\). The dark-matter temperature is then lower by a factor of about \(12.8\) in the presence of self-number-changing interactions.


This factor is already large enough to indicate a parametrically important change in the free-streaming properties of the produced dark matter. Since the characteristic velocity scales schematically as \(v \sim p/m_\chi\), a colder momentum distribution implies a correspondingly smaller free-streaming length and therefore a weaker suppression of small-scale structure. In this sense, the effect of mediator cannibalization can qualitatively change whether the resulting fermionic dark matter behaves as effectively warm or effectively cold. A precise translation into structure-formation bounds would, however, require the computation of the linear matter power spectrum for the full non-thermal distribution~\cite{Ballesteros:2020adh}, which goes beyond the scope of this work.


Note that the maximum cooling, and therefore the largest possible modification of the dark-matter free-streaming properties, is not arbitrary, but is fixed by the total energy injected into the dark sector. Initially, the energy density stored in the mediator sector is negligible. During freeze-in, Higgs decays transfer a finite amount of energy into the dark sector, and once this injection becomes negligible the subsequent evolution is governed by covariant energy-momentum conservation,
\begin{equation}
	\dot{\rho}_\varphi + 3H(\rho_\varphi + P_\varphi)=0 \,.
\end{equation}
As long as the mediator remains relativistic, one has \(P_\varphi \simeq \rho_\varphi/3\), so that
\begin{equation}
	\dot{\rho}_\varphi + 4H\rho_\varphi = 0 \,,
\end{equation}
and therefore the comoving relativistic energy density
\begin{equation}
	U_0 \equiv a^4 \rho_\varphi
\end{equation}
is conserved.\footnote{By contrast, \(a^3\rho_\varphi\) is the total energy in a comoving volume, but it is not conserved in the relativistic regime because the fluid performs \(P\,dV\) work during the expansion.} This is illustrated in~\Cref{fig:rho_evol}. Self-number-changing reactions may redistribute the injected energy among mediator particles and drive the system toward chemical equilibrium, but they do not change the conserved quantity \(U_0\).

For a relativistic gas,
\begin{equation}
	\rho_\varphi = z_\varphi \frac{\pi^2}{30} g_\varphi T_\varphi^4\,,
\end{equation}
which implies
\begin{equation}
	T_\varphi = \frac{1}{a}\left( \frac{30}{z_\varphi g_\varphi \pi^2} U_0 \right)^{1/4}\,.
\end{equation}
The lowest possible temperature is attained once chemical equilibrium is reached, corresponding to \(z_\varphi = 1\). In that limit, the minimum temperature is therefore fixed by
\begin{equation}
	T_\varphi = \frac{1}{a}\left( \frac{30}{g_\varphi \pi^2} U_0 \right)^{1/4}\,.
\end{equation}

\section{Summary}
\label{sec:summary_ch3}

In this chapter we investigated several realizations of cannibal dark matter produced through freeze-in, with particular emphasis on whether the portal interaction responsible for populating the dark sector can also give rise to observable signatures in indirect detection, long-lived particle searches, or structure-formation probes. In all cases, we derived and solved the coupled Boltzmann equations for the relevant number densities and temperatures, consistently including decays, annihilations, elastic scatterings, and \(3\leftrightarrow2\) reactions. This framework allows for a consistent treatment of heat exchange between the visible and dark sectors, as well as the subsequent evolution of particle abundances through freeze-in and freeze-out. In the final part of the chapter, we also went beyond the moment-based description and considered the phase-space distribution of dark matter itself, in order to assess the impact of warm-dark-matter bounds in scenarios where the mediator, rather than the dark matter particle, undergoes cannibal evolution.

The first scenario considered a self-interacting real scalar dark matter candidate in a spontaneously broken \(\mathbb{Z}_2\) theory. In this setup, the breaking of the stabilizing symmetry generates the \(3\leftrightarrow2\) processes required for cannibalization, but at the same time induces dark matter decay. This leads to a strong tension: the portal coupling must remain small enough to satisfy lifetime and indirect-detection bounds, yet a sufficiently large coupling is needed for efficient freeze-in production. As a result, the viable parameter space is severely restricted, and freeze-in by itself is generally unable to produce the observed dark matter abundance. This scenario therefore requires either an additional production channel or a modified cosmological history.

The second scenario involved a complex scalar dark matter candidate stabilized by a \(\mathbb{Z}_3\) symmetry. In this case, freeze-in can successfully account for the observed relic abundance while remaining compatible with existing constraints. A notable feature of this setup is the non-trivial role of \(2\to3\) processes, which can significantly enhance the dark-sector yield and thereby alter the final abundance. Unlike the broken \(\mathbb{Z}_2\) case, the dark matter is stable and no late decays are induced. However, because the interaction with the Standard Model proceeds only through a very weak Higgs portal, the model remains experimentally difficult to probe with current searches.

The third scenario extended the \(\mathbb{Z}_3\) setup by introducing an unstable Higgs-like mediator \(\phi\). Here the phenomenology depends strongly on the mass hierarchy between the mediator and the dark matter. For \(m_s>m_\phi\), annihilation into mediators is kinematically open and can lead to an excessive mediator abundance during Big Bang nucleosynthesis, resulting in stringent cosmological bounds. In this regime, both dark matter self-interactions and dark matter--mediator interactions are important in determining the mediator lifetime and the viability of the model. The same region can also lead to sizable \(s\)-wave annihilation rates and therefore to potentially observable indirect-detection signals. By contrast, for \(m_s<m_\phi\), the mediator population is efficiently depleted into dark matter at early times, which substantially relaxes the cosmological bounds from late-time energy injection. In the GeV-scale regime, this opens interesting prospects for long-lived particle searches, in particular at experiments such as MATHUSLA.

A further result of this chapter is that cannibal dynamics need not be carried by the dark matter particle itself in order to leave an imprint on structure formation. This was illustrated in the warm-dark-matter analysis, where the mediator \(\varphi\) undergoes cannibal evolution while the dark matter is instead given by a feebly coupled fermion \(\chi\) produced through \(\varphi\to\chi\bar\chi\). In this case, the relevant quantity is no longer only the final relic abundance, but also the full phase-space distribution of the produced dark matter. Since the fermion does not possess sufficiently strong self-interactions to maintain local thermal equilibrium, its unintegrated Boltzmann equation must be solved directly. This allows one to determine the effective dark-matter temperature.

The main lesson of this analysis is that the presence or absence of self-number-changing reactions in the mediator sector affects not only the total abundance of the produced dark matter, but also the shape of its momentum distribution. When cannibal self-interactions are efficient, the mediator approaches chemical equilibrium and redistributes the available energy among a larger number of particles. As a consequence, the dark matter produced from mediator decays inherits a colder spectrum. In the opposite case, where self-interactions are absent and the mediator remains chemically underpopulated, the mediator bath is hotter and more dilute. Fewer mediators are then available to decay, suppressing the overall normalization of the dark-matter distribution, but each mediator carries on average more energy, so that the daughter particles inherit a harder spectrum. This leads to a significantly larger effective dark-matter temperature and therefore to a stronger suppression of small-scale structure. In this sense, the cannibal dynamics of the parent sector can substantially modify the resulting structure-formation bounds even when the dark matter particle itself does not participate in the equilibration of the dark sector.



These scenarios show that freeze-in production in cannibal dark sectors can reproduce the observed relic density while leading to a rich and highly model-dependent phenomenology. Depending on the particle content and symmetry structure of the dark sector, the same framework can be essentially invisible, severely constrained, or potentially testable in future experiments. More generally, our results highlight that interactions confined entirely within the dark sector can have a major impact not only on the relic abundance and late-time signatures, but also on the momentum distribution of dark matter and the associated structure-formation constraints.

All results presented in this chapter were derived under the standard assumption of radiation domination during dark matter production. As discussed in Chapter~\ref{ch:2}, however, this assumption need not hold in the early Universe. In the next chapter we therefore relax instantaneous reheating and study cannibal dark matter produced via freeze-in in a non-standard matter-dominated cosmological background, with particular emphasis on low reheating temperatures. As we shall see, such scenarios naturally allow for larger portal couplings and lead to more promising experimental signatures while preserving the characteristic features of cannibal dynamics.

%% file: Chapters/Chapter4.tex

\chapter{Production of cannibal dark matter during reheating}\label{ch:4} 

\begin{chapterpublication}
	This chapter is largely based on the publication
	\textit{Freezing-in Cannibals with Low-reheating Temperature}~\cite{Bernal:2025osg} by N.~Bernal, E.~Cervantes, K.~Deka and A.~Hryczuk,
	DOI~[\href{https://link.springer.com/article/10.1007/JHEP09(2025)083}{10.1007/JHEP09(2025)083}].
	
	\vspace{0.5em}
	
	The author of this thesis contributed to the project in three main directions.
	First, the author contributed to developing the conceptual connection between non-standard reheating dynamics and the self-interacting dark matter framework produced via freeze-in, as introduced in the previous chapter.
	Second, the author carried out the numerical implementation and parts of the phenomenological analysis, including the development of the Boltzmann solvers, the execution of parameter scans, and the incorporation of cosmological and experimental constraints.
	Third, the author contributed to the preparation of the manuscript, including its structure, the presentation of the analytical results, and the overall exposition.
	
\end{chapterpublication}

The previous chapter presented a phenomenological study of cannibal DM produced via the freeze-in mechanism in three distinct scenarios. One of the underlying assumptions was that the expansion of the Universe is dominated by SM radiation during freeze-in. As discussed in~\Cref{ch:2}, this need not be the case. In fact, the present-day DM abundance depends not only on particle-physics dynamics, but also on the cosmological history of the Universe. Even though the earliest stages of cosmic evolution remain largely unknown, it is commonly assumed that the Universe was radiation dominated from the end of inflation until matter--radiation equality. Reheating is typically taken to be instantaneous and to occur at a temperature well above the characteristic scales relevant for DM production. However, this assumption is not justified~\cite{Allahverdi:2020bys}. 

Cosmic reheating, i.e. the epoch during which the inflaton transfers its energy density to SM particles and generates the SM thermal bath~\cite{Dolgov:1989us,Traschen:1990sw,Kofman:1994rk,Kofman:1997yn}, is in general a continuous process, and
as we discussed in~\Cref{sec:reheating}, it may conclude at temperatures only slightly above the scale of Big Bang nucleosynthesis.
A non-instantaneous reheating history can significantly affect DM production in freeze-in scenarios~\cite{Bhattiprolu:2022sdd,Arcadi:2024wwg,Lebedev:2024mbj}. 

\section{Rate of expansion beyond radiation domination}

\begin{figure}[t!]
	\centering
	\includegraphics[width=0.7\textwidth]{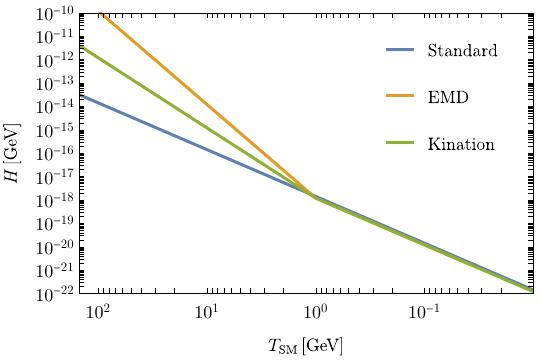}
	\caption{Evolution of the Hubble rate as a function of the temperature of the Standard Model plasma for standard cosmology (blue), early matter domination (EMD, orange) and kination (green) with $T_\text{rh} = 1~\text{GeV}$. Above $T_\text{rh}$, the following hierarchy is satisfied: $H_\text{kin.}>H_\text{EMD}>H_\text{standard}$.} 
	\label{fig:pH}
\end{figure}

The standard cosmological expansion scenario assumes that the Universe is dominated by SM radiation, $\rho_\text{rad}$, at all relevant
stages of DM production. Non-standard expansion histories assume that the Universe might be dominated
by a different fluid,\footnote{In principle it can be a set of different fluids, but we remain agnostic about the matter content and discuss the simplest scenario with a single fluid here.} in particular an inflaton field, here denoted as $\phi$. 
The dynamics of reheating can be parameterized in terms of the
effective equation-of-state parameter $\omega$ of the inflaton field, which is defined as $\omega\equiv P_\phi/\rho_\phi$.
If the inflaton oscillates in a potential $V(\phi)\propto \phi^n$,
its averaged equation of state is given by
$\omega = (n-2)/(n+2)$~\cite{Turner:1983he},
implying that its energy density scales as
\begin{equation}
    \rho_\phi(a) \propto a^{-3(1+\omega)}.
\end{equation}
The commonly assumed quadratic potential corresponds to $\omega=0$,
while $\omega=1/3$ arises for quartic oscillations and
$\omega=1$ describes a phase denoted as kination.
In scenarios with $\omega>1/3$, the inflaton energy density redshifts
faster than radiation and reheating can occur without significant entropy injection into the SM, 
while, for $\omega\le 1/3$, the inflaton must decay or annihilate efficiently
in order for the SM bath to eventually dominate. 

The evolution of the SM temperature during reheating can be parametrized
in a model-independent way as
\begin{equation}
    T(a) = T_{\rm rh} \left(\frac{a_{\rm rh}}{a}\right)^{\alpha},
\end{equation}
where $T_{\rm rh}$ is defined by $\rho_\phi(T_{\rm rh})=\rho_\text{rad}(T_{\rm rh})$,
and the exponent $\alpha$ encodes the microphysics of reheating,
including the inflaton equation of state and the nature of its decay products.
In the standard perturbative decay of a quadratic inflaton,
one recovers $\alpha=3/8$, while alternative decay channels,
annihilation-dominated reheating, or non-standard potentials
lead to different values of $\alpha$.
Requiring that radiation eventually dominates imposes the consistency condition
\begin{equation}
    \alpha \le \frac{3}{4}(1+\omega),
\end{equation}
which excludes regions of parameter space where reheating never completes. 

\begin{figure}[t!]
	\centering
	\includegraphics[width=0.65\textwidth]{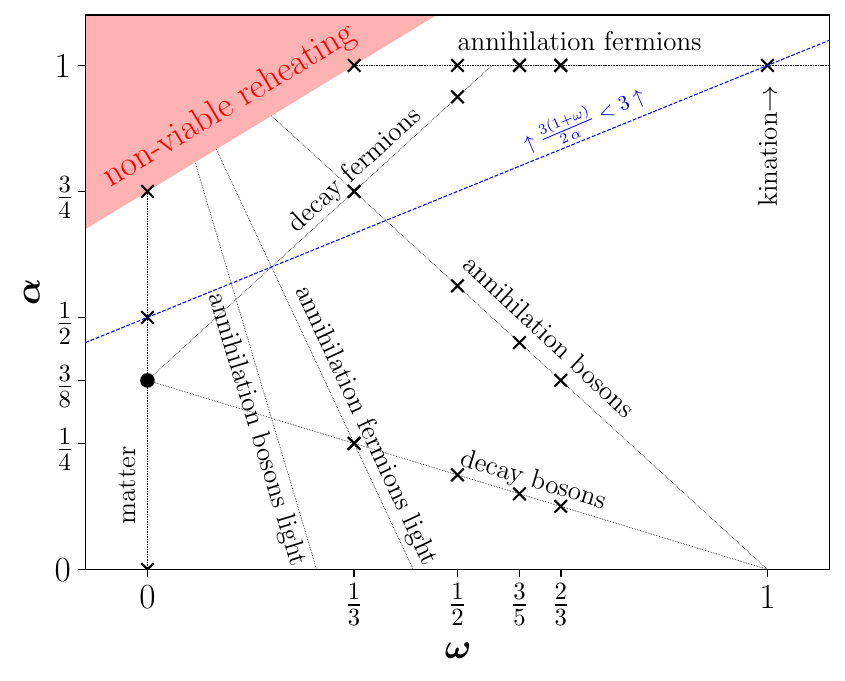}
	\caption{Overview of the reheating scenarios considered. The black dot denotes the conventional setup, in which the inflaton redshifts as non-relativistic matter and decays into Standard Model particles with a constant decay width, whereas the black crosses indicate the alternative cases discussed in the main text. The red region in the upper-left corner does not lead to a viable reheating history. The area above the blue dotted line satisfies \( \tfrac{3(1+\omega)}{2\alpha} < 3 \). Figure from~ \cite{Bernal:2024yhu}.}
	\label{fig:dif_cosmos}
\end{figure}

In this work we will focus on the case where 
the inflaton oscillates as \textit{non-relativistic matter}, i.e., in a quadratic potential with $\omega = 0$ and $\alpha = 3/8$. A complete description of the allowed reheating scenarios in the $(\omega,\alpha)$ space is shown in~\Cref{fig:dif_cosmos} taken from~\cite{Bernal:2024yhu}.

%

Since the DM energy density is subdominant, the background evolution can first be determined independently from the inflaton and radiation sectors. This background is then used to study the evolution of the DM number density and temperature in the subsequent sections.
Assuming perturbative decay with a $100\%$ branching ratio into SM particles,\footnote{In principle, the inflaton could also decay directly into DM. However, even a very small branching ratio into DM can yield a sizeable contribution to the relic abundance during reheating, potentially dominating over production from SM states. Since our aim is to isolate the portal-mediated freeze-in mechanism, we do not consider this possibility here.} and instantaneous thermalization due to SM gauge interactions, the energy densities obey the Boltzmann equations~\cite{Chung:1998rq,Giudice:2000ex}\footnote{Strictly speaking, non-perturbative and non-linear reheating dynamics cannot be fully captured by these Boltzmann equations. Here we focus on the late stages of reheating, where the linear regime provides a good approximation~\cite{Bassett:2005xm,Allahverdi:2010xz,Amin:2014eta,Lozanov:2019jxc,Barman:2025lvk}.}
\begin{align}
	& \frac{d\rp}{dt} + 3 H \rp = -\Gamma \rp\,, \label{eq:BEinf}\\
	& \frac{ds}{dt} + 3 H s = +\frac{\Gamma}{T}\rp\,, \label{eq:BEsm}
\end{align}
where $\rp$ denote the inflaton energy density, $s$ is the SM entropy density and $\Gamma$ denotes the total inflaton decay width. The entropy density is related to the SM temperature $T$ through
\begin{equation}
	s(T) = \frac{2\pi^2}{45}\, \gss(T)\, T^3\,,
\end{equation}
with $\gss(T)$ the effective numbers of relativistic degrees of freedom contributing to entropy and energy density. The Hubble expansion rate is given in this case by
\begin{equation} \label{eq:Hubble}
	H^2 = \frac{\rp + \rR}{3 M_P^2}\,,
\end{equation}
with $\rho_\text{rad}$ is given in Eq.~\eqref{eq:rho_rad}.

\begin{figure}[t!]
	\centering
	\includegraphics[width=0.55\textwidth]{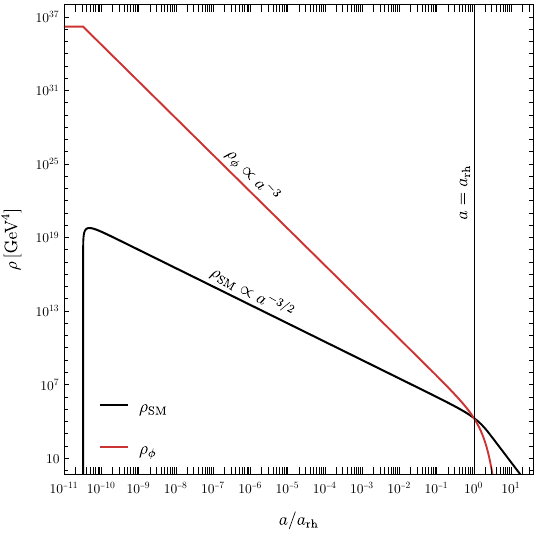}
	\includegraphics[width=0.55\textwidth]{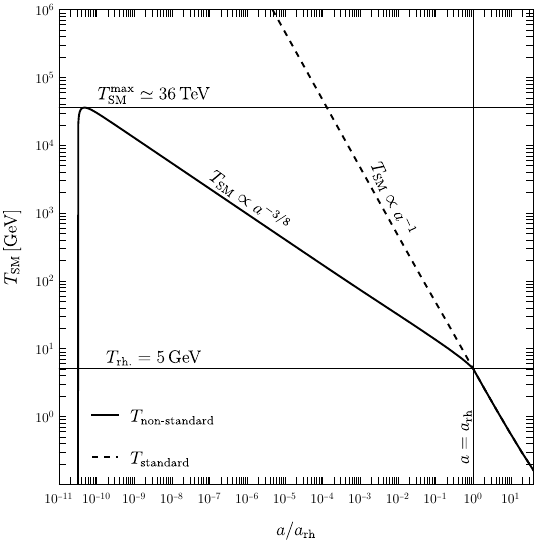}
	\caption{Evolution of the energy density of the inflaton field $\rp$ and the SM radiation $\rR$ (top) and the SM temperature $T$ (down), as a function of the cosmic scale factor $a$, for $\Trh = 5$~GeV.}
	\label{fig:energy_dens_SMtemp}
\end{figure}

Instead of $\Gamma$, it is customary to introduce the reheating temperature $\Trh$, defined by the condition $\rp(\Trh) = \rR(\Trh)$. We also define the corresponding scale factor $\arh \equiv a(\Trh)$. The system of Eqs.~\eqref{eq:BEinf}--\eqref{eq:BEsm} admits approximate analytical solutions. During reheating ($a \ll \arh$), the decay term in Eq.~\eqref{eq:BEinf} can be neglected, allowing one to first solve for $\rp$ and subsequently for $\rR$. After reheating ($a \gg \arh$), the source term in Eq.~\eqref{eq:BEsm} becomes negligible.
The SM temperature evolves approximately as
\begin{equation}
	T(a) \simeq \Trh \times
	\begin{dcases}
		\left(\frac{\arh}{a}\right)^{3/8} & \text{for } a \leq \arh\,,\\[6pt]
		\left(\frac{\gss(\Trh)}{\gss(T)}\right)^{1/3} \frac{\arh}{a} & \text{for } a \geq \arh\,,
	\end{dcases}
\end{equation}
while the Hubble rate scales as
\begin{equation} \label{eq:H}
	H(T) \simeq \frac{\pi}{3 M_P}\times
	\begin{dcases}
		\sqrt{\frac{\gs(\Trh)}{10}}\, \frac{T^4}{\Trh^2} & \text{for } T \geq \Trh\,,\\[6pt]
		\sqrt{\frac{\gs(T)}{10}}\, T^2 & \text{for } T \leq \Trh\,.
	\end{dcases}
\end{equation}
At late times, once reheating has completed and radiation domination is established, entropy conservation implies $T \propto 1/a$ and $H \propto T^2/M_P$.
An illustrative example of the evolution of $\rp$ and $\rR$ (top) and of the SM temperature (down) as functions of the scale factor is shown in Fig.~\ref{fig:energy_dens_SMtemp}, for $\Trh = 5\,\text{GeV}$.
To preserve the successful predictions of BBN, reheating must complete at temperatures $\Trh > T_\text{BBN} \simeq 4\,\text{MeV}$~\cite{Sarkar:1995dd,Kawasaki:2000en,Hannestad:2004px,Barbieri:2025moq}.\footnote{Reheating temperatures in the MeV--GeV scale can challenge conventional baryogenesis. Alternative mechanisms such as non-thermal leptogenesis~\cite{Zhang:2023oyo}, Affleck--Dine leptogenesis~\cite{Affleck:1984fy}, or Mesogenesis~\cite{Elahi:2021jia} may nevertheless remain viable.}

\section{Freeze-in production during reheating} \label{sec:FI_reh}
We investigate the freeze-in production of DM in the presence of both non instantaneous reheating and dark-sector self-interactions. Besides numerically solving the cBE governing the DM yield and temperature, Eqs.~\eqref{eq:N}~and~\eqref{eq:T_evolution_a}, we also solve and the evolution of the SM bath stemming from the reheating process by solving Eq.~\eqref{eq:BEinf}. 
This allows us to consistently track portal interactions together with $2\to 3$ cannibal processes in a time-dependent cosmological background.
We find that these effects can significantly shift the viable parameter space and open regions that would otherwise be excluded under the assumptions of instantaneous reheating or negligible self-interactions.

We take as a benchmark model the potential introduced in~\Cref{sec:Z3}, in particular Eqs.~\eqref{HP_Z3}~and~\eqref{eq:VS}, i.e., a complex scalar field $S$ as DM candidate
charged under a $\mathbb{Z}_3$ symmetry with a Higgs Portal interaction. A related analysis of \(\mathbb{Z}_3\)-symmetric complex scalar dark matter beyond the standard radiation-domination can be found in~\cite{Mitra:2025cmo}, where the DM relic is determined by thermal freeze-out, with important annihilations and semi-annihilations with the SM mediated by an effective operator.


Unlike~\Cref{sec:Z3}, here we will denote the mass of the DM candidate as \(m\) instead of \(m_s\), and its temperature $T'$ instead of $T_S$. As in~\Cref{sec:Z3}, we assume no CP violation, so that the real and imaginary parts of the complex field contribute equally to \(\Omega_{\rm DM}\).\footnote{As discussed in the previous chapter, this assumption could be relaxed, in which case the two components would in general contribute differently to the final relic abundance. A genuine asymmetric-DM scenario~\cite{Kaplan:2009ag}, however, would require that the relic density be set predominantly by a surviving asymmetry after the symmetric component has been depleted. We do not consider that possibility here.}

If the reheating temperature is low, contributions from fermionic and eventual hadronic $2\leftrightarrow 2$ processes cannot be neglected~\cite{Lebedev:2024mbj}.
For low reheating temperatures, production takes place at $T\!\sim\!{\cal O}(\mathrm{MeV{-}GeV})$,\footnote{The inflaton continuously injects entropy into the SM plasma, diluting the DM density at earlier stages when heavier states are present; reproducing the observed relic abundance therefore requires a larger effective production rate. In other words, it is required to increase the portal coupling, which at the same time enhances the relative impact of these fermionic and hadronic $2\leftrightarrow2$ channels.} where heavy SM states are absent from the thermal bath and Higgs-decay production is negligible. The total FI yield is then controlled by scatterings among the light degrees of freedom that remain abundant, i.e.\ leptons at $T\!\sim\!\mathrm{MeV}$ and, around and below the QCD crossover, hadronic states instead of quarks and gluons~\cite{Lebedev:2024mbj}. In both the high and low $\Trh$ cases, the production that occurs before the electroweak phase transition is much smaller than the two other modes, and therefore we set the initial condition at $a_I = a_\text{ewpt}$ in this work, where $T(a_\text{ewpt}) = 150$~GeV~\cite{Heeba:2018wtf}. Note that we also fix the inflationary scale such that the maximal temperature reached by the SM plasma during reheating is much higher than $T_\text{ewpt}$, ensuring that all SM species remain in thermal equilibrium at all times, that is, during and after reheating.


\subsection{Production from fermionic and hadronic collisions}
\label{subsec:FI_reh2}

The dilution of the DM energy density due to the accelerated expansion during reheating ceases once the Universe reaches the reheating temperature $\Trh$. For reheating temperatures of order of a few GeV or higher, the dominant production channel is the Higgs decay process $h \to S^* S$. Its contribution to the zeroth and second moment are presented in~\Cref{subsec:Higgs_decay}. On the other hand, if reheating ends at temperatures in the MeV scale, Higgs-induced production occurs during a phase of substantial SM entropy dilution and achieving the observed relic density then requires enhanced DM production from the thermal plasma, i.e., larger values of $\lhs$ are needed. This increases the production rates from both fermionic and hadronic SM states, the latter becoming relevant only after the QCD phase transition.
Notably, larger values of $\lhs$ may also induce kinetic and chemical equilibrium between the dark and SM sectors transitioning the DM candidate from a FIMP to a WIMP~\cite{Silva-Malpartida:2023yks,Belanger:2024yoj}. For sufficiently low reheating temperatures, interaction rates can decouple \textit{before} reheating completes, requiring a consistent treatment of the temperature evolution to accurately capture kinetic decoupling.

The DM temperature is affected by both annihilations and elastic scatterings with bath particles. Annihilations contribute to both $C_0$ and $C_2$. For instance, $C_0^{S^*S\leftrightarrow \bar f f}$ can be expressed in terms of the thermal average $\braket{\sigma_{S^*S\to \bar f f}\,\text{v}}$, as detailed in~\cite{Edsjo:1997bg}. The hadronic contribution to the zeroth moment reads
\begin{equation}\label{eq:C0hadrons}
	C_0^{S^*S\to \text{hadrons}}
	=
	-\int_p\int_{\tilde p}
	\sigma\, \text{v}_{S^*S\to \text{hadrons}}
	f_s(p)\, f_s(\tilde p)\,.
\end{equation}
Note that here $\text{v}$ denotes the relative velocity of $S^*$ and $S$ prior to collision, not the SM Higgs VEV, while
\begin{equation}
	\sigma\,\text{v}_{S^*S\to \text{hadrons}}
	\equiv
	\frac{\lhs^2 v_h^2}{\sqrt{s}}
	\frac{1}{(s-m_h^2)^2 + m_h^2\Gamma_h(m_h)^2}
	\Gamma_{\text{hadrons}}(\sqrt{s})\,,
\end{equation}
with $\Gamma_h(m_h) = 4.042\,\text{MeV}$~\cite{Djouadi:2005gi}. 

The second-moment contribution $C_2^{S^*S\leftrightarrow \bar f f}$ involves the temperature-weighted average $\braket{\sigma_{S^*S\to \bar f f}\,\text{v}}_2$, defined in~\cite{Binder:2017rgn},
\begin{equation}
	C_2^{S^*S\to \text{hadrons}}
	=
	-\int_p\int_{\tilde p}
	\frac{p^2}{E}\,
	\sigma\,\text{v}_{S^* S \to \text{hadrons}}\,
	f_s(p)\, f_s(\tilde p)\,.
\end{equation}

These expressions apply in the deconfined phase prior to the QCD transition, which we assume to occur instantaneously at $T_\text{QCD} = 200\,\text{MeV}$.\footnote{We model the QCD transition as an instantaneous crossover at $T_{\rm QCD}=200\,\mathrm{MeV}$ for simplicity. In reality, the QCD transition is a smooth crossover occurring over $T\simeq 150$--$200\,\mathrm{MeV}$, leading to a gradual change in the effective relativistic degrees of freedom and in the relevant annihilation channels. A more refined treatment would interpolate continuously between quark and hadron degrees of freedom and modify the production rate near this temperature. Since the temperature interval of the crossover is narrow and DM production in the parameter region considered is not sharply localized at $T\sim T_{\rm QCD}$, this approximation does not qualitatively affect our results.} Below this temperature, quarks are confined into hadrons. We therefore use the decay width $\Gamma_\text{hadrons}$ from~\cite{Winkler:2018qyg}.

Elastic scatterings also influence the temperature evolution by driving $T' \to T$. 
Immediately after the electroweak phase transition, hard scattering with Higgs bosons can be estimated using Eq.~\ref{eq:S_med_scatter}.
A second source of kinetic equilibration arises from scattering with fermions, which we approximate as
\begin{equation} \label{eq:C2_el_scatter_ferm}
	C_2^{S f\leftrightarrow Sf}
	\simeq
	\frac{N}{N_\text{eq}}\,
	\frac{\lhs^2\, m_f^2\, N_c}{64 \pi^5\, m_h^4}\,
	T'^2\, T^2\,
	(T - T')\,
	\left(m_f^2 + 3 T T'\right)
	e^{-m/T'}\,
	e^{-m_f/T}\,,
\end{equation}
where $N = n a^3$ and $N_\text{eq} = n_\text{eq} a^3$ are the comoving number density and comoving number density in equilibrium, respectively. Additionally, $N_c=1$ for leptons and $N_c=3$ for quarks. A detailed derivation is provided in Appendix~\ref{sec:scatter_ferm}.

\section{Results} \label{sec:Results}

The interplay between the FI mechanism, self-equilibration and the evolving Hubble rate \textit{during} and \textit{after} reheating gives rise to a non-trivial thermal history. To clarify the underlying dynamics, we first examine two benchmark scenarios that reproduce the observed DM relic abundance. These examples correspond to representative cases in which the freeze-out of DM self-interactions occurs either during reheating or after its completion. We then extend the analysis by performing a broad parameter scan.
For any process to be relevant, its interaction rate must compete with the Hubble expansion rate. During reheating, the latter scales as $H \propto T^4/(M_P \Trh^2)$, while after reheating it follows the standard radiation-dominated behavior $H \propto T^2/M_P$; see Eq.~\eqref{eq:H}. At high temperatures, DM production is dominated by Higgs decays. As the temperature drops, this channel becomes Boltzmann suppressed, while a significant fraction of the produced DM is diluted during reheating.

For sufficiently low $\Trh$, DM production from annihilations of light SM states via an off-shell Higgs becomes increasingly relevant. Although this channel is typically subdominant compared to Higgs decay, it can compensate for the DM dilution. At early times, when the DM number density is small but with enough kinetic energy, the $2 \to 3$ process dominates. As the DM abundance increases, the inverse $3 \to 2$ process increases as well, thermalizing the dark sector.

\subsection{Freeze-out after reheating} 
\label{subsec:trh1GeV}

\begin{figure}
	\def\sepf{0.496}
	\centering
	\includegraphics[width=\sepf\columnwidth]{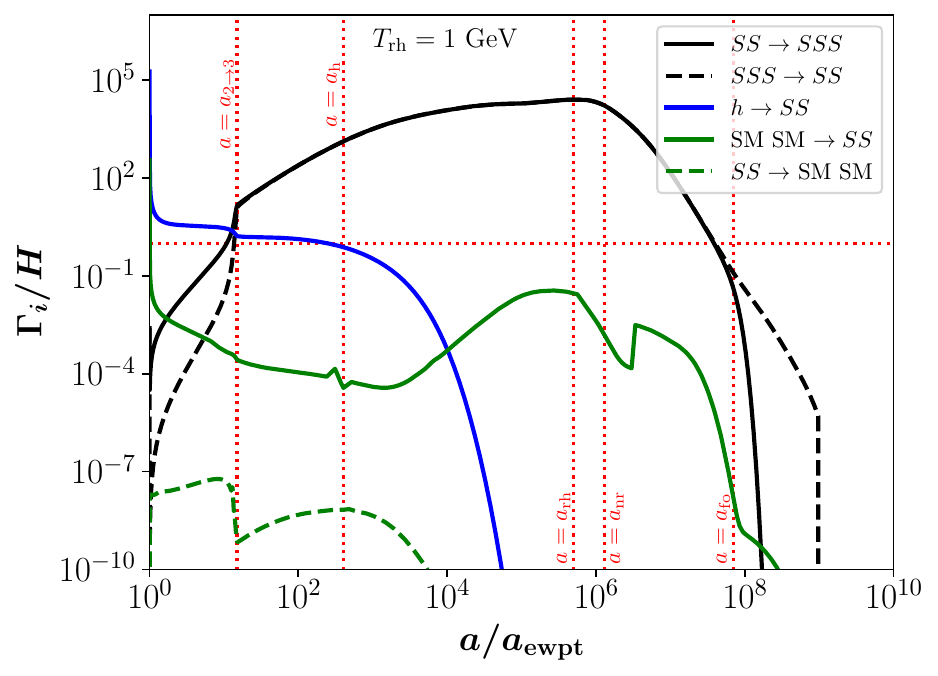}
	\includegraphics[width=\sepf\columnwidth]{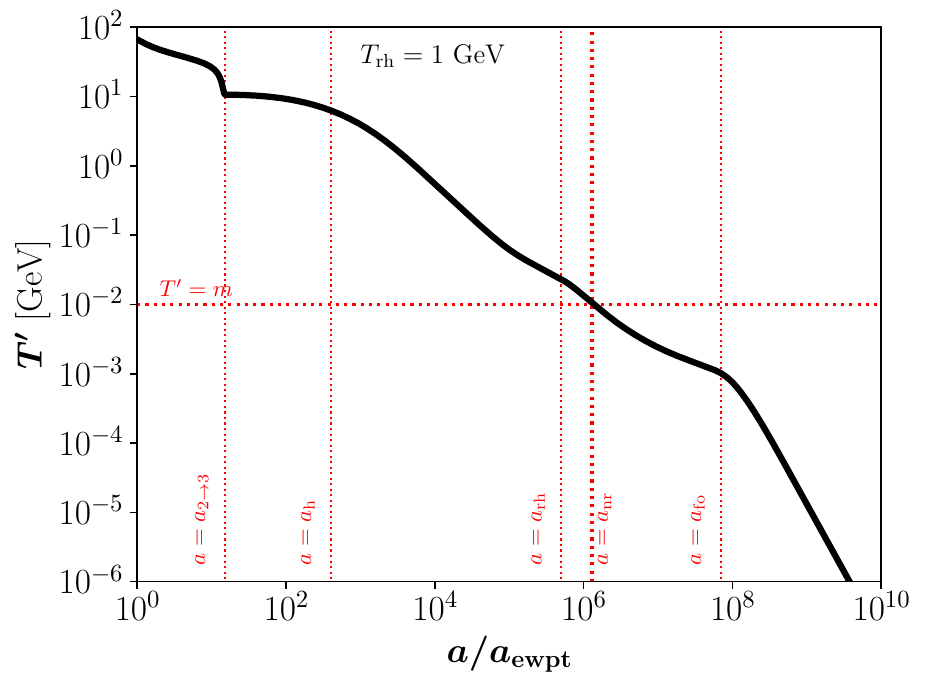}
	\includegraphics[width=\sepf\columnwidth]{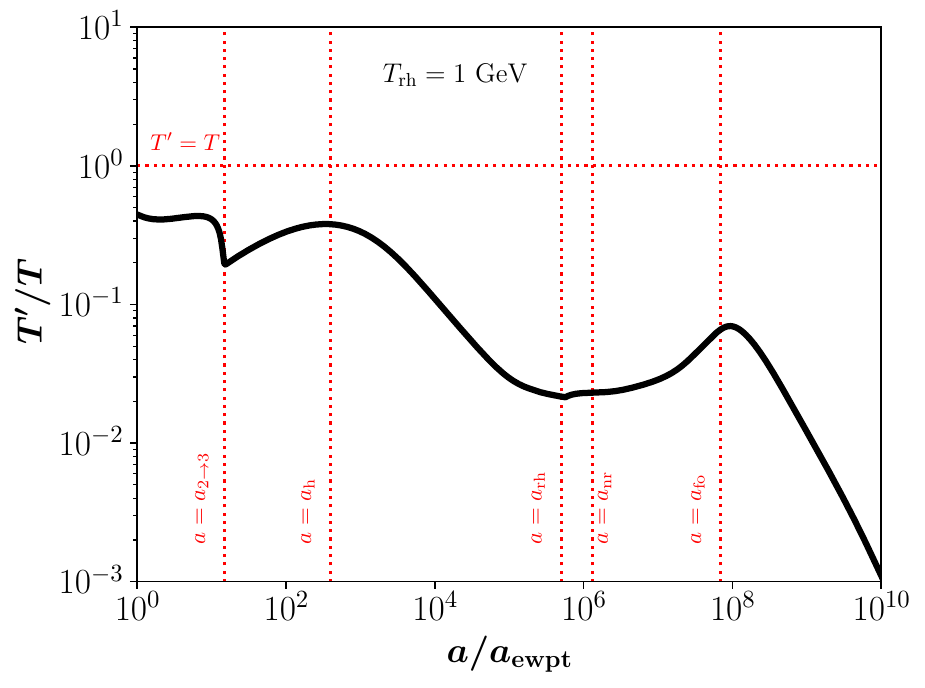}
	\includegraphics[width=\sepf\columnwidth]{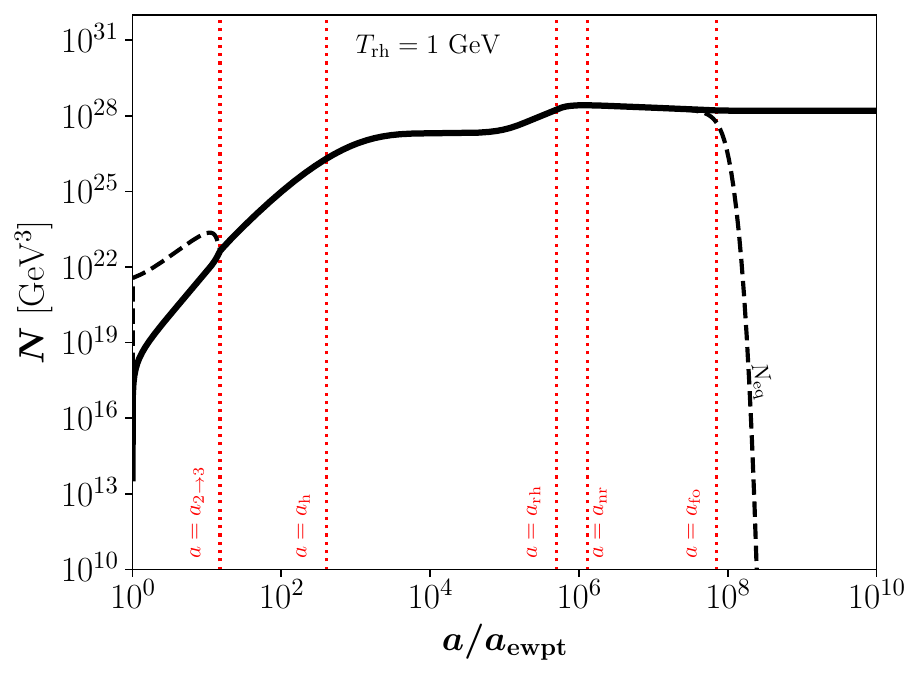}
	\caption{Results for $\Trh = 1$~GeV, with $m = 10$~MeV, $\lhs = 1.8 \times 10^{-7}$ and $\ls = 0.01$. Top left: Interaction rates normalized to the Hubble rate. Top right: Evolution of the dark-sector temperature. Bottom left: Ratio between dark-sector and SM temperatures. Bottom right: Evolution of the comoving number density.}
	\label{fig:Trh1GeV}
\end{figure}

In~\Cref{fig:Trh1GeV} we show the results for $\lhs = 1.8 \times 10^{-7}$ and $\Trh = 1$~GeV, with $m = 10$~MeV and $\ls = 0.01$. Here DM is produced mainly through Higgs decays. The top-left panel displays the ratio of the relevant interaction rates to the Hubble rate. The top-right panel shows the evolution of the dark-sector temperature. The bottom-left panel presents the ratio $T'/T$, while the bottom-right panel shows the evolution of the comoving DM number density.

At early times, Higgs decays (solid blue curve in the top-left panel) set the initial DM abundance. As the number density builds up, the $2 \to 3$ self-interaction (thick black curve) becomes relevant, which leads to a decrease in $T'$. Eventually the rate $\Gamma_{3\to 2}$ (dashed black curve) also increases, leading to chemical equilibrium within the dark sector ($N = N_\text{eq}$). Higgs-induced production continues until its rate falls below the Hubble rate (second vertical dotted red line at $a = a_h$). During this period, annihilations of SM particles (solid green curve) remain subdominant. After this stage, DM is no longer efficiently sourced and remains relativistic for some time, implying $T' \propto a^{-1}$. Close to the end of reheating (third vertical dotted red line at $a = a_{\rm rh}$), SM annihilations give a small additional contribution. At this point the rate of expansion transitions from $H \propto a^{-3/2}$ to $H \propto a^{-2}$. When the DM temperature drops to $T' \simeq m$ (at $a = a_{\rm nr}$), it becomes non-relativistic and its number density becomes Boltzmann suppressed. Entropy conservation within the dark sector leads to a slower decrease of $T'$, approximately $T' \sim 1/\log(a^3)$~\cite{1992ApJ...398...43C}. As a result, $T'/T$ increases for a period due to cannibalization. When the $3 \to 2$ process decouples (last vertical dotted red line at $a = a_{\rm fo}$), the DM temperature scales as $T' \propto a^{-2}$, as expected for a non-relativistic decoupled species, while the SM temperature scales as $T \propto a^{-3/8}$ during reheating and as $T \propto a^{-1}$ afterward.

The different stages are also visible in the bottom-right panel, where the thick black line shows the comoving number density $N = n a^3$. For comparison, the dashed black line indicates the equilibrium value $N_\text{eq} = n_\text{eq} a^3$.

\subsection{Freeze-out during reheating} 
\label{subsec:trh10MeV}

\begin{figure}
	\def\sepf{0.496}
	\centering
	\includegraphics[width=\sepf\columnwidth]{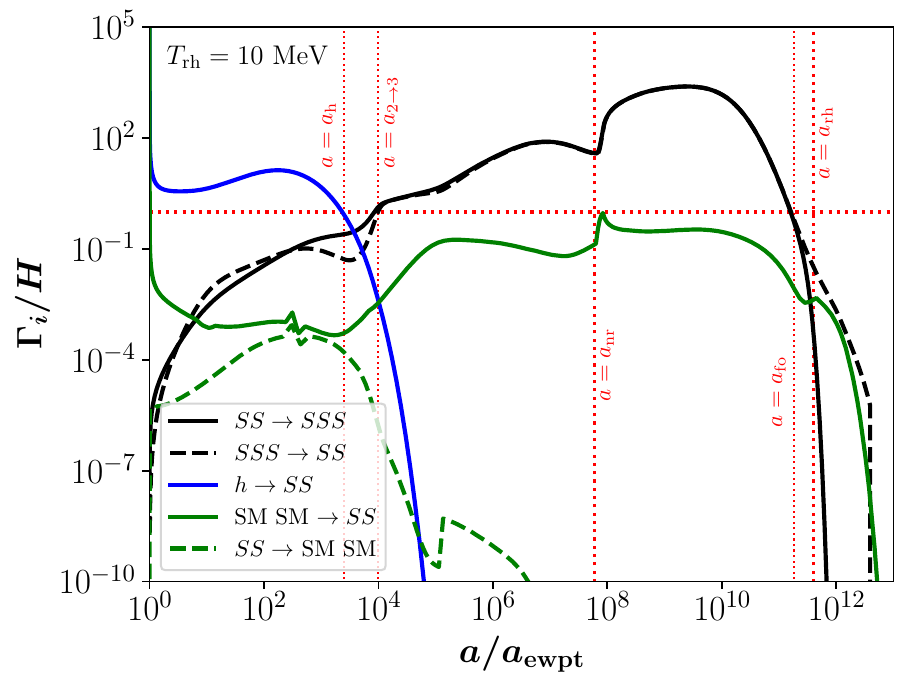}
	\includegraphics[width=\sepf\columnwidth]{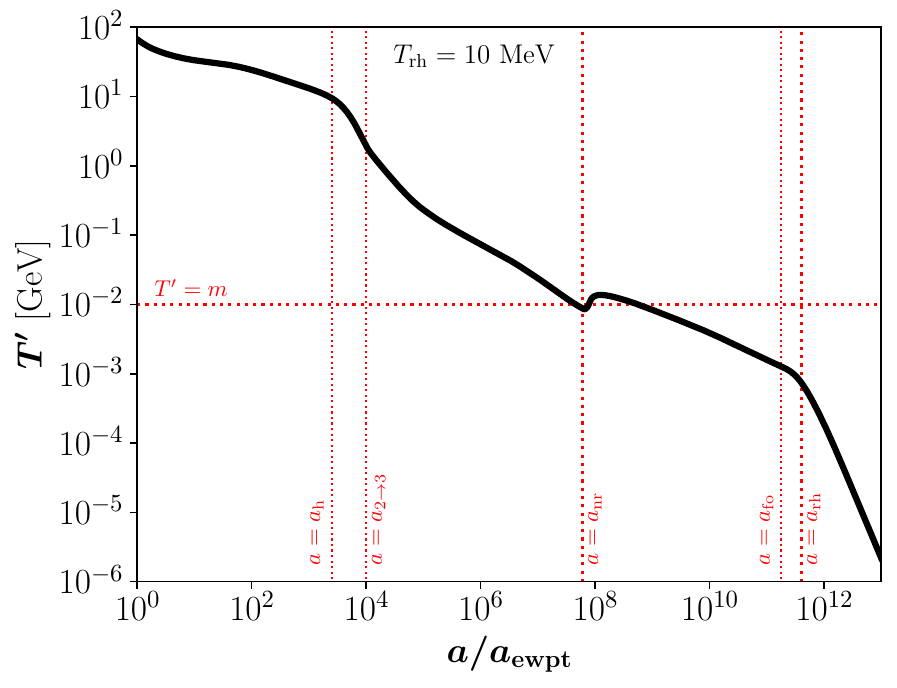}
	\includegraphics[width=\sepf\columnwidth]{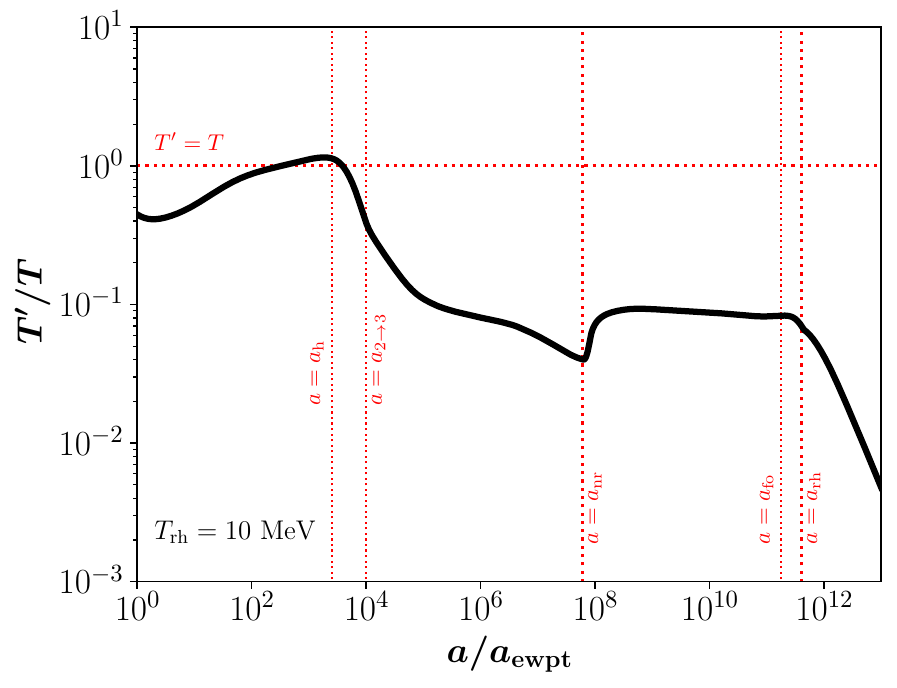}
	\includegraphics[width=\sepf\columnwidth]{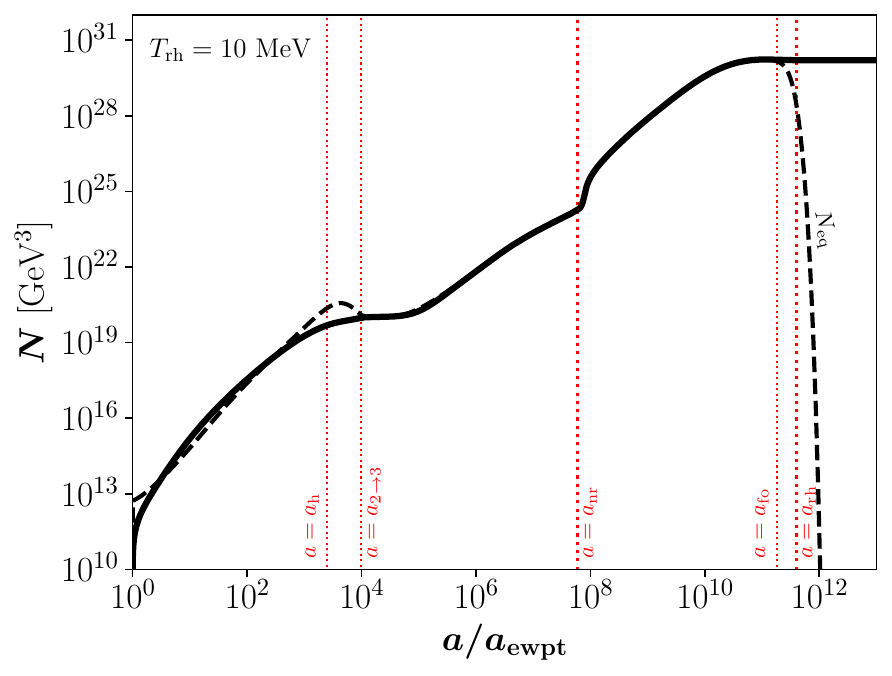}
	\caption{Same as Fig.~\ref{fig:Trh1GeV}, but for $\Trh = 10$~MeV and $\lhs = 4.5 \times 10^{-4}$.}
	\label{fig:Trh10MeV}
\end{figure}

Let us now consider $\Trh = 10$~MeV. The results are shown in Fig.~\ref{fig:Trh10MeV}. We keep the same values of $m$ and $\ls$ as in Section~\ref{subsec:trh1GeV} with $\lhs = 4.5 \times 10^{-4}$. 
Because reheating ends at such a low temperature, more DM dilutes than in the previous case. To reproduce the observed relic abundance, a substantially larger portal coupling is therefore required. In this setup, self-interactions are initially subdominant with respect to both Higgs decays and the Hubble expansion rate, as seen in the top-left panel. Higgs decays dominate DM production and gradually raise the ratio $T'/T$, shown in the bottom-left panel. This continues until the Higgs decay rate drops below the Hubble rate (first vertical red dotted line at $a = a_h$).

After this point, the $2 \to 3$ self-production process becomes briefly efficient. This leads to a decrease in $T'$. The effect ends soon afterward (second vertical red dotted line at $a_{2\to3}$), when the inverse $3 \to 2$ reactions become efficient and bring the number-changing processes into equilibrium. At the same time, SM annihilations ($2\,\text{SM} \to 2\,S$) start to contribute, so that DM is never completely source-free.
Eventually DM becomes non-relativistic and its number density undergoes Boltzmann suppression and $T'$ follows the usual logarithmic scaling with $a$. Around this epoch, SM annihilation rates experience a mild enhancement near the hadronization scale. This provides an additional source of DM and induces a small increase in $T'$ and in $T'/T$.

The ratio $T'/T$ remains approximately constant until freeze-out takes place (fourth vertical red dotted line at $a_{\rm fo}$). Reheating ends shortly afterward, marked by the final vertical red line at $a = a_{\rm rh}$. The bottom-right panel shows the evolution of the comoving number density $N$. A larger DM population must be generated compared to the $\Trh = 1$~GeV case in order to compensate for the entropy dilution.\footnote{Since we fix $\ls = 10^{-2}$ in Sections~\ref{subsec:trh1GeV} and~\ref{subsec:trh10MeV}, the $2 \to 3$ reactions never dominate the overall production. For $\ls$ close to unity, as discussed in Section~\ref{subsec:param-regions} and shown in Fig.~\ref{fig:ls_vs_relic}, the impact of self-interactions becomes more pronounced and modifies the evolution of both $T'$ and $N$ until $3 \to 2$ processes take over.}

\subsection{Parameter regions} 
\label{subsec:param-regions}

\begin{figure}[t!]
	\centering
	\includegraphics[width=0.65\textwidth]{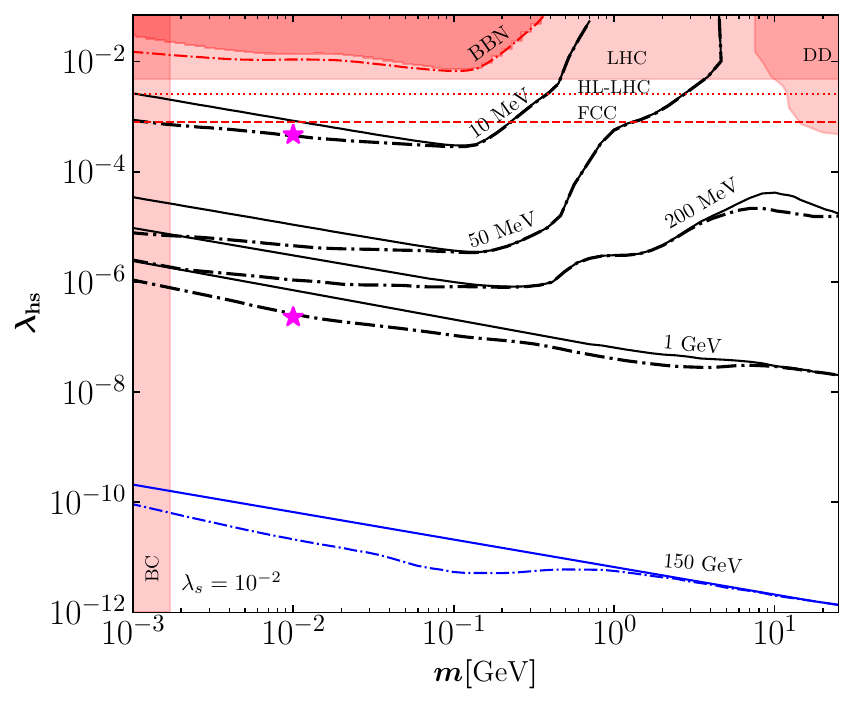}
	\includegraphics[width=0.65\textwidth]{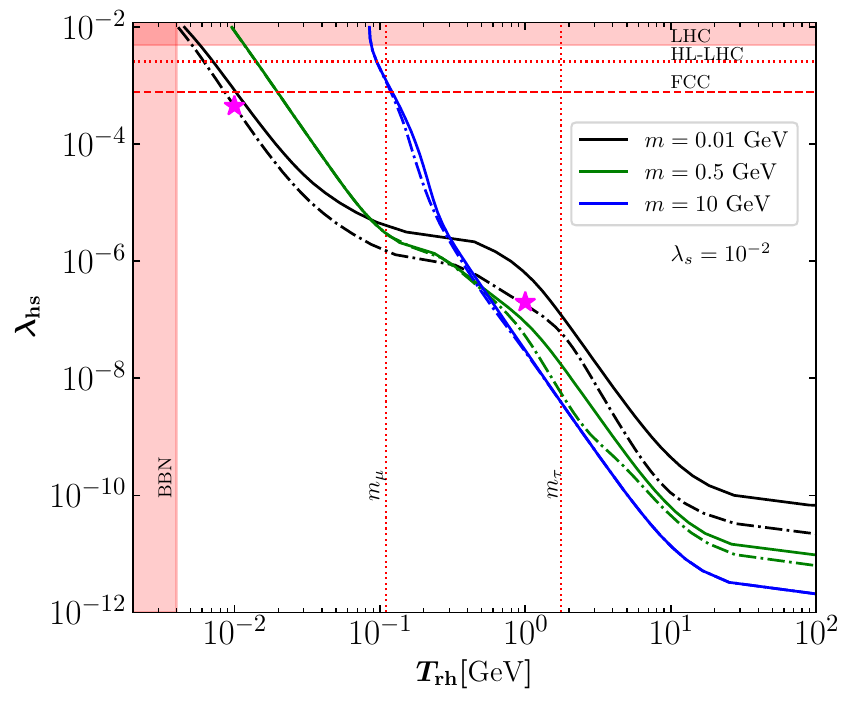}
	\caption{Top panel: Regions reproducing the observed DM relic abundance in the plane $[m,\, \lhs]$ for different reheating temperatures $\Trh$. Down panel: Same requirement shown in the $[\Trh,\, \lhs]$ plane for different DM masses. Dot-dashed lines correspond to $\ls = 10^{-2}$, while solid lines correspond to the case without self-interactions ($\ls = g_s = 0$).}
	\label{fig:sol_lhs_ms_Trh_FIMPs}
\end{figure}

Having discussed representative benchmark cases, we now turn to a broader exploration of the parameter space. In the top panel of Fig.~\ref{fig:sol_lhs_ms_Trh_FIMPs}, we show contours of constant reheating temperature $\Trh$ in the $[m,\, \lhs]$ plane. Dot-dashed lines correspond to scenarios including DM self-interactions with $\ls = 10^{-2}$, while solid lines represent the case without self-interactions, $\ls = g_s = 0$. The two magenta stars indicate the benchmark points analyzed in Sections~\ref{subsec:trh1GeV} and~\ref{subsec:trh10MeV}.

For large reheating temperatures, in particular $\Trh = 150$~GeV, the contour (blue line) matches the standard radiation-dominated expectation, where DM production takes place after reheating. The absence of self-interactions is the typical infrared FI. As $\Trh$ decreases, a larger fraction of DM production occurs during reheating. Since entropy dilution is significant in this regime, larger production rates are required. For intermediate reheating temperatures, such as $\Trh \sim 200$~MeV, self-interactions become relevant even for DM masses in the range $m \sim 2$--$20$~GeV. In this region, annihilations involving heavier fermions ($\tau$, $c$, and $b$) are efficient.

For $\Trh$ between 10~MeV and 100~MeV, entropy dilution is very strong. DM produced from Higgs decays and heavy fermion annihilations is largely diluted. Production then relies mainly on muon annihilations. As a result, once the DM mass exceeds the muon threshold ($m \sim 100$~MeV), the required portal coupling increases sharply to compensate for the exponential decrease of muon number density.

As in the previous chapter, $\sigma_{2\to 2}$ is constrained by the Bullet Cluster (1E~0657--56)~\cite{Randall:2008ppe}. For $\ls = 10^{-2}$, the excluded region is shown in red (BC) in the top panel.
Additionally, the decay $h \to S^* S$ contributes to the Higgs invisible width,
\begin{equation}
	\Gamma_{h\to S^*S} = \frac{\lhs^2\, v^2}{32\pi\, m_h} 
	\sqrt{1-\frac{4m^2}{m_h^2}}\,.
\end{equation}
This is constrained by LHC limits on the invisible branching ratio, $\text{BR}_{\text{inv}} \leq 10\%$~\cite{ATLAS:2023tkt, CMS:2023sdw}, which set an upper bound on $\lhs$. This exclusion appears as the red horizontal band. Future projections from the HL-LHC ($\text{BR}_{\text{inv}} \leq 1.9\%$) and FCC ($\text{BR}_{\text{inv}} \leq 0.2\%$) are shown as dotted and dashed red lines~\cite{Dawson:2022zbb}.

The BBN limit on reheating~\cite{Sarkar:1995dd, Kawasaki:2000en, Hannestad:2004px, Barbieri:2025moq} is shaded in red. Direct detection experiments such as LZ~\cite{LZ:2024zvo} and Xenon1T~\cite{XENON:2018voc} constrain the portal coupling for DM masses above roughly 1~GeV. These bounds are indicated in the upper-right red region (DD). Limits from DM-electron scattering are negligible due to the small electron Yukawa coupling. We also note proposals to probe reheating temperatures in the MeV range at $e^+e^-$ colliders~\cite{Barman:2024nhr, Barman:2024tjt}.

The bottom panel of Fig.~\ref{fig:sol_lhs_ms_Trh_FIMPs} shows contours of fixed DM mass in the $[\Trh,\, \lhs]$ plane. The structure follows the same reasoning: the required portal coupling is determined by which SM states dominate production. For clarity, we only mark the muon and tau thresholds, as the charm and bottom thresholds lie close to the tau one.

\begin{figure}[t!]
	\centering
	\includegraphics[width=0.57\textwidth]{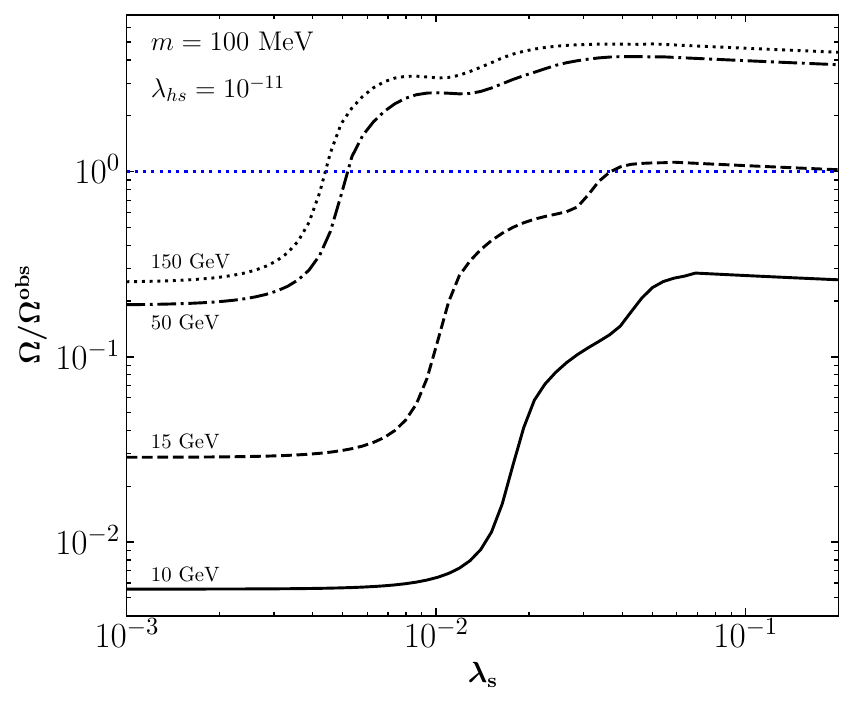}
	\includegraphics[width=0.55\textwidth]{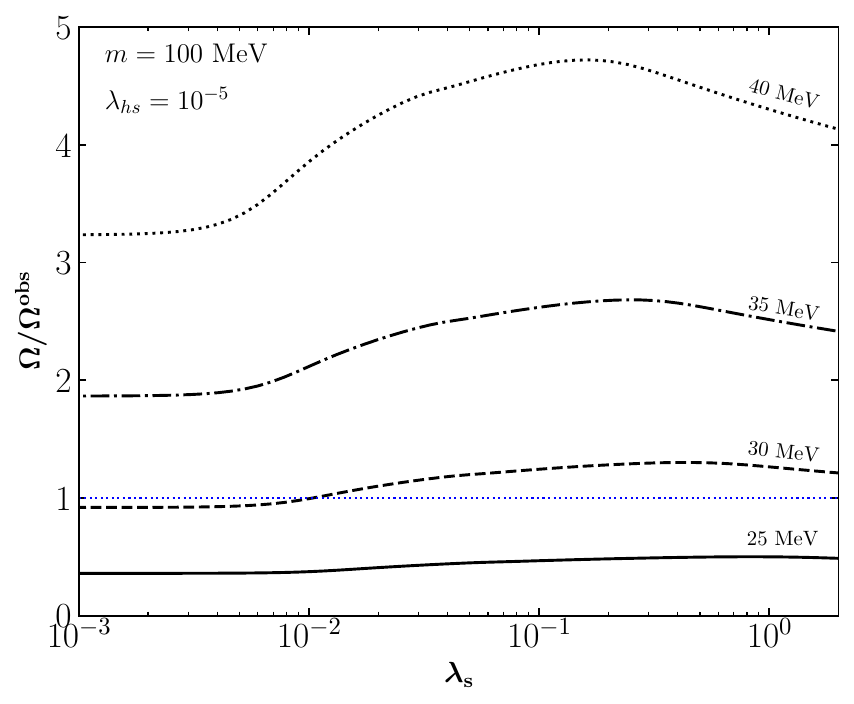}
	\caption{Ratio of the relic abundance obtained in this framework to the observed value as a function of $\ls$ for $m = 100$~MeV and different reheating temperatures. Top: $\lhs = 10^{-11}$. Down: $\lhs = 10^{-5}$.}
	\label{fig:ls_vs_relic}
\end{figure}

In Fig.~\ref{fig:ls_vs_relic}, we show the ratio of the predicted relic abundance to the observed value as a function of $\ls$, fixing $m = 100$~MeV and considering two representative portal couplings. Contours correspond to different reheating temperatures.
In the top panel ($\lhs = 10^{-11}$), the small portal coupling requires large reheating temperatures to reproduce the correct relic density. In this regime, $\ls$ can significantly modify the abundance. Large values of $\ls$ enhance production through $2 \to 3$ processes and can increase the relic abundance by more than an order of magnitude. Once $\ls$ is large enough for rapid self-thermalization, the curves flatten and form a plateau.
In the bottom panel ($\lhs = 10^{-5}$), the larger portal coupling allows smaller reheating temperatures. Here the role of $\ls$ is reduced. DM production from the portal is already efficient, and increasing $\ls$ quickly drives the dark sector toward equilibrium. As $\ls$ approaches unity, $3 \to 2$ processes become very efficient and start to reduce the relic abundance, since the already large DM population undergoes stronger cannibal depletion.

\subsection{Prospects for future colliders}

\begin{figure}[t!]
	\centering
	\includegraphics[width=0.75\textwidth]{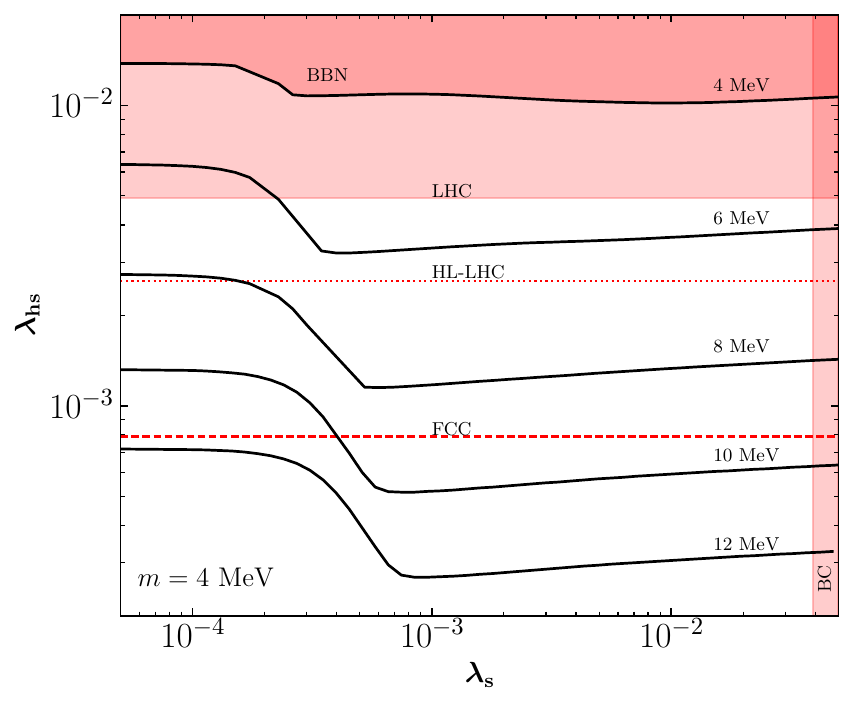}
	\caption{Reheating temperature $\Trh$ required to reproduce the observed DM relic abundance in the plane $[\ls,\,\lhs]$, for $m = 4$~MeV.} 
	\label{fig:ls_vs_lhs}
\end{figure}

In Fig.~\ref{fig:ls_vs_lhs}, we show the parameter space for a light DM candidate with mass $m = 4$~MeV in the $[\ls,\, \lhs]$ plane. The contours correspond to different reheating temperatures $\Trh$ that reproduce the observed relic abundance. The red-shaded region above the $\Trh = 4$~MeV contour is excluded by BBN. The wide horizontal red band is ruled out by current LHC limits on invisible Higgs decays. The vertical red region on the right is excluded by bounds on DM self-interactions from the Bullet Cluster. Projected sensitivities of the HL-LHC ($\text{BR}_{\text{inv}} \leq 1.9\%$) and FCC ($\text{BR}_{\text{inv}} \leq 0.2\%$) on the invisible Higgs decay are indicated by dotted and dashed red lines, respectively~\cite{Dawson:2022zbb}.

As an example, consider the contour corresponding to $\Trh = 6$~MeV. For very small $\ls$, the required portal coupling $\lhs$ lies in the region already excluded by collider bounds. However, increasing $\ls$ enhances the DM yield through $2 \to 3$ processes, which lowers the value of $\lhs$ needed to match the relic density. This shift moves the model into a region that can be probed by future collider experiments. This example illustrates how self-interactions, together with low reheating temperatures, can make light DM scenarios viable that would otherwise be excluded.

\section{Summary} 
\label{sec:summary_ch4}

In this chapter, we have presented a detailed analysis of FI production of DM in the presence of non-instantaneous reheating and dark-sector self-interactions. We focused on a $\mathbb{Z}_3$-symmetric scalar DM candidate with cannibal processes. Our study is based on the numerical solution of coupled Boltzmann equations describing the evolution of the DM number density and temperature, together with the background dynamics of the SM bath during reheating, which also required a dedicated Boltzmann solver for the energy density of the inflaton and SM radiation. This framework reveals a thermal history that differs significantly from the standard picture with instantaneous reheating and negligible self-interactions. The final relic abundance depends sensitively on the reheating temperature $\Trh$, the Higgs portal coupling $\lhs$, and the self-coupling $\ls$.

We paid particular attention to low reheating temperatures, $\Trh \simeq 10$~MeV, where entropy dilution is significant. To reproduce the observed relic abundance, larger portal couplings $\lhs$ are required. In this regime, production from SM annihilations becomes important, and the freeze-out of self-interactions can take place during reheating itself. The evolution of $T'$ is then shaped by both ongoing energy injection from the SM bath and threshold effects, such as hadronization.

Our parameter scans over $m$, $\lhs$, $\ls$, and $\Trh$ lead to several conclusions. The portal coupling needed to obtain the correct relic abundance increases rapidly as $\Trh$ decreases. Self-interactions are especially relevant for light DM. The $2 \to 3$ processes can substantially enhance the DM yield, allowing successful freeze-in with smaller values of $\lhs$. Their impact is strongest when $\lhs$ is small and $\Trh$ is large. For larger $\lhs$, portal production is already efficient, and increasing $\ls$ mainly affects the timing of self-thermalization. If $\ls$ is large enough, strong $3 \to 2$ processes can even reduce the relic abundance by depleting an initially overproduced DM population. The viable parameter space is constrained by several bounds: self-interaction limits from the Bullet Cluster, invisible Higgs decay constraints from the LHC, the lower bound on $\Trh$ from BBN, and direct detection experiments. We find that current LHC bounds are already competitive with BBN and direct detection in parts of the parameter space. Future facilities such as the HL-LHC and FCC are expected to probe even larger regions. This is in contrast to conventional freeze-in scenarios with high $\Trh$, where the required portal couplings are typically of order $\lhs \sim 10^{-11}$ and remain beyond experimental reach.

We also identify regions in which the combination of sizable self-interactions and low reheating temperatures opens up parameter space for light DM with $m \sim \mathcal{O}(\text{MeV})$. Without self-interactions, these scenarios would require portal couplings that are already excluded. A sufficiently large $\ls$ enhances DM production through number-changing processes and allows smaller, experimentally viable values of $\lhs$. These regions offer prospects for future collider searches and provide indirect sensitivity to the dynamics of low-temperature reheating.

Overall, our analysis shows that a realistic assessment of freeze-in DM requires a consistent treatment of reheating and dark-sector self-interactions. Their interplay can modify both the relic abundance and the experimental signatures in a significant way, and it opens up phenomenological possibilities that are absent in more simplified cosmological settings.

%% file: Chapters/Chapter5.tex

\chapter{Inducing an inverse first order phase transition}\label{ch:5} 

\begin{chapterpublication}
	This chapter is based on publication in preparation with F. Kahlhoefer, J. Matuszak and R. Santiago.
	
	\vspace{0.5em}
	
	This research project originated during the author’s internship at KIT in the group of F. Kahlhoefer. The conceptual development, numerical implementation, and analysis presented in this chapter were carried out largely by the author of this thesis, with feedback and discussion from the collaborators listed above. In particular, all numerical results and figures presented in this chapter were produced by the author of this thesis.
	
	The corresponding work is currently in preparation for publication.
\end{chapterpublication}

So far we have studied the production of dark sectors via the freeze-in mechanism under the assumption that one of the produced dark-sector particles constitutes the observed dark matter abundance and must satisfy $\Omega_c h^2 = 0.120 \pm 0.001$~\cite{Planck:2018vyg}. In general, freeze-in can populate dark sector particles that are not dark matter.

In this chapter we consider such a scenario, focusing in particular on the case in which the produced particles are scalars that thermalize within the dark sector and form their own thermal bath. In this case, the scalar field may develop a non-vanishing vacuum expectation value, which acts as an order parameter for the thermodynamic state of the system and allows one to track the onset of phase transitions.

As previously discussed, kinetic equilibrium between the Standard Model and the dark sector requires interactions mediated by operators with sufficiently large couplings. The absence of any experimental signal from dark-sector particles suggests instead that such couplings, if present, are very small. In that regime, dark-sector particles are slowly produced from the Standard Model bath through decays or annihilations. During this process, the dark-sector's energy density increases, potentially leading to non-negligible thermal corrections to its effective potential. 

A cosmological phase transition occurring as the dark-sector's entropy increases has been discussed in~\cite{Barni:2025ced, Buen-Abad:2023hex, Barni:2025gnm, Dent:2024bhi}, while the hydrodynamics of an inverse first-order phase transition was studied in~\cite{Barni:2024lkj}. In such a transition, the system evolves from a broken phase to a symmetric one. The main goal of this chapter is to describe this process beyond the relativistic approximation used in~\cite{Dent:2024bhi}.

Among the most extensively studied scenarios of cosmological first-order phase transitions (see, e.g.,~\cite{Balan:2025uke,Bringmann:2023iuz,Bringmann:2026xcx}) are those in which a dark sector is in kinetic equilibrium with the SM and therefore remains in the symmetric phase at high temperatures. As the Universe cools, the system then undergoes a phase transition to the broken phase. In contrast, here we consider a dark sector that is only very weakly coupled to the Standard Model and is populated gradually via freeze-in. The implications of the absence of kinetic equilibrium in related contexts have been discussed in~\cite{Ertas:2021xeh,Li:2025nja}, where it was shown that a hot dark sector, defined by $\xi \equiv T_{\rm ds}/T > 1$, can increase the energy stored and thereby potentially strengthen the gravitational-wave signal associated with a first-order phase transition.


The occurrence of a phase transition does not rely on a specific symmetry, but on the structure of the effective potential. What is needed is a metastable minimum separated from the true vacuum by a barrier, which can arise from tree-level interactions, radiative corrections, or thermal effects. In fact, a first-order phase transition can already occur in vacuum, provided that the potential contains two non-degenerate minima separated by a barrier. In cosmological applications, the surrounding plasma modifies this structure through thermal corrections to the scalar effective potential. At sufficiently high temperature these corrections often favor the symmetric phase, whereas the transition to the broken phase occurs only later, as the Universe expands and the plasma becomes dilute enough. Thermal effects may thus shape or induce the transition. Symmetries nevertheless remain important for model building, since they determine the allowed interactions of the dark sector and may protect the stability of a dark-matter candidate.

In this work we remain agnostic about the ultraviolet completion and adopt a minimal Abelian U(1) dark sector as a concrete realization. This choice is convenient because gauge-boson thermal corrections can naturally generate the barrier required for a FOPT, while keeping the setup simple enough to isolate the effect of gradual dark-sector population through freeze-in, where our main interest is the description of an inverse transition.

From a physical point of view, the mechanism studied in this chapter can be summarized as follows. The dark sector is initially only sparsely populated, in the broken phase. As freeze-in proceeds, energy is gradually injected into the hidden sector through rare production processes. If the dark-sector interactions are sufficiently efficient internally, this injected energy is redistributed among its degrees of freedom enforcing a thermal shape and can be described by an effective dark-sector temperature and fugacity. The build-up of this plasma modifies the scalar effective potential and may eventually favor a symmetric configuration over the broken one. In this way, the gradual population of the hidden sector can induce an inverse phase transition, even though the Standard Model bath itself is cooling as the Universe expands.


A further subtlety concerns the meaning of chemical equilibrium in the broken phase. In strict thermal equilibrium, a chemical potential is associated with an exactly conserved charge. In a broken phase, however, the propagating degrees of freedom do not carry a relevant conserved gauge charge, so there is no fundamental equilibrium chemical potential of that type. Nevertheless, if the system is close to local thermal equilibrium while particle-number-changing reactions are still inefficient, departures from chemical equilibrium can be conveniently parametrized by an effective fugacity, or equivalently by an effective chemical potential for particle number. This should be understood as a kinetic description of an under- or over-populated distribution, rather than as a chemical potential associated with an exact conserved charge. Such a situation arises naturally in freeze-in, where $n\ll n_{\rm eq}$. A fully consistent treatment is therefore formulated in real-time thermal field theory. In practice, we will show that number-changing reactions in the model become efficient enough to drive the dark sector close to full chemical equilibrium.

Finally, although the gravitational-wave phenomenology of a first-order transition also depends on bubble-wall dynamics and plasma hydrodynamics, these questions lie beyond the scope of this chapter. Our focus is on the thermal evolution of the dark sector and on the conditions under which an inverse first-order phase transition can occur.

The chapter is organized as follows. We first introduce the minimal U(1) dark-sector setup and the assumptions underlying its population through freeze-in. We then discuss the thermodynamic description of the plasma, with particular emphasis on the role of chemical equilibration and on the effective-potential treatment used throughout this work. Finally, we combine these ingredients to determine under which conditions the gradual build-up of the hidden sector can induce an inverse first-order phase transition.


\section{The model}\label{sec:model}

We consider an extension of the SM with a complex scalar \(\Phi\) charged under a dark \(\mathrm{U}(1)\) symmetry and the associated gauge boson \(A'_\mu\). The Lagrangian reads
\begin{equation}
	\begin{split}
		\mathcal{L} =  |D_\mu \Phi|^2 - \frac{1}{4}F'_{\mu\nu}F'^{\mu\nu}
		+ \mu^2\Phi^*\Phi - \lambda(\Phi^*\Phi)^2
		+ \frac{\epsilon}{2\cos\theta_W} B_{\mu\nu}F'^{\mu\nu}\,.
	\end{split}
\end{equation}
Here \(\Phi\) carries dark charge \(Q_\Phi=+1\), \(F'_{\mu\nu} = \partial_\mu A'_\nu - \partial_\nu A'_\mu\) is the field-strength tensor of the dark gauge boson, $B_{\mu\nu} = \partial_\mu B_\nu - \partial_\nu B_\mu$
is the field-strength tensor of the SM \(\mathrm{U}(1)_Y\) gauge field \(B_\mu\), and $\theta_{W}$ is the Weinberg angle. Furthermore, \(\lambda\) is a dimensionless quartic coupling, \(\mu\) is the bare mass parameter of \(\Phi\), which we take to satisfy \(\mu^2<0\) so that \(\langle \Phi \rangle \neq 0\) at zero temperature, and \(\epsilon\) denotes the kinetic-mixing parameter.

At zero temperature $\Phi$ obtains a VEV $v_\phi$, hence we expand the field as $\Phi\to (\phi + v_\phi + i\varphi)/\sqrt{2}$, where $\phi$, $\varphi$ and $v_\phi$ are real. The potential after symmetry breaking is
\begin{equation}\label{eq:model}
\begin{split}
    \mathcal{L}=&\frac{1}{2}\partial_\mu\phi\partial^\mu\phi+\frac{1}{2}\partial_\mu\varphi\partial^\mu\varphi - \frac{1}{4}F'_{\mu\nu}F'^{\mu\nu}-\frac{1}{2}m_\phi^2\phi^2+\frac{1}{2}m_{A'}^2 A'^2_{\mu}
    \\&-g A'_\mu(\varphi\partial^\mu\phi-\phi\partial^\mu\varphi-v_\phi\partial^\mu\varphi) + \frac{g^2}{2}\phi^2 A'^2_\mu + g^2v_\phi\phi A'^2_\mu
    \\&-\lambda v_\phi \phi^3 - \lambda v_\phi\varphi^2\phi - \frac{\lambda}{4}\phi^2\varphi^2-\frac{\lambda}{4}\phi^4-\frac{\lambda}{4}\varphi^4
    \\&+\frac{\epsilon}{2\cos\theta_w}B_{\mu\nu}F'^{\mu\nu}\,,
\end{split}
\end{equation}
where $g$ is the gauge coupling associated with the $\text{U}(1)$ symmetry, and the bare masses of the various fields depend on the VEV as
\begin{equation}\label{eq:masses}
    m_\phi^2(v_\phi) = 3\lambda v_\phi^2-\mu^2\,,\qquad m_\varphi^2(v_\phi) = \lambda v_\phi^2 -\mu^2 = 0\,,\qquad m_{A'}^2(v_\phi)=g^2 v_\phi^2\,.
\end{equation}

Although a Higgs-portal operator, \(|H|^2|\Phi|^2\), is gauge invariant and would in general provide an additional production channel for the dark sector, we do not include it in this chapter. We work in the limit in which its contribution is subdominant compared with kinetic-mixing-induced production. This should be viewed as a simplifying assumption adopted in order to isolate the phenomenology of kinetic-mixing-dominated freeze-in. If present with an unsuppressed coupling, Higgs-mediated production would typically be dominated by Higgs decays and would occur mainly at temperatures of order the electroweak scale, \(T\sim 100\,\mathrm{GeV}\). By contrast, dark photons are produced through inverse decays, \(\bar f f \to A'\), and the corresponding rate is sensitive to the dark-photon mass, which we take in the MeV range. The relevant production therefore takes place at lower temperatures, opening a parametrically different regime of dark-sector population.


%
\subsection{Production and equilibration of $A'$ and $\phi$}\label{subsec:equilibration}

\paragraph{Bremsstrahlung.}

We assume instantaneous reheating to a Standard Model temperature \(T_{\rm SM}^{\rm rh}=150\,\mathrm{GeV}\), so that the evolution considered in this chapter begins once reheating is complete and the SM bath has thermalized. The subsequent cosmological evolution takes place during radiation domination.


%
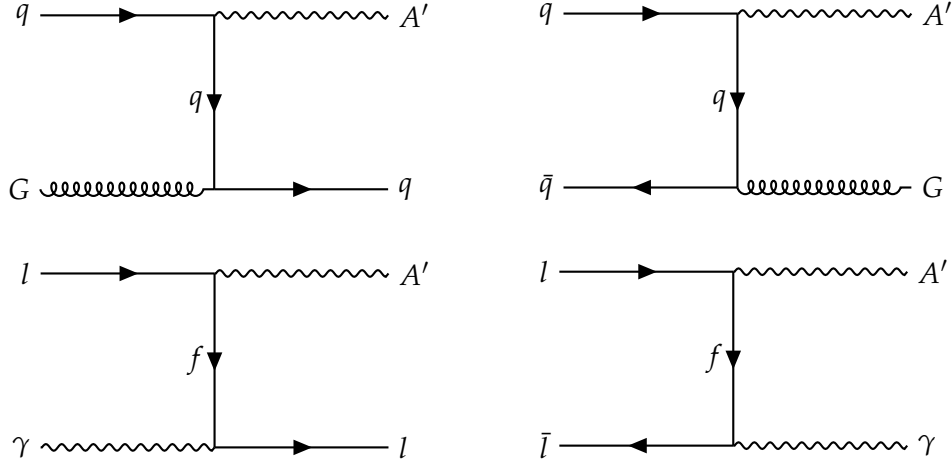
\begin{figure}[t!]
	\centering
	
	\begin{tikzpicture}
		\begin{feynman}[large,baseline=(current bounding box.center)]
			\vertex [dot] (a) [dot];
			\vertex [below=2.3cm of a] (b) [dot];
			
			\vertex [left=2.3cm of a] (i1) {\(q\)};
			\vertex [left=2.3cm of b] (i2) {\(G\)};
			\vertex [right=2.3cm of b] (f1) {\(q\)};
			\vertex [right=2.3cm of a] (f2) {\(A'\)};
			
			\diagram*{
				(i1) -- [fermion] (a),
				(a)  -- [fermion,edge label'=\(q\)] (b),
				(i2) -- [gluon] (b),
				(b)  -- [fermion] (f1),
				(a)  -- [photon] (f2),
			};
		\end{feynman}
	\end{tikzpicture}
	\hspace{1.0cm}
	\begin{tikzpicture}
		\begin{feynman}[large,baseline=(current bounding box.center)]
			\vertex [dot] (a) [dot];
			\vertex [below=2.3cm of a] (b) [dot];
			
			\vertex [left=2.3cm of a] (i1) {\(q\)};
			\vertex [left=2.3cm of b] (i2) {\(\bar q\)};
			\vertex [right=2.3cm of b] (f1) {\(G\)};
			\vertex [right=2.3cm of a] (f2) {\(A'\)};
			
			\diagram*{
				(i1) -- [fermion] (a),
				(a)  -- [fermion,edge label'=\(q\)] (b),
				(i2) -- [anti fermion] (b),
				(b)  -- [gluon] (f1),
				(a)  -- [photon] (f2),
			};
		\end{feynman}
	\end{tikzpicture}
	
	\vspace{0.5cm}
		\begin{tikzpicture}
		\begin{feynman}[large,baseline=(current bounding box.center)]
			\vertex [dot] (a) [dot];
			\vertex [below=2.3cm of a] (b) [dot];
			
			\vertex [left=2.3cm of a] (i1) {\(l\)};
			\vertex [left=2.3cm of b] (i2) {\(\gamma\)};
			\vertex [right=2.3cm of b] (f1) {\(l\)};
			\vertex [right=2.3cm of a] (f2) {\(A'\)};
			
			\diagram*{
				(i1) -- [fermion] (a),
				(a)  -- [fermion,edge label'=\(f\)] (b),
				(i2) -- [photon] (b),
				(b)  -- [fermion] (f1),
				(a)  -- [photon] (f2),
			};
		\end{feynman}
	\end{tikzpicture}
	\hspace{1.0cm}
	\begin{tikzpicture}
		\begin{feynman}[large,baseline=(current bounding box.center)]
			\vertex [dot] (a) [dot];
			\vertex [below=2.3cm of a] (b) [dot];
			
			\vertex [left=2.3cm of a] (i1) {\(l\)};
			\vertex [left=2.3cm of b] (i2) {\(\bar l\)};
			\vertex [right=2.3cm of b] (f1) {\(\gamma\)};
			\vertex [right=2.3cm of a] (f2) {\(A'\)};
			
			\diagram*{
				(i1) -- [fermion] (a),
				(a)  -- [fermion,edge label'=\(f\)] (b),
				(i2) -- [anti fermion] (b),
				(b)  -- [photon] (f1),
				(a)  -- [photon] (f2),
			};
		\end{feynman}
	\end{tikzpicture}
	\caption{Top: representative quark-exchange diagrams for $A'$ production at high temperature: $qG\to qA'$ (left, $u$-channel quark exchange) and $q\bar q\to GA'$ (right, $t$-channel quark exchange). Down: same as top with leptons and SM photons.}
\end{figure}

The first regime of production corresponds to SM temperatures that satisfy $T\gg m_{A'}$. At such temperatures, bremsstrahlung collisions (Compton-like scattering) and annihilation processes are the leading production channels:
\begin{equation}\label{eq:bremstrahlung}
\ell \gamma \to \ell A', 
\qquad
\bar\ell \ell \to \gamma A', 
\qquad
q G \to G A',
\qquad
\bar q q \to G A'.
\end{equation}

Here $q$, $G$, and $\gamma$ denote SM quarks, gluons, and photons, respectively.\footnote{There are also annihilation channels $\bar ff,\bar qq\to A'A'$. However, the corresponding matrix elements are suppressed as $\epsilon^2$.} These channels are collinearly enhanced: the squared matrix elements contain $t$- and/or $u$-channel propagators that become singular in the collinear limit ($t\to 0$ and/or $u\to 0$). 
For instance, in the relativistic limit, the matrix element for the process $qG\to q A'$ is given by
\begin{equation}\label{eq:M_brem_rel}
	\overline{|\mathcal{M}|^2}(qG\to qA')
	\;\simeq\;
	\mathcal{C}_{qg}\,g_s^2\,g_q^2
	\left(-\frac{s}{u}-\frac{u}{s}\right),
	\qquad (s>0,\;u<0),
\end{equation}
where $g_q = \epsilon e Q_q$ is the effective coupling of the dark photon to the quark \(q\), $g_f=\epsilon e Q_f$ is the effective coupling of the dark photon and $\mathcal{C}_{qg}$ is an order-one spin/color averaging factor. Here $g_s\simeq 1.21$ is the gauge coupling of quarks and gluons,
\footnote{At finite temperature, the QCD coupling runs with the renormalization scale \(\mu\), not directly with \(T\). For relativistic \(2\to2\) processes in a plasma, one may take \(\mu\sim\pi T\), which induces only a mild temperature dependence because the running is logarithmic. Over the range relevant here, \(g_s\) therefore remains of order unity. Thermal corrections mainly enter through screening masses that regulate collinear singularities, rather than through an \(\mathcal{O}(1)\) shift of \(g_s\). Treating \(g_s\) as approximately constant is thus adequate for our purposes~\cite{Blaizot:2003iq}.}. 

Processes involving color states are modified after hadronization at \(T\sim T_{\rm QCD}\). In the benchmark studied in this chapter, the dark sector possesses a MeV-scale vacuum expectation value, which yields a MeV-scale dark-photon mass.
The bremsstrahlung channels are therefore already open throughout the much earlier quark--gluon phase, and the dominant contribution is accumulated at temperatures well above the QCD crossover.
By the time the SM plasma reaches hadronization, the additional contribution to the yield is already subleading.
For this reason, the detailed dynamics of the QCD transition can be neglected at the level of the present analysis.

%

Finally, although Eq.~\eqref{eq:M_brem_rel} shows the squared matrix element in the relativistic limit, in the numerical analysis we used \texttt{CalcHEP}~\cite{Belyaev:2012qa} to obtain the full tree-level analytical expression. The infrared/collinear enhancement was regulated by introducing thermal screening masses for the relevant SM states. This should be understood as an effective infrared regulator, rather than as a fully resummed finite-temperature treatment.

\paragraph{Coalescence.}
Yet another relevant channel of production corresponds to inverse decays of SM fermions,
\begin{equation}
    \bar f f \to A'\,,
\end{equation}
whose production peaks when the temperature of the SM is of the order of the dark photon's mass.\footnote{An additional contribution arises from the process $\bar f f \to A' \to A' \phi$, however, $\bar f f \to A'$ dominates parametrically due to kinematics and phase-space power counting~\cite{Xu:2025wlq}. Additionally, this channel is closed in the symmetric phase.} 
%
%
%
%
%

The produced dark photons subsequently generate dark Higgs bosons via the gauge interactions
\begin{equation}\label{eq:phi-A}
    \frac{g^2}{2}\,\phi^2 A_\mu'^2, 
    \qquad 
    g^2 \phi_b\,\phi A_\mu'^2,
    \qquad 
    \lambda \phi_b\,\phi^3 .
\end{equation}
Since we are interested in FOPT, we consider large gauge couplings ($g\gtrsim 0.1$) to generate a potential barrier, therefore, we assume that as soon as the dark photons are produced, they instantly thermalize the dark sector, meaning we assume a Bose-Einstein distribution for $\phi$ and $A'$. 



\paragraph{Resonant production from inverse decays}
%
The SM photon, $\gamma$, acquires a thermal mass due to its interactions with the SM plasma that is parametrically $m_\gamma \sim e T$, therefore, as the universe cools, there is a period in time when $m_{A'} = m_\gamma$ and the coalescence production is enhanced due to the resonance. Such a process can provide the necessary energy to induce a FOPT and therefore we consider such a process in detail here.

Neglecting in-medium effects for the moment, the squared matrix element 
summed over initial and final spins is
\begin{equation}
    \sum |\mathcal{M}_{\bar f f}|^2 
    = 16\pi \alpha_{\rm eff} m_{A'}^2 
    \left( 1 + 2 \frac{m_f^2}{m_{A'}^2} \right),
\end{equation}
which also determines the vacuum decay width
\begin{equation}
    \Gamma_{A' \to \bar f f}
    =
    \frac{\alpha_{\rm eff}}{3} m_{A'}
    \left( 1 + 2 \frac{m_f^2}{m_{A'}^2} \right)
    \sqrt{1 - \frac{4 m_f^2}{m_{A'}^2}}\,,
\end{equation}
where $\alpha_{\rm eff} \equiv N_c^f Q_f^2 \epsilon^2 \alpha $.

Working in the Maxwell–Boltzmann approximation and focusing on the 
relativistic limit $m_f \ll m_{A'}$, the collision term can be written as
\begin{equation}
    C_0^{\bar f f \to A'}
    =
    \frac{3}{2\pi^2}\,
    \Gamma_{A' \to \bar f f}\,
    m_{A'}^2\, T\, 
    K_1\!\left(\frac{m_{A'}}{T}\right),
\end{equation}
in the continuum limit without resonance.

In the thermal plasma, inverse decays proceed through an
in-medium SM photon propagator, which will be enhanced around $m_{A'} = m_\gamma$. The collision term
receives corrections from the longitudinal and transverse
polarization modes of the photon.

Separating the contributions according to the polarization
of the produced vector boson with four-momentum
$(\omega,\vec q)$, the collision term can be written as~\cite{Fradette:2014sza}
\begin{equation}\label{eq:C0_ff_to_A}
C_0^{\bar f f \to A'}
=
\frac{3m_{A'}}{2\pi^2}
\Gamma_{A'\to\bar f f}
\int_{m_{A'}}^\infty d\omega
\sqrt{\omega^2-m_{A'}^2}\,
e^{-\omega/T}
\left[
\frac{1}{3}
\frac{m_{A'}^4}{|m_{A'}^2-\Pi_L|^2}
+
\frac{2}{3}
\frac{m_{A'}^4}{|m_{A'}^2-\Pi_T|^2}
\right],
\end{equation}
where $\Pi_{L,T}(\omega,|\vec q|,T)$ are the longitudinal
and transverse photon polarization functions in the plasma.

The denominators originate from the resummed in-medium
propagator and encode possible resonant enhancement.
In particular, a resonance occurs when
\begin{equation}
\mathrm{Re}\,\Pi_{L,T}(\omega,T) = m_{A'}^2 .
\end{equation}
Since the plasma frequency scales as $\omega_p^2 \sim \alpha T^2$,
this condition is typically satisfied at temperatures
$T \sim m_{A'}/\sqrt{\alpha}$, i.e. parametrically larger than
$m_{A'}$. In general, for a one-component ultra-relativistic plasma,
the expressions for the real parts of the polarization tensors are given by \cite{Braaten:1993jw},
\begin{eqnarray}
{\rm Re}\Pi_T(\omega) &=& \omega_p^2\frac{3\omega^2}{2\vec{q}^2}\left( 1-\frac{m_V^2}{\omega^2}\frac{\omega}{2|\vec{q}|}
\log\frac{\omega +|\vec{q}|}{\omega -|\vec{q}|}\right), \nonumber
\\
{\rm Re}\Pi_L(\omega) &=& 3\omega_p^2\frac{m_V^2}{\vec{q}^2}\left( \frac{\omega}{2|\vec{q}|}
\log\frac{\omega +|\vec{q}|}{\omega -|\vec{q}|} - 1 \right), \label{eq:RePi}
\end{eqnarray}
where all the factors of $|\vec{q}|$ can be replaced with $\sqrt{\omega^2-m_{A'}^2}$. 

In the limit of vanishing plasma density,
$\Pi_{L,T}\to 0$, the expression in brackets tends to unity,
and the collision term reduces to the continuum result of~\eqref{eq:C0_ff_to_A}
reproducing the non-resonant expressions discussed above.


In principle, both quarks and leptons contribute to the coalescence channel. However, quarks are confined into hadrons across the QCD transition. For MeV-scale dark-photon masses, the resonance condition $m_{A'} \approx m_\gamma$ is reached around $T \sim T_{\rm QCD}$, so a reliable treatment would require modeling hadronization effects. In this work we avoid that complication and include only the leptonic contribution, with the understanding that this yields a lower estimate for the total production rate.


\paragraph{Chemical equilibrium}
Interactions between $\phi$ and $A'$ lead to $\mu_\phi = \mu_{A'}$, however, they alone do not guarantee a vanishing chemical potential. To account for chemical equilibration, we consider the self-number changing reactions for $\phi$, $3\phi\leftrightarrow2\phi$, which are analogous to the real scalar candidate with broken $\mathbb{Z}_2$ symmetry studied in~\Cref{sec:real_scalar}. 

Additionally, the broken phase allows for interactions of the form 
\begin{equation}\label{eq:cannibal_w_A}
	\begin{split}
		3\phi\leftrightarrow2A' ,\qquad 2A'\phi\leftrightarrow2A',\qquad 2\phi A' \leftrightarrow \phi A',\qquad 2A'\phi\leftrightarrow 2\phi\,,
	\end{split}
\end{equation} 
which are in fact typically orders of magnitude larger than $3\phi\leftrightarrow 2\phi$. This is due to the size of the gauge coupling $g$, which is usually orders of magnitude larger than $\lambda$ in FOPT studies.

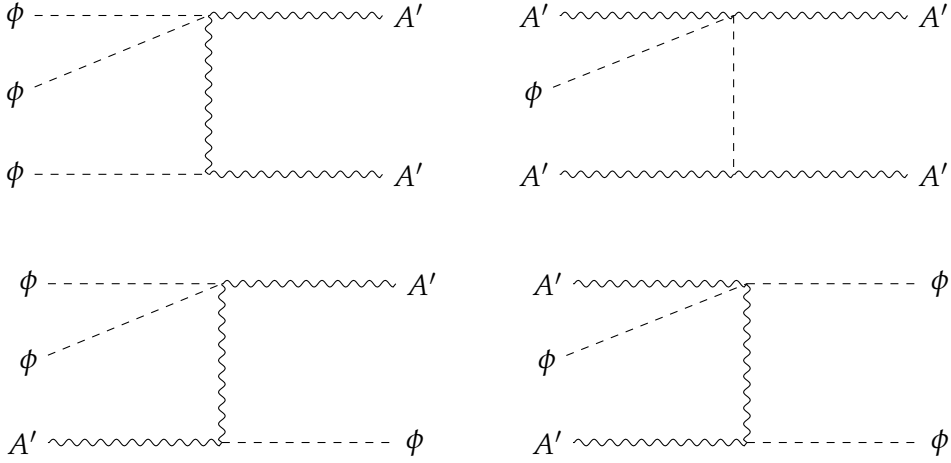
\begin{figure}[t!]
	\centering
	\begin{tikzpicture}[baseline=(in)]
		\begin{feynman}
			\vertex [dot] (a);  
			\vertex [below=0.6cm of a] (in);
			\vertex [below=1.5cm of in] (b);
			\vertex [left =2.3cm of a] (i1) {$\phi$};  
			\vertex [below=1.05cm of i1] (ii2){$\phi$};
			\vertex [left =2.3cm of b] (i2){$\phi$};
			\vertex [right=2.3cm of b] (f1){$A'$};
			\vertex [right=2.3cm of a] (f2){$A'$};
			\diagram* {
				(i1) -- [scalar](a)
				-- [photon] (f2),
				(a) -- [photon] (b),
				(i2) -- [scalar] (b),
				(b) -- [photon] (f1),
				(ii2) -- [scalar] (a),
			};
		\end{feynman}
	\end{tikzpicture} 
	\hspace{0.8cm}
	\begin{tikzpicture}[baseline=(in)]
		\begin{feynman}
			\vertex [dot] (a);  
			\vertex [below=0.6cm of a] (in);
			\vertex [below=1.5cm of in] (b);
			\vertex [left =2.3cm of a] (i1) {$A'$};  
			\vertex [below=1.05cm of i1] (ii2){$\phi$};
			\vertex [left =2.3cm of b] (i2){$A'$};
			\vertex [right=2.3cm of b] (f1){$A'$};
			\vertex [right=2.3cm of a] (f2){$A'$};
			\diagram* {
				(i1) -- [photon](a)
				-- [photon] (f2),
				(a) -- [scalar] (b),
				(i2) -- [photon] (b),
				(b) -- [photon] (f1),
				(ii2) -- [scalar] (a),
			};
		\end{feynman}
	\end{tikzpicture}
	
	\vspace{0.8cm}
	\begin{tikzpicture}[baseline=(in)]
		\begin{feynman}
			\vertex [dot] (a);  
			\vertex [below=0.6cm of a] (in);
			\vertex [below=1.5cm of in] (b);
			\vertex [left =2.3cm of a] (i1) {$\phi$};  
			\vertex [below=1.05cm of i1] (ii2){$\phi$};
			\vertex [left =2.3cm of b] (i2){$A'$};
			\vertex [right=2.3cm of b] (f1){$\phi$};
			\vertex [right=2.3cm of a] (f2){$A'$};
			\diagram* {
				(i1) -- [scalar](a)
				-- [photon] (f2),
				(a) -- [photon] (b),
				(i2) -- [photon] (b),
				(b) -- [scalar] (f1),
				(ii2) -- [scalar] (a),
			};
		\end{feynman}
	\end{tikzpicture} 
	\hspace{0.8cm}
	\begin{tikzpicture}[baseline=(in)]
		\begin{feynman}
			\vertex [dot] (a);  
			\vertex [below=0.6cm of a] (in);
			\vertex [below=1.5cm of in] (b);
			\vertex [left =2.3cm of a] (i1) {$A'$};  
			\vertex [below=1.05cm of i1] (ii2){$\phi$};
			\vertex [left =2.3cm of b] (i2){$A'$};
			\vertex [right=2.3cm of b] (f1){$\phi$};
			\vertex [right=2.3cm of a] (f2){$\phi$};
			\diagram* {
				(i1) -- [photon](a)
				-- [scalar] (f2),
				(a) -- [photon] (b),
				(i2) -- [photon] (b),
				(b) -- [scalar] (f1),
				(ii2) -- [scalar] (a),
			};
		\end{feynman}
	\end{tikzpicture}
	
\caption{Some of the $g$-leading order diagrams of the self-number changing reaction processes in Eq.~\eqref{eq:cannibal_w_A}. There are diagrams with more vertices and/or different channels. 
}
\label{fig:feynman_cannibal_w_A}
\end{figure}

%
%

In the symmetric phase, $3\leftrightarrow 2$ reactions involving $A'$ and $\phi$ are absent at tree level. Higher-multiplicity channels of the form $4\leftrightarrow 2$ remain open, e.g.
\[
2A'\Phi^*\Phi \leftrightarrow 2A'\,,
\]
but they are expected to be more suppressed than the broken-phase $3\leftrightarrow 2$ processes due to the larger final-state phase space and larger power counting in the couplings. Hence, their efficiency must be assessed separately.

%
%

At this stage it is important to distinguish between two physically different uses of the chemical potential. On the one hand, charge conservation constrains chemical potentials, but does not by itself set them to zero. For instance, in the symmetric phase one has
\begin{equation}
	\mu_{\Phi^*} = - \mu_\Phi \, ,
\end{equation}
which follows from exact U(1) charge conservation. Here ``exact'' means that the charge is protected by the underlying symmetry of the theory and is conserved by the microscopic dynamics. On the other hand, this relation still allows \(\mu_\Phi \neq 0\). The vanishing of the chemical potential is instead a consequence of efficient number-changing reactions that enforce chemical equilibrium. If these reactions are slow, a non-zero fugacity may persist irrespective of whether the system is in the symmetric or broken phase. In that case, the relevant quantity is not an exact conserved charge, but an approximately conserved particle number, in the sense that number-changing processes are present but may be inefficient, so that the total number density does not approach its equilibrium value.

\paragraph{Reduction to an effective two–equation system.}

%
%
%
%
%
%

In principle one should evolve the full set of Boltzmann equations
for the number and energy densities of both $A'$ and $\phi$,
\begin{equation}
	\begin{split}
		\dot n_{A'}+3H n_{A'} &= C_0^\text{FI} + C_0[A']^{2A'\leftrightarrow 2\phi} + C_0[A']^{3\leftrightarrow 2}\,,
		\\\dot n_{\phi}+3H n_{\phi} &= C_0[\phi]^{2A'\leftrightarrow 2\phi} +C_0[\phi]^{3\leftrightarrow 2},
	\end{split}
\end{equation}
together with the corresponding temperature equations. Note that here $C_0^\text{FI}$ contains the coalescence as well as the 2-to-2 scattering production. 

The dynamics simplifies considerably when the internal dark–sector
interaction rate satisfies
\begin{equation}
	\Gamma_{2A'\to 2\phi}
	\;\gg\; H\,.
	\label{eq:gamma_condition}
\end{equation}
In this regime the gauge interactions enforce the same temperature and chemical potential for the system of two particles, long before cosmic expansion or freeze-in production can significantly modify the distributions.

In fact, the processes
$2A'\leftrightarrow2\phi$ do not inject or remove energy from the dark
sector but merely redistribute it between $A'$ and $\phi$.
In that case, the dark sector can be described by a common temperature
$T_{\rm ds}$ and chemical potential $\mu$, such that
\begin{equation}
	T_{A'} = T_\phi = T_\text{ds},
	\qquad
	\mu_{A'} = \mu_\phi = \mu.
\end{equation}
The individual number and energy densities then become functions of
$(T_\text{ds},\mu)$,
\begin{equation}
	n_\psi = n_\psi(T_{\rm ds},\mu),
	\qquad
	\rho_\psi = \rho_\psi(T_{\rm ds},\mu),
\end{equation}
with $\psi\in\{A',\phi\}$.

In this limit the only true source term for the dark sector correspond to the freeze-in production channels discussed above. It is therefore sufficient to evolve the total dark–sector densities,
\begin{equation}
	n \equiv n_{A'}+n_\phi,
	\qquad
	\rho \equiv \rho_{A'}+\rho_\phi,
\end{equation}
according to
\begin{equation}\label{eq:n_and_rho}
	\begin{split}
		\dot n + 3H n &= C_0^\text{FI} + C_0^{3\leftrightarrow 2},
		\\ \dot \rho + 3H (\rho  + P) &= C_E^\text{FI}\,,
	\end{split}
\end{equation}
where $C^\text{FI}$ is the collision operator corresponding to the freeze-in mechanism. The individual densities $n_{A'}$ and $n_\phi$ are then obtained
algebraically from their equilibrium ratios at $(T_{\rm ds},\mu)$.

In practice, since $A'$ and $\phi$ remain relativistic during the
bremmstrahlung production stage, their equilibrium abundances are of the
same parametric order and the dark sector behaves effectively as a
single fluid.

\section{Effective potential of the dark sector}\label{sec:effective_potential}

Since number-changing interactions need not be efficient from the outset, the dark sector can in principle depart from chemical equilibrium. It is convenient to parameterize the phase-space distributions by a time-dependent fugacity \(z\equiv e^{\mu/T_{\rm ds}}\), even though \(\mu\) is not associated with an exact conserved charge. Rather, \(z\) should be understood as a convenient parametrization of departures from chemical equilibrium associated with approximate particle-number conservation, as discussed for instance in~\cite{Calzetta:2008iqa}.


In such a situation, the imaginary-time (Matsubara) formalism discussed in~\Cref{sec:Thermal_corr}, which assumes thermal equilibrium (or, at most, a chemical potential associated with an exactly conserved charge) and is formulated in terms of a thermal partition function, is not the most general starting point. We therefore adopt a real-time quasiparticle description, in which medium effects are expressed directly in terms of the actual occupation numbers \(f_\psi\), where $\psi\in\{\phi,\,A' \}$. We do not attempt to discuss the real-time formalism in this thesis, since this would take us too far from the main focus of the chapter. Instead, we use the thermal corrections derived from this formalism. In the limit of full equilibrium, this description reproduces the standard thermal results obtained in the Matsubara formalism. Interested readers may consult~\cite{Lundberg:2020mwu,Bellac:2011kqa,Landsman:1986uw} for complete discussions.


As we show below, in the U(1) model studied here the number-changing reactions involving \(A'\) are sufficiently efficient to drive the system rapidly toward chemical equilibrium, so that \(z=1\) during the relevant stages of the evolution. The usefulness of the present framework is thus that it allows this equilibration to be established dynamically, rather than imposed from the outset.

\paragraph{Pressure representation of the medium.}

Within this quasiparticle description, we denote collectively by \(\psi\) the fluctuating fields around the homogeneous background \(\phi_b\), with field-dependent masses \(m_\psi(\phi_b)\) and single-particle energies \(E_\psi(\phi_b)=\sqrt{p^2+m_\psi^2(\phi_b)}\).

For a homogeneous state characterized by occupation numbers \(f_\psi\), the pressure of a quasiparticle species \(\psi\) is given by~\cite{Kolb:1990vq}
\begin{equation}
	P_\psi(\phi_b)
	=
	g_\psi\int_p \frac{p^2}{3E_\psi}\,f_\psi,
	\label{eq:p_general}
\end{equation}
where \(g_\psi\) counts the physical degrees of freedom. 
This expression follows from the spatial components of the energy--momentum tensor and is valid for arbitrary occupation numbers \(f_\psi\) in a homogeneous and isotropic state. With the Maxwell--Boltzmann ansatz, it reduces to the pressure formula introduced in Eq.~\eqref{eq:thermoquantities}.

For a homogeneous and isotropic medium, we identify the thermal contribution to the effective potential as minus the pressure~\cite{Lundberg:2020mwu,Bellac:2011kqa,Landsman:1986uw},
\begin{equation}
	V_{\rm med}(\phi_b)=-P_{\rm med}(\phi_b)
	=-\sum_\psi P_\psi(\phi_b).
\end{equation}
This relation provides a practical way of constructing the medium correction directly from the occupation numbers, without invoking a thermal partition function.

Parameterizing the non-equilibrium distributions as
\begin{equation}
	f_\psi
	=
	\frac{1}{z^{-1}e^{E(\phi_b)/T_\psi}\pm 1},
	\qquad
	z=e^{\mu/T_\psi}\,,
\end{equation}
where $T_\psi$ corresponds to the temperature of the species $\psi$. We obtain
\begin{equation}\label{eq:Vmed1}
	V_{\rm med}(\phi_b;T,z)
	=
	-\sum_\psi g_\psi
	\int_p
	\frac{p^2}{3E}
	\left(z^{-1}e^{E/T_\psi}\pm 1\right)^{-1}.
\end{equation}

Equivalently,
\begin{equation}\label{eq:Vmed2}
	V_{\rm med}(\phi_b;T,z)
	=
	\frac{T_\psi^4}{2\pi^2}
	\sum_\psi g_\psi(\pm)
	\int_0^\infty dy\,y^2
	\ln\!\left(
	1\pm z\,e^{-\sqrt{y^2+x_\psi^2(\phi_b)}}
	\right),
\end{equation}
with $x_\psi(\phi_b)\equiv m_\psi(\phi_b)/T_\psi$. Eq.~\eqref{eq:Vmed2} is usually found in the literature~\cite{Quiros:1999jp,Bringmann:2023iuz}.
For \(z=1\), this reduces to the standard equilibrium expression for the thermal effective potential as derived in the Matsubara formalism in~\Cref{sec:Thermal_corr}.

There is no closed form for Eqs.~\eqref{eq:Vmed1} and~\eqref{eq:Vmed2}. However, a useful semi-analytic representation is obtained by expanding the distribution function as
\begin{equation}\label{eq:dist_expansion}
	f_\psi
	=
	\frac{1}{z^{-1}e^{E/T_\psi}\pm1}
	=
	\frac{z\,e^{-E/T_\psi}}{1\pm z\,e^{-E/T_\psi}}
	=
	\sum_{k=1}^{\infty}(\mp1)^{k-1}z^k e^{-kE/T_\psi}.
\end{equation}

This series converges pointwise whenever \(|z\,e^{-E/T_\psi}|<1\). For massive species (\(E\ge m_\psi\)) and \(0<z\le 1\), this condition is automatically satisfied for all momenta. If one allows \(z>1\), convergence for all \(p\) requires \(ze^{-m_\psi/T_\psi}<1\), equivalently \(\mu<m_\psi\). For exactly massless bosons the \(p\to0\) region is infrared sensitive due to Bose enhancement.

During freeze-in, and in the absence of self-number-changing reactions, one has \(z\le 1\), for which the fugacity expansion is well defined. If self-number-changing reactions later enforce \(z=1\), the only delicate regime is the infrared bosonic sector. In the symmetric phase, however, the relevant dispersion relation includes thermal corrections, $E=\sqrt{p^2+m_{\rm eff}^2}$,
so that \(E>0\) for all \(p\) as long as \(m_{\rm eff}\neq0\). The expansion remains pointwise convergent, although convergence can become slow when \(m_{\rm eff}/T_\psi\ll1\) and \(z\simeq1\).

We can expand the pressure as
\begin{equation}
	\int_p \frac{p^2}{3E}
	\left(e^{(E-\mu)/T_\psi}\pm1\right)^{-1}
	=
	\begin{cases}
		\frac{m_\psi^2T_\psi^2}{2\pi^2}
		\sum_{k=1}^{\infty}\frac{z^k}{k^2}K_2(km_\psi/T_\psi),
		& \text{for bosons},\\[1.2ex]
		\frac{m_\psi^2T_\psi^2}{2\pi^2}
		\sum_{k=1}^{\infty}\frac{z^k}{k^2}(-1)^{k-1}K_2(km_\psi/T_\psi),
		& \text{for fermions}.
	\end{cases}
\end{equation}

In fact, any thermodynamic quantity can be expanded in powers of \(z\),
\begin{equation}
	\mathcal O = \sum_{k=1}^{\infty} z^k\,\mathcal O_k^{\rm eq}.
\end{equation}
In particular,
\begin{equation}\label{eq:thermodynamic_k}
	\begin{split}
		n_k^{\rm eq}
		&=
		(\pm1)^{k-1}\frac{g}{2\pi^2}\frac{m_\psi^2T_\psi}{k}K_2(y),\\
		p_k^{\rm eq}
		&=
		(\pm1)^{k-1}\frac{T_\psi}{k}n_k^{\rm eq},\\
		\rho_k^{\rm eq}
		&=
		(\pm1)^{k-1}\frac{g}{2\pi^2}\frac{m_\psi^2T_\psi^2}{k^2}\left(3K_2(y)+yK_1(y)\right),\\
		s_k^{\rm eq}
		&=
		\frac{\rho_k^{\rm eq}+p_k^{\rm eq}}{T_\psi}
		=
		\frac{\partial p_k^{\rm eq}}{\partial T_\psi},
	\end{split}
\end{equation}
with
\begin{equation}
	y\equiv \frac{k\,m_\psi}{T_\psi}.
\end{equation}

\paragraph{Daisy resummation.}

The need for daisy resummation reflects the infrared sensitivity of soft bosonic modes when \(\Pi_\psi/m_\psi^2(\phi_b)\gtrsim 1\). Although this is often discussed in Matsubara language in terms of the zero mode, the same physics appears in real time as an enhancement of static soft modes in the retarded propagator~\cite{Carrington:1997sq}. Hence, we include
\begin{equation}
V_{\rm daisy}
=
-\frac{T_\psi}{12\pi}
\sum_{\psi=\phi,\varphi,A'_L}
g_\psi
\left[
(m_\psi^2+\Pi_\psi)^{3/2}
-
(m_\psi^2)^{3/2}
\right],
\end{equation}
where the longitudinal gauge mode $A'_L$ is treated separately,
as usual in thermal field theory.

The one-loop thermal masses $\Pi_\psi$ in the model are given by 
\begin{equation}\label{eq:Pi}
	\begin{split}
		\Pi_\phi &= 3\lambda\,\mathcal{I}_\phi + \lambda\,\mathcal{I}_\varphi + 3g^2\,\mathcal{I}_{A'}\,,\\
		\Pi_\varphi&=\lambda\,\mathcal{I}_\phi + 3\lambda\,\mathcal{I}_\varphi + 3g^2\mathcal{I}_{A'}\,,
		\\ \Pi_{A'}&=2g^2\left(\mathcal{I}_{\phi}+\mathcal{I}_{\varphi}\right)\,,
	\end{split}
\end{equation}
where
\begin{equation}
	\mathcal{I}_\psi = \int_p \frac{f_\psi}{E}.
\end{equation} 
The derivation is provided in~\Cref{ap:thermal_masses}.
%
%
In other words, the Daisy resummation is essential (cf. Eq.~\eqref{eq:resummed_propagator}) when $\Pi_\psi/m_\psi$ becomes large  (particularly near the phase transition when $m_\psi\to 0$). In fact, $z<1$ can maintain $\Pi_\psi<m_\psi$ in broader regions of the phase space.


\begin{figure}[t]
	\centering
	\setlength{\tabcolsep}{8pt}
	
	\begin{tabular}{ccccc}
		
		\begin{tikzpicture}
			\begin{feynman}
				\diagram [horizontal=i to f, scale=0.85] {
					i -- [scalar] v1 -- [scalar] f,
					v1 -- [scalar, loop, min distance=1.9cm] v1,
				};
			\end{feynman}
		\end{tikzpicture}
		&
		\begin{tikzpicture}
			\begin{feynman}
				\diagram [horizontal=i to f, scale=0.85] {
					i -- [scalar] v1 -- [scalar] f,
					v1 -- [scalar, loop, min distance=1.9cm, edge label'={\small $\varphi$}] v1,
				};
			\end{feynman}
		\end{tikzpicture}
		&
		\begin{tikzpicture}
			\begin{feynman}
				\diagram [horizontal=i to f, scale=0.85] {
					i -- [scalar] v1 -- [scalar] f,
					v1 -- [photon, loop, min distance=1.9cm, edge label'={\small $A_\mu$}] v1,
				};
			\end{feynman}
		\end{tikzpicture}
		
		\\[-2pt]
		{\small (a)} & {\small (b)} & {\small (c)} 
	\end{tabular}
	
	\caption{One-loop contributions to
		the thermal self-energy $\Pi_\phi$ for a scalar field $\phi$. Panel (a) correspond to 
		a $\phi^4$ interaction, (b) to a scalar portal
		$\phi^2\varphi^2$, and (e) gauge
		interactions (schematically).}
	\label{fig:ring_jll}
\end{figure}
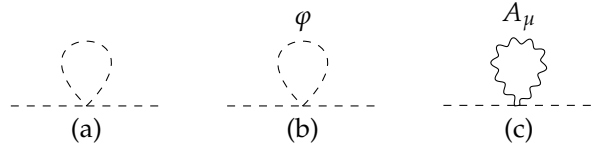

\paragraph{The effective potential}

We can now write the finite-temperature, finite-fugacity effective potential corresponding to the Lagrangian in Eq.~\eqref{eq:model} as
\begin{equation}\label{eq:Veff}
    V_\text{eff}(\phi_b;T,z)= V_\text{tree}(\phi_b)+ V_\text{CW}(\phi_b) + V_\text{ct}(\phi_b) + V_\text{med}(\phi_b;T,z)+ V_\text{daisy}(\phi_b;T,z)\,.
\end{equation}

In the Landau gauge and the \(\overline{\rm MS}\) renormalization scheme, the Coleman--Weinberg contribution takes the form
\begin{equation}
  V_{\rm CW}(\phi_b) = \sum_{\psi=\phi,\varphi,A'} \pm g_\psi \frac{m_\psi^4(\phi_b)}{64\pi^2}
  \left(\log\frac{m_\psi^2(\phi_b)}{\Lambda^2} - k_\psi\right),
  \label{CWpotential}
\end{equation}
where the $+$ ($-$) sign applies to bosons (fermions), $g_\psi$ are the degrees of freedom of particle $\psi$, $m_\psi(\phi_b)$ are the background field-dependent masses, and $k_\psi = 3/2$ for scalars and fermions and $k_\psi = 5/6$ for gauge bosons. We take the renormalization scale $\Lambda$ to be the VEV at zero temperature, $v_\phi$. The field-dependent masses in the CW potential are computed with renormalized parameters, i.e., using the expressions from eq.~\eqref{eq:masses} with the replacement $v_\phi \to \phi_b$. 

The counter terms potential are chosen such that the renormalization condition leads to the tree-level values of dark Higgs mass and VEV at zero temperature, i.e.,
\begin{equation}
	V_\text{ct}(\phi_b) = -\frac{\delta\mu^2}{2}\phi_b^2+\frac{\delta\lambda}{4}\phi_b^4\,.
\end{equation}
with 
\begin{equation}
	\begin{split}
		\delta\mu^2 &= \left.\left(\frac{2}{3\phi_b}\frac{dV_\text{CW}(\phi_b)}{d\phi_b}-\frac{1}{2}\frac{d^2V_\text{CW}}{d\phi_b^2} \right)\right\vert_{\phi_b = v_\phi}\qquad \text{and}
		\\ 	\delta\lambda &= \left.\left(\frac{1}{2\phi_b^3}\frac{dV_\text{CW}(\phi_b)}{d\phi_b}-\frac{1}{2\phi_b^2}\frac{d^2V_\text{CW}}{d\phi_b^2} \right)\right\vert_{\phi_b = v_\phi}\,.
	\end{split}
\end{equation}

Note that, even though the model does not contain fermions, we decided to present the definition of the effective potential in full generality. A similar study with a U(1) gauge theory in the context of FOPT with charged fermions as DM candidates is studied in~\cite{Bringmann:2023iuz}.

\subsection{Evolution of the effective potential}\label{subsec:evol}

Tracking the free-energy-density evolution during freeze-in could, in principle, be formulated within a real-time non-equilibrium framework. More fundamentally, such an approach would begin with the Schwinger--Keldysh formalism, from which effective kinetic equations, such as the Boltzmann equation~\eqref{eq:fBE} or the corresponding moment equations in local thermal equilibrium, can be obtained under suitable assumptions. One could then attempt to reconstruct the thermodynamic evolution by inserting the resulting distribution functions into \(V_{\rm eff}\).

In practice, however, this route is not particularly convenient once the problem is formulated directly in terms of the effective potential. In particular, it is not straightforward to implement the Daisy-resummed contribution consistently at the level of the unintegrated kinetic description.

More importantly, during a first-order phase transition the system cannot in general be described as a single homogeneous phase.
When the metastable minimum satisfies
\begin{equation}
	V_{\rm eff}(\phi_{\rm old})>V_{\rm eff}(\phi_{\rm new}),
\end{equation}
there is a free-energy difference
\begin{equation}
	\Delta V_{\rm eff}=V_{\rm old}-V_{\rm new},
\end{equation}
which drives bubble nucleation and expansion.
The associated pressure difference, \(-\Delta V_{\rm eff}\), acts between the bubble wall and the surrounding plasma.
This phase-conversion dynamics is not encoded in a single-phase Boltzmann equation for the distribution functions alone.

A convenient parametrization is instead to track the fraction of the volume that has converted to the new phase,
\begin{equation}
	\rho=(1-f)\rho_{\rm old}+f\rho_{\rm new}+\rho_{\rm wall},
\end{equation}
where \(f\) denotes the fraction of the Universe in the new vacuum and is related to the false-vacuum survival probability through \(f=1-P_f\) ($P_f$ is the fraction in the false vacuum defined in Eq.~\eqref{eq:false_vacuum_fraction}), namely
\begin{equation}\label{eq:false_vacuum_f}
	f(t)
	=
	1-\exp\!\left[
	-\frac{4\pi}{3}
	\int_{t_c}^{t}dt'\,
	\Gamma(t')\,a^3(t')\,R^3(t,t')
	\right].
\end{equation}
In the following we neglect \(\rho_{\rm wall}\), which is expected to be subdominant~\cite{Bringmann:2023iuz,Bringmann:2023opz,Li:2025nja}.

The corresponding evolution equation for the total energy density is then
\begin{equation}\label{eq:rho_evol_phases}
	\begin{split}
		&(1-f)\left[\dot\rho_{\rm old}+3H(\rho_{\rm old}+P_{\rm old})\right]
		\\&+f\left[\dot\rho_{\rm new}+3H(\rho_{\rm new}+P_{\rm new})\right]
		\\
		&
		+\dot f\,(\rho_{\rm new}-\rho_{\rm old})
		=
		j\,,
	\end{split}
\end{equation}
with $j$ a source term.
The effect of the potential difference is encoded in the difference between the phase energy densities, since
\(\rho_{\rm new}-\rho_{\rm old}\) contains the vacuum contribution associated with \(-\Delta V_{\rm eff}\).
In this way, Eq.~\eqref{eq:rho_evol_phases} accounts for reheating from the release of vacuum energy during phase conversion.

For this reason, and in contrast to the treatment in~\Cref{ch:3,ch:5}, it is more convenient here to work directly with the energy--momentum tensor rather than with the unintegrated Boltzmann equation.
The point is not that the two descriptions are inconsistent, but that once the thermodynamic quantities are encoded in \(V_{\rm eff}\), the continuity equation provides the more economical framework, while phase conversion can be incorporated explicitly through the \(f\)-dependent terms.\footnote{This assumes that the background field tracks the minimum of \(V_{\rm eff}\) adiabatically, so that coherent oscillations and the associated kinetic contribution \(\dot\phi_b^2/2\) can be neglected.}

We describe the dark sector through the energy--momentum tensor \(T^{\mu\nu}\), supplemented by an external source \(J^\nu\),
\begin{equation}
	\nabla_\mu T^{\mu\nu}=J^\nu.
\end{equation}
For a homogeneous background with no net momentum injection, \(J^\nu=(j,0,0,0)\), and in LTE this reduces to
\begin{equation}\label{eq:continuity_energy_source}
	\dot\rho+3H(\rho+P)=j,
\end{equation}
where \(\rho\equiv T^{00}\) and \(P\equiv T^{ii}/(3a^2)\).

In the absence of phase conversion, the source term \(j\) can be identified with the energy moment of the collision operator, i.e.\ with the right-hand side of the energy-density Boltzmann equation.
In the present setup this corresponds to the freeze-in energy injection, \(j=C_E^{\rm FI}\).
By contrast, the energy transfer associated with the transition itself is not part of \(j\); it is already accounted for by the phase-conversion term \(\dot f(\rho_{\rm new}-\rho_{\rm old})\) in Eq.~\eqref{eq:rho_evol_phases}. 

If one starts from the Boltzmann equation and takes its energy moment, one obtains the continuity equation with a source given by the energy-weighted collision term. This identifies \(j\) with the freeze-in energy injection \(C_E^{\rm FI}\). The vacuum and Daisy-resummed contributions are not additional collision terms; rather, they are already incorporated in the thermodynamic quantities \(\rho\) and \(P\) derived from the effective potential. In~\Cref{sec:moments_fBE} we discuss that one can identify the Daisy-resummed terms in the fBE moments when the mass varies in time.

In scale-factor derivatives ($d/dt = aH\,d/da$) the continuity equation results in
\begin{equation}\label{eq:continuity_energy_source_a}
	\rho' = \frac{j}{aH} - \frac{3}{a}(\rho+P).
\end{equation}

For the remainder of this subsection and to keep the notation clean, \(T\) denotes the dark-sector temperature, assuming the dark sector as internally thermalized (due to the large gauge coupling) and characterized by a single temperature.

With the dark sector internally thermalized and adiabatic tracking of the minum,
\begin{equation}\label{eq:min_condition}
	\left.\partial_\phi V_{\rm eff}(\phi;T,z)\right|_{\phi=\phi_{\min}(T,z)} = 0\,,
\end{equation}
where $T$ encodes the temperatures of all species within the dark sector.

%
%
%
%

Using $(T,z)$ as independent variables, the total energy density can be written as
\begin{equation}\label{eq:rho_from_V}
	\rho(T,z) = V(T,z) - T\left(\frac{\partial V}{\partial T}\right)_{z}.
\end{equation}
(Equivalently, $\rho = -P + Ts + \mu n$ with $\mu=T\ln z$). Note that we introduce the shorthand $V = V_{\rm eff}(\phi_{\rm min};T, z)$. 



\paragraph{(i) First derivative of $\rho$ with respect to $a$}
We start from $\rho(\phi,T,z)=V(\phi;T,z)-T\,\partial_T V(\phi;T,z)$.
Taking an $a$-derivative gives
\begin{equation}\label{eq:rho_prime_general}
	\rho' = \rho_\phi\,\phi' + \rho_T\,T' + \rho_z\,z',
\end{equation}
with
\begin{align}
	\rho_\phi &= \partial_\phi V_{\rm eff} - T\,\partial_{\phi T}V_{\rm eff}
	= V_\phi - T V_{\phi T}, \\
	\rho_T &= \partial_T V_{\rm eff} - \partial_T\!\left(T\,\partial_T V_{\rm eff}\right)
	= V_T - (V_T + T V_{TT}) = -T V_{TT},\\
	\rho_z &= \partial_z V_{\rm eff} - T\,\partial_{Tz}V_{\rm eff}
	= V_z - T V_{Tz},
\end{align}
where $V_\phi\equiv \partial_\phi V_{\rm eff}$, $V_{TT}\equiv \partial_T^2 V_{\rm eff}$,
etc., all evaluated at the instantaneous field value. We, now evaluate on the minimum (Eq.~\eqref{eq:min_condition}), where $V_\phi=0$,
\begin{equation}\label{eq:rho_prime_on_minimum_phi_prime}
	\rho' = -T V_{\phi T}\,\phi' \;-\; T V_{TT}\,T' \;+\; (V_z - T V_{Tz})\,z'.
\end{equation}

\paragraph{(ii) Expressing $\phi'$ in terms of $(T',z')$}
We now take the derivative of the minimum condition $\partial_\phi V_\text{eff}(\phi_{\min};T,z)=0$ with respect to $a$,
\begin{equation}\label{eq:min_condition_prime}
	0=(V_\phi)' = V_{\phi\phi}\,\phi' + V_{\phi T}\,T' + V_{\phi z}\,z'\,
\end{equation}
Hence
\begin{equation}\label{eq:phi_prime_solution}
	\phi' = -\frac{V_{\phi T}\,T' + V_{\phi z}\,z'}{V_{\phi\phi}}.
\end{equation}

\paragraph{(iii) Final expression for $\rho'$ purely in terms of $(T',z')$}
We insert Eq.~\eqref{eq:phi_prime_solution} into Eq.~\eqref{eq:rho_prime_on_minimum_phi_prime}.
Using symmetry of mixed partials $V_{\phi T}=V_{T\phi}$, one obtains
\begin{equation}\label{eq:rho_prime_final_Tz}
	\rho'
	=
	\left[
	T\frac{(V_{\phi T})^2}{V_{\phi\phi}}
	- T V_{TT}
	\right]T'
	\;+\;
	\left[
	T\frac{V_{\phi T}V_{\phi z}}{V_{\phi\phi}}
	+ V_z - T V_{Tz}
	\right]z'.
\end{equation}

\paragraph{(iv) Using energy continuity to solve for $T'$ and $z'$}
We combine Eq.~\eqref{eq:rho_prime_final_Tz} with Eq.~\eqref{eq:continuity_energy_source_a},
\begin{equation}\label{eq:energy_eq_Tz}
	\left(
	T\frac{(V_{\phi T})^2}{V_{\phi\phi}}
	- T\,V_{TT}
	\right)T'
	\;+\;
	\left(
	T\frac{V_{\phi T}V_{\phi z}}{V_{\phi\phi}}
	+ V_z - T\,V_{Tz}
	\right)z'
	=
	\frac{j}{aH} - \frac{3}{a}\left(-T\,V_T \right),
\end{equation}
where we used $\rho + P = -T\,V_T$. Then
\begin{equation}\label{eq:T_ds_evol_from_Veff}
	T' = \frac{j/(aH) + 3T\,V_T/a - \left(
		T\frac{V_{\phi T}V_{\phi z}}{V_{\phi\phi}}
		+ V_z - T\,V_{Tz}
		\right)z'}{T\frac{(V_{\phi T})^2}{V_{\phi\phi}}
		- T\,V_{TT}}
\end{equation}

To close the system for $(T,z)$ we consider the number density of the system 
with its corresponding nBE and source term $C_0$ in Eq.~\eqref{eq:n_and_rho}.
Notice that $n(T,z)$ can be obtained from $P(T,z)$ with the relation $n=\frac{z}{T}P_z$,
providing a second linear relation between $T'$ and $z'$.
To connect the effective potential with the number density evolution, first notice that 
\begin{equation}\label{eq:n_from_P_z}
	n(T,z) = \frac{z}{T}\left(\frac{\partial P}{\partial z}\right)_T
		\equiv \frac{z}{T}\,P_z = -\frac{z}{T}\left(\frac{\partial V}{\partial z}\right)_T.
\end{equation}
Since $n=n(T,z)$,
\begin{equation}\label{eq:n_prime_chainrule}
	n' = n_T\,T' + n_z\,z',
\end{equation}
with
\begin{align}
	n_T \equiv \left(\frac{\partial n}{\partial T}\right)_z
	&=
	\left(\frac{\partial}{\partial T}\right)_z\left(\frac{z}{T}P_z\right)
	=
	-\frac{z}{T^2}P_z + \frac{z}{T}P_{zT},
	\label{eq:nT_from_P}
	\\
	n_z \equiv \left(\frac{\partial n}{\partial z}\right)_T
	&=
	\left(\frac{\partial}{\partial z}\right)_T\left(\frac{z}{T}P_z\right)
	=
	\frac{1}{T}P_z + \frac{z}{T}P_{zz}.
	\label{eq:nz_from_P}
\end{align}
Eq.~\eqref{eq:n_prime_chainrule} gives the linear relation between $T'$ and $z'$,
\begin{equation}\label{eq:number_eq_Tz}
		\left(\frac{z}{T^2}V_z - \frac{z}{T}V_{zT}\right)T'
		+
		\left(-\frac{1}{T}V_z - \frac{z}{T}V_{zz}\right)z'
		=
		\frac{C_0}{aH} - \frac{3}{a}n,
\end{equation}
or
\begin{equation}\label{eq:z_from_Veff}
	z' = -T\frac{C_0/(aH) + 3z\,V_z/(aT)-\left(\frac{z}{T^2}V_z - \frac{z}{T}V_{zT}\right)T'}{V_z + z\,V_{zz}}\,.
\end{equation}
%

In the study~\cite{Li:2025nja} it is stressed that a correct phenomenological prediction from a FOPT requires a description of the thermal history encompassing the reheating of the plasma.

In principle the system of ODEs, Eqs.~\eqref{eq:T_ds_evol_from_Veff} and~\eqref{eq:z_from_Veff}, are equivalent to the fBE description in~\Cref{sec:moments_fBE}, Eqs.~\eqref{eq:T_evolution*} and~\eqref{eq:z_evolution*}, as long as the system is away from a FOPT. Near the transition, the equation for the temperature contains corrections from the potential barrier $\Delta V_{\rm eff}$, as shown in Eq.~\eqref{eq:rho_evol_phases}, where the reheating of the plasma is manifest once the transition completes. We provide a derivation of such equations with a thermal mass in~\Cref{sec:moments_fBE}.

The formalism introduced above allows for a genuinely dynamical fugacity. As shown in the next section, however, in the present U(1) model the \(A'\)-assisted number-changing reactions are so efficient that the dark sector is rapidly driven to \(z=1\).

\section{Results}\label{sec:ch5_results}

As a benchmark point to study an inverse FOPT, we fix
\begin{equation}\label{eq:BM_FOPT}
	\lambda = 10^{-3}\,,\qquad v_\phi = 100\,\text{MeV}\,,\qquad g = 0.3\qquad\text{and}\qquad  \epsilon = 3\times 10^{-11}\,,
\end{equation}
which yields \(m_\phi = 1.82\,\text{MeV}\) and \(m_{A'} = 30\,\text{MeV}\). The parameters $\lambda$, \(v_\phi\) and $g$ are chosen such that the dark-sector spectrum lies at the MeV scale and the resonance condition \(m_{A'}\simeq m_\gamma(T)\) is reached at relatively low SM temperatures. The kinetic mixing \(\epsilon\) is then selected to illustrate the regime in which the resonant energy injection is strong enough to induce an inverse transition, but not so large that the system remains in the symmetric phase throughout the resonance. In this sense, the benchmark is not meant to be generic, but rather to provide a representative example of a multi-transition thermal history.

\begin{figure}[t!]
	\centering
	\includegraphics[width=0.8\textwidth]{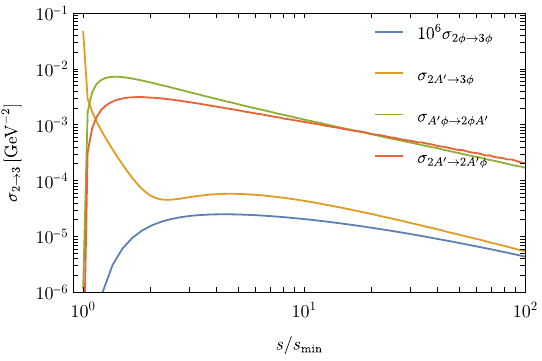}
	\caption{Cross sections for self-number-changing \(2\to 3\) reactions for the benchmark point in Eq.~\eqref{eq:BM_FOPT}. The blue curve corresponds to \(10^6 \sigma_{2\phi\to 3\phi}\), illustrating the strong suppression of this channel relative to the others. The horizontal axis is normalized to \(s_{\rm min}=(\sum_i m_i)^2\), where the \(m_i\) are the masses of the initial-state particles.}
	\label{fig:sigma_2_to_3}
\end{figure}

\subsection{Chemical equilibration}\label{subsec:chem_Eq}

The relatively large gauge coupling enhances self-number-changing reactions involving \(A'\), as discussed in~\Cref{subsec:equilibration}.
We first discuss chemical equilibration for the benchmark point in Eq.~\eqref{eq:BM_FOPT}, and then argue that this behavior is generic whenever the hierarchy \(g\gg\lambda\), required for a FOPT, is realized.
To this end, we evaluate the \(2\to 3\) cross sections for the processes in Eq.~\eqref{eq:cannibal_w_A} using \texttt{CalcHEP}~\cite{Belyaev:2012qa}.

As shown in~\Cref{fig:sigma_2_to_3}, the cross sections involving at least one \(A'\) are parametrically enhanced.
This follows from the squared matrix elements associated with the diagrams in~\Cref{fig:feynman_cannibal_w_A}, which scale as
\(\vert\mathcal{M}\vert^2\sim\lambda^3 m_{A'}^4/m_{\phi}^6 = g^4/(8v_\phi^2)\).
Suppressing these interactions would require either lowering \(g\), which removes the barrier responsible for the FOPT, or increasing \(v_\phi\) substantially.
By comparison, for \(3\phi\leftrightarrow 2\phi\), the matrix element in Eq.~\eqref{eq:mat_el_real} scales as \((\sqrt{3\lambda^3}/m_\phi)^2 = \lambda^2/(2v_\phi^2)\).
Hence, as long as \(g\gg \lambda\), the squared matrix elements involving \(A'\) are enhanced by a factor \(g^4/(4\lambda^2)\) relative to \(3\phi \leftrightarrow 2\phi\), explaining the strong suppression of that channel in~\Cref{fig:sigma_2_to_3}.

These reactions are only present in the broken phase, where the trilinear interaction \(g^2v_\phi \phi A'_\mu A'^\mu\) is non-vanishing.
When the symmetry is restored, this term disappears, and the leading processes capable of enforcing vanishing chemical potentials are instead of the \(4\leftrightarrow 2\) type.
As in the broken phase, one expects the dominant channels to involve at least one \(A'\), for instance \(A'\Phi\Phi^\ast\Phi \to A'\Phi\).
However, once the symmetry is restored, the particles become effectively massless and only thermal masses regulate the amplitudes.
A proper treatment then requires controlling the collinear and logarithmic enhancements of the \(t\)- and \(u\)-channel contributions through thermal screening, which lies beyond the capabilities of \texttt{CalcHEP}.

We estimate the efficiency of the \(4\leftrightarrow 2\) processes by using the broken-phase \(3\leftrightarrow 2\) reactions as a proxy in the symmetric phase, while multiplying their rate by an overall suppression factor of \(10^{-5}\). This factor is not meant to represent a precise matching to the underlying symmetric-phase matrix elements; rather, it is introduced as a deliberately conservative stress test, intended to mimic a substantial reduction of the true number-changing rate once the cubic interaction vanishes and thermal screening effects become important. Even under this strongly suppressed prescription, we find essentially no departure from chemical equilibrium. This indicates that our conclusion does not depend sensitively on the precise value of the symmetric-phase rate. The same hierarchy \(g\gg \lambda\) that is needed to generate a barrier in a \(\mathrm{U}(1)\) model therefore also tends to guarantee chemical equilibrium in the symmetric phase.

We stress that the factor \(10^{-5}\) is not a quantitative estimate of the true symmetric-phase \(4\leftrightarrow2\) rate. It is introduced as a deliberately strong suppression of the proxy interaction in order to test the robustness of the conclusion that chemical equilibrium is maintained. 


It is natural to ask whether this link between first-order dynamics and fast number-changing reactions extends beyond the present U(1) setup. A similar pattern may arise in models with discrete symmetries.
For instance, in \(\mathbb{Z}_3\)-symmetric scalar theories, a tree-level cubic interaction of the form \(\kappa(\Phi^3+{\Phi^\dagger}^3)\) can both help generate a vacuum barrier and mediate number-changing processes already in the symmetric phase.
This suggests that rapid equilibration may be common in at least some classes of first-order transitions, although establishing this requires an explicit calculation of the relevant cross sections on a model-by-model basis.
A systematic taxonomy of which phase-transition models genuinely require a dynamical treatment of the fugacity is outside of the scope of the present work.

\subsection{Evolution and reheating}\label{subsec:Reheating}

A fully consistent treatment of reheating during a first-order phase transition would track the phase-conversion fraction
\(f(t)=1-e^{-I(t)}\), such that the energy continuity equation contains the explicit conversion term
\(\dot f\,(\rho_{\rm new}-\rho_{\rm old})\) of Eq.~\eqref{eq:rho_evol_phases}.
Implementing this requires the nucleation rate
\(\Gamma(T)\propto e^{-S_3(T)/T}\), and repeated evaluations of the bounce action and bubble profile along the thermal history.

As a first approximation, we instead treat each transition as instantaneous.
The system is evolved along the metastable branch until a nucleation or percolation criterion is reached, after which the evolution is continued in the new phase with thermodynamic variables fixed by energy conservation.
In this approximation, the energy released into the plasma during a standard transition is taken to be \(\kappa_{\rm heat}\Delta\rho_{\rm ds}\), with \(\kappa_{\rm heat}\simeq 1\) at the transition point.
This neglects any fraction of the available energy stored in bubble walls, but it captures the leading effect of vacuum-energy release on the post-transition state.

Assuming that the available latent heat is fully converted into plasma energy, we model the phase transition as instantaneous compared with the Hubble time. The scale factor therefore changes negligibly during the conversion, \(a_{\rm old}\simeq a_{\rm new}\), and the post-transition state is obtained by matching the total dark-sector energy density across the transition hypersurface,
\begin{equation}
	\rho_{\rm old}=\rho_{\rm new}\,.
\end{equation}
Equivalently,
\begin{equation}\label{eq:rho_radiation}
	\rho_{\rm new}^{\rm rad}
	=
	\rho_{\rm old}^{\rm rad}
	+\rho_{\rm old}^{\rm vac}
	-\rho_{\rm new}^{\rm vac}\,,
\end{equation}
where \(\rho=-P+T\partial_T P = V-T\partial_T V\).
Here \(\rho^{\rm vac}=V_{T=0}\) denotes the vacuum contribution, while \(\rho^{\rm rad}\) contains the thermal part of the effective potential. 
The reheating temperature of the dark sector is obtained by solving Eq.~\eqref{eq:rho_radiation} at the transition point.

We note that for a relativistic plasma undergoing adiabatic cosmological evolution one has \(\rho \propto a^{-4}\), whereas non-relativistic matter scales approximately as \(a^{-3}\). In the present treatment, however, the reheating step is taken to occur at essentially fixed scale factor, so these distinctions do not affect the matching at leading order.

Before discussing the temperature evolution in detail, we note that solving the coupled ODE system for \((T_{\rm ds},z)\) shows that \(z\to 1\) already around the electroweak epoch, \(T\sim 100\,\mathrm{GeV}\).
The dark sector therefore remains in chemical equilibrium throughout the stages relevant for the present analysis, and it is sufficient to solve the temperature equation~\eqref{eq:T_ds_evol_from_Veff} with \(z=1\).
We also considered the case of low-temperature and instantaneous reheating of the SM. We found again rapid chemical equilibration, so that typically \(\Gamma_{\rm cannibal}>H>\Gamma_{\rm FI}\), largely independently of the initial condition.

In principle, an inverse first-order transition of the dark sector already occurs at high SM temperatures (\(T\sim 150\,\mathrm{GeV}\)), in which case the full \((T_{\rm ds},z)\) system should be evolved from the outset.
However, any gravitational-wave signal produced at such early times is strongly suppressed relative to the SM radiation background, which still dominates the total energy density near the electroweak epoch.
For this reason, we focus on the later stage of the evolution, where the dark-sector dynamics becomes more distinctive.

\begin{figure}[t!]
	\centering
	\includegraphics[width=0.8\textwidth]{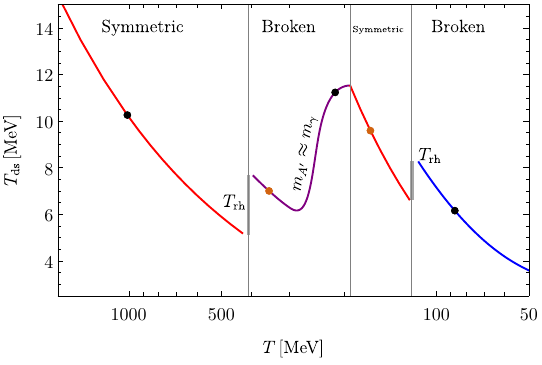}
	\caption{Evolution of \(T_{\rm ds}\) as a function of the SM temperature \(T\).
		At high temperature the dark sector is in the symmetric phase (red).
		Below \(T\simeq 500\,\mathrm{MeV}\), the system undergoes a transition to the broken phase; the associated release of vacuum energy reheats the dark plasma, producing the jump visible in purple.
		At later times, resonant \(\bar l l\to A'\) production increases the dark-sector energy density and drives the system back to the symmetric phase, inducing an inverse first-order transition (red).
		As the Universe continues to expand and cool, the dark sector eventually undergoes a second transition to the broken phase, again accompanied by reheating (blue).
		The scattered points indicate the benchmark temperatures for which the effective potentials shown in \Cref{fig:Veff_plots} are evaluated.}
	\label{fig:temp_tot_zoom}
\end{figure}

\begin{figure}[t!]
	\centering
	\includegraphics[width=0.8\textwidth]{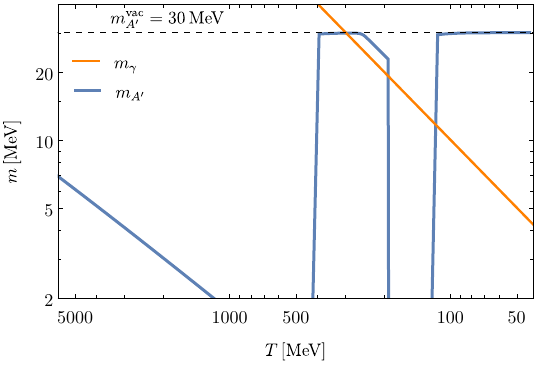}
	\caption{Longitudinal thermal mass of the dark photon (blue) and thermal mass of the SM photon (orange).}
	\label{fig:masses}
\end{figure}

The resulting evolution of \(T_{\rm ds}\) is shown in~\Cref{fig:temp_tot_zoom}.
Around the electroweak scale, bremsstrahlung production of dark photons, cf.~Eq.~\eqref{eq:bremstrahlung}, provides the dominant freeze-in channel.
At lower temperatures this contribution becomes negligible, and around \(T\sim 200\,\mathrm{MeV}\) coalescence becomes the dominant source. The vertical gray-thick lines in~\Cref{fig:temp_tot_zoom} indicate the reheating temperatures associated with the standard transitions into the broken phase.
These reheating jumps arise because the transition releases vacuum energy into the dark plasma.
By contrast, the inverse transition from broken to symmetric phase does not produce an analogous temperature jump.
In the present setup, this inverse transition is driven by resonant energy injection from the SM bath, which temporarily restores the symmetric minimum.
Unlike the standard transitions into the broken phase, the inverse transition is not accompanied by latent-heat release into the plasma. Instead, it is driven by continuous energy injection from the SM bath, which restores the symmetric minimum. For this reason, it appears as a smooth change of branch rather than as a reheating event.

The absence of a reheating jump during the inverse transition does not by itself imply a stronger signal than in the corresponding standard transition.
The dominant GW source still depends on the partition of energy between bubble walls and plasma motion, and therefore on the wall velocity and the friction exerted by the surrounding medium. This requires a careful computation of the wall velocity that can be performed, e.g., with the \texttt{WallGo} routine~\cite{vandeVis:2025plm}, and that goes beyond the scope of this work.


Note that at high $T_{\rm ds}$ the longitudinal dark-photon mass is generated thermally through interactions with the dark plasma.
Using Eq.~\eqref{eq:Pi}, one finds in the high-temperature limit
\((m_{A'}^L)^2=g^2 T_{\rm ds}^2/3\).
After the transition to the broken phase, the dark photon instead acquires a mass from the dark Higgs vacuum expectation value, and \(m_{A'}^L\) remains approximately constant at \(m_{A'}^L\simeq 30\,\mathrm{MeV}\).
For a finite interval, roughly around \(T\sim 200\,\mathrm{MeV}\), this mass coincides with the thermal photon mass in the SM plasma,
\(m_\gamma = 2\sqrt{\pi\alpha}\,T/3\),
which leads to resonant dark-photon production, as shown in~\Cref{fig:masses}.

\begin{figure}[t!]
	\def\sepf{0.496}
	\centering
	\includegraphics[width=\sepf\columnwidth]{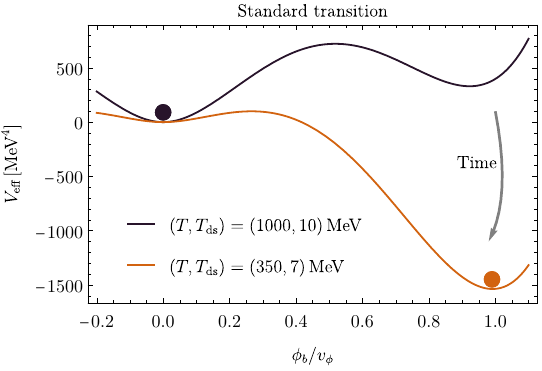}
	\includegraphics[width=\sepf\columnwidth]{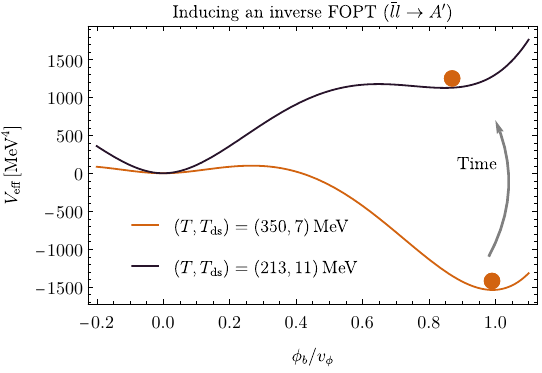}
	\includegraphics[width=\sepf\columnwidth]{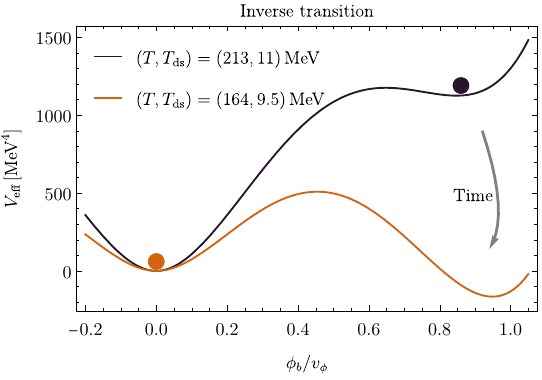}
	\includegraphics[width=\sepf\columnwidth]{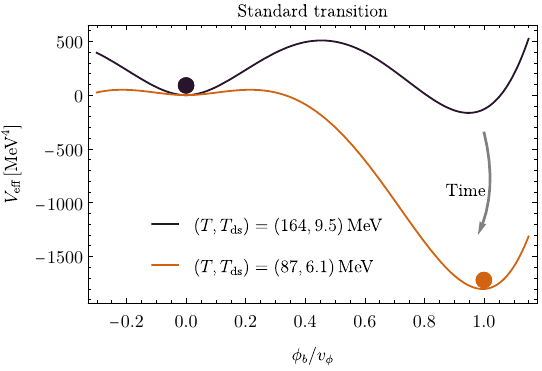}
	\caption{Effective potential evaluated at the benchmark points highlighted in~\Cref{fig:temp_tot_zoom}. The ordering in time is from left to right and top to bottom. Arrows indicate the direction of the evolution, and the filled circles denote the vacuum occupied by the system. The sequence shows that the phase history is not monotonic: after entering the broken phase, resonant energy injection temporarily restores the symmetric minimum and induces an inverse first-order transition, before the system finally returns to the broken phase at later times.}
	\label{fig:Veff_plots}
\end{figure}

\begin{figure}[t!]
	\centering
	\includegraphics[width=0.8\textwidth]{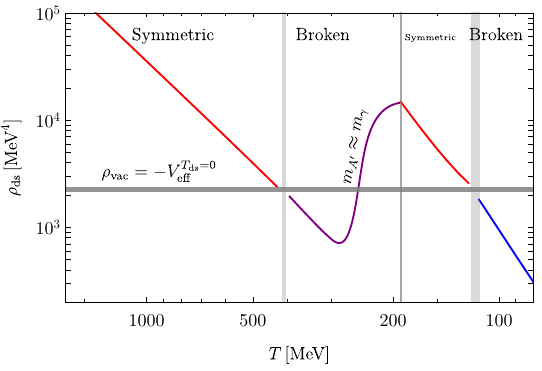}
	\caption{Energy density of the dark sector as a function of the SM temperature. The color coding of the different phases follows~\Cref{fig:temp_tot_zoom}. We also show the vacuum contribution and indicate the transition thresholds with gray vertical lines. The standard transitions into the broken phase are accompanied by reheating jumps, reflecting the transfer of vacuum energy into the plasma, whereas the inverse transition does not show an analogous discontinuity. The vacuum contribution is negative because the effective potential is normalized such that \(V_{\rm eff}(\phi_b=0,T_{\rm ds}\gg m_\phi)=0\) in the symmetric phase, so that \(V_{\rm eff}(\phi_b=v_\phi)<0\) in the broken phase.}
	\label{fig:rho_plot}
\end{figure}

\begin{figure}[t!]
	\centering
	\includegraphics[width=0.8\textwidth]{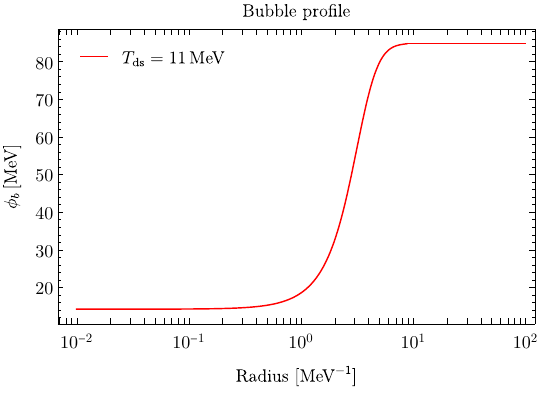}
	\caption{Bounce solution for the inverse first-order transition. The field approaches the stable minimum near the center of the bubble and asymptotes to the metastable minimum at large radius.}
	\label{fig:bubble_profile}
\end{figure}

The effective potential for four cases is depicted in~\Cref{fig:Veff_plots}. Taken together, these benchmark potentials illustrate that the phase history is not monotonic: the system first leaves the symmetric phase, is then driven back into it by resonant energy injection, and finally re-enters the broken phase once the plasma cools further. The panels show which minimum is occupied at each stage, when the broken minimum appears or disappears, and how the temporary restoration of the symmetric minimum gives rise to the inverse transition.
The filled circles identify the vacuum occupied by the system at each benchmark point.

The corresponding evolution of the energy density is shown in~\Cref{fig:rho_plot}.
This figure makes the energy bookkeeping explicit.
The standard transitions into the broken phase are accompanied by a transfer of vacuum energy into the plasma, visible as reheating jumps, while the inverse transition proceeds without an analogous discontinuity.
We also show the vacuum contribution \(\rho_{\rm vac}\), together with the transition thresholds marked by the gray vertical lines.

We also display the bounce profile for the inverse transition in~\Cref{fig:bubble_profile}.
It exhibits the standard critical-bubble behavior: the field is close to the true vacuum in the interior and approaches the false vacuum at large radius, with the transition region identifying the bubble wall.
Its main purpose here is to make explicit the field configuration corresponding to the bounce action \(S_3\).


Qualitatively, the gravitational-wave signal from the inverse transition is expected to resemble the intermediate (phase-2) peak in Ref.~\cite{Dent:2024bhi}, as both arise from a symmetry-restoring first-order transition triggered after the sector is reheated from an initially broken phase. In that analysis, the inverse transition occurs at a lower Standard Model temperature than the final standard transition and therefore peaks at a lower observed frequency, with the phase-2 peak lying about one order of magnitude below the phase-3 peak in the maximally separated case. The later transitions can be stronger because they take place at smaller Hubble values and are therefore less redshift suppressed.
In the present U(1) model this suggests that the inverse transition should populate lower frequencies than the final transition, potentially moving the signal toward the PTA band for sufficiently low transition temperatures.
A quantitative prediction, however, requires a dedicated computation of the nucleation temperature, the transition rate \(\beta/H\), and the efficiency factors governing the partition of energy between bubble walls and the plasma, whose full evaluation we leave for future work.

\section{Summary}\label{sec:ch5_summary}

In this chapter we developed a framework to study the thermal evolution of a dark sector populated via freeze-in, with the aim of describing when such a sector can undergo an inverse first-order phase transition. The main question was how the effective potential evolves while the hidden sector is continuously populated by the Standard Model bath, and whether departures from chemical equilibrium have to be followed explicitly through a time-dependent fugacity.

We applied this framework to a U(1) model with a complex scalar and a dark photon. In this example, the main result is that self-number-changing reactions involving the dark photon are very efficient. As a consequence, the hidden sector is driven rapidly toward chemical equilibrium, so that \(z\to 1\) well before the phase transitions of interest. This means that the formalism developed here does allow for a dynamical fugacity, but in the benchmark studied in this chapter the relevant transitions take place after chemical equilibrium has already been established.

Within this regime, the dark sector exhibits a non-monotonic phase history. The system starts in the broken phase. As freeze-in proceeds, resonant dark-photon production injects energy into the hidden sector and modifies the effective potential until the symmetric minimum becomes favored, triggering an inverse first-order phase transition. At later times, the dark sector cools again because of the expansion of the Universe, and the usual transition back to the broken phase takes place. In this way, a sector populated out of equilibrium can undergo a broken \(\to\) symmetric \(\to\) broken sequence even though the visible sector keeps cooling throughout.

An important qualitative feature is that the inverse transition does not show a reheating jump in the dark-sector temperature. In the setup studied here, the transition is driven by the gradual build-up of energy in the hidden sector rather than by the release of latent heat, and therefore it appears in the thermal history as a continuous change from one branch to another.

A broader lesson suggested by this analysis is that, in gauge theories of this kind, the same interactions that help generate a first-order barrier can also make number-changing reactions very efficient. This tends to reduce the importance of a dynamical fugacity at the time of the transition. Whether the same conclusion holds more generally in models with a different structure of interactions, for example theories with discrete symmetries and tree-level cubic terms, remains to be understood. Since the analysis presented here is based on a benchmark rather than on a full parameter scan, it will also be important to determine how generic this multi-step thermal history is. Finally, while the bubble profile was presented explicitly, a quantitative prediction for the gravitational-wave signal still requires a dedicated computation of the bounce action, wall dynamics, and efficiency factors.

%% file: Chapters/Chapter6.tex
\chapter{Conclusions and outlook}\label{ch:conclusions}

In this thesis, we studied the production and evolution of hidden sectors populated through the freeze-in mechanism. In the first part of the thesis, the focus was on scenarios in which the dark sector contains a dark-matter candidate undergoing cannibal dynamics. In the final part, the emphasis shifted to more general hidden sectors populated via freeze-in, whose relevance lies not necessarily in providing dark matter, but in exhibiting a non-trivial thermal history, phase structure, or phase-space dynamics. The central theme throughout is that freeze-in should not be viewed as a purely passive production mechanism. Even when the coupling to the Standard Model is feeble, interactions within the hidden sector can qualitatively modify the cosmological evolution, the relic abundance, the momentum distribution of the produced particles, and the resulting phenomenology.

A first general conclusion is that self-interactions within the hidden sector can play a decisive role in determining whether freeze-in remains viable. In the models studied here, number-changing reactions such as \(2\leftrightarrow3\) modify the relation between the portal coupling and the final relic abundance, and in some cases they open viable regions of parameter space that would otherwise be excluded. At the same time, the precise impact of these reactions depends on the symmetry structure of the dark sector. In particular, broken \(\mathbb{Z}_2\) models are strongly constrained by the loss of stability induced by symmetry breaking, while \(\mathbb{Z}_3\)-symmetric models provide a more robust realization of cannibal freeze-in.

The analysis of models with unstable mediators further illustrates that dark-sector phenomenology is highly sensitive to the particle content and mass hierarchy. Depending on whether the mediator is lighter or heavier than the dark matter, late-time cosmological constraints, indirect-detection signatures, and long-lived particle signals can all be drastically altered. More broadly, this demonstrates that interactions confined entirely within the hidden sector may leave observable imprints even when the portal to the Standard Model is extremely weak. A particularly interesting example of this arises in the warm-dark-matter study discussed in Chapter~\ref{ch:3}, where the dark matter itself does not undergo cannibalization, but is instead produced from the decay of a cannibalising mediator. In this case, the relevant imprint is not primarily the relic density, but the phase-space distribution inherited by the dark matter from the mediator sector. We found that the presence or absence of self-number-changing reactions in the mediator bath substantially modifies both the normalization and the shape of the dark-matter distribution. In particular, efficient mediator self-interactions drive the bath toward chemical equilibrium and redistribute the available energy among a larger number of particles, resulting in a colder and more sharply peaked dark-matter spectrum. In the absence of such reactions, the mediator sector remains more dilute and hotter, so that the produced dark matter inherits a broader and harder momentum distribution. This shows that cannibal dynamics can influence structure-formation constraints even when the dark matter particle itself does not participate directly in the cannibal phase.

A second main conclusion is that the cosmological background during dark-matter production is of central importance. In the standard radiation-dominated picture, the portal couplings required for freeze-in are typically so small that direct tests are extremely challenging. By contrast, when reheating is non-instantaneous and the reheating temperature is low, the relation between production efficiency and relic abundance changes substantially. In this regime, larger portal couplings become possible, and the resulting parameter space can be significantly more accessible to collider, direct-detection, and long-lived particle searches. This shows that the phenomenology of freeze-in dark matter cannot be assessed reliably without specifying the thermal history of the visible sector.

The final part of the thesis extended these ideas to the dynamical evolution of an effective potential in a dark U(1) sector populated via freeze-in. Unlike the previous chapters, this setup was not introduced as a dark-matter model, but as an example of a hidden sector with a non-trivial phase structure. This led to a particularly interesting result: although the framework was formulated to allow departures from chemical equilibrium through a non-trivial fugacity \(z\), the dark-photon-assisted number-changing reactions turn out to be so efficient that the dark sector is rapidly driven to \(z=1\). In this sense, chemical equilibration is not an assumption in the model, but rather an outcome of the dynamics. This observation suggests a broader connection between the interactions that help generate a first-order barrier and those that enforce fast equilibration.

Within this equilibrated regime, the hidden sector can nevertheless exhibit a non-monotonic phase history. In the benchmark model studied here, energy injection from the Standard Model bath temporarily restores the symmetric phase and induces an inverse first-order phase transition before the system later returns to the broken phase. An important qualitative feature is that this inverse transition does not exhibit the reheating jump characteristic of standard transitions into the broken phase. Instead, it is driven by continuous energy injection rather than by latent-heat release. This provides a concrete example of how freeze-in dynamics and phase-transition dynamics can become deeply intertwined.

This interplay is particularly relevant in models with a first-order phase transition, since the sizable gauge coupling required to realize such a transition can simultaneously drive rapid number-changing reactions and chemical equilibration.

Taken together, the results of this thesis show that hidden sectors populated via freeze-in can exhibit a much richer cosmological behavior than is often assumed. Their dynamics depend not only on the portal interaction with the Standard Model, but also on the internal interaction structure of the hidden sector, the phase-space evolution of the produced states, and the background cosmological history.
The lesson is that a realistic treatment must follow the evolution of the hidden sector beyond the minimal assumption of a non-interacting bath.

Several directions for future work remain open. In the warm-dark-matter scenario studied in Chapter~\ref{ch:3}, the next natural step would be to go beyond the effective-temperature estimate and perform a full structure-formation analysis based on the computed phase-space distribution. This would require propagating the non-thermal distribution through the corresponding linear matter power spectrum and, ultimately, confronting it with Lyman-\(\alpha\) data through dedicated simulations. 

In the case of the dark U(1) model, an important extension would be to determine whether the inverse first-order phase transition identified in this work can lead to an observable gravitational-wave signal. This would require computing the bounce action and the nucleation history in the non-equilibrium background considered here. More broadly, it would be worthwhile to investigate whether the rapid chemical equilibration found in this model is a generic feature of scalar fields featuring FOPTs, or whether it is specific to the gauge structure and interaction pattern of the dark U(1) setup. In this context, other realizations, such as \(\mathbb{Z}_3\)-symmetric models, would provide a useful comparison.

%% file: Appendices/AppendixA.tex

\chapter{Collision operators for FI and cannibal reactions} 
\label{AppendixA} 

\section{Matrix elements}\label{ap:Matrix_elem_self}

We collect here the matrix elements relevant for the number-changing reactions discussed in the main text.

We begin with the real scalar model of Section~\ref{sec:real_scalar}.
For the process $\varphi_1\varphi_2\leftrightarrow\varphi_3\varphi_4\varphi_5$, the tree-level amplitude can be written as
\begin{equation}
	\begin{split}
		\label{eq:mat_el_real}
		i\mathcal{M}_{3\varphi\leftrightarrow 2\varphi}
		=-ig^3\Bigg(&
		\frac{1}{SS_{34}}+\frac{1}{SS_{35}}+\frac{1}{SS_{45}}
		+\frac{1}{T_{15}S_{34}}+\frac{1}{T_{14}S_{35}}+\frac{1}{T_{13}S_{45}}
		\\&
		+\frac{1}{T_{25}S_{34}}+\frac{1}{T_{24}S_{35}}+\frac{1}{T_{23}S_{45}}
		+\frac{1}{T_{14}T_{23}}+\frac{1}{T_{15}T_{23}}+\frac{1}{T_{13}T_{24}}
		\\&\hspace{12em}
		+\frac{1}{T_{14}T_{25}}+\frac{1}{T_{13}T_{25}}+\frac{1}{T_{15}T_{24}}
		\Bigg)
		\\&
		-ig\lambda\Bigg(
		\frac{1}{S}+\frac{1}{S_{34}}+\frac{1}{S_{35}}+\frac{1}{S_{45}}
		+\frac{1}{T_{13}}+\frac{1}{T_{14}}+\frac{1}{T_{15}}
		+\frac{1}{T_{23}}+\frac{1}{T_{24}}+\frac{1}{T_{25}}
		\Bigg)\,,
	\end{split}
\end{equation}
where we defined $S_{ij}=s_{ij}-m_\varphi^2$, $T_{ij}=t_{ij}-m_\varphi^2$, and $S\equiv S_{12}$. In the broken phase the cubic coupling satisfies $g=\sqrt{3\lambda}\,m_\varphi$. It is therefore convenient to factor out $\sqrt{3\lambda^3}/m_\varphi$ and work with dimensionless Mandelstam variables $\tilde s_{ij}=s_{ij}/m_\varphi^2$ and $\tilde t_{ij}=t_{ij}/m_\varphi^2$.

For the $\mathbb{Z}_3$-symmetric complex scalar in Section~\ref{sec:Z3}, two independent amplitudes contribute. For
$S^\ast S\leftrightarrow SSS$ (denoted $\mathcal{M}^{(1)}$),
\begin{equation}
	\begin{split}
		i\mathcal{M}^{(1)}
		=&-ig_s^2\!\left(
		\frac{1}{S_{45}T_{13}}+\frac{1}{S_{35}T_{14}}+\frac{1}{S_{34}T_{15}}
		\right)
		\\&
		-ig_s\lambda_s\!\left(
		\frac{1}{S_{34}}+\frac{1}{S_{35}}+\frac{1}{S_{45}}
		+\frac{1}{T_{13}}+\frac{1}{T_{14}}+\frac{1}{T_{15}}
		\right)\,,
	\end{split}
\end{equation}
while for $SS\leftrightarrow S^\ast S^\ast S$ (denoted $\mathcal{M}^{(2)}$),
\begin{equation}
	\begin{split}
		i\mathcal{M}^{(2)}
		=&-ig_s^2\!\left(
		\frac{1}{SS_{34}}+\frac{1}{T_{14}T_{23}}+\frac{1}{T_{13}T_{24}}
		\right)
		\\&
		-ig_s\lambda_s\!\left(
		\frac{1}{S}+\frac{1}{S_{34}}
		+\frac{1}{T_{13}}+\frac{1}{T_{14}}
		+\frac{1}{T_{23}}+\frac{1}{T_{24}}
		\right)\,,
	\end{split}
\end{equation}
with $S_{ij}=s_{ij}-m_s^2$ and analogous definitions for $T_{ij}$.

\section{$3\varphi\leftrightarrow 2\varphi$ collision integrals}\label{ap:C0self}

The zeroth-moment collision contribution associated with the $3\varphi\leftrightarrow2\varphi$ reaction is obtained by integrating the full collision operator over phase space,
\begin{equation}
	\begin{split}
		\braket{C_{3\varphi\leftrightarrow 2\varphi}}
		&\equiv \frac{g_\varphi}{n_\varphi}\int\frac{d^3\vec p}{(2\pi)^3}\,C_{3\varphi\leftrightarrow 2\varphi}
		\\
		&=\frac{1}{n_\varphi}\frac{1}{2!}\frac{1}{3!}
		\int d\Pi_1\cdots d\Pi_5\,
		\vert\tilde{\mathcal{M}}_{3\varphi\leftrightarrow 2\varphi}\vert^2
		\left(f_1f_2-f_3f_4f_5\right)
		\\
		&=\frac{1}{12\,n_\varphi}\int d\Pi_1\cdots d\Pi_5\,
		\vert\tilde{\mathcal{M}}_{3\varphi\leftrightarrow 2\varphi}\vert^2\,
		e^{-(E_1+E_2)/T_\varphi}\left[
		\left(\frac{n_\varphi}{n_\varphi^{\rm eq}}\right)^2
		-\left(\frac{n_\varphi}{n_\varphi^{\rm eq}}\right)^3
		\right]
		\\
		&= s\,Y_\varphi\,\braket{\sigma_{3\varphi\to 2\varphi}v^2}\,s\,
		\left(Y_\varphi^{\rm eq}-Y_\varphi\right)\,,
	\end{split}
	\label{eq:C0_self}
\end{equation}
where we have defined $\braket{C} \equiv C_0/n_i$ and
where $g_\varphi$ denotes the number of internal degrees of freedom (here $g_\varphi=1$). In the second equality we used the symmetry factors for identical particles, and in the third equality we rewrote the product of distribution functions via
$f_3f_4f_5=(n_\varphi/n_\varphi^{\rm eq})\,f_1f_2$.

To evaluate the three-body phase-space integral, it is convenient to proceed in two steps: first boost to a ``laboratory'' frame defined as the center-of-mass frame of two final-state particles (e.g.\ $\varphi_3\varphi_4$), and then boost to the overall center-of-mass frame~\cite{Byckling:1971vca}. In this intermediate frame one may write
$p_1+p_2-p_5=(\sqrt{s_{34}},\vec0\,)^{\mathrm{T}}$, with
\begin{equation}
	s_{34}\equiv(p_3+p_4)^2=(p_1+p_2-p_5)^2=s+m_\varphi^2-2\sqrt{s}\,E_5\,.
\end{equation}
By construction,
$p_3^{\rm lab}=(\sqrt{s_{34}}/2,\vec p_3^{\,\rm lab})^{\mathrm{T}}$
and
$p_4^{\rm lab}=(\sqrt{s_{34}}/2,-\vec p_3^{\,\rm lab})^{\mathrm{T}}$.
The transformation from this lab frame to the total CM frame can be represented as a boost along the $z$ axis followed by a rotation about the $y$ axis,
\begin{equation}
	\mathcal{B}_z(\gamma)=
	\begin{pmatrix}
		\gamma & 0 & 0 & \gamma\beta \\
		0 & 1 & 0 & 0 \\
		0 & 0 & 1 & 0 \\
		\gamma\beta & 0 & 0 & \gamma
	\end{pmatrix},
	\qquad
	\mathcal{R}_y(\alpha)=
	\begin{pmatrix}
		1 & 0 & 0 & 0 \\
		0 & \cos\alpha & 0 & \sin\alpha \\
		0 & 0 & 1 & 0 \\
		0 & -\sin\alpha & 0 & \cos\alpha
	\end{pmatrix},
\end{equation}
so that
$p_i^{\rm cm}=\mathcal{R}_y(\alpha+\pi)\,\mathcal{B}_z\!\left(\frac{E_{12}}{\sqrt{s_{34}}}\right)p_i^{\rm lab}$,
with
$E_{12}=(s+s_{34}-m_\varphi^2)/(2\sqrt{s})$ and $\beta=\sqrt{1-1/\gamma^2}$.
After these changes of variables, the thermal average can be expressed as
\begin{equation}\label{eq:3to2_average}
	\begin{split}
		\braket{\sigma_{3\varphi\to 2\varphi}v^2}
		&=\frac{1}{2!3!}\frac{1}{(n_\varphi^{\rm eq})^3}
		\int d\Pi_1\cdots d\Pi_5\,
		\vert\tilde{\mathcal{M}}_{3\varphi\leftrightarrow2\varphi}\vert^2\,
		e^{-(E_1+E_2)/T_\varphi}
		\\
		&=\frac{1}{12}\frac{1}{(n_\varphi^{\rm eq})^3}\frac{3}{8(2\pi)^4}
		\int d\Pi_1\,d\Pi_2\,e^{-(E_1+E_2)/T_\varphi}
		\int d\tilde E_5\,d\phi_4\,dx_4\,dx_5\,
		J\,\vert\mathcal{M}(\tilde s_{ij},\tilde t_{ij})\vert^2\,,
	\end{split}
\end{equation}
where $\tilde E_5\equiv E_5/m_\varphi$ and the integration ranges are
$\tilde E_5\in\big[1,(s-3m_\varphi^2)/(2m_\varphi\sqrt{s})\big]$,
$x_{4,5}(=\cos\theta_{4,5})\in[-1,1]$, and $\phi_4\in[0,2\pi)$.
The Jacobian is
\begin{equation}
	J=\sqrt{\tilde E_5^{\,2}-1}\,
	\sqrt{1-\frac{4m_\varphi^2}{m_\varphi^2-2\tilde E_5 m_\varphi\sqrt{s}+s}}\,.
\end{equation}
We evaluate the remaining four-dimensional integral with a Monte-Carlo routine.
For the initial-state phase space it is useful to trade variables for
$E_+\equiv E_1+E_2$ and $s$, which yields~\cite{Edsjo:1997bg}
\begin{equation}
	d\Pi_1\,d\Pi_2
	=\frac{1}{(2\pi)^4}\frac{1}{8}\,dE_+\,dE_-\,ds
	=\frac{1}{(2\pi)^4}\frac{p_{12}}{2}\sqrt{\frac{E_+^2-s}{s}}\,dE_+\,ds\,,
\end{equation}
with $p_{12}=\tfrac{1}{2}\sqrt{s-4m_\varphi^2}$.
In the non-relativistic regime one obtains the characteristic mass scaling
\begin{equation}\label{eq:ap_m_scaling}
	\braket{\sigma_{3\varphi\to2\varphi}v^2}
	\sim \lambda^3\,\frac{m_\varphi^4}{(n_\varphi^{\rm eq})^3}
	\sim \frac{\lambda^3}{m_\varphi^5}\,.
\end{equation}

On the other hand, the second moment of the collision operator takes the form
\begin{equation}
	\begin{split}
		3T_\varphi n_\varphi \braket{C_{3\varphi\leftrightarrow 2\varphi}}_2
		=&-\frac{1}{3!}\int d\Pi_1\cdots d\Pi_5\,
		\frac{\vec p_1^{\,2}}{E_1}\, f_1f_2\,
		\vert\tilde{\mathcal{M}}_{3\varphi\leftrightarrow 2\varphi}\vert^2
		\\
		&+\frac{1}{2!2!}\int d\Pi_1\cdots d\Pi_5\,
		\frac{\vec p_3^{\,2}}{E_3}\, f_1f_2\,
		\vert\tilde{\mathcal{M}}_{3\varphi\leftrightarrow 2\varphi}\vert^2
		\\
		&-\frac{1}{2!2!}\int d\Pi_1\cdots d\Pi_5\,
		\frac{\vec p_3^{\,2}}{E_3}\, f_3f_4f_5\,
		\vert\tilde{\mathcal{M}}_{3\varphi\leftrightarrow 2\varphi}\vert^2
		\\
		&+\frac{1}{3!}\int d\Pi_1\cdots d\Pi_5\,
		\frac{\vec p_1^{\,2}}{E_1}\, f_3f_4f_5\,
		\vert\tilde{\mathcal{M}}_{3\varphi\leftrightarrow 2\varphi}\vert^2\,,
	\end{split}
	\label{eq:C2_self}
\end{equation}
where $\braket{C}_2 = C_2/(3n_iT_i)$.
The contributions proportional to $1/3!$ can be combined into
\begin{equation}
	\braket{\sigma_{3\varphi\to2\varphi}v^2\,\vec p^{\,2}/E}\,
	n_\varphi^2\,(n_\varphi-n_\varphi^{\rm eq})\,,
\end{equation}
where
\begin{equation}
	\braket{\sigma_{3\varphi\to2\varphi}v^2\,\vec p^{\,2}/E}
	=\frac{g_\varphi^3}{(n_\varphi^{\rm eq})^3}\frac{1}{3!}
	\int d\Pi_1\,d\Pi_2\,
	\frac{\vec p_1^{\,2}}{E_1}\,f_1f_2\,
	4F\,\sigma_{2_\varphi\to3_\varphi}\,,
\end{equation}
and $F=\sqrt{(s/2-m_\varphi^2)^2-m_\varphi^4}$.
The terms proportional to $1/(2!2!)$ yield
\begin{equation}
	\frac{1}{2!2!}\left(\frac{n_\varphi}{n_\varphi^{\rm eq}}\right)^2
	\int d\Pi_1\cdots d\Pi_5\,
	f_1f_2\,\frac{\vec p_3^{\,2}}{E_3}\,
	\vert\tilde{\mathcal{M}}_{3\varphi\leftrightarrow2\varphi}\vert^2
	\left(1-\frac{n_\varphi}{n_\varphi^{\rm eq}}\right)\,.
	\label{eq:C2_2}
\end{equation}
The remaining three-body integral,
$\int d\Pi_3\,d\Pi_4\,d\Pi_5\,(\vec p_3^{\,2}/E_3)\vert\tilde{\mathcal{M}}_{3\leftrightarrow2}\vert^2$,
is not Lorentz invariant because of the factor $\vec p_3^{\,2}/E_3$. Upon boosting to the overall center-of-mass frame, the energy transforms as~\cite{Arcadi:2019oxh}
\begin{equation}\label{eq:Energy_to_CM}
	E_3\to E_3\cosh\eta+\vec p_3^{\,z}\sinh\eta\,,
	\qquad
	\vec p_3^{\,z}=\sqrt{E_3^2-m_\varphi^2}\cos\theta_3\,,
\end{equation}
where $\eta$ denotes the rapidity. This motivates introducing
\begin{equation}\label{eq:sigma_tilde}
	\begin{split}
		4F\,\tilde\sigma_{2\varphi\to3\varphi}(\eta)
		=\frac{1}{2!2!}\int d\Pi_3\,d\Pi_4\,d\Pi_5\,
		\frac{\vec p_3^{\,2}}{E_3}\,
		\vert\tilde{\mathcal{M}}_{3\varphi\leftrightarrow2\varphi}\vert^2\,,
	\end{split}
\end{equation}
and defining its thermal average as
\begin{equation}
	\begin{split}
		\braket{\tilde\sigma_{2\varphi\to3\varphi}v}
		=\frac{g_\varphi^2}{(2\pi)^4(n_\varphi^{\rm eq})^2}
		&\int_{3m_\varphi}^{\infty} dE_{\rm cm}\,
		\sqrt{\frac{E_{\rm cm}^2}{4}-m_\varphi^2}\,E_{\rm cm}^2
		\\
		&\times\int_{0}^{\infty} d\eta\,\sinh^2\eta\,
		\exp\!\left(-\frac{E_{\rm cm}}{T_\varphi}\cosh\eta\right)\,
		4F\,\tilde\sigma_{2\varphi\to3\varphi}(\eta)\,.
	\end{split}
\end{equation}
With these definitions, the second moment can be expressed in terms of comoving variables as
\begin{equation}
	\begin{split}\label{eq:C2_real_cannibal_scalar}
		3T_\varphi \braket{C_{3\varphi\leftrightarrow2}}_2
		= s^2Y_\varphi\left(Y_\varphi-Y_\varphi^{\rm eq}\right)
		\left(
		\braket{\sigma_{3\varphi\to2\varphi}v^2\,\vec p^{\,2}/E}
		-\braket{\tilde\sigma_{3\varphi\to2\varphi}v^2}
		\right)\,.
	\end{split}
\end{equation}

\subsection{$3\leftrightarrow 2$ collision integrals for complex DM}\label{ap:C_Z3}
%
We consider the number-changing process $SS\leftrightarrow S^*S^*S$. The corresponding collision term for $S^*$ reads
\begin{equation}
	\begin{split}
		C_{SS\leftrightarrow S^*S^*S}[S^*]=\frac{1}{2E_{S^*}g_{S^*}}&\int\Bigg(
		(1+f_{S^*})\vert\tilde{\mathcal{M}}_{12\to \underline{S^*}45}\vert^2\left(\frac{1}{2!}d\Pi_1d\Pi_2f_1f_2\right)d\tilde\Pi_4d\tilde\Pi_5
		\\&-f_{S^*}\vert\tilde{\mathcal{M}}_{\underline{S^*}45\to 12}\vert^2 d\Pi_4d\Pi_5 f_4 f_5 \left(\frac{1}{2!}d\tilde\Pi_1d\tilde\Pi_2 \right)
		\Bigg)\,,
	\end{split}
\end{equation}
while for $S$ it is
\begin{equation}
	\begin{split}
		C_{SS\leftrightarrow S^*S^*S}[S] =\frac{1}{2E_{S}g_{S}} &\int\Bigg( 
		-f_S\vert\tilde{\mathcal{M}}_{\underline{S}2\to 345}\vert^2 d\Pi_2f_2\left(\frac{1}{2!}d\tilde\Pi_3d\tilde\Pi_4d\tilde\Pi_5\right)
		\\&+(1+f_S)\vert\tilde{\mathcal{M}}_{12\to 34\underline{S}}\vert^2\left(\frac{1}{2!}d\Pi_1d\Pi_2 f_1f_2\right)\left(\frac{1}{2!}d\tilde\Pi_3d\tilde\Pi_4\right)
		\\&-f_S\vert\tilde{\mathcal{M}}_{34\underline{S}\to12}\vert^2\left(\frac{1}{2!}d\Pi_3d\Pi_4f_3f_4\right)\left(\frac{1}{2!}d\tilde\Pi_1d\tilde\Pi_2\right)
		\\&+(1+f_S)\vert\tilde{\mathcal{M}}_{345\to \underline{S}2}\vert^2\left(\frac{1}{2!}d\Pi_3d\Pi_4d\Pi_5f_3f_4f_5\right)d\tilde\Pi_2
		\Bigg)\,.
	\end{split}
\end{equation}
Assuming CP conservation, the matrix elements satisfy
$\mathcal{M}_{S^*S\leftrightarrow SSS}=\mathcal{M}_{SS^*\leftrightarrow S^*S^*S^*}$ and
$\mathcal{M}_{SS\leftrightarrow S^*S^*S}=\mathcal{M}_{S^*S^*\leftrightarrow SSS^*}$.
We denote the former by $\mathcal{M}^{(1)}$ and the latter by $\mathcal{M}^{(2)}$.
We further assume $f_S=f_{S^*}$, implying $n_S=n_{S^*}$ and total number density $n=n_S+n_{S^*}$.
Neglecting Bose enhancement factors and setting $g_S=1$, we obtain
\begin{equation}
	\int \frac{d^3\vec p_{s^*}}{(2\pi)^3}C[S^*] = \frac{1}{2}\int d\Pi_1\dots d\Pi_5\,\left(\frac{1}{3}\vert\tilde{\mathcal{M}}^{(1)}\vert^2 + \frac{1}{2}\vert\tilde{\mathcal{M}}^{(2)}\vert^2\right)\,(f_1 f_2-f_3 f_4 f_5)\,.
\end{equation}
The integrated collision term $\int \frac{d^3\vec p_s}{(2\pi)^3}\,C[S]$ takes the same form. The corresponding second moment is
\begin{equation}
	\begin{split}
		\int \frac{d^3\vec p_1}{(2\pi)^3} \frac{\vec p^{\,2}_1}{E_1}C[S^*] = &-\frac{1}{3}\int d\Pi_1\dots d\Pi_5\vert\tilde{\mathcal{M}}^{(1)}\vert^2\frac{\vec p_1^{\,2}}{E_1}(f_1f_2-f_3f_4f_5) 
		\\&+\frac{1}{2}\int d\Pi_1\dots d\Pi_5 \vert\tilde{\mathcal{M}}^{(1)}\vert^2\frac{\vec p^{\,2}_3}{E_3}(f_1f_2-f_3f_4f_5)
		\\&-\frac{1}{2}\int d\Pi_1\dots d\Pi_5\vert\tilde{\mathcal{M}}^{(2)}\vert^2\frac{\vec p_1^{\,2}}{E_1}(f_1f_2-f_3f_4f_5)
		\\&+\frac{3}{4}\int d\Pi_1\dots d\Pi_5\vert\tilde{\mathcal{M}}^{(2)}\vert^2\frac{\vec p_3^{\,2}}{E_3}(f_1f_2-f_3f_4f_5)\,,
	\end{split}
\end{equation}
which can be written as
\begin{equation}
	\begin{split}
		\frac{3}{2}\int &d\Pi_1\dots d\Pi_5\vert\tilde{\mathcal{M}}\vert^2 \frac{\vec p^{\,2}_3}{E_3}(f_1f_2-f_3f_4f_5) 
		\\&- \int d\Pi_1 \dots d\Pi_5\vert\tilde{\mathcal{M}}\vert^2\frac{\vec p_1^{\,2}}{E_1}(f_1f_2-f_3f_4f_5)\,,
	\end{split}
\end{equation}
with $\vert\tilde{\mathcal{M}}\vert^2=\frac{1}{3}\vert\tilde{\mathcal{M}}^{(1)}\vert^2 + \frac{1}{2}\vert\tilde{\mathcal{M}}^{(2)}\vert^2$.
As in Section~\ref{ap:C0self}, the thermal averages in Eq.~\eqref{eq:C2_real_cannibal_scalar} keep the same structure; the only modification is the replacement of the matrix element and the corresponding Boltzmann factors.

\section{Freeze-in collision integrals}\label{ap:CFI}
\subsection{Higgs decay}\label{subsec:Higgs_decay}
The zeroth-moment collision term for Higgs decay is
\begin{equation}
	n_\chi\braket{C_{h\leftrightarrow\chi\chi}} = \int d\Pi_1d\Pi_2 d\Pi_3\vert \tilde{\mathcal{M}}_{h\leftrightarrow\chi \chi}\vert^2
	\left(f_1-f_2f_3\right)\,,
\end{equation}
where $\chi$ denotes either $\varphi$ or $S$, and we label momenta as $h_1\leftrightarrow \chi_2\chi_3$. Since $\mathcal{M}$ is constant, it can be taken outside the integration. Neglecting inverse decay, we obtain
\begin{equation}\label{eq:ap_Higgs_decay_C0}
	\begin{split}
		n_\chi\braket{C_{h\to\chi\chi}}&=\frac{c^2}{16\pi^2}\int d\Pi_1 f_1 \int \frac{d^3\vec p_2}{E_2}
		\frac{d^3\vec p_3}{E_3}
		\delta^{(4)}(p_1-p_2-p_3)
		\\&=c^2\frac{m_h}{16\pi^3}\sqrt{1-\frac{4m_\chi^2}{m_h^2}}\, T K_1(m_h/T)\,,
	\end{split}
\end{equation}
where $c$ denotes the relevant coupling. For interactions with $\varphi$, $c=\lambda_{h\varphi}v$, while for $S$ it is $c=A_{\phi s}\theta$. We note an overall factor-of-two difference with respect to some conventions~\cite{Lebedev:2019ton}, which originates from symmetry-factor choices.

The second moment is
\begin{equation}
	3T_\chi n_\chi \braket{C_{h\to \chi\chi}}_2 = \int d\Pi_1d\Pi_2 d\Pi_3\, \vert \tilde{\mathcal{M}}_{h\to\chi\chi}\vert^2 \frac{\vec p_2^{\,2}}{E_2}f_1\,.
\end{equation}
Since $\vec p_2^{\,2}/E_2$ is not Lorentz invariant, we boost it to the center-of-mass frame,
\begin{equation}
	E_2 \to E_2\,\cosh\eta+\vec p_2^{\,z}\sinh\eta\,,
\end{equation}
and express the boost in terms of $z=\cosh\eta=E_h/m_h$. This gives
\begin{equation}\label{eq:ap_CMboost}
	\frac{\vec p_2^{\,2}}{E_2} \to E_2 z + \vert \vec p_2\vert \cos\theta_2 \sqrt{z^2-1} - \frac{m_\chi^2}{E_2 z + \vert \vec p_2\vert \cos\theta_2 \sqrt{z^2-1}}\,.
\end{equation}
The resulting expression for $3T_\chi n_\chi\braket{C_{h\to\chi\chi}}_2$ is
\begin{equation}\label{eq:ap_Higgs_decay_C2}
	\begin{split}
		&\frac{c^2}{4(2\pi)^2}\int d\Pi_1f_1\int d\cos\theta_2\,\frac{2\pi \vert \vec p_2\vert}{E_2}\left.\left(E_2\,z - \frac{m_\chi^2}{E_2 z + \vert \vec p_2\vert \cos\theta_2 \sqrt{z^2-1}}\right)\right\vert_{E_2=m_h/2}
		\\&=\frac{c^2 m_h^2}{32\pi^3}\sqrt{1-\frac{4m_\chi^2}{m_h^2}}\Bigg(T K_2(m_h/T)
		\\&\hspace{8em}-\frac{2m_\chi^2}{\sqrt{m_h^2-4m_\chi^2}}\int_1^\infty dz\,e^{-m_h z/T}\log\left(\frac{m_h z + 2\vert\vec p_2\vert\sqrt{z^2-1}}{m_h z - 2\vert\vec p_2\vert\sqrt{z^2-1}}\Bigg)\right)\,.
	\end{split}
\end{equation}
In the hierarchical limit $m_\chi\ll m_h$, the logarithmic term is negligible.

\subsection{Three-body Higgs decay}\label{ap:triple_higgs_decay}
For three-body Higgs decay, the zeroth moment for the mediator $\phi$ is
\begin{equation}
	g_\phi n_\phi \braket{C_{h\leftrightarrow \phi S S^* }} = \int d\Pi_1d\Pi_2 d\Pi_3 d\Pi_4\vert \tilde{\mathcal{M}}_{h\leftrightarrow \phi S S^* }\vert^2
	\left(f_1-f_2f_3f_4\right)\,,
\end{equation}
with momenta labeled as $h_1\leftrightarrow \phi_2S_3S^*_4$. Neglecting inverse decay gives
\begin{equation}
	n_\phi \braket{C_{h\to \phi S S^* }} = \frac{\lambda_{\phi s}^2\theta^2}{256\pi^5}\int d\Pi_1\,f_1 \int \frac{d^3\vec p_2}{E_2}
	\frac{d^3\vec p_3}{E_3} \frac{d^3\vec p_4}{E_4}
	\delta^{(4)}(p_1-p_2-p_3-p_4)\,.
\end{equation}
The three-body phase-space integral can be evaluated as in Section~\ref{ap:C0self}. Writing
$s_{34}=(p_3+p_4)^2=(p_1-p_2)^2=m_h^2+m_\phi^2-2m_hE_2$, one obtains
\begin{equation}
	8\pi^2  \int_{m_\phi}^{E_2^\text{max}} dE_2\, \vert \vec p_2\vert^{2} \sqrt{1-\frac{4m_s^2}{m_h^2+m_\phi^2-2m_hE_2}}\,,
\end{equation}
where
\begin{equation}
	\begin{split}
		E_2^\text{max} &= \sqrt{(\vec q_2^\text{ max})^2+m_\phi^2}\qquad \text{and} 
		\\\vert \vec q_2^\text{ max}\vert &= \sqrt{m_h^4+(m_\phi^2-4m_s^2)^2-2m_h^2(m_\phi^2+4m_s^2)}/(2m_h)\,,
	\end{split}
\end{equation}
with $\vert\vec p_2^\text{ max}\vert$ the maximal momentum of $\phi$. In the limit $m_\phi,\,m_s\ll m_h$ the integral can be evaluated analytically, yielding $\pi^2 m_h^2$. The zeroth moment then becomes
\begin{equation}
	\begin{split}
		n_\phi \braket{C_{h\to \phi SS^*}}&\simeq  \frac{\lambda_{\phi s}^2\theta^2}{1024\pi^5}\,m_h^4 \int_1^\infty dz\,\sqrt{z^2-1}\exp(-m_h z/T)
		\\&=\frac{\lambda_{\phi s}^2\theta^2}{1024\pi^5}m_h^3 T K_1(m_h/T)\,. 
	\end{split}
\end{equation}

The second moment is
\begin{equation}
	3T_\phi n_\phi \braket{C_{h\to \phi S S^* }}_2 = \int d\Pi_1d\Pi_2 d\Pi_3 d\Pi_4\vert \tilde{\mathcal{M}}_{h\to \phi S S^* }\vert^2\,\frac{\vec p_2^{\,2}}{E_2}\,f_1\,.
\end{equation}
Proceeding as for the two-body decay, we boost $\vec p_2^{\,2}/E_2$ to the center-of-mass frame. Here we already work in the massless limit and neglect the logarithmic contribution. This yields
\begin{equation}
	3T_\phi n_\phi \braket{C_{h\to \phi S S^* }}_2 \simeq  \frac{\lambda_{\phi s}^2\theta^2}{3072\pi^5}m_h^4 T K_2(m_h/T)\,.
\end{equation}
The expression for $S$ is obtained analogously.
%
\subsection{Higgs annihilation}
\label{sec:Higgs_ann}
The zeroth moment is
\begin{equation}
	n_\varphi \braket{C_{hh\leftrightarrow \varphi\varphi}} = \frac{1}{2!}\int d\Pi_1 d\Pi_2 d\Pi_3 d\Pi_4 \vert\tilde{\mathcal{M}}_{hh\leftrightarrow\varphi\varphi}\vert^2\left(f_1f_2 -f_3 f_4\right)\,,
\end{equation}
with momentum assignment $h_1 h_2\leftrightarrow \varphi_3\varphi_4$. As in the decay calculation, we neglect the inverse process. Using the standard relation to the $2\to2$ cross section~\cite{Edsjo:1997bg}, the collision integral can be written as
\begin{equation}\label{eq:integral_hh_to_phiphi}
	n_\varphi \braket{C_{hh\to \varphi\varphi}}= \frac{T}{32\pi^4}\int_{4m_h^2}^\infty \,ds\,\sqrt{s}(s-4m_h^2)K_1(\sqrt{s}/T)\,\sigma_{hh\to\varphi\varphi}\,.
\end{equation}

The second moment is
\begin{equation}
	3T_\varphi n_\varphi\braket{C_{hh\to\varphi\varphi}}_2 = \frac{1}{2!}\int d\Pi_1\dots d\Pi_4\,\vert\tilde{\mathcal{M}}_{hh\leftrightarrow\varphi\varphi}\vert^2\frac{\vec p_3^{\,2}}{E_3}\,f_1 f_2\,.
\end{equation}
We evaluate this expression in the center-of-mass frame, boosting $\vec p_3^{\,2}/E_3$ as in Eq.~\eqref{eq:ap_CMboost}. In the present case $z=E_+/\sqrt{s}$, with $E_+=E_1+E_2$. Performing the two-body final-state phase-space integration yields
\begin{equation}
	\begin{split}
		&\int d\Pi_3 d\Pi_4\,\vert\tilde{\mathcal{M}}_{hh\to\varphi\varphi}\vert^2\,\left(E_3 z+\vert \vec p_3\vert\cos\theta_3\sqrt{z^2-1}-\frac{m_\varphi^2}{E_3 z + \vert \vec p_3\vert \cos\theta_3\sqrt{z^2-1}}\right)
		\\&=\frac{\vert\mathcal{M}_{hh\to\varphi\varphi}\vert^2}{8\pi}\frac{\vert\vec p_3\vert}{\sqrt{s}}\left(\sqrt{s}z- \frac{m_\varphi^2}{\vert\vec p_3\vert}\frac{1}{\sqrt{z^2-1}}\log\left(\frac{\sqrt{s} z + \vert\vec p_3\vert\sqrt{z^2-1}}{\sqrt{s} z - \vert\vec p_3\vert\sqrt{z^2-1}}\right)\right)\,.
	\end{split}
\end{equation}
%

\subsection{Production from electroweak states}\label{ap:EW}
Freeze-in production from electroweak initial states is important because, at high energies, the gauge-boson propagators lead to only mild suppression. We compute the relevant matrix elements with \texttt{CalcHEP 3.8.10}~\cite{Belyaev:2012qa} and use their high-energy limits in the cBE. The full expressions are lengthy; below we report the corresponding cross sections in the large-$s$ limit (and for $m_\phi\to0$):
\begin{equation}
	\begin{split}
		\sigma_{hh\to h\phi}&= \frac{\lambda_h^2\theta^2}{16\pi s}\,,
		\\
		\sigma_{Z h\to Z\phi}&= \theta^2 \frac{s-2m_Z^2-6m_h^2+12m_Z^2\log \left( \frac{s}{m_Z^2}\right)}{144\pi v^4}\,,
		\\
		\sigma_{W^{\pm}Z\to W^\pm \phi} &= \theta^2\frac{480m_W^4-72m_W^2m_Z^2 + 24m_Z^4}{864 m_Z^2 \pi v^4}\,,
		\\
		\sigma_{W^{+}W^{-}\to h \phi} &= \theta^2\frac{s-2m_W^2+8m_W^2\log\left(\frac{s}{m_W^2}\right)}{144\pi v^4}\,,
		\\
		\sigma_{W^{+}W^{-}\to Z \phi} &= \theta^2\frac{m_W^2(8m_W^2+m_Z^2)}{18m_Z^2\pi v^4}\,,
		\\
		\sigma_{W^{\pm} h\to W^{\pm} \phi} &=\theta^2\frac{12m_W^2\log \left(\frac{s}{m_W^2}\right)-6m_h^2-2m_W^2+s}{144\pi v^4} \,,
		\\
		\sigma_{ZZ\to h\phi} &=\theta^2\frac{s+3m_h^2 - 2m_Z^2 + 8m_Z^2\log\left(\frac{s}{m_Z^2}\right)}{144\pi v^4}\,.
	\end{split}
\end{equation}

\section{Dark Matter - mediator interactions}\label{ap:C_DM_med}
%
\subsection{Production-annihilation}
The collision operator for $\phi\phi\leftrightarrow SS^*$ is
\begin{equation}
	\begin{split}
		C_{\phi\phi\leftrightarrow SS^*}[S]=\frac{1}{2E_Sg_S}\int\Bigg(&-f_S\vert\tilde{\mathcal{M}}_{\underline S S_2^*\to\phi_3 \phi_4}\vert^2 \,d\Pi_2 f_2\left(\frac{1}{2!}d\Pi_3 d\Pi_4 \right)
		\\&+\vert\tilde{\mathcal{M}}_{\underline S S_2^*\leftarrow\phi_3 \phi_4}\vert^2 d\Pi_2\left(\frac{1}{2!}d\Pi_3d\Pi_4\,f_3f_4\right) \Bigg)\,.
	\end{split}
\end{equation}
We allow for dilution in the mediator sector as well. The corresponding zeroth-moment thermal average becomes
\begin{equation}
	\begin{split}
		&\braket{C_{\phi\phi\leftrightarrow SS^*}[S]}=\\
		&-\left(\frac{n_S}{n_S^{\text{eq}}}\right)^2\frac{\lambda_{\phi s}^2}{2!n_S}\frac{T_S}{512\pi^5}\int_{\max(4m_\phi^2,4m_s^2)}^\infty ds\,\frac{\sqrt{s-4m_\phi^2}\sqrt{s-4m_s^2}}{\sqrt{s}}K_1(\sqrt{s}/T_S)
		\\&+\left(\frac{n_\phi}{n_\phi^{\text{eq}}}\right)^2\frac{\lambda_{\phi s}^2}{2!n_S}\frac{T_\phi}{512\pi^5}\int_{\max(4m_\phi^2,4m_s^2)}^\infty ds\,\frac{\sqrt{s-4m_\phi^2}\sqrt{s-4m_s^2}}{\sqrt{s}}K_1(\sqrt{s}/T_\phi)\,.
	\end{split}
\end{equation}
In the relativistic regime, the lighter state can be treated as approximately massless. To capture the leading non-relativistic correction, we expand $\sqrt{s-4m^2}\simeq \sqrt{s}-2m^2/\sqrt{s}$ with $m=\min(m_\phi,m_s)$, and approximate
\begin{equation}
	\begin{split}
		\int_{4M^2}^\infty ds\,\frac{\sqrt{s-4m_\phi^2}\sqrt{s-4m_s^2}}{\sqrt{s}}K_1(\sqrt{s}/T_\text{ds})\approx &4M^2 T_\text{ds} K_1(M/T_\text{ds})^2
		\\&-e^{-2M/T_\text{ds}}\frac{m^2 T_\text{ds}^2}{M}\,,
	\end{split}
\end{equation}
where $M=\max(m_\phi,m_s)$ and $T_\text{ds}$ stands for either $T_\phi$ or $T_S$. Finally, the collision term for $\phi$ follows from number conservation,
$\braket{C_{\phi\phi\leftrightarrow SS^*}[\phi]} = -2\frac{n_S}{n_\phi}\braket{C_{\phi\phi\leftrightarrow SS^*}[S]}$.
We use this approximation in the cBE system, Eq.~\eqref{eq:system_med_DM}.

\subsection{Scattering}
We next consider elastic scattering between the $S$ and $\phi$ fluids. Following the parametrization of~\cite{Aboubrahim:2023yag}, we write
\begin{equation}
	C_\text{scatter}[S] = \frac{1}{128\pi^3 E_1 g_S\vert\vec p_1\vert}\int_{m_S}^\infty dE_3\int_{\text{max}(m_\phi,E_3-E_1+m_\phi)}^\infty dE_2\,\Pi(E_1,E_2,E_3)\mathcal{P}(f_1,\dots,f_4)\,,
\end{equation}
where momenta are assigned as $S_1 \phi_2\leftrightarrow S_3 \phi_4$. The kinematic factor $\Pi$ is
\begin{equation}
	\Pi(E_1,E_2,E_3) = \lambda_{\phi s}^2\left(k_+ - k_-\right)
	\Theta(k_+-k_-)\,,
\end{equation}
with $k_+=\text{min}\!\left(\vert\vec p_1\vert + \vert\vec p_3\vert,\, \vert \vec p_2\vert + \vert\vec p_4\vert \right)$,
$k_-=\text{max}\!\left(\big|\vert \vec p_1\vert - \vert \vec p_3\vert\big|,\, \big|\vert\vec p_2 \vert - \vert \vec p_4\vert\big|\right)$,
and $\Theta$ the Heaviside step function. The functional $\mathcal{P}$ collects the distribution functions,
\begin{equation}
	\mathcal{P}(f_1,f_2,f_3,f_4)= f_3 f_4 - f_1 f_2\,.
\end{equation}
Since the matrix element is constant, we estimate the collision operator in the relativistic limit, taking $E_j\simeq|\vec p_j|$ for all external legs.

In this limit, the second moment becomes (setting $g_S=1$)
\begin{equation}
	\begin{split}
		\braket{C_\text{scatter}[S]}_2 &\simeq \frac{1}{n_S\,3 T_S}\frac{1}{2\pi^2}\int_0^\infty dE_1\,E_1^2\,C_\text{scatter}
		\\&=\frac{\lambda_{\phi s}^2}{n_S\,3 T_S\,256\pi^5}\int_0^\infty dE_1\,E_1 \int_0^\infty dE_3\int_{\text{max}(0,E_3-E_1)}^\infty dE_2\,\Pi\,\mathcal{P}\,.
	\end{split}
\end{equation}
The integrand can be written as
\begin{equation}
	\Pi\,\mathcal{P}=\,(E_1 + E_2 - |E_1 - E_3| - |E_2 - E_3|)\,
	\Theta(E_1 + E_2 - |E_1 - E_3| - |E_2 - E_3|)\,(f_3 f_4-f_1f_2)\,,
\end{equation}
and, for equilibrium distributions,
\begin{equation}
	f^\text{eq}_3f^\text{eq}_4-f^\text{eq}_1f^\text{eq}_2
	=\left(e^{-E_3/T_S}e^{-(E_1-E_3)/T_\phi}-e^{-E_1/T_S}\right)e^{-E_2/T_\phi}\,.
\end{equation}
The integral over $E_2$ can be carried out analytically by splitting the domain into $E_2<E_3$ and $E_2>E_3$. One finds
\begin{equation}
	\begin{split}
		\int_{\text{max}(0,E_3-E_1)}^{E_3} dE_2&\,(E_1 + 2E_2 - E_3 -|E_1-E_3|)\,e^{-E_2/T_\phi} 
		\\&+ \int_{E_3}^\infty dE_2\,(E_1 + E_3 -|E_1-E_3|)\,e^{-E_2/T_\phi}
		\\&=2e^{-E_3/T_\phi}T_\phi^2\left(e^{\frac{E_1+E_3-|E_1-E_3|}{2T_\phi}}-1\right)\,,
	\end{split}
\end{equation}
and the remaining $E_3$ integration yields
\begin{align}
	&2T_\phi^2\int_0^\infty dE_3\,e^{-E_3/T_\phi}\left(e^{\frac{E_1+E_3-|E_1-E_3|}{2T_\phi}}-1\right)\left(e^{-E_3/T_S}e^{-(E_1-E_3)/T_\phi}-e^{-E_1/T_S}\right)\notag\\
	&=2T_\phi^2\frac{e^{-E_1(1/T_\phi+1/T_S)}(e^{E_1/T_S}T_S^2-e^{E_1/T_\phi}\left(E_1(T_\phi-T_S)+T_S^2 \right))}{T_\phi-T_S}\,.
\end{align}
Finally, performing the $E_1$ integral with measure $\int dE_1\,E_1$ gives
\begin{equation}\label{eq:S_med_scatter}
	\braket{C_\text{scatter}[S]}_2\simeq
	\left(\frac{n_S}{n_S^\text{eq}}\right)\left(\frac{n_\phi}{n_\phi^\text{eq}}\right)
	\frac{\lambda_{\phi s}^2}{n_S\,3T_S\,128\pi^5}\,T_\phi^2T_S^2\,(T_\phi-T_S)\,
	e^{-m_S/T_S}e^{-m_\phi/T_\phi}\,.
\end{equation}
The exponential factors are included to mimic the onset of Boltzmann suppression outside the relativistic regime. To include scattering in the $\phi$ temperature equation, we use energy conservation, which implies
$\braket{C_\text{scatter}[\phi]} = -\frac{3T_S n_S}{3T_\phi n_\phi}\,\braket{C_\text{scatter}[S]}$.

\section{Dark matter scattering with fermions} \label{sec:scatter_ferm}
Here we denote the temperature of the dark sector as $T'$ and the mass of the DM as $m$ aligned with the notation of~\Cref{ch:4}.
The squared matrix element for the elastic scattering of DM with a SM fermion reads
\begin{equation}
	\vert\mathcal{M}_{Sf\to Sf}\vert^2
	=
	N_c\, \lhs^2\, m_f^2\,
	\frac{4 m_f^2 - t}{(t - m_h^2)^2}\,,
\end{equation}
where $N_c$ denotes the number of colors. For the temperatures of interest, $T' \leq T < m_h$, and given the hierarchies $m \ll m_h$ and $m_f \ll m_h$, the momentum transfer is suppressed relative to $m_h^2$. We therefore approximate
\(
(t - m_h^2)^2 \simeq m_h^4.
\)

We again evaluate the collision operator in the relativistic limit, while retaining the Boltzmann suppression factors $e^{-m/T'} e^{-m_f/T}$ to account for low-temperature effects. Writing
\begin{equation}
	4 m_f^2 - t
	=
	2 m_f^2
	+
	2 E_2 E_4
	-
	2 \vert \vec p_2 \vert \vert \vec p_4 \vert \cos\alpha\,,
\end{equation}
with momenta labeled as $S_1 f_2 \to S_3 f_4$, the constant part of the numerator gives
\begin{equation}\label{eq:ap1}
	C_2^{S f\leftrightarrow S f}
	\supset
	\frac{N}{N_\text{eq}}
	\frac{\lhs^2\, m_f^4\, N_c}{64 \pi^5\, m_h^4}\,
	T'^2\, T^2\,
	(T - T')\,
	e^{-m/T'} e^{-m_f/T}.
\end{equation}

We next consider the term with angular dependence,
\begin{align}
	C_2^{S f\leftrightarrow S f}
	&\supset
	\frac{2 N_c \lhs^2 m_f^2}{m_h^4}
	\int d\Pi_1 \cdots d\Pi_4\,
	\frac{\vec p_1^{\,2}}{E_1}
	\left(
	E_2 E_4
	-
	\vert \vec p_2 \vert \vert \vec p_4 \vert \cos\alpha
	\right)
	\nonumber\\
	&\qquad\times
	\delta^{(4)}\!\left(\sum_f p_f - \sum_i p_i\right)
	\left(f_3 f_4 - f_1 f_2\right).
\end{align}

We first integrate over $p_1$ and $p_2$,
\begin{align}
	&\int d\Pi_1 d\Pi_2\,
	\frac{\vec p_1^{\,2}}{E_1}
	\left(
	E_2 E_4
	-
	\vert \vec p_2 \vert \vert \vec p_4 \vert \cos\alpha
	\right)
	\delta^{(4)}\!\left(\sum_f p_f - \sum_i p_i\right)
	\nonumber\\
	&\qquad=
	\int d\cos\alpha\,
	\frac{1}{8\pi}
	\frac{\vert \vec p_1 \vert}{\sqrt{s}}
	\left(\frac{\vec p_1^{\,2}}{E_1}\right)^{\!\text{cm}}
	\left(
	E_1 E_4
	-
	\vert \vec p_1 \vert \vert \vec p_4 \vert \cos\alpha
	\right)
	\nonumber\\
	&\qquad\simeq
	\frac{(E_3 + E_4)\, s}{16\pi}.
\end{align}

In the above, we have boosted to the center-of-mass (CM) frame and performed the angular integration. In the relativistic limit we take $E_i \simeq \vert \vec p_i \vert$. The non-invariant quantity transforms as
\begin{equation}
	\left(\frac{\vec p_1^{\,2}}{E_1}\right)^{\!\text{cm}}
	=
	E_1 z
	+
	\vert \vec p_1 \vert \cos\alpha \sqrt{z^2 - 1}
	-
	\frac{m^2}{E_1 z + \vert \vec p_1 \vert \cos\alpha \sqrt{z^2 - 1}},
\end{equation}
where $z \equiv (E_3 + E_4)/\sqrt{s}$. We further approximate
\(
(\vec p_1/E_1)^{\text{cm}} \simeq E_1 (z + \sqrt{z^2 - 1}\cos\alpha)
\)
and use $\sqrt{z^2 - 1} \ll 3z$.

The remaining phase-space elements reduce to
\begin{equation}
	d\Pi_3\, d\Pi_4
	=
	\frac{1}{64 \pi^4}\,
	dE_3\, dE_4\, ds\,,
\end{equation}
with
\begin{equation}
	s
	=
	m^2 + m_f^2
	+
	2 E_3 E_4
	-
	2 \vert \vec p_3 \vert \vert \vec p_4 \vert \cos\theta.
\end{equation}

Carrying out the remaining integrations yields the dominant term
\begin{align}
	\int dE_3 dE_4\,
	e^{-E_3/T'} e^{-E_4/T}
	\int_0^{4E_3 E_4} ds\,
	\frac{(E_3+E_4)s}{32\pi}
	=
	\frac{3}{2\pi}
	T'^3 T^3 (T+T').
\end{align}

For the back-reaction term (suppressing overall factors),
\begin{align}
	\int dE_3 dE_4\,
	E_3 e^{-E_3/T'} e^{-E_4/T}
	\int_0^{4E_3 E_4} ds\,
	\frac{s}{32\pi}
	=
	\frac{3}{\pi}
	T'^4 T^3.
\end{align}

Restoring all prefactors and combining both pieces, we obtain
\begin{equation}\label{eq:ap2}
	C_2^{S f\leftrightarrow S f}
	\supset
	\frac{N}{N_\text{eq}}
	\frac{3 \lhs^2 m_f^2 N_c}{64 \pi^5 m_h^4}\,
	T'^3 T^3
	(T - T')
	e^{-m/T'} e^{-m_f/T}.
\end{equation}

Adding the contributions from Eqs.~\eqref{eq:ap1} and~\eqref{eq:ap2}, we finally arrive at
\begin{equation}
	C_2^{S f\leftrightarrow S f}
	\simeq
	\frac{N}{N_\text{eq}}
	\frac{\lhs^2 m_f^2 N_c}{64 \pi^5 m_h^4}\,
	T'^2 T^2
	(T - T')
	\left(m_f^2 + 3 T T'\right)
	e^{-m/T'} e^{-m_f/T},
\end{equation}
which reproduces Eq.~\eqref{eq:C2_el_scatter_ferm}.

\section{Vacuum stability of a scalar mixing with the Higgs}\label{ap:vac_stab}
%
The interactions between the Higgs and the mediator in~\Cref{ch:3} are described by the scalar potential
\begin{equation}\label{V(h,phi)}
	V(H,\phi) =  \mu_h^2\,H^\dagger H + \lambda_h(H^\dagger H)^2 +(B_{h\phi}\phi + \lambda_{h\phi}\phi^2)H^\dagger H 
	+\lambda_1\phi +\frac{1}{2}\mu_\phi^2\phi^2 + \frac{g_\phi}{3!}\phi^3 + \frac{\lambda_\phi}{4!}\phi^4 \,,
\end{equation}
where $H$ denotes the SM $\text{SU}(2)_\text{L}$ Higgs doublet. In unitarity gauge we write
$H = \frac{1}{\sqrt{2}}(0,h)^\intercal$, and after electroweak symmetry breaking we shift
$h\to h+v$ with $v\simeq 246\,\text{GeV}$.
We keep the linear term $\lambda_1\phi$. In many singlet-scalar setups this term is set to zero, since a field shift
$\phi\to \phi+\phi_0$ can be used to remove the tadpole~\cite{Fradette:2017sdd}. Equivalently, one may expand around
$\phi\to \phi+w$ and impose the minimization condition such that the linear term in the shifted field vanishes. For the moment we allow $\lambda_1\neq 0$.

The stationary conditions at $(h,\phi)=(v,w)$ are
\begin{equation}\label{extremum_cond}
	\begin{split}
		\left.\frac{\partial V}{\partial \phi}\right\vert_{h=v,\phi = w} &= \frac{g_\phi}{2}w^2 + \frac{\lambda_\phi}{6}w^3 
		+ \mu_\phi^2 w + v^2w\lambda_{h\phi} + \frac{1}{2}v^2 B_{h\phi}+\lambda_1=0\,,
		\\
		\left.\frac{\partial V}{\partial h}\right\vert_{h=v,\phi = w} &= \lambda_hv^3 + v\mu_h^2 + vwB_{h\phi} + vw^2\lambda_{h\phi}=0\,.
	\end{split}
\end{equation}
Solving these equations for $\lambda_h$ and $\lambda_1$ gives
\begin{equation}\label{lambdas}
	\begin{split}
		\lambda_h&= -\frac{\mu_h^2-w B_{h\phi}-w^2\lambda_{h\phi}}{v^2}\,,
		\\
		\lambda_1 &= -\left(\frac{g_\phi}{2}w^2 + \frac{\lambda_\phi}{6}w^3 
		+ \mu_\phi^2 w + v^2w\lambda_{h\phi} + \frac{1}{2}v^2 B_{h\phi} \right)\,.
	\end{split}
\end{equation}
With these substitutions, $(v,w)$ is a critical point and the tadpole term vanishes. In particular, if one wishes to have a solution with $w=0$, the first condition in Eq.~\eqref{extremum_cond} requires
$v^2 B_{h\phi}/2 + \lambda_1 = 0$. Conversely, imposing $(v^2B_{h\phi}/2 + \lambda_1)\phi=0$ ensures that $w=0$ is a stationary point. For simplicity we adopt $w=0$, i.e.
$v^2 B_{h\phi}/2 + \lambda_1=0$, which matches the convention used in the literature~\cite{Fradette:2017sdd}. In that case, the $\phi$-stationarity condition reduces to
\begin{equation}\label{extremum_cond2}
	\frac{g_\phi}{2}w^2 + \frac{\lambda_\phi}{6}w^3 + \mu_\phi^2 w + v^2w\lambda_{h\phi}=0\,.
\end{equation}

The scalar mass matrix is obtained from the second derivatives of the potential evaluated at $(v,w)$,
\begin{equation}
	M^2 = 
	\begin{pmatrix}
		-2\left(\mu_h^2+w(B_{h\phi}+w\lambda_{h\phi})\right) & v\left(B_{h\phi}+2w\lambda_{h\phi}\right)
		\\
		v\left(B_{h\phi}+2w\lambda_{h\phi}\right) & g_\phi w+\frac{w^2\lambda_\phi}{2}+\mu_\phi^2 + v^2\lambda_{h\phi}
	\end{pmatrix}\,,
\end{equation}
which is diagonalized by the rotation
\begin{equation}\label{eq:rot_matrix}
	\mathcal{O} =
	\begin{pmatrix}
		\cos\theta&\sin\theta\\
		-\sin\theta&\cos\theta
	\end{pmatrix}\,,
\end{equation}
so that $\mathcal{O}^\intercal M^2\mathcal{O}=\text{diag}(m_h^2,m_\phi^2)$. The mixing angle is fixed by the off-diagonal entry,
\begin{equation}\label{eq:mixing_angle}
	\sin 2\theta  =\frac{2v(B_{h\phi} +2w\lambda_{h\phi})}{m_\phi^2-m_h^2}\,.
\end{equation}

A nonzero singlet vev $w\neq 0$ does not break any symmetry, so it does not change the qualitative physics. One must still check whether Eq.~\eqref{extremum_cond2} admits additional stationary points and whether they correspond to minima. Besides $w=0$, Eq.~\eqref{extremum_cond2} yields
\begin{equation}
	w_\pm = \frac{-3g_\phi\pm \sqrt{3}\sqrt{3g_\phi^2 - 8\lambda_\phi \mu_\phi^2 - 8v^2\lambda_\phi\lambda_{h\phi}}}{2\lambda_\phi}\,.
\end{equation}
If the discriminant is negative,
\begin{equation}
	0>3g_\phi^2 - 8\lambda_\phi \mu_\phi^2 - 8v^2\lambda_\phi\lambda_{h\phi}\,,
\end{equation}
then $w_\pm$ are absent. This condition can be rewritten as
\begin{equation}
	3g_\phi^2 - 8\lambda_\phi \mu_\phi^2 - 8v^2\lambda_\phi\lambda_{h\phi}
	= 3g_\phi^2-4(m_\phi^2+m_h^2)\lambda_\phi + 4(m_h^2-m_\phi^2)\lambda_\phi\cos 2\theta\,.
\end{equation}
In the small-mixing limit $\theta\ll1$, this simplifies to
\begin{equation}
	3g_\phi^2-8m_\phi^2\lambda_\phi<0
	\quad\Rightarrow\quad
	\frac{3g_\phi^2}{8m_\phi^2}<\lambda_\phi\,.
\end{equation}
Finally, we take $\lambda_\phi$ and $g_\phi$ to be small so that $\phi$ does not chemically thermalize the mediator sector. With these assumptions, minimizing the full potential in Eq.~\eqref{eq:full_potential} yields $\braket{S}=0$ (for $k<8/3$) and $w=0$.

\chapter{Thermal masses for the U(1) model}\label{ap:thermal_masses}

Here we derive the thermal-mass formulas in Eq.~\eqref{eq:Pi} of~\Cref{ch:5}.

For \(\Phi=(\phi+i\varphi)/\sqrt{2}\), the interaction terms relevant for the thermal masses are
\begin{equation}
	\mathcal L_{\rm int}\supset
	-\frac{\lambda}{4}\phi^4
	-\frac{\lambda}{2}\phi^2\varphi^2
	-\frac{\lambda}{4}\varphi^4
	+\frac{g^2}{2}A'_\mu A'^\mu(\phi^2+\varphi^2)
	+gA'^\mu(\varphi\partial_\mu\phi-\phi\partial_\mu\varphi).
\end{equation}

In the real-time formalism, two-point functions are arranged into a \(2\times2\) contour matrix. For the present one-loop tadpole calculation, it is sufficient to use the \(11\) component, i.e.\ the time-ordered propagator on the forward branch of the contour.~\cite{Lundberg:2020mwu,Bellac:2011kqa,Landsman:1986uw}. Its thermal part for a bosonic quasiparticle \(\psi\) is
\begin{equation}
	iD^{11}_{\psi,\rm th}(p)
	=
	2\pi\,\delta(p^2-m_\psi^2)\,f_\psi(|p^0|).
\end{equation}
Performing the \(p^0\)-integration then gives
\begin{equation}
	\int \frac{d^4p}{(2\pi)^4}\, iD^{11}_{\psi,\rm th}(p)
	=
	\int_p \frac{f_\psi(E_\psi)}{E_\psi}
	\equiv \mathcal I_\psi,
\end{equation}
with \(E_\psi=\sqrt{p^2+m_\psi^2}\). For a momentum-independent one-loop tadpole, the thermal self-energy takes the generic form
\begin{equation}
	i\Pi = S\,(iV)\int \frac{d^4p}{(2\pi)^4}\, iD^{11}_{\psi,\rm th}(p),
\end{equation}
where \(V\) is the corresponding four-point vertex and \(S\) is the symmetry factor.

The momentum-independent scalar self-energies are obtained from tadpole diagrams. From
\begin{equation}
	-\frac{\lambda}{4}\phi^4=-\frac{6\lambda}{4!}\phi^4,
\end{equation}
the \(\phi^4\) vertex is \(-i6\lambda\). The one-loop tadpole therefore gives
\begin{equation}
	i\Pi_\phi^{(\phi)}
	=
	\frac{1}{2}(-i6\lambda)
	\int \frac{d^4p}{(2\pi)^4}\, iD^{11}_{\phi,\rm th}(p),
\end{equation}
so that
\begin{equation}
	\Pi_\phi^{(\phi)}=3\lambda\,\mathcal I_\phi.
\end{equation}

Similarly, the mixed interaction \(-\lambda\phi^2\varphi^2/2\) gives the vertex \(-i2\lambda\), hence
\begin{equation}
	i\Pi_\phi^{(\varphi)}
	=
	\frac{1}{2}(-i2\lambda)
	\int \frac{d^4p}{(2\pi)^4}\, iD^{11}_{\varphi,\rm th}(p),
\end{equation}
which yields
\begin{equation}
	\Pi_\phi^{(\varphi)}=\lambda\,\mathcal I_\varphi.
\end{equation}

The gauge seagull \(g^2A'^2\phi^2/2\) gives the vertex \(ig^2 g_{\mu\nu}\), so that the tadpole contribution is proportional to \(g_{\mu\nu}D^{\mu\nu}_{A'}\). In the quasiparticle description of a massive vector, this contraction counts the three physical polarizations, yielding
\begin{equation}
	\Pi_\phi^{(A')}=3g^2\,\mathcal I_{A'}.
\end{equation}

Therefore,
\begin{equation}
	\Pi_\phi=3\lambda\,\mathcal I_\phi+\lambda\,\mathcal I_\varphi+3g^2\,\mathcal I_{A'}.
\end{equation}
Exchanging \(\phi\leftrightarrow\varphi\) gives
\begin{equation}
	\Pi_\varphi=\lambda\,\mathcal I_\phi+3\lambda\,\mathcal I_\varphi+3g^2\,\mathcal I_{A'}.
\end{equation}

For the gauge boson, the Debye mass is obtained from the static longitudinal polarization tensor,
\begin{equation}
	\Pi_{A'}\equiv \Pi^{00}_{A'}(0,\mathbf{0}).
\end{equation}
At one loop, this receives both bubble and seagull contributions from the charged scalars. In the static limit, these combine into the standard Debye-mass expression, which for each real scalar degree of freedom reads
\begin{equation}
	\Pi^{00}_{\psi}(0,\mathbf{0})=2g^2\,\mathcal I_\psi.
\end{equation}
Summing over \(\psi=\phi,\varphi\), one finds
\begin{equation}
	\Pi_{A'}=2g^2(\mathcal I_\phi+\mathcal I_\varphi).
\end{equation}

In the relativistic limit and for fugacity \(z=1\), one has $\mathcal I_\psi=T_\psi^2/12$ for bosons, while for fermions $\mathcal I_\psi=T_\psi^2/24.$
%
\section{Moments of the fBE with varying mass}\label{sec:moments_fBE}

\subsection{Energy density moment}\label{subsec:evol_temp}
In this subsection of the appendix we comment that an evolution equation for the energy density can be obtained from Eq.~\eqref{eq:fBE} by integrating both sides by $\sum_\psi g_\psi\int_p E$ and taking into account that the mass varies over time. The result is
\begin{equation}\label{eq:energy_density_1}
	\dot \rho+3H(\rho+P)=n\langle 1/E\rangle\frac{1}{2}\frac{d(m^2)}{dt}+g\int_p E\,C\equiv n\langle 1/E\rangle\frac{1}{2}\frac{d(m^2)}{dt}+j\,.
\end{equation}
The presence of the varying mass $(m^2)'$ is due to the dispersion relation dependence on this mass: $E(a) = \sqrt{p^2+m(a)^2}$. In~\Cref{subsec:varying_mass} we provide a discussion of the physical interpretation of the varying mass. In terms of the scale factor ($d/dt = aH\,d/da$) we obtain
\begin{equation}
	\rho' = \frac{j}{aH}+\langle n/(2E)\rangle (m^2)'-\frac{3}{a}(\rho+P)\,.
\end{equation}

In the remainder of the appendix, we will denote the temperature in question simply as $T$ to keep the notation clean. Then 
\begin{equation}
	\rho' = \frac{\partial \rho}{\partial T}T'+ \frac{\partial \rho}{\partial z}z' + \frac{\partial \rho}{\partial m^2}(m^2)'+\frac{\partial \rho}{\partial a}\,.
\end{equation}
Note that for simplicity we dropped the subscript "ds" in the temperature. 

Here $\rho$ does not depend explicitly on $a$ (see Eq.~\eqref{eq:thermodynamic_k}). Hence
\begin{equation}\label{eq:T_evolution*}
	T' = \frac{j/(aH)-3(\rho+P)/a - (\frac{\partial \rho}{\partial z} + \beta)z'}{\frac{\partial \rho}{\partial T} + \alpha}\,.
\end{equation}

In this case, the curvature of the potential evaluated at the minimum encodes the effects of the thermal mass, there $\alpha$ and $\beta$ are defined as
\begin{equation}
	\alpha\equiv  -\frac{z}{2T}n_z\left(V_{\phi\phi T}-\frac{V_{\phi\phi\phi}V_{\phi T}}{m^2}\right)\qquad\text{and}\qquad \beta\equiv  -\frac{z}{2T}n_z\left(V_{\phi\phi z}-\frac{V_{\phi\phi\phi}V_{\phi z}}{m^2}\right)\,,
\end{equation}\label{eq:alpha beta}
where we used the following useful relations:
\begin{equation}
	\left\langle \frac{n}{2E} \right\rangle = -P_{m^2}\,\qquad\text{and}\qquad \rho_{m^2}+P_{m^2} = -\frac{z}{2T}n_z\,.
\end{equation}
%

\subsection{Pressure moment}\label{subsec:evol}
We previously discussed in detail how to obtain an equation for the temperature from the effective potential (\Cref{ch:5}) and from the energy density equations (\Cref{subsec:evol_temp}). One can insist in the phase-space picture of the EOM for the temperature and the number density in the pressure moment. Here we will discuss this case, reconciling the thermal mass with the scenario discussed in~\Cref{ch:5}.
As noted in~\Cref{ch:2} one can obtain a set of cBE for the comoving number density, $N = n/a^3$, and the temperature, $T_{\rm ds}$ of the dark Higgs via the pressure or energy density moments. 

Here we will briefly discuss the pressure moment beyond the Maxwell-Boltzmann approximation. Recall that 'velocity' dispersion, $\tilde T$, from the pressure is obtained by integrating both sides of Eq.~\eqref{eq:fBE} over $g\int_p p^2/E$ to derive an equation for $P$, and then identifying $\tilde T\equiv \langle p^2/(3E)\rangle = P/n$ (the thermal average is defined as $\langle X\rangle \equiv (g_\psi/n) \int_p X\,f$). This velocity dispersion coincides with the temperature that appears in the distribution when such a distribution is the Maxwell-Boltzmann one (cf. Eq.~\eqref{eq:thermodynamic_k}, $k=1$). However, when the distribution is Bose-Einstein (or Fermi-Dirac), the parametric (thermodynamic) temperature of the distribution is not exactly $P/n$, as was already extensively discussed in~\Cref{subsec:cBE_pressure}. This is an important distinction, as the parametric temperature is the one entering the effective potential in Eq.~\eqref{eq:Veff}. 

The equation for $\tilde T(\equiv P/n)$ is
\begin{equation} \label{eq:Ttilde}
	\tilde T' = - \frac{2\,  \tilde T}{a} + \frac{1}{3\, a} \left\langle\frac{p^4}{E^3}\right\rangle+ \frac{a^2}{3\, H\, N}\, C_2 - \frac{N'}{N} \tilde T - \frac{1}{6}\left\langle\frac{p^2}{E^3}\right\rangle \frac{d(m^2)}{da},
\end{equation}
where here $'=\frac{d}{da}$, and $C_2 = g\int_p p^2\,C/E$ as defined in Eq.~\eqref{eq:C2_def}. To obtain an equation for the actual thermodynamic temperature, we first observe that 
\begin{equation}
	\tilde T' = (P/n)' =  -\frac{n'}{n^2}P + \frac{P'}{n} = -\frac{n'}{n^2}P +\frac{1}{n}\left(P_T T' +P_z z' + P_{m^2} (m^2)'  \right)\,.
\end{equation}
The notation is $A_x= \partial A/\partial x$. 

Therefore, the equation for temperature $T$ is
\begin{equation}\label{eq:T}
	\begin{split}
		T' &= \frac{1}{P_T}\left(n\,\tilde T ' + \frac{P}{aN}(aN'-3N)- \frac{n\,T_\phi}{z}z' - \frac{3P-\rho}{2m^2}(m^2)'\right)
		\\&=\frac{1}{P_T}\Bigg(n \left(- \frac{2\,  \tilde T}{a} + \frac{1}{3\, a} \left\langle\frac{p^4}{E^3}\right\rangle+ \frac{a^2}{3\, H\, N}\, C_2 - \frac{N'}{N} \tilde T \right) 
		\\&\qquad\qquad + \frac{P}{aN}(aN'-3N)- \frac{nT_\phi}{z}z' - \left(\frac{3P-\rho}{2m^2}+\frac{n}{6}\left\langle\frac{p^2}{E^3}\right\rangle \right)\frac{d(m^2)}{da}\Bigg)
	\end{split}
\end{equation}
where we used $P_z =nT_\phi/z $ and $P_{m^2} = (3P-\rho)/(2m^2)$. The effective (screened) mass squared is given by the second derivative of the potential with respect to the background field evaluated at the minimum: 
\begin{equation}
	m^2 =\left.\frac{\partial^2V_\text{eff}(\phi_b;T,z)}{\partial\phi_b^2}\right\vert_{\phi_b=\phi_\text{min}}\,,
\end{equation}
and it encodes the interactions with the Goldstone mode and the dark photon. The derivative with respect to the scale factor is
\begin{equation}\label{eq:dm2}
	\frac{d(m^2)}{da} = \left.\left(\frac{\partial m^2}{\partial T}\right)\right\vert_{z} T' +\left.\left(\frac{\partial m^2}{\partial z}\right)\right\vert_{T} z' +\left(\frac{\partial m^2}{\partial \phi_b}\right)\phi_b'\,.
\end{equation}
Along the minimization trajectory $\phi_b = \phi_\text{min}(T,z)$ we use $\partial_{\phi_b} V_\text{eff}= 0$ to eliminate $\phi_b'$,
\begin{equation}
	0=\frac{d}{da}\partial_{\phi_b}V_\text{eff} =(\partial_{\phi_b\phi_b}V_\text{eff})\,\phi_\text{min}' + (\partial_{\phi_b T}\,V_\text{eff})T'+(\partial_{\phi_b z}\,V_\text{eff}) z'\,.
\end{equation}
We treat $m^2$ as a function of the \textit{free} background $\phi_b$ to evaluate the corresponding partial derivatives, followed by imposing the minimization condition ($\phi_b=\phi_\text{min}$) \textit{at the end}. The equation for the temperature evolution is
\begin{equation}\label{eq:Tv2}
	\begin{split}
		T' &=\frac{1}{1+\kappa_1}\frac{1}{P_T}\Bigg(n \left(- \frac{2\,  \tilde T}{a} + \frac{1}{3\, a} \left\langle\frac{p^4}{E^3}\right\rangle+ \frac{a^2}{3\, H\, N}\, C_2 - \frac{N'}{N} \tilde T \right) 
		+ \frac{P}{aN}(aN'-3N) + \kappa_2z'\Bigg)\,,
	\end{split}
\end{equation}
where
\begin{equation}
	\begin{split}
		\kappa_1 &\equiv \frac{1}{P_T}\left(\frac{3P-\rho}{2m^2}+\frac{n}{6}\left\langle\frac{p^2}{E^3}\right\rangle\right) \left(V_{\phi_b\phi_bT}-\frac{V_{3\phi_b}V_{\phi_bT}}{m^2}\right)
		\quad\text{and\qquad} 
		\\\kappa_2 &\equiv  -\frac{1}{P_T}\left(\frac{3P-\rho}{2m^2}+\frac{n}{6}\left\langle\frac{p^2}{E^3}\right\rangle\right) \left(V_{\phi_b\phi_bz}-\frac{V_{3\phi_b}V_{\phi_bz}}{m^2}\right) - \frac{nT}{zP_T}\,.
	\end{split}
\end{equation}
A useful relation for $\langle p^2/E^3\rangle$ is
\begin{equation}
	m^2\left\langle \frac{p^2}{E^3}\right\rangle = \left\langle\frac{p^2}{E} \right\rangle -\left\langle\frac{p^4}{E^3} \right\rangle = 3\tilde T-   \left\langle\frac{p^4}{E^3} \right\rangle\,.
\end{equation}

\subsection{Fugacity evolution}\label{subsec:fugacity}

We now turn our attention to obtain an equation for the fugacity $z$ from the number density Boltzmann equation. Note that we can use Eq.~\eqref{eq:dist_expansion} and Eq.~\eqref{eq:thermoquantities} to obtain
\begin{equation}\label{eq:zprime}
	z' = \frac{N' - \sum_k z^k\left(N_k^\text{eq} \right)' }{\sum_k k\,N_k^\text{eq}z^{k-1}}\,,
\end{equation}
where the derivative of $N_k^\text{eq}$ is 
\begin{equation}
	\left(N_k^\text{eq}\right)' = \frac{g}{4\pi^2}a^2 m\left(\frac{6}{k} m\,(T+a\,T')\,K_2(k\,m/T)  +\frac{a}{T}  (2m^2\,T'  - T\,(m^2)')K_1(k\,m/T)\right)\,.
\end{equation}

Using the $k$th value for the energy density in Eq.~\eqref{eq:thermodynamic_k}, the derivative of the co-moving number of particles can be expressed as
\begin{equation}
	\left(N_k^\text{eq}\right)' = \frac{3}{a} N_k^\text{eq}+\frac{a^3\,k\,T'}{T^2}\rho_k^\text{eq}+\frac{\,(m^2)'}{2m^2\,T}(3N_k^\text{eq}\,T-k\,a^3 \rho_k^\text{eq}) \,.
\end{equation}
Rewriting Eq.~\eqref{eq:zprime} in terms of known thermodynamic quantities, we obtain:
\begin{equation}\label{eq:preliminary_z'}
	z' =\frac{1}{N_z}\left(N' - \frac{3}{a}N - \frac{a^3 T'}{T^2}z\,\rho_z + \frac{a^3}{2m^2T}\frac{d(m^2)}{da}\sigma\right)\,,
\end{equation}
with $N_z= \sum_k k\,z^{k-1}N_k^\text{eq} $ and
\begin{equation}
	\begin{split}
		N = \sum_k z^k N_k^\text{eq}\,,\qquad
		z\,\rho_z =\sum_k z^k\,k\,\rho_k^\text{eq} \,,\qquad  \sigma\equiv \sum_k z^k\,k\,(\rho_k^\text{eq}-3p_k^\text{eq}) = z\frac{\partial (\rho-3p)}{\partial z}\,.
	\end{split}
\end{equation}
Cf. Eq.~\eqref{eq:thermodynamic_k}. 

After substituting Eq.~\eqref{eq:dm2}, we find
\begin{equation}\label{eq:z_evolution*}
	z' = \frac{1}{1-\eta_2}\frac{1}{N_z}\left(N'-\frac{3}{a}N + \frac{a^3}{T}\left(-\frac{z\,\rho_z}{T}+ \frac{\sigma}{2m^2}\eta_1\right)T' \right)\,,
\end{equation}
where we define
\begin{equation}
	\eta_1 \equiv \frac{\partial^3V_\text{eff}}{\partial^2\phi_b\partial T}-\frac{1}{m^2}\frac{\partial^3 V_\text{eff}}{\partial\phi_b^3}\frac{\partial^2V_\text{eff}}{\partial\phi_b\partial T}\,,\qquad  
	\eta_2 \equiv \frac{\sigma}{2m^2\,T\,n_z}\left(\frac{\partial^3V_\text{eff}}{\partial^2\phi_b\partial z}-\frac{1}{m^2}\frac{\partial^3 V_\text{eff}}{\partial\phi_b^3}\frac{\partial^2V_\text{eff}}{\partial\phi_b\partial z}\right)\,.
\end{equation}\label{eq:etas}

\section{Making sense of a varying mass: the thermodynamic interpretation}\label{subsec:varying_mass}

To gain physical insight of the role of the varying mass term in Eq.~\eqref{eq:T} recall that the equation for energy density, Eq.~\eqref{eq:energy_density_1}, can also be obtained from Eq.~\eqref{eq:fBE} by integrating both sides by $\sum_\psi g_\psi\int_p E$,
\begin{equation}\label{eq:energy_density}
	\dot \rho+3H(\rho+P)=n\langle 1/E\rangle\frac{1}{2}\frac{d(m^2)}{dt}+g\int_p E\,C
\end{equation}
The first term in the right hand side looks like a 'source term'. To show that this performs thermodynamic work and therefore the first and second law of thermodynamics are satisfied, we first recall that the entropy density (for bosons) is defined as 
\begin{equation}
	s=-\sum_\psi g_\psi\int_p\left(f_\psi\log f_\psi-(1+ f_\psi)\log (1+f_\psi)\right)\,,
\end{equation}
and the useful identity
\begin{equation}
	\frac{d}{df_\psi}\left(f_\psi\log f_\psi-(1+ f_\psi)\log (1+f_\psi)\right)=\log\frac{f_\psi}{1+f_\psi} =-\frac{E-\mu}{T}\,.
\end{equation}
In the last equality we assumed the Bose-Einstein distribution. 

Hence, using Eq.~\eqref{eq:fBE}
\begin{equation}
	\dot s=-\sum_\psi g_\psi\int_p \log\frac{f_\psi}{1+f_\psi}\partial_t f_\psi = -\sum_{\psi}g_\psi\int_p \log\frac{f_\psi}{1+f_\psi} (H\,p\,\partial_p\,f_\psi+C) 
\end{equation}
and after replacing $\log\frac{f_\psi}{1+f_\psi}=-\frac{E-\mu}{T}$ we get
\begin{equation}\label{eq:entropy_eq}
	\dot s +3Hs=\frac{1}{T}\sum_\psi g_\psi\int_p E\,C-\frac{\mu}{T}\sum_\psi g_\psi\int_p C\,. 
\end{equation}
That is, the entropy equation derived from Eq.\eqref{eq:fBE} does not receive contributions from a varying mass, and hence entropy is conserved, as it should be. 

Now, we can replace the r.h.s of Eq.~\eqref{eq:entropy_eq} with the energy and number density Eqs.~\eqref{eq:energy_density} and~\eqref{eq:nBE}, i.e., we identify the terms on the r.h.s with the energy density and number density evolutions, respectively. This leads to the familiar first law of thermodynamics,
\begin{equation}
	T\, d(a^3 s) = d(a^3\rho)+p\,d(a^3)-a^3n\frac{1}{2}\langle1/E\rangle\,d(m^2)-\mu \,d(a^3n)\,.
\end{equation}
In other words,
\begin{equation}
	T\,dS= dU+p\,dV - G\,d(m^2)-\mu \,dN\,,
\end{equation}
where $G=\frac{N}{2}\langle 1/E\rangle $. 

This is the first law of thermodynamics with an extra term $-G\,d(m^2)$. This term can be understood as parametric work (due to the internal changes in degrees of freedom of the system). When $m^2$ changes, each particle's energy $E=\sqrt{p^2+m^2}$ changes, even if momentum is kept fix. So the system must either absorb or release energy as its mass parameter shifts.
That absorbed or released energy shows up as the $G\,d(m^2)$ parametric work.